\documentclass[11pt,twoside,a4paper,cmspaper,final,collab]{cms-tdr}
\def\svnVersion{398c27d-D}\def\svnDate{2026/09/12}\def\cmsCernNoTag{CERN-EP-2026-222}\def\cmsCernDate{\today}\def\cmsMessage{Submitted to SciPost Physics}
\begin{document}\cmsNoteHeader{MLG-25-002}

\newcommand{\zee}{$\PZ\to\Pe\Pe$\xspace}
\newlength\cmsTabWidth
\ifthenelse{\boolean{cms@external}}
{
    \setlength{\cmsTabWidth}{0.9\textwidth}
    \newcommand{\cmsTableMaybe}[1]{\resizebox{\cmsTabWidth}{!}{#1}}
}{
        \setlength{\cmsTabWidth}{\textwidth}
    \newcommand{\cmsTableMaybe}[1]{#1}
}
\newcommand{\cmsTableAlways}[1]{\resizebox{\cmsTabWidth}{!}{#1}}
\newlength\cmsTabSkip\setlength{\cmsTabSkip}{1.5ex}
\cmsNoteHeader{MLG-25-002}
\title{Calibration of electromagnetic shower features in the CMS calorimeter with machine-learning techniques}

\date{\today}

\abstract{
 Monte Carlo simulations are used extensively in high-energy particle physics analyses. However, an incomplete description of the underlying physics can lead to significant discrepancies between observables measured in simulated and collision data. To mitigate potential biases arising from such mismodelling, it is essential to calibrate simulations to data. This paper presents two novel calibration methods based on machine-learning techniques: a reweighting approach, which employs a classifier to learn the ratio of probability density functions between simulation and data; and a normalising-flow approach, which learns a high-dimensional transformation to map simulation to data. Compared to traditional calibration methods, both approaches offer continuous, unbinned corrections across high-dimensional feature spaces, enabling improved global agreement with data. The techniques are demonstrated in the context of correcting the features of simulated electromagnetic showers in the CMS calorimeters, using proton-proton collision data collected during 2022 at $\sqrt{s} = 13.6$\TeV, corresponding to an integrated luminosity of 26.7\fbinv. The strengths and limitations of the two methods are compared.
}

\hypersetup{%
pdfauthor={CMS Collaboration},%
pdftitle={Calibration of electromagnetic shower features in the CMS calorimeter with machine-learning techniques},%
pdfsubject={CMS},%
pdfkeywords={CMS, ECAL, ML, calibration}} 

\maketitle 

\tableofcontents

\newpage
\section{Introduction}\label{sec:introduction}
Monte Carlo (MC) simulations are a crucial ingredient for data analysis across many scientific disciplines. 
In high-energy particle physics experiments, they are used to simulate the hard scattering process in a collision event,
as well as the subsequent showering and interaction of the final-state particles with the detector material. 
The output is a collection of simulated collision events that, after applying the same reconstruction algorithms, 
can be compared to collision data for parameter estimation and hypothesis testing. The MC simulations typically provide a good description of the data. 
However, deviations can arise when the MC generation algorithms lack a full description of the underlying physics. 
This ranges from missing higher-order terms in the perturbative expansion of a scattering amplitude to mismodelling of the radiation damage within a region of the detector. 
In statistical analyses, sizeable deviations between simulation and data can introduce measurement biases. 
If left uncorrected, these biases must be accounted for by systematic uncertainties, 
thereby reducing the precision of a measurement or the sensitivity of a search for new physics.
Conversely, correcting these biases allows the associated systematic uncertainties to be reduced, leading to improved measurement precision and search sensitivity.

Final-state particle identification (PID) algorithms are used extensively in particle physics analyses, 
providing an output score to enable the identification of particles of interest from unwanted backgrounds. The algorithms typically leverage machine learning (ML) to interpret the complex patterns of energy deposits (hits) left by particles as they traverse the detector. The input features to these algorithms encode information about how final state particles shower and interact with the detector medium, which depends on challenging-to-model physics processes and a detailed knowledge of the detector geometry and operational conditions. As a result, it is common to observe sizeable deviations between MC simulation and data in the PID output distributions.

Calibrating the PID output distribution in MC simulation is not a straightforward task. The typical approach is to correct the efficiency of a fixed selection on the PID output by calculating the binned probability density ratios between collision data and simulation. These so-called ``scale factors" can be derived in bins of one or more observables, \eg the particle transverse momentum $\pt$ and pseudorapidity $\eta$, and are applied as a multiplicative event weight associated with the calibrated object. 
This method faces problems when the density ratio between data and simulation varies substantially within a given bin, or when dealing with a high-dimensional space, because of the finite number of samples. Such drawbacks motivate the use of calibration methods that are continuous (unbinned) in the observable space, and can handle a high number of dimensions. Furthermore, methods that correct the entire distribution facilitate the use of PID variables in subsequent stages of an analysis, \eg in event classifiers that distinguish signal from background.

The identification of particles whose signature includes electromagnetic (EM) showers is a critical component for many measurements in CMS, 
\eg in the Higgs boson to diphoton decay channel ($\PH \to \PGg\PGg$), 
and mismodelling of the EM shower signature is often one of the leading experimental systematic uncertainties~\cite{CMS:2021kom}.  
Such PID algorithms rely on input features describing the shape and isolation properties of the EM shower. 
This paper presents the application of two novel ML calibration methods designed to correct these properties of simulated EM showers in CMS.
The methods target corrections in a high-dimensional feature space constructed primarily from measurements with the CMS electromagnetic calorimeter (ECAL)~\cite{CMS:2008xjf,CMS:2023gfb,CMS:1997ysd}.
The calibration of signatures in other CMS subdetectors, such as the tracker~\cite{CMS:2014pgm}, is beyond the scope of this work;
consequently, the application of these corrections to fully reconstructed electrons and photons is left to future analyses.

A sample of unbiased EM showers (``probes") is obtained by applying the tag-and-probe technique~\cite{CMS:2010svw} to \zee decays in both simulation and data.
The probes are then used to train the ML calibration algorithms,
which correct the high-dimensional PID input feature space in simulation to resemble that observed in data. 
The final performance of the calibration is evaluated by propagating these corrections through the PID algorithm, 
and comparing the agreement between data and simulation in the PID output distribution.

While both ML calibration methods detailed in this paper are faced with the same task, they operate in fundamentally different ways.
The ``reweighting method" approximates the probability density ratio between simulation and data distributions using an ML classifier to discriminate between samples drawn from the two sources~\cite{Rogozhnikov:2016bdp,Cranmer:2015bka,Andreassen:2019nnm,Nachman:2021opi,Diefenbacher:2020rna,CMS:2024jdl}. 
The resulting classifier output can be converted into a weight that, as with the scale factor approach, acts as a multiplicative correction to the total event weight. 
This technique has advantages over the standard binned approach, since the weights are fully continuous and can be derived for a high-dimensional space. 
Instead, the ``normalising-flow method" directly changes the values of the input features by learning a mapping between the simulation and data distributions~\cite{Pollard:2021fqv,nonparametric,3619be1e75774bb18b26af7d2f580a53,Golling:2022nkl,Raine:2022hht,Hallin:2021wme,Algren:2023qnb,Bright-Thonney:2023sqf,Golling:2023mqx,papamakarios2021normalizing_review}. This technique also has benefits over traditional approaches as the mapping is derived for continuous, high-dimensional distributions.
Both methods have been shown to perform well in simple calibration examples with pseudo-data~\cite{Cranmer:2015bka, Daumann:2024kfd}.
This paper documents the first application of such methods to CMS collision data,
including the subsequent challenges that arise in this environment.
The normalising-flow method detailed here was used for EM shower calibration in a CMS $\PH \to \PGg\PGg$ analysis at $\sqrt{s} = 13.6$\TeV~\cite{CMS:2025fwn},
published prior to this paper,
demonstrating the practical applicability of these techniques for a physics measurement.

Several ML-based techniques for continuous calibration in particle physics have been explored in the literature.
The ATLAS Collaboration has investigated the use of optimal transport maps, 
estimated with input-convex neural networks~\cite{DBLP:journals/corr/AmosXK16} 
to calibrate jet flavour tagging classifiers~\cite{ATLAS_bjet_OT_calibration,Pollard:2021fqv}. 
During the LHC data-taking period from 2016--2018, CMS developed a chained quantile regression technique, in which boosted decision trees (BDTs) are used to estimate conditional quantiles of the data and simulated distributions~\cite{CMS:2022wpo}. 
Other approaches aimed at minimising discrepancies between data and simulation in PID outputs include adversarial training~\cite{Stein2022JetTaggingAdversarial} and domain adaptation techniques~\cite{CMS_2025_tau_corrections}.
The core techniques underlying the calibration methods presented here have also been applied in other contexts in particle physics. 
In particular, the reweighting approach has been used by CMS to transform simulated samples to alternative configurations, 
thereby avoiding the computational cost of generating new samples~\cite{CMS:2024jdl}.
More generally, the central idea of the reweighting method, \ie using a classifier to estimate the probability density ratio between two distributions,
has been exploited by ATLAS and CMS in several applications, 
including using the \textsc{OmniFold} procedure for unbinned differential cross section measurements~\cite{Andreassen:2019cjw,ATLAS:2024xxl,CMS:2025sws}, 
and recently in the first neural simulation based inference analysis performed at the LHC~\cite{ATLAS:2024jry}.

The paper is organised as follows. Section~\ref{sec:cms} provides a brief description of the CMS detector. In Section~\ref{sec:data}, the data and simulation samples used to train and evaluate the performance of the calibration methods are detailed, and Section~\ref{sec:selection} explains the object selection procedure used to isolate the EM showers. A description of the calibration methods is given in Section~\ref{sec:calibration_overview}, where the mathematical formulation used in the rest of the paper is defined. The performance of the methods is shown in Section~\ref{sec:results}. Following this, the complementary strengths and limitations of the two methods are compared in Section~\ref{sec:discussion}. Finally, the paper is summarised in Section~\ref{sec:summary}.

The code repositories used to implement the reweighting and normalising-flow methods are publicly available in Refs.~\cite{icenet,oneflow}.

\section{The CMS detector}\label{sec:cms}
The CMS apparatus~\cite{CMS:2008xjf,CMS:2023gfb} is a multipurpose, nearly hermetic detector, designed to trigger on~\cite{CMS:2020cmk,CMS:2016ngn,CMS:2024aqx} and identify electrons, muons, photons, and (charged and neutral) hadrons~\cite{CMS:2020uim,CMS:2018rym,CMS:2014pgm}. Its central feature is a superconducting solenoid of 6\unit{m} internal diameter, providing a magnetic field of 3.8\unit{T}. Within the solenoid volume are a silicon pixel and strip tracker, a lead tungstate crystal ECAL, and a brass and scintillator hadron calorimeter (HCAL), each composed of a barrel and two endcap sections. 
Forward calorimeters extend the pseudorapidity coverage provided by the barrel and endcap detectors. Muons are reconstructed using gas-ionisation detectors embedded in the steel flux-return yoke outside the solenoid. 
More detailed descriptions of the CMS subdetectors (excluding the ECAL, which is described below), together with definitions of the coordinate system and the relevant kinematic variables, can be found in Refs.~\cite{CMS:2008xjf,CMS:2023gfb}.

The ECAL, which is the most relevant subdetector for this paper, consists of 75\,848 lead tungstate crystals, which provide coverage in pseudorapidity $\abs{\eta} < 1.48 $ in a barrel region (EB) and $1.48 < \abs{\eta} < 3.0$ in two endcap regions (EE). 
The crystals are arranged in a quasi-projective geometry, with their front faces tilted by $3^\circ$ with respect to the radial direction to minimise inter-crystal gaps aligned with particle trajectories.
The EB (EE) crystals have front-face dimensions of $2.2{\times}2.2\cm^2$ ($2.86{\times}2.86\cm^2$), corresponding to a transverse granularity of $0.0175\times 0.0175$ in $(\eta,\phi)$ space in the EB, increasing progressively to a maximum value of $0.05{\times}0.05$ in the EE. 
These dimensions are comparable to the Moli\`ere radius of lead tungstate, such that the transverse distribution of energy among neighbouring crystals provides powerful discrimination between different shower topologies.
The EB (EE) crystals have a length of 23\cm (22\cm), corresponding to 26 (25) radiation lengths, 
which is sufficient to limit the longitudinal leakage of high-energy EM showers to an acceptable level.
The crystals are arranged in a single layer, and therefore provide no longitudinal shower shape information.
Preshower detectors consisting of two planes of silicon sensors interleaved with a total of three radiation lengths of lead are located in front of each EE detector.  

Events of interest are selected using a two-tiered trigger system. The first level, composed of custom hardware processors, uses information from the calorimeters and muon detectors to select events at a rate of around 100\unit{kHz} within a fixed latency of 4\mus~\cite{CMS:2020cmk}. The second level, known as the high-level trigger (HLT), consists of a farm of processors running a version of the full event reconstruction software optimised for fast processing, and reduces the event rate to a few kHz before data storage~\cite{CMS:2016ngn,CMS:2024aqx}. 

\section{Data samples and simulated events}\label{sec:data}
The calibration algorithms were trained using a subset of the proton-proton (pp) collision data collected in 2022 at $\sqrt{s} = 13.6$\TeV, corresponding to an integrated luminosity of 26.7\fbinv~\cite{CMS-PAS-LUM-22-001}.
In data, \zee events are collected using a single-electron trigger with a $\pt$ threshold of 30\GeV~\cite{CMS:2016ngn,CMS:2024aqx,CMS:2020uim,CMS-DP-2024-052}. 

{\tolerance=1200
The MC simulation sample for \zee is generated at next-to-leading order in quantum chromodynamics with \MGvATNLO (v2.9.9)~\cite{Alwall:2014hca}, interfaced with \PYTHIA (v8.306)~\cite{Bierlich:2022pfr} for the simulation of the parton shower, fragmentation, and hadronisation, using the CP5 underlying event tune~\cite{CMS:2019csb}.
Parton distribution functions are taken from the NNPDF 3.1 next-to-next-to-leading order set~\cite{NNPDF:2017mvq}.
The \PYTHIA generator is also used to simulate additional pp interactions occurring in the same or neighbouring bunch crossings (pileup). 
Simulated events are weighted to match the pileup distribution observed in data. 
The average number of interactions per bunch crossing in 2022 data is 46, assuming a total inelastic pp cross section of 80\,mb at $\sqrt{s} = 13.6$\TeV. The response of the CMS detector is simulated using the \GEANTfour package~\cite{Agostinelli:2002hh}.
\par}

Background contributions from top quark-antiquark ($\ttbar$) production and diboson $(\PV\PV)$ production are negligible compared to the contribution from \zee decays. This was explicitly checked using dedicated simulation samples, generated with \MGvATNLO (v2.9.13) and interfaced with \PYTHIA for $\ttbar$, and with \PYTHIA directly for $\PV\PV$. 
After the event reconstruction and selection described in Section~\ref{sec:selection}, 
these backgrounds contribute around 0.1\% inclusively, and less than 1\% in the tails of the distributions relevant for the calibration. 
Backgrounds from quark- or gluon-initiated jets that are misidentified as electrons are also found to be negligible in the selected phase space. 
As a result of the high purity of \zee events in the selected phase space, 
no backgrounds were considered in either the training or the evaluation of the calibration algorithms for this application.

\section{Event reconstruction and selection}\label{sec:selection}
The primary vertex (PV) is taken to be the vertex corresponding to the hardest scattering in the event,
evaluated using tracking information alone, 
as described in Section 9.4.1 of Ref.~\cite{CMS-TDR-15-02}.
The standard global event reconstruction used in CMS measurements, also called particle-flow (PF) event reconstruction~\cite{CMS:2017yfk}, aims to reconstruct and identify each individual particle in an event, with an optimised combination of all subdetector information. 
In this paper, the methods aim to calibrate the features of EM showers in the CMS ECAL,
which are mostly relevant for the PF reconstruction of photons and electrons ($\Pe/\Pgg$).
The EM showers are reconstructed using the algorithm described below.

Energy deposits (clusters) in the ECAL form the basis of the reconstruction algorithm~\cite{CMS:2020uim}. 
Clusters are built iteratively around a ``seed" ECAL crystal, identified as a local energy maximum above a given threshold. 
This is performed by aggregating crystals that share at least one corner in common with a crystal already in the cluster, and have an energy well above the electronics noise level. 
The clusters are then dynamically merged into ``superclusters" (SCs) to ensure good containment of the EM shower originating from the same initial particle. 
The energy of the SC is estimated by summing the energy of each crystal, 
after calibrating for response variations in time~\cite{CMS:2020uim}.
The spatial position of the SC in pseudorapidity and azimuthal angle ($\eta, \phi$), 
is defined as the average position of the individual crystals,
weighted by the logarithm of the crystal energy deposit.

The SC energy is corrected to account for imperfect shower containment in the operational ECAL crystals and losses in the clustering procedure. 
The correction is performed using a parametric multivariate regression, derived using simulation.
After applying the regression, the relative energy resolution for EM showers with $\pt$ between 1 and 100\GeV, and with $\abs{\eta} < 1.0$, improves from about 1.3 to 0.9\%.
For EM showers in the outer EE ($2.0 < \abs{\eta} < 2.5$), the relative energy resolution improves from approximately 3.1 to 2.4\%, for the same $\pt$ range.
Residual corrections to account for the differences between data and simulation in the energy scale and resolution are then applied.
In accordance with the calibration methods described in this paper,
these corrections are derived using electrons from \zee decays.
The residual corrections are described in more detail in Ref.~\cite{CMS:2025fwn}.

All SCs are required to pass a minimum $\pt$ threshold of 25\GeV, and lie within $\abs{\eta}<2.5$, excluding the barrel-endcap transition region $(1.44<\abs{\eta}<1.57)$.
Each SC must also pass the standard preselection criteria based on shower shape and isolation variables used in CMS $\PH\to\PGg\PGg$ analyses (provided in Ref.~\cite{CMS:2025fwn}), with the exception of an electron-based veto, which is inverted.
This ensures that the calibration methods are trained on a sample of EM showers that are typical of those used in CMS physics analyses.

A sample of unbiased EM showers (``probes'') is obtained from both simulation and data using the tag-and-probe technique~\cite{CMS:2010svw}, applied to \zee decays.
In addition to the SC selection criteria described above, 
the tag is required to have $\pt > 40$\GeV, satisfy tight isolation requirements, and match an object that passes the HLT threshold. 
These stricter conditions on the tag ensure that the calibration algorithms, trained using the probes, remain insensitive to the tag definition. 
In simulation, the tag-and-probe candidates are further required to match the generator-level electrons from the $\PZ$ boson decay.
Both electrons from the \zee decay can serve as the tag,
providing a maximum of two tag-and-probe candidates per event.
To suppress background contributions, the invariant mass of the tag-and-probe pair $m_{\Pe\Pe}$, is required to be between 80 and 100\GeV.

\begin{figure}[htb!]
    \centering
    \includegraphics[width=0.49\linewidth]{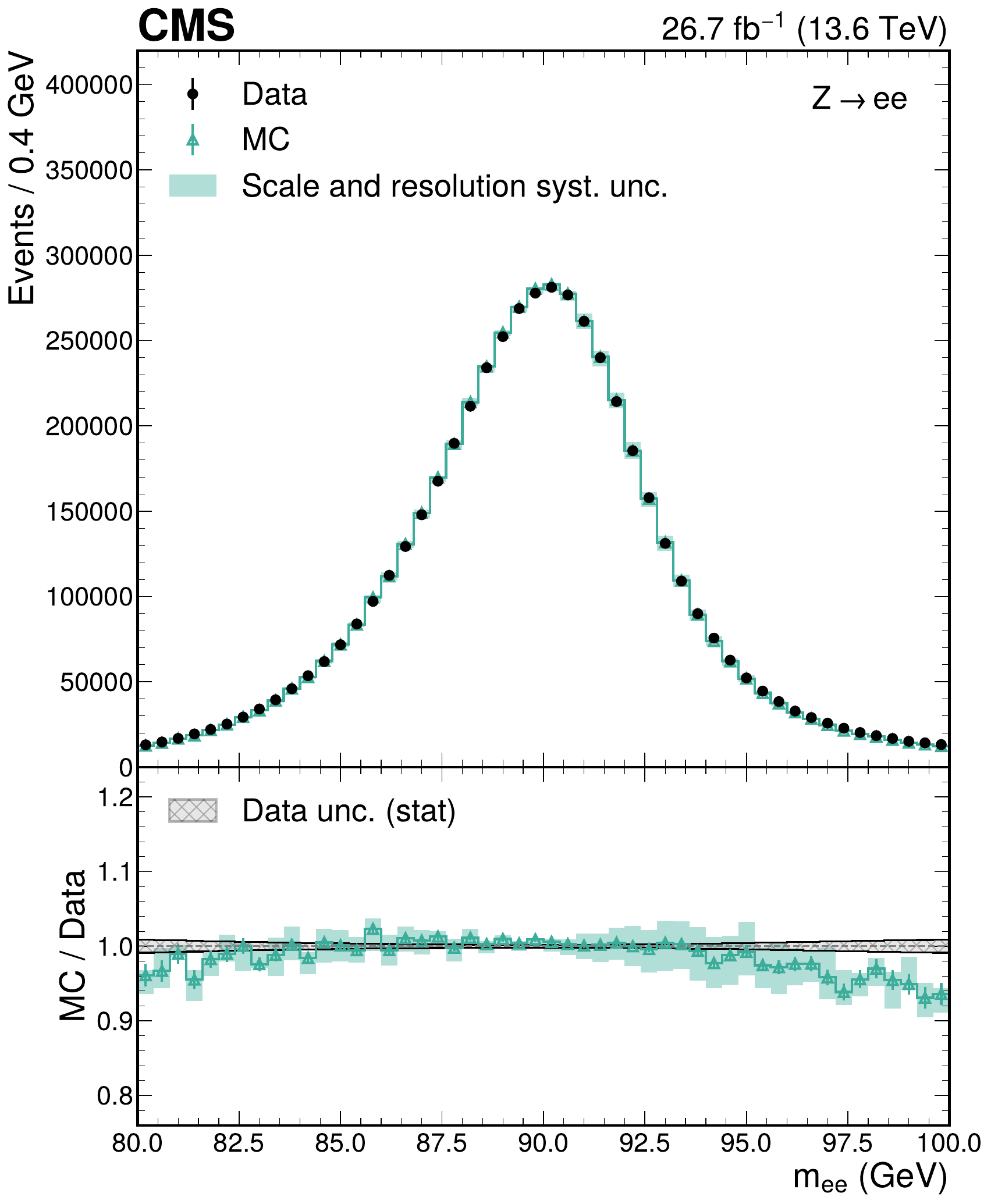}
    \includegraphics[width=0.49\linewidth]{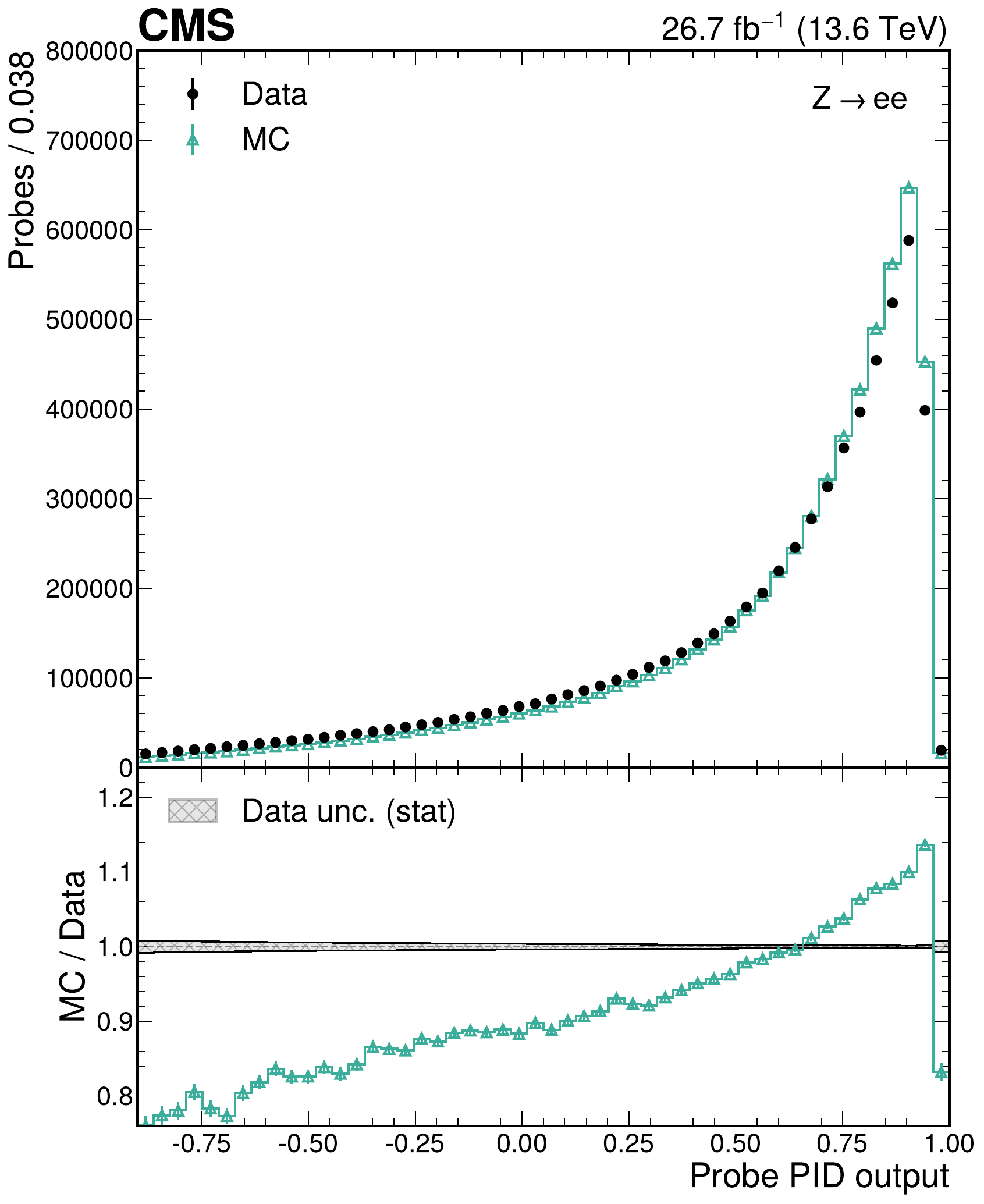}
    \caption{
    Invariant mass spectrum for tag-and-probe candidates from \zee events (left) and the PID output distribution for the probe EM showers (right) in data (black points) and simulation (teal histogram with triangular points). 
    The statistical uncertainties in the simulation and data are shown by the error bars on the points.
    For the left-hand figure, 
    the invariant mass for events in which both electrons pass the tag selection is included only once.
    The systematic uncertainty arising from the residual energy scale and resolution corrections is shown by the teal band in the left-hand figure. 
    The lower panels show the ratio of simulation to data, 
    where the statistical uncertainty in the data is indicated by the grey hatched band.}
    \label{fig:selection}
\end{figure}

The $m_{\Pe\Pe}$ spectrum is shown for simulation and data in Fig.~\ref{fig:selection} (left).
In this and all subsequent figures,
the ratio panels show simulation divided by data, 
with data chosen as the reference process since the calibration methods are designed to correct simulation to data. 
To avoid double counting, the $m_{\Pe\Pe}$ for events in which both electrons pass the tag selection are included only once in this figure.
The simulation histogram is normalised to have the same yield as data.
Good agreement is observed across the bulk of the distribution within uncertainties, 
where the uncertainty in the simulation includes both a statistical component and the systematic uncertainties arising from the residual energy scale and resolution corrections.
This level of agreement demonstrates the effectiveness of the applied corrections.
A slight underprediction is visible in the tails of the distribution, suggesting a small mismodelling of the resolution in simulation;
however, this effect does not have a significant impact on the calibration procedures.
Furthermore, the observed agreement supports the conclusion that contributions from background processes in the selected phase space are small,
validating the exclusion of background samples from the calibration algorithm training.

The targeted PID algorithm in this work is a boosted decision tree (BDT)~\cite{CMS:2020uim,CMS-DP-2024-052}.
The discriminating features rely primarily on measurements from the CMS calorimeters, 
while information from the tracker enters only through variables describing the isolation of the EM shower.
After training, the BDT score is used to distinguish EM showers originating from prompt $\Pe/\Pgg$'s produced at the PV
from those originating from non-prompt particles, such as the collimated diphoton decays of neutral mesons inside hadronic jets.
The BDT is implemented in \textsc{XGBoost}~\cite{DBLP:journals/corr/ChenG16}, and is trained on prompt and non-prompt photons from simulated $\PGg$~+~jet events. 
Probe EM showers from \zee decays used to train the calibration algorithms are required to pass a minimal threshold requirement on the PID score to remove the most background contaminated part of the phase space.

The PID input features include the shower shape and isolation variables listed in Table~\ref{tab:input_features}. In addition, the energy and $\eta$ of the SC as well as $\rho$, defined as the median of the transverse energy density in the event, are included to account for the dependence of the discriminating features on these kinematic and event-level features. The shower shape and isolation features often have significant mismodelling in the simulation. Sources of the mismodelling include the modelling of noise, the imperfect description of the amount of material upstream of the CMS ECAL in simulation, inactive detector regions, and the time evolution of the detector response, in particular caused by radiation damage~\cite{CMS:2020uim}.
The accumulation of these differences leads to a sizeable discrepancy between data and simulation in the PID output distribution. This is evident in Fig.~\ref{fig:selection} (right), which shows the uncorrected PID output distributions for the selected probes in simulation and data.
Using this uncorrected PID output in a physics analysis would lead to a measurement bias, which must be covered by introducing sizeable systematic uncertainties.
The calibration methods described in this paper aim to correct the shower shape and isolation input features, to ultimately obtain a good agreement between data and simulation in the PID output.

\begin{table*}[ht!]
    \centering
    \topcaption{Features used in the calibration algorithms. The features are split into two subsets: kinematic and event-level features, $x_k$, and shower shape and isolation features, $x_s$. 
    The final three columns show whether the feature is used as an input for each model,
    as indicated by a checkmark.
    The labels S1, S2, and NF refer to the Stage 1 reweighting model,
    the Stage 2 reweighting model, and the normalising-flow model, respectively,
    as described in Section~\ref{sec:calibration_overview}.
    The checkmarks with a ``c" subscript highlight the conditional features for the normalising-flow model. 
    A more detailed description of each feature is provided in Refs.~\cite{CMS:2020uim,CMS-DP-2024-052}.}
    \centering
    \renewcommand{\arraystretch}{1.2}
   \cmsTableAlways{
        \begin{tabular}{rp{12cm}ccc}
            Feature &\multicolumn{1}{c}{Description} & S1 & S2 & NF \\
            \hline 
            \multicolumn{5}{c}{Kinematic and event-level features, $x_k$} \\[\cmsTabSkip]
            Probe \pt & Transverse momentum & \checkmark & \checkmark & \checkmark$_c$ \\
            Probe $\eta$ & Pseudorapidity & \checkmark & \checkmark & \checkmark$_c$ \\
            Probe $\phi$ & Azimuthal angle & \checkmark & \checkmark & \checkmark$_c$ \\
            $\rho$ & Median energy density per unit $(\eta,\phi)$ area in the event (sensitive to pileup) & \checkmark & \checkmark & \checkmark$_c$ \\[\cmsTabSkip]
            \multicolumn{5}{c}{Shower shape and isolation features, $x_s$} \\[\cmsTabSkip]
            Probe $R_9$ & Ratio of energy sum in 3~x~3 grid of crystals surrounding SC seed, to total SC energy. & \NA & \checkmark & \checkmark \\
            Probe $S_4$ & Ratio of energy sum in 2~x~2 grid containing most energetic crystals, to energy in 5~x~5 grid surrounding SC seed. & \NA& \checkmark & \checkmark \\
            Probe $\eta$-width & Lateral spread of shower in $\eta$ direction, calculated as standard deviation in crystal $\eta$ values, weighted by the logarithm of the crystal energy. & \NA& \checkmark & \checkmark \\
            Probe $\phi$-width & Lateral spread of shower in $\phi$ direction, calculated as standard deviation in crystal $\phi$ values, weighted by the logarithm of the crystal energy. & \NA & \checkmark & \checkmark \\
            Probe $\sigma_{i{\eta}i{\eta}}$ & Second moment of log-weighted distribution of crystal energies in $\eta$, calculated in 5~x~5 matrix surrounding SC seed, rescaled to units of crystal size. & \NA & \checkmark & \checkmark \\
            Probe $\sigma_{i{\eta}i{\phi}}$ & Covariance of log-weighted distribution of crystal energies in $\eta$ and $\phi$, calculated in 5~x~5 matrix surrounding SC seed, rescaled to units of crystal size. & \NA & \checkmark & \checkmark \\
            Probe H/E & Ratio of energy deposited in the HCAL in a cone of radius $\Delta R = \sqrt{(\Delta \eta)^2 + (\Delta \phi)^2} = 0.15$ around the SC direction, to the energy of the SC. & \NA & \NA & \checkmark \\
            Probe $\mathcal{I}^{\mathrm{HCAL}}_{\text{PF, cluster}}$ & Sum of transverse energy of PF clusters in the HCAL, falling inside a cone of radius $\Delta R = 0.3$ around the SC. & \NA & \NA & \checkmark \\
            Probe $\mathcal{I}_{\mathrm{PF,ch}}$ & Sum of transverse energy of PF charged hadrons falling inside a cone of radius $\Delta R = 0.3$ around the SC, measured with respect to the selected PV. & \NA & \checkmark & \checkmark \\
            Probe $\mathcal{I}^{\mathrm{WV}}_{\mathrm{PF,ch}}$ & Sum of transverse energy of PF charged hadrons falling inside a cone of radius $\Delta R = 0.3$ around the SC, measured with respect to the interaction vertex that provides the largest isolation sum. This vertex is referred to as the ``worst" vertex (WV). The variable is effective in isolating misidentified photon candidates originating from jets coming from a vertex other than the PV. & \NA & \checkmark & \checkmark \\
            Probe $\mathcal{I}^{\mathrm{ECAL}}_{\text{PF, cluster}}$ & Sum of transverse energy of PF clusters in the ECAL, falling inside a cone of radius $\Delta R = 0.3$ around the SC. & \NA & \checkmark & \checkmark \\
            Probe $\mathcal{I}^{\text{hollow}}_{\text{track}}(\Delta R = 0.3)$ & Sum of transverse energy of all tracks in a hollow cone with a smaller (larger) radius of $\Delta R = 0.04$ ($\Delta R = 0.3$) around the SC. & \NA & \checkmark & \checkmark \\
            Probe $\mathcal{I}^{\text{solid}}_{\text{track}}(\Delta R = 0.4)$ & Sum of transverse energy of all tracks in a solid cone of radius $\Delta R = 0.4$ around the SC. & \NA & \checkmark & \checkmark \\
            Probe $\sigma_{\mathrm{RR}}$ (EE-only) & Standard deviation of the shower spread in the $(x,y)$ plane of the preshower detector. & \NA & \checkmark & \checkmark \\
            Probe $E_{\mathrm{ES}}/E_{\text{raw}}$ (EE-only) & Energy deposited in the preshower detector divided by the total energy of the SC. & \NA & \checkmark & \checkmark \\
        \end{tabular}
    }
    \label{tab:input_features}
\end{table*}

After selection, we obtain approximately 10 million and 19 million probe EM showers from data and simulation, respectively.
The samples are split into three components: 60\% are used for training the calibration algorithms, 10\% are used as validation samples to ensure that no overfitting occurs and for optimising the hyperparameters of the models, and 30\% are used as the test samples to evaluate the final performance of the methods. 
Only the test samples are used for the results in Section~\ref{sec:results}.

\section{Calibration methods}\label{sec:calibration_overview}
The events are characterised by a set of features $x \in \Omega$, where $\Omega$ represents the corresponding feature space of all physically allowed values of $x$.
Simulated events are distributed according to a probability density function (pdf) $p_{\mathrm{MC}}(x)$, whereas events from collision data are distributed according to $p_{\text{data}}(x)$.
Typically there are discrepancies between the two pdfs.
The act of calibration refers to modifying the simulation such that the resulting pdf approximates the one observed in data,
\begin{equation}
    p_{\text{MC}}(x) \longrightarrow \widetilde{p}_{\text{MC}}(x) \approx p_{\text{data}}(x).
\end{equation}
As the densities are intractable and therefore not directly accessible, the calibrations are derived using samples of events. 
The simulated sample $\mathcal{X}_{\mathrm{MC}} = \big\{(x_i,w^{\text{nom}}_i)\big\}_{i=1,\ldots,N_{\mathrm{MC}}}$, 
where $N_{\mathrm{MC}}$ is the total number of simulated events,
includes per-event weights $w^{\text{nom}}_i$,
such that the expectation value of an arbitrary observable $\mathcal{O}$ is calculated according to,
\begin{equation}
    \mathbb{E}(\mathcal{O}) = \frac{\sum_i w^{\text{nom}}_i\mathcal{O}_i}{\sum_i w^{\text{nom}}_i}.
\end{equation}
Note that $w^{\text{nom}}_i$ can take negative values to account for the cancellation of terms in higher-order calculations or to avoid double-counting in jet matching and merging algorithms. However, negative weights often lead to problems during the training of ML algorithms. The two methods presented in this paper adopt different strategies for handling negative weights, which are described in more detail in Sections~\ref{sec:reweighting} and \ref{sec:norm_flow}.
The simulated sample is compared to a data sample, $\mathcal{X}_{\text{data}} = \big\{x_i\big\}_{i=1,\ldots,N_{\text{data}}}$, 
such that data events effectively carry a unit weight.
Before calibration, the simulated sample is normalised to the data sample, $\sum^{N_{\mathrm{MC}}}_i w^{\text{nom}}_i = N_{\text{data}}$,
where $N_{\text{data}}$ is the total number of data events.

The calibration methods are trained with the ultimate aim to correct the PID output distribution for probe EM showers from \zee decays.
Rather than simply correcting the one-dimensional PID output distribution,
the methods provide a more general calibration by correcting the high-dimensional PID input feature space.
The corrections are then propagated through the PID algorithm,
and the performance is evaluated by comparing the agreement between the corrected simulation and data in the PID output.
The variables listed in Table~\ref{tab:input_features} define the features $x$, which are input to the calibration methods.
These are separated into two subsets: $x_s$ refer to features related to the probe shower shape and isolation, and $x_k$ refer to the probe kinematic features and global event variables.
The uncorrected distributions of the $x_k$ and $x_s$ features can be found in Figs.~\ref{fig:results_kinematic_pt_rho}--\ref{fig:results_shower_5} in Section~\ref{sec:results},
where the performance of the calibration methods is shown.

For the application detailed in this paper, the methods provide a correction for the features related to the EM shower and detector response, \ie $x_s$, 
that is conditional on the features related to the hard scattering or underlying event, \ie $x_k$.
Crucially, the methods are not learning to correct $x_k$, and are designed to leave $p_{\mathrm{MC}}(x_k)$ unchanged.
The resulting conditional calibrations can then be evaluated and applied to an independent data set, where the probability density over $x_k$ is different.
This conditional property is achieved in different ways for the two methods, as described in Sections~\ref{sec:reweighting} and \ref{sec:norm_flow}.

The validity of the calibration algorithms is, in principle, guaranteed only if identical MC generators and tuning configurations are used for both training and deployment.
Although the calibration is learnt to be conditional on $x_k$,
and the \GEANTfour component (common to all full simulation samples) largely determines the modelling of $x_s$,
residual dependencies on the MC configuration can remain.
For example, differences in the modelling of radiation emitted by final-state particles after the hard scattering process may alter the radiation pattern around the EM shower,
thereby affecting $x_s$, in particular the isolation-related features.
When the MC configuration used for training differs from that used in deployment,
these effects must be carefully checked to ensure the calibration algorithms remain valid.

\subsection{Reweighting method}\label{sec:reweighting}
The reweighting method is based on the procedures described in Refs.~\cite{Rogozhnikov:2016bdp,Cranmer:2015bka,Andreassen:2019nnm,Nachman:2021opi,Diefenbacher:2020rna,CMS:2024jdl}.
The ideal reweighting function to morph simulation to data is defined by the probability density ratio,
\begin{equation}
    p_{\text{data}}(x)/p_{\mathrm{MC}}(x).
\end{equation}  
An ML classifier trained to distinguish between simulation and data can be used to approximate this ratio. 
For a classifier $f(x): \mathbb{R}^d \to [0,1]$, trained on a $d$-dimensional feature space using a weighted binary cross-entropy (wBCE) loss
\begin{equation}\label{eq:bce_loss}
    L_{\mathrm{wBCE}}[f] = - {\sum_{i \in \mathcal{X}_{\text{data}}} \ln f(x_i)} - \sum_{i \in \mathcal{X}_{\mathrm{MC}}} w^{\text{nom}}_i\ln(1-f(x_i)),
\end{equation}
the ``likelihood ratio trick"~\cite{Andreassen:2019nnm} provides a reweighting function $w(x)$ that approximates the density ratio,
\begin{equation}
    w(x) = \frac{f(x)}{1-f(x)} \approx \frac{p_{\text{data}}(x)}{p_{\mathrm{MC}}(x)}.
\end{equation}
The reweighting function can then be applied as a multiplicative correction to the nominal event weight to morph simulation to data: $w^{\mathrm{corr}}_i = w(x_i)w^{\text{nom}}_i$.
By using ML for this task,
we are able to account for the large dimensionality of $\Omega$, and provide a weight that is continuous in $x$.

In this application, the aim is to derive a calibration for $x_s$ that is conditional on $x_k$, \ie we want to learn the ratio of conditional densities $p_{\text{data}}(x_s|x_k)/p_{\mathrm{MC}}(x_s|x_k)$.
An ML classifier that is trained on the full set of $(n+m)$ input features, $f(x_s,x_k) : \mathbb{R}^{n+m} \to [0,1]$, will learn to approximate the joint density ratio
\begin{equation}\label{eq:joint_density_classifier}
    \frac{f(x_s,x_k)}{1-f(x_s,x_k)} \approx \frac{p_{\text{data}}(x_s,x_k)}{p_{\mathrm{MC}}(x_s,x_k)} = \frac{p_{\text{data}}(x_s|x_k)p_{\text{data}}(x_k)}{p_{\mathrm{MC}}(x_s|x_k)p_{\mathrm{MC}}(x_k)},
\end{equation}
where the joint densities are expressed in terms of the conditional and marginal densities in the final expression. 
To isolate the ratio of conditional densities, $p_{\mathrm{MC}}(x_k)$ must be initially reweighted to match $p_{\text{data}}(x_k)$ prior to training.
This prevents the classifier from learning discrepancies between simulation and data arising from kinematic or pileup mismodelling,
such as missing higher-order effects in the hard-scattering process or an imperfect description of the underlying event.

The initial reweighting is performed by first training a classifier $f_{\mathrm{S1}}(x_k) : \mathbb{R}^{m} \to [0,1]$, to distinguish between simulation and data using only $x_k = \{\pt,\eta,\phi,\rho\}$. The associated reweighting function, referred to as Stage 1 (S1),
\begin{equation}
    w_{\mathrm{S1}}(x_k) = \frac{f_{\mathrm{S1}}(x_k)}{1-f_{\mathrm{S1}}(x_k)} \approx \frac{p_{\text{data}}(x_k)}{p_{\mathrm{MC}}(x_k)}.
\end{equation}
is applied as a multiplicative correction to the nominal event weight. The reweighted MC is then used to train a classifier on the full $(n+m)$ input features, $f_{\mathrm{S2}}(x_s,x_k) : \mathbb{R}^{n+m} \to [0,1]$.
By applying this two-stage procedure, $p_{\mathrm{MC}}(x_k)$ is initially reweighted to $p_{\text{data}}(x_k)$, and these terms will cancel in the ratio of Eq.~\ref{eq:joint_density_classifier}. 
The resulting reweighting function, referred to as Stage 2 (S2), approximates the ratio of conditional densities,
\begin{equation}
    w_{\mathrm{S2}}(x_s,x_k) = \frac{f_{\mathrm{S2}}(x_s,x_k)}{1-f_{\mathrm{S2}}(x_s,x_k)} \approx \frac{p_{\text{data}}(x_s|x_k)}{p_{\mathrm{MC}}(x_s|x_k)}.
\end{equation}
In this manner, the conditionality is amortised into the trained S2 model.
This means that when applying the reweighting method to calibrate $x_s$ for an independent data set,
one would only apply the S2 reweighting function $w_{\mathrm{S2}}(x)$.
Despite the fact that the weight correction applies globally to an event, the initial S1 reweighting means that in theory $p_{\mathrm{MC}}(x_k)$ should remain unchanged after the application of $w_{\mathrm{S2}}(x)$. 
This property, referred to as kinematic invariance, is explicitly demonstrated and discussed in Section~\ref{sec:results_kinematic}.

The training of the S2 model was found to be significantly more stable when events with negative weights were excluded. 
Negative weights are associated to the kinematic properties of an event rather than the EM shower formation or detector response.
Consequently, while the marginal density $p_{\mathrm{MC}}(x_k)$ may differ between positively and negatively weighted events, 
the conditional density $p_{\mathrm{MC}}(x_s|x_k)$ is expected to be the same.
A potential caveat is that a small fraction of negative weights may originate from the parton shower,
which can affect the radiation pattern surrounding the EM shower and therefore mildly violate the assumption that $p_{\mathrm{MC}}(x_s|x_k)$ is the same for both positive and negative weights.
This effect was explicitly checked by comparing the $x_s$ distributions for positively and negatively weighted events after aligning their $p_{\mathrm{MC}}(x_k)$ marginal distributions via a dedicated reweighting.
Any residual differences were found to be negligible compared to the discrepancies between simulation and data,
for this application.

\begin{figure}[htb!]
    \centering
    \includegraphics[width=.9\linewidth]{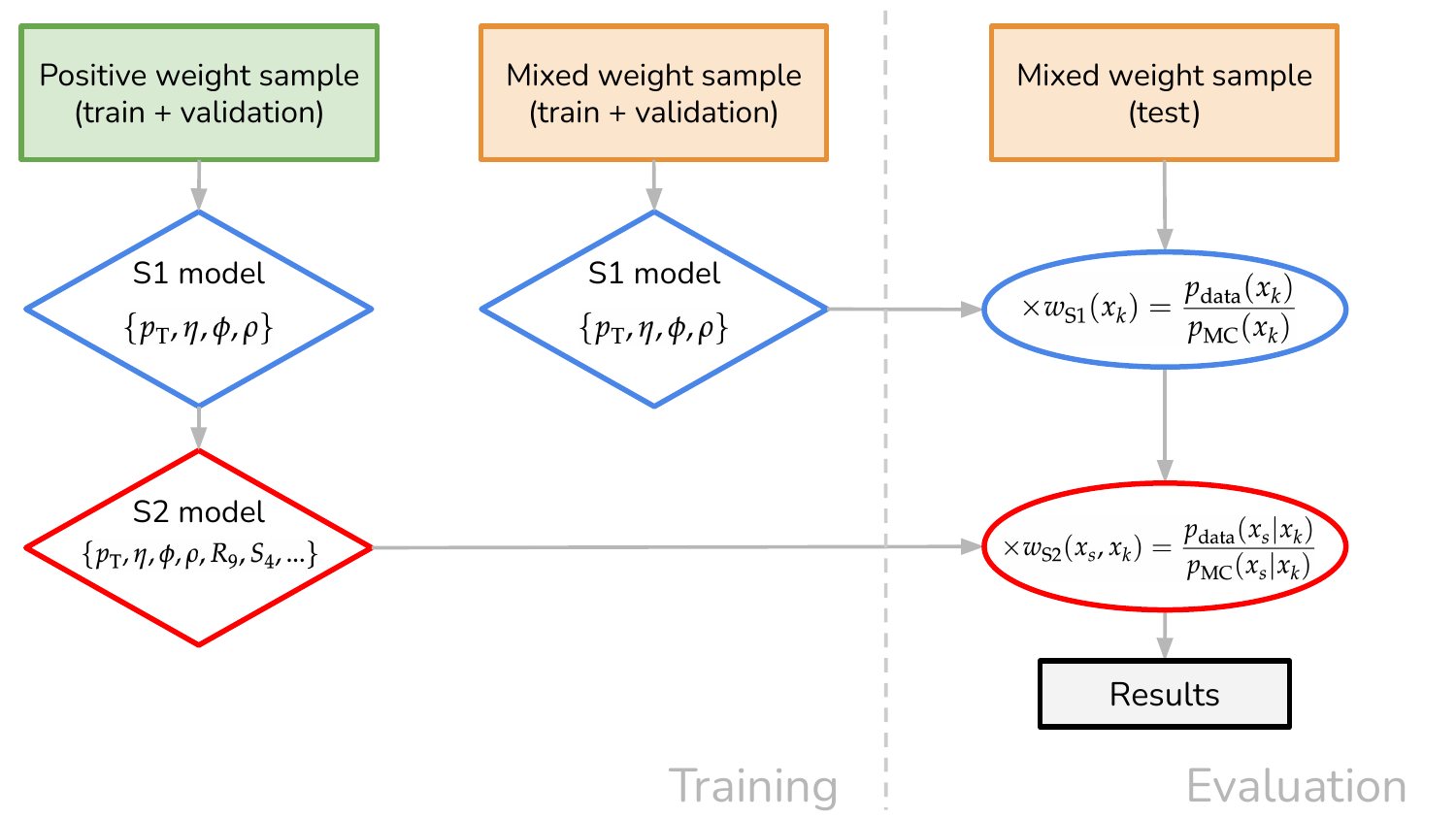}
    \caption{Schematic diagram of the two-staged model training and evaluation steps.
    Only positively weighted events are used in the training of the S2 model. This necessitates an S1 model trained using only positively weighted events during the initial training process. For evaluation, a separate S1 model is required that is trained on both positively and negatively weighted events.}
    \label{fig:reweighting_flow}
\end{figure}

Based on the discussion above, all events with negative weights are excluded during the training of both the S1 and S2 models to enhance the training stability.
For the evaluation (as discussed in Section~\ref{sec:results}), a separate S1 model is trained to align the $x_k$ distributions for the simulation sample containing both positive and negative weights. 
The S2 model trained on only positively weighted events is then applied.
A diagram summarising the training and evaluation stages of the reweighting method is provided in Fig.~\ref{fig:reweighting_flow}.

Although the S2 model is trained to be conditional on the probe $\eta$,
separate classifier pairs, $(f_{\mathrm{S1}},f_{\mathrm{S2}})$, were trained for each region of the CMS ECAL: the EB, the positive-$z$ endcap (EE$+$), and the negative-$z$ endcap (EE$-$). 
This choice improved the training stability and was found to provide a more accurate description of variations in both $x_s$ and $x_k$ across the different subdetectors.
Each classifier is implemented using the \textsc{IceBoost} model from the \textsc{IceNet} software package~\cite{icenet}. \textsc{IceBoost} builds on gradient BDTs from \textsc{XGBoost}~\cite{DBLP:journals/corr/ChenG16}, and integrates the automatic differentiation framework from \textsc{PyTorch}~\cite{paszke2019pytorch} to facilitate the construction of custom terms in the loss function.
It also supports multiple procedures for estimating the Hessian matrix during training.
The Hessian matrix contains the second derivatives of the loss function with respect to the model output predictions,
encoding the local curvature of the loss and helping to determine the size of the model update at each training iteration.
In this application, a stochastic Hutchinson estimator~\cite{Hutchinson01011990} was employed to estimate the Hessian matrix for all classifiers (details can be found in the software documentation~\cite{icenet}).
Unlike the standard \textsc{XGBoost} model, the \textsc{IceBoost} model can accommodate both positive and negative weights in the training, which is necessary for the additional evaluation-stage S1 model discussed in Section~\ref{sec:results}.
We note that neural network classifiers could also be employed for this task; however BDTs were found to offer greater training stability and were more straightforward to optimise.
The S1 and S2 models were trained using NVIDIA Tesla V100 GPUs requiring up to 32 GB of VRAM, and took around 1--2 hours and 6--10 hours to train, respectively, depending on the detector region. 
Each model was initially trained with a maximum of 400 estimators, \ie up to 400 boosted decision trees.
During training, the performance on the validation data set was monitored after each additional estimator was added.
The model configuration corresponding to the number of estimators that minimised the total validation loss was then selected.

The application of the reweighting method to a real-world scenario presents a number of challenges that are not apparent in simple studies with pseudo-data.
For example, CMS collision data include many imperfections and irregularities, ranging from inactive detector regions to the unwanted effects of data compression techniques that convert raw hits to analysis-ready events. 
The challenges faced are discussed in the following paragraphs,
along with the method enhancements designed to overcome a number of these challenges and improve the calibration performance.

Stage-2 models trained using only the wBCE loss (Eq.~\ref{eq:bce_loss}) were found to fluctuate significantly in the level of kinematic invariance achieved.
A regularisation term $\mathrm{R}_{\mathrm{SWD}}$ based on the sliced Wasserstein distance (SWD)~\cite{rabin2011wasserstein,swd_wasserstein_barycenters} was included in the loss function to penalise deviations in $x_k$ between the nominal and corrected simulation.
This helps to stabilise the training and favour S2 models with good kinematic invariance.
The $\mathrm{R}_{\mathrm{SWD}}$ term, calculated over the kinematic features $x_k$, is added to the wBCE loss according to,
\begin{equation}
    L_{\text{tot}} = L_{\mathrm{wBCE}} + \beta\,\mathrm{R}_{\mathrm{SWD}},
\end{equation}
where $\beta$ is a hyperparameter controlling the influence of the SWD term.
The full Wasserstein distance~\cite{inbook_villani} is a measure of the difference between two pdfs. 
However, the computation becomes extremely expensive at high dimensionality.
In the SWD approach, $p_{\mathrm{MC}}(x_k)$ is projected onto many one-dimensional (1D) directions (``slices"), 
where the directions are uniformly sampled from the unit hypersphere in the 4-dimensional $x_k$ feature space. 
The 1D earth mover's distance (EMD)~\cite{inbook_villani}, \ie the 1-Wasserstein distance, between the uncorrected and corrected simulation is then calculated for each slice, using a subset of the full simulation sample.
The EMD quantifies the minimum ``work'' required to transform one distribution into another, 
with smaller values indicating greater similarity between the distributions.
In this calculation, the corrected simulation is extracted by applying the S2 weight correction obtained at that iteration in the training cycle. 
The $\mathrm{R}_{\mathrm{SWD}}$ term in the loss is then calculated as the average of the 1D Wasserstein distances across all slices,
\begin{equation}
   \mathrm{R}_{\mathrm{SWD}} = \frac{1}{N_{\text{slices}}}\sum_{j=1}^{N_{\text{slices}}} {\mathrm{EMD}}(\Pi_{\theta_j}\,\chi^{(s)}_{\mathrm{MC}},\Pi_{\theta_j}\,\widetilde{\chi}^{(s)}_{\mathrm{MC}}),
\end{equation}
where $N_{\text{slices}}$ is the number of slices, 
$\Pi_{\theta_j} = \theta^{\mathrm{T}}_jx_k$ is the projection operator for slice $j$, such that $\theta_j$ is a unit vector sampled from the 4-dimensional unit hypersphere,
and this is applied to $\chi^{(s)}_{\mathrm{MC}}$ and $\widetilde{\chi}^{(s)}_{\mathrm{MC}}$ which are the uncorrected and corrected simulation subsamples, respectively.
The number of slices and the size of the simulation sample used in the $\mathrm{R}_{\mathrm{SWD}}$ calculation are provided in Table~\ref{tab:hyperparam}.
It was found that 500 slices with 500k simulated events provided a stable estimate of $\mathrm{R}_{\mathrm{SWD}}$,
while remaining within the memory constraints of the GPUs used in training.

\begin{table*}[ht!]
    \centering
    \topcaption{Summary of hyperparameters used in the reweighting models. 
    The first block lists the \textsc{XGBoost}-specific parameters, where information on the meaning of each parameter can be found in the \textsc{XGBoost} documentation~\cite{DBLP:journals/corr/ChenG16}.
    The parameters not listed are set to the \textsc{XGBoost} (v2.1.4) default values.
    The second block contains additional custom parameters, which are described in the text.}
    \centering
    \renewcommand{\arraystretch}{1.2}
   \cmsTableMaybe{
        \begin{tabular}{rccc{c}@{\hspace*{5pt}}ccc}
            & \multicolumn{3}{c}{S1 models} && \multicolumn{3}{c}{S2 models} \\ \cline{2-4}\cline{6-8}
            & EB & EE+ & EE- && EB & EE+ & EE- \\
            \cline{2-8} 
            Learning rate & 0.05 & 0.05 & 0.05 && 0.05 & 0.1 & 0.1 \\
            Maximum tree depth & 9 & 9 & 9 && 8 & 8 & 8 \\
            $\PGg$ (minimum loss reduction for split) & 1 & 1 & 1 && 1 & 1 & 1 \\
            Maximum delta step & 1 & 1 & 1 && 1 & 1 & 1 \\
            Maximum number of estimators (trees) & 400 & 400 & 400 && 400 & 400 & 400 \\
            L1 regularisation & 0.05 & 0.05 & 0.05 && 0.05 & 0.05 & 0.05 \\
            L2 regularisation & 2 & 2 & 2 && 2 & 2 & 2 \\[\cmsTabSkip]
            $\beta$ (SWD) & \NA & \NA & \NA && 0.01 & 0.0025 & 0.0025 \\
            Number of SWD slices & \NA & \NA & \NA && 500 & 500 & 500 \\
            Number of MC events per SWD slice & \NA & \NA & \NA && 500k & 500k & 500k \\
            $\sigma_0$ (Gaussian noise regularisation) & \NA & \NA & \NA && 0.1 & 0 & 0 \\
            $\tau$ (temperature scaling) & 1 & 1 & 1 && 0.9 & 1.1 & 1.1 \\
            $w_{\text{ceil}}$ (weight clipping) & 100 & 100 & 100 && 100 & 100 & 100 \\
        \end{tabular}
    }
    \label{tab:hyperparam}
\end{table*}

Some of the input features to the PID algorithm exhibit discontinuities. For example, the quantities related to the isolation of the EM shower have a peak at zero, followed by a gap, and then a smoothly falling tail. Additional discontinuities existed in some features due to the numerical precision in which the values were stored in the data sets; a too low precision leads to spikes in the distribution at discrete values. 
In principle, the reweighting method can have difficulty in learning a density ratio that is smoothly continuous in $x$ when there are sizeable gaps in the training data feature space.
To mitigate this effect, a data smoothing technique was introduced in the model training~\cite{tran2023one}.
In each training iteration, the samples $\mathcal{X}_{\mathrm{MC}}$ and $\mathcal{X}_{\text{data}}$ are modified by adding a Gaussian noise term to the input features,
\begin{equation}
    x_i \longrightarrow x'_i = x_i\sqrt{1-\sigma^2(t)} + y_i\sigma^2(t),
\end{equation}
where $y_i \sim \mathcal{N}^d(0,1)$, and $\mathcal{N}^d(0,1)$ is the $d$-dimensional unit Gaussian function. The $\sigma^2(t)$ term encodes the magnitude of the Gaussian noise and is scheduled to decrease during the training cycle according to,
\begin{equation}
    \sigma^2(t;\sigma_0,\theta,\lambda,T) = \sigma^2_0 \Bigg(\frac{\sigmoid(\frac{\theta}{\lambda})-\sigmoid(\frac{\theta}{\lambda}\frac{t}{T})}{\sigmoid(\frac{\theta}{\lambda})-\sigmoid(0)}\Bigg),
\end{equation}
where $t$ is the current iteration (number of estimators) in the training cycle, and $T$ is the total number of estimators used during the full training. The $\sigma_0$ hyperparameter was varied for each model and set to the values provided in Table~\ref{tab:hyperparam}, whilst $\theta$ was fixed to 3, and $\lambda$ was fixed to 0.9 and 1.1 for the EB and EE models, respectively.
In the early stages of this work, only samples with reduced numerical precision in some input features were available. Under these conditions, the data smoothing technique proved essential for achieving good performance and producing smooth reweighting functions across the full feature space. 
With the inclusion of higher-precision samples, the benefit of the data smoothing technique diminished, such that only the EB S2 model showed a small improvement when trained with a non-zero Gaussian noise magnitude ($\sigma_0=0.1$).

Certain input features exhibited large initial discrepancies between simulation and data, particularly those related to the CMS HCAL (as shown in Fig.~\ref{fig:results_shower_3}). 
To correct these features via reweighting requires a function that varies significantly from unity.
However, the application of such a function will result in a large variance in event weights, thereby reducing the statistical power of the simulation sample.
To avoid this scenario, the two HCAL-related features were dropped from the training of $f_{\mathrm{S2}}$, and were therefore not corrected in the calibration.
This option was only possible as the two features exhibit low importance in the PID feature ranking, and therefore do not contribute much towards the PID output shape for this task.
However, if this is not possible, then features with large initial discrepancies should be corrected beforehand, using a method that does not alter the event weights, \eg the normalising-flow method detailed in Section~\ref{sec:norm_flow}. 
The reweighting method could then be applied to correct any residual discrepancies.

\begin{figure}[htb!]
    \centering
    \includegraphics[width=1\linewidth]{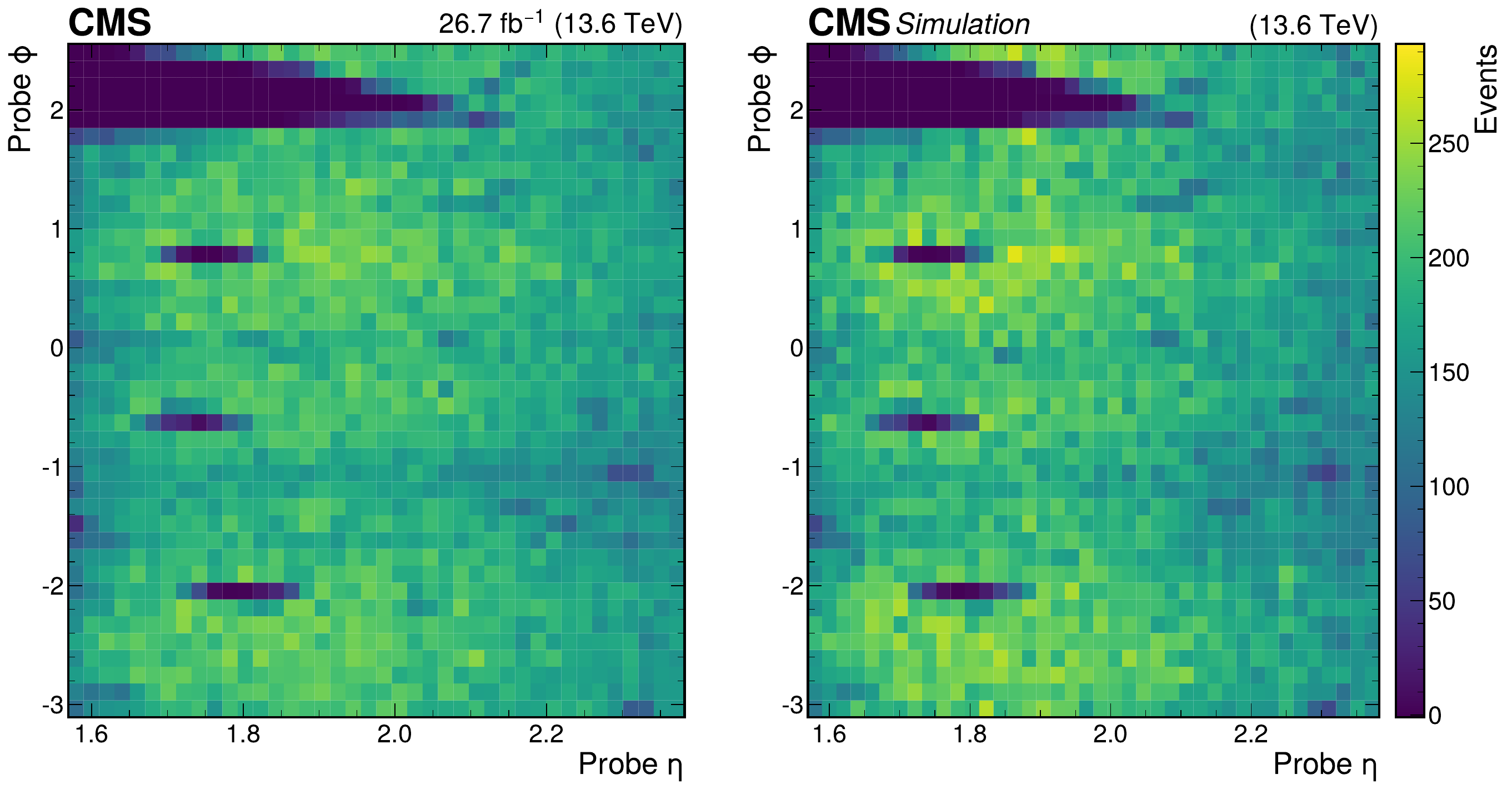}
    \caption{Heatmap for probes in data (left) and simulation (right) in the $(\eta,\phi)$ plane for the EE$+$ subdetector.
    The hole from the power cooling issue can be seen in the upper-left corner of the plots.
    The simulation is normalised to have the same yield as data.}
    \label{fig:heatmap_eep}
\end{figure}

The reweighting method was found to be somewhat unstable when there exist sizeable ``holes" in the feature space due to inactive detector regions. 
The largest hole originates from a power cooling issue in September 2022, where about 7\% of the ECAL channels in EE$+$ were disabled~\cite{CMS-DP-2024-052}. 
Figure~\ref{fig:heatmap_eep} shows the density of probes in $(\eta,\phi)$ space for the EE$+$ subdetector, for data (left) and simulation (right). Whilst some attempt has been made to account for the inactive detector regions in simulation, the boundaries are often misaligned with data. This can lead to small fluctuations (spikes) in the learned weight corrections at the hole boundaries. An additional strategy was tested in which sharp boundaries were imposed around the inactive detector regions through dedicated selection criteria applied to both data and simulation; however, this yielded no significant reduction in the weight correction fluctuations at the hole boundaries.
No further attempt was made to fix the issue of inactive detector regions,
and we leave this to future investigations.

The training hyperparameters were selected for each model using a grid scan procedure~\cite{JMLR:v13:bergstra12a}. 
Models were trained across a range of values for key parameters: $\beta$, $\sigma_0$, maximum depth of trees, and the learning rate, 
and the performance was subsequently assessed on the validation sample. 
The density of points in the grid scan was limited by the available computing resources, and therefore finer scans could potentially yield further improvements in the performance. 
Two additional hyperparameters are applied only at the evaluation stage, and not during training. 
The first corresponds to a temperature scaling of the model output introduced to obtain a ``well-calibrated" classifier~\cite{DBLP:journals/corr/GuoPSW17},
where in this context ``well-calibrated'' refers to the property that the classifier output accurately represents the true density ratio. 
Each classifier $f$ terminates with a sigmoid function, ensuring that the output represents a probability bounded between zero and one. 
Temperature scaling replaces this sigmoid function, introducing an additional temperature parameter $\tau$, with
\begin{equation}\label{eq:temperature_scaling}
    \sigmoid\Big(\frac{z}{\tau}\Big) = \frac{1}{1+e^{-z/\tau}}.
\end{equation}
The second hyperparameter corresponds to a weight clipping procedure, designed to limit the magnitude of the weight corrections obtained in the reweighting method. A hyperparameter $w_{\text{ceil}}$ is introduced such that the final weight correction is given by $\mathrm{min}(w_{\mathrm{S2}},w_{\text{ceil}})$.
These evaluation-stage hyperparameters are also scanned over in the grid-scan procedure.

Multiple metrics were established to identify the hyperparameter configurations that performed well. A reduced $\chi^2$ metric was defined to quantify the level of agreement between 1D histograms from data and simulation,
\begin{equation}\label{eq:chi2_reduced}
    \chi^2/n_{\mathrm{dof}} = \frac{1}{n_{\text{bins}}-1} \sum_i \frac{(d_i-s_i)^2}{\sigma_{\mathrm{d},i}^2+\sigma_{\mathrm{s},i}^2}.
\end{equation}
The terms $d_i$ and $s_i$ represent the data and simulation yields in bin $i$, respectively, where the simulation is calculated as the sum of weights in that bin. The simulation weights are modified by the relevant weight corrections from the trained S1 and S2 models. The terms $\sigma_{\mathrm{d},i}$ and $\sigma_{\mathrm{s},i}$ are the statistical uncertainties in the data and simulation yields in bin $i$, respectively.
These uncertainties are calculated according to the standard Poisson uncertainty for the data, and the weighted Poisson uncertainty for the simulation,
 such that $\sigma^2_{\mathrm{s},i}$ corresponds to the sum of squared weights.
Finally, $n_{\text{bins}}$ is the number of bins in the histogram. One of the hyperparameter optimisation metrics employed was the $\chi^2/n_{\mathrm{dof}}$ evaluated over the PID output distribution.

An additional $\chi^2$-based metric was defined to quantify the degree of kinematic invariance achieved by the S2 model. This was obtained by replacing the data yields in Eq.~\ref{eq:chi2_reduced} with the uncorrected simulation sample yields, and by removing the $\sigma_{\mathrm{d},i}$ term in the denominator to avoid double-counting the statistical uncertainty in the simulation. The $\chi^2/n_{\mathrm{dof}}$ was computed separately for each of the four $x_k$ variables and summed, with smaller values indicating S2 models that better preserve the $x_k$ distributions.

\begin{figure}[htb!]
    \centering
    \includegraphics[width=.49\linewidth]{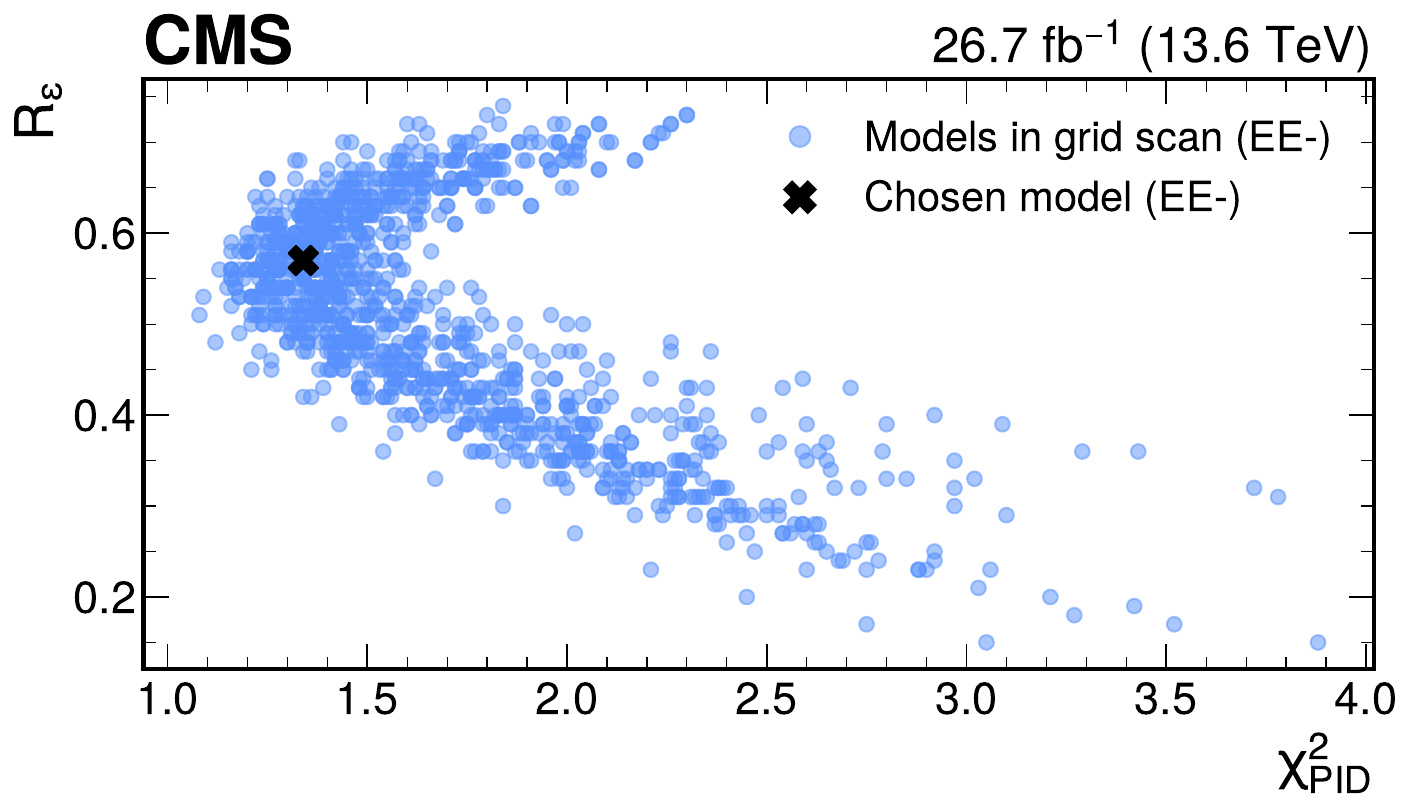}
    \includegraphics[width=.49\linewidth]{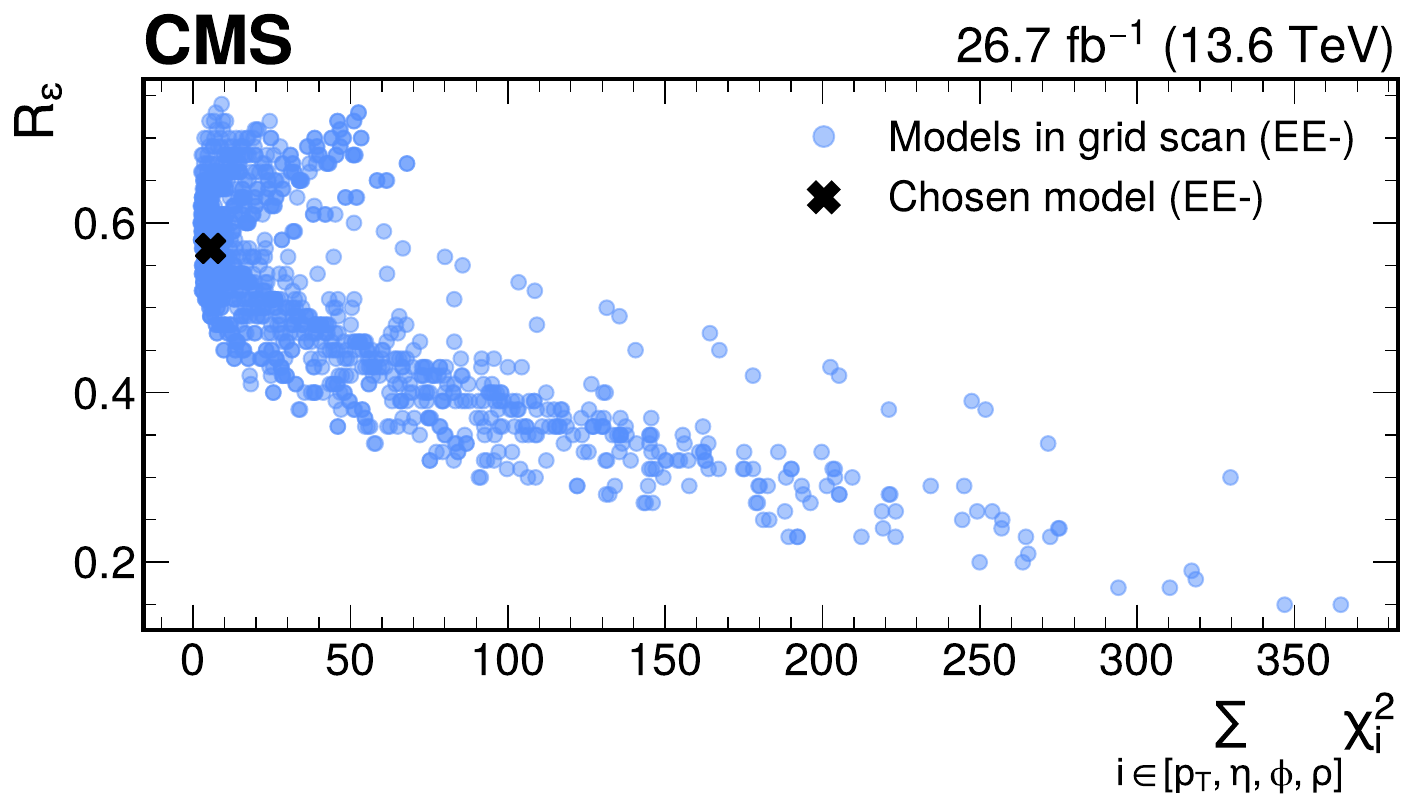}
    \includegraphics[width=.49\linewidth]{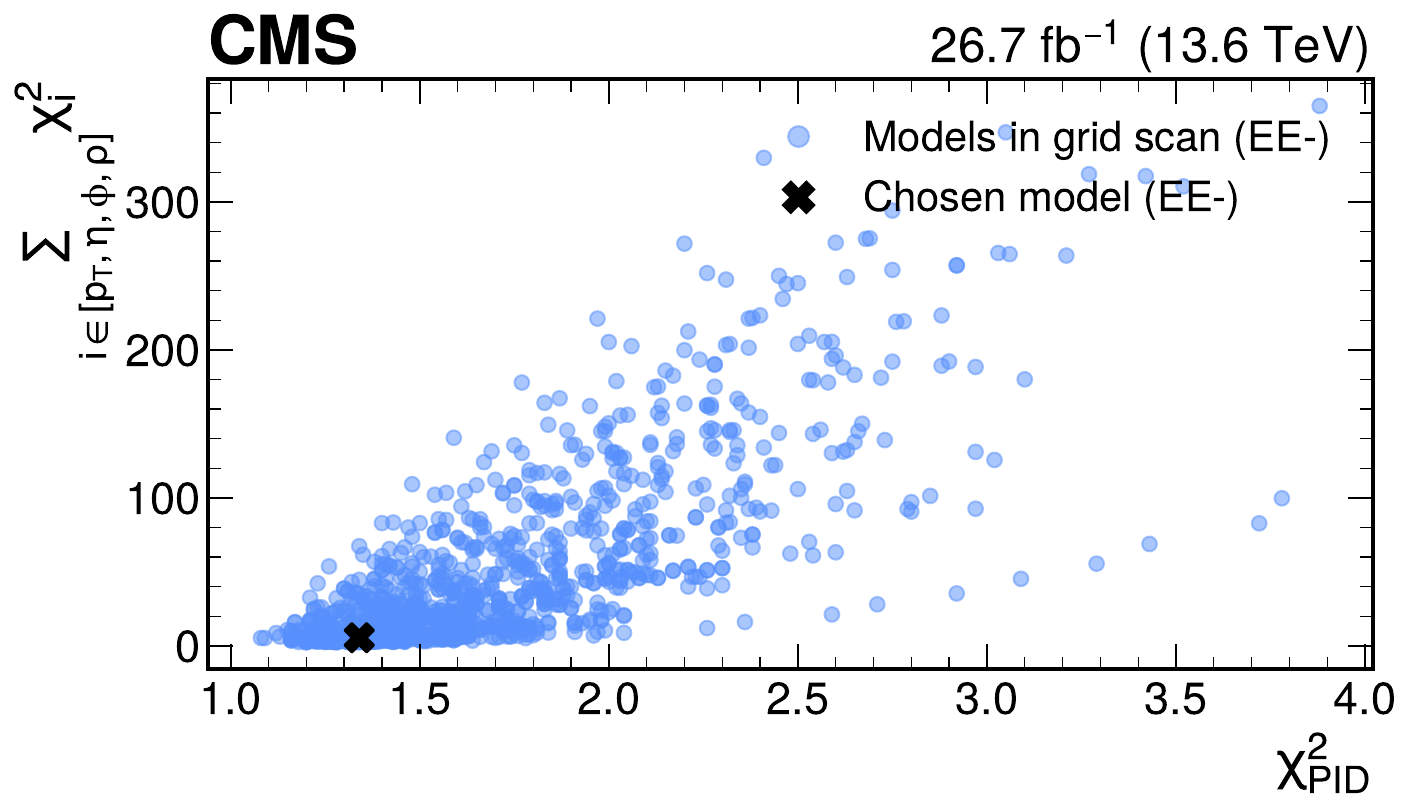}
    \caption{Scatter plots showing the performance of the different S2 EE$-$ models obtained in the grid-scan procedure. The performance is shown for different pairs of the following metrics: the $\chi^2/n_{\mathrm{dof}}$ for the PID output, the sum of $\chi^2/n_{\mathrm{dof}}$ values over the $x_k$ features evaluated between the nominal and corrected simulation, and $R_\varepsilon$. The black cross indicates the chosen S2 EE$-$ model.}
    \label{fig:rwgt_gridscan}
\end{figure}

The reweighting method increases the variance of event weights, and thereby reduces the statistical power of the simulation sample. The effective sample size (ESS) $\varepsilon$ of a weighted sample is defined as, 
\begin{equation}
    \varepsilon = \frac{(\sum_i w_i)^2}{\sum_i w_i^2}.
\end{equation}
To quantify the reduction in the statistical power, the ESS is computed before and after applying the S2 reweighting, and their ratio is defined as, 
\begin{equation}\label{eq:rwgt_statpower}
    R_\varepsilon = \frac{\varepsilon(S1{\times{S2}})}{\varepsilon(S1)}.
\end{equation}
Values of $R_\varepsilon$ close to unity correspond to models that preserve the statistical power of the sample. 
Figure~\ref{fig:rwgt_gridscan} shows the performance of the S2 EE$-$ models obtained during the grid-scan procedure,
where each point corresponds to a different hyperparameter configuration. 
Results are shown for pairs of the following performance metrics: 
the $\chi^2/n_{\mathrm{dof}}$ calculated over the PID output, 
the sum of $\chi^2/n_{\mathrm{dof}}$ values across the $x_k$ features evaluated between the nominal and corrected simulation, 
and $R_\varepsilon$. 
A bias variance trade-off~\cite{hastie_09_elements-of.statistical-learning} exists between $R_\varepsilon$ and the PID output $\chi^2/n_{\mathrm{dof}}$ (upper left) when $R_\varepsilon$ is above a value of around 0.55: lower $\chi^2/n_{\mathrm{dof}}$ values (reduced bias) are associated with lower $R_\varepsilon$ values (increased variance). The final model should be selected from the Pareto front~\cite{inbook_pareto}, depending on the requirements of the application. 
In this work, the model indicated by the black cross was chosen. 
Although this model may not appear optimal in any single 2D projection, it performs well consistently across all metrics. 
The final set of hyperparameters used for each model is provided in Table~\ref{tab:hyperparam}. 

\begin{figure}[htb!]
    \centering
    \includegraphics[width=.325\linewidth]{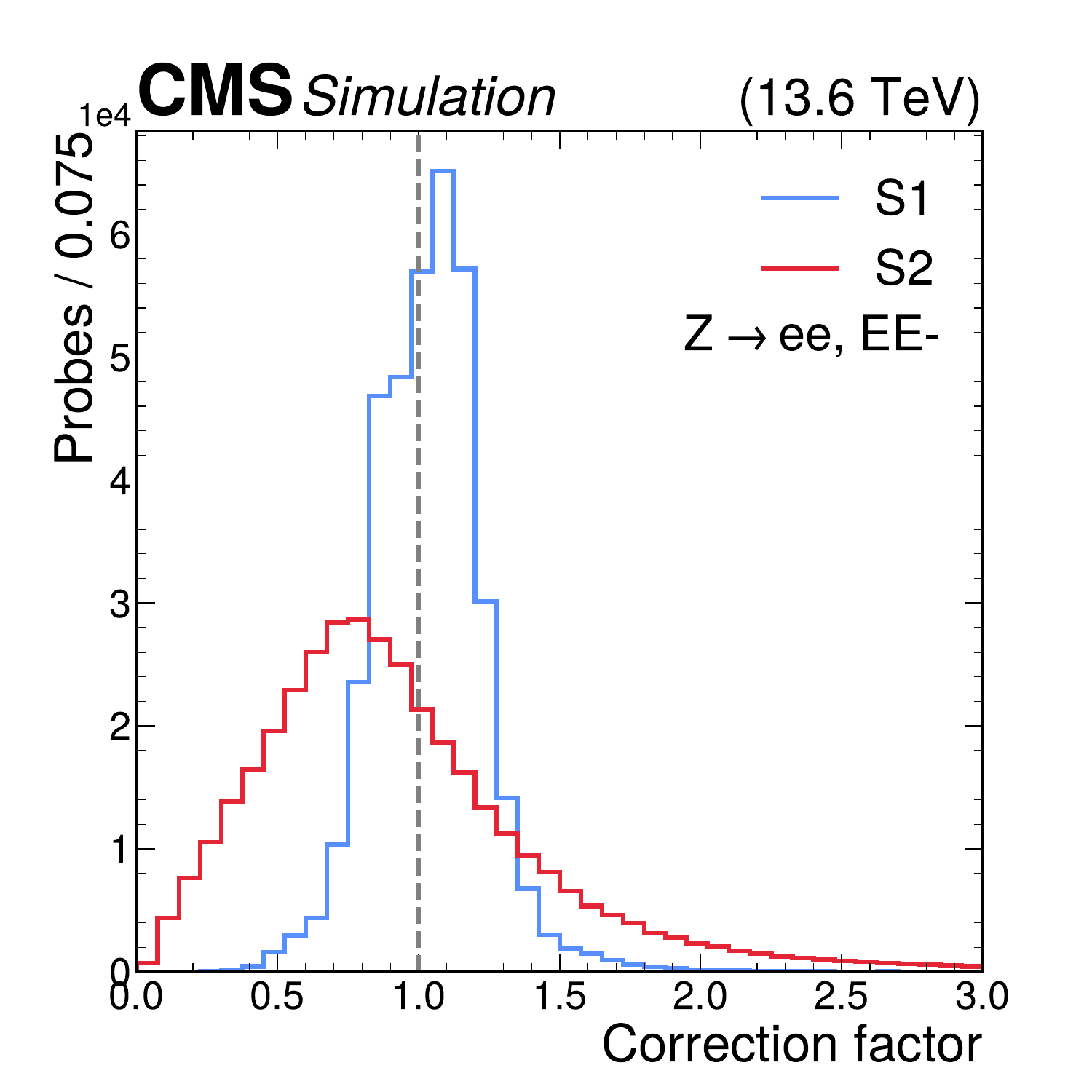}
    \includegraphics[width=.325\linewidth]{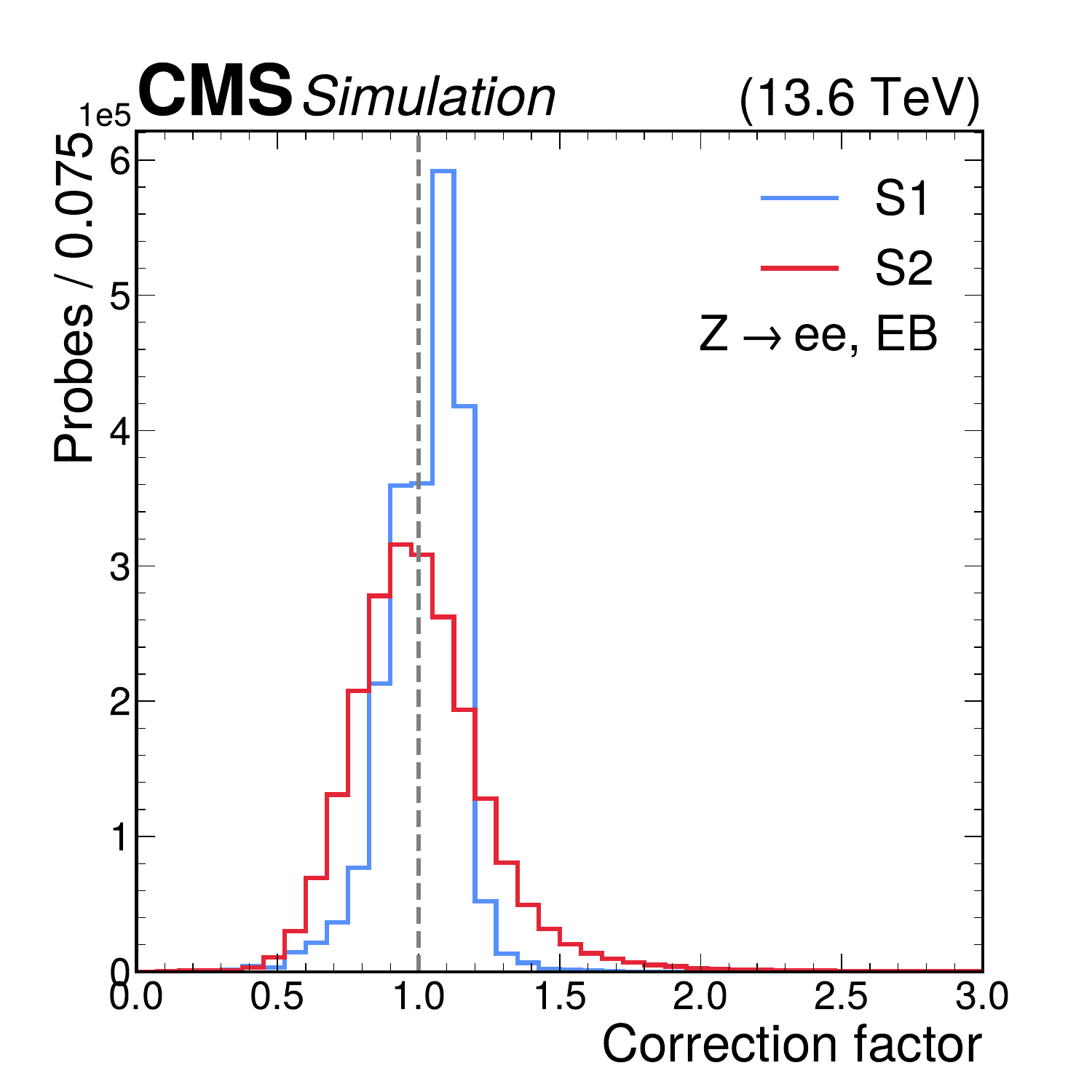}
    \includegraphics[width=.325\linewidth]{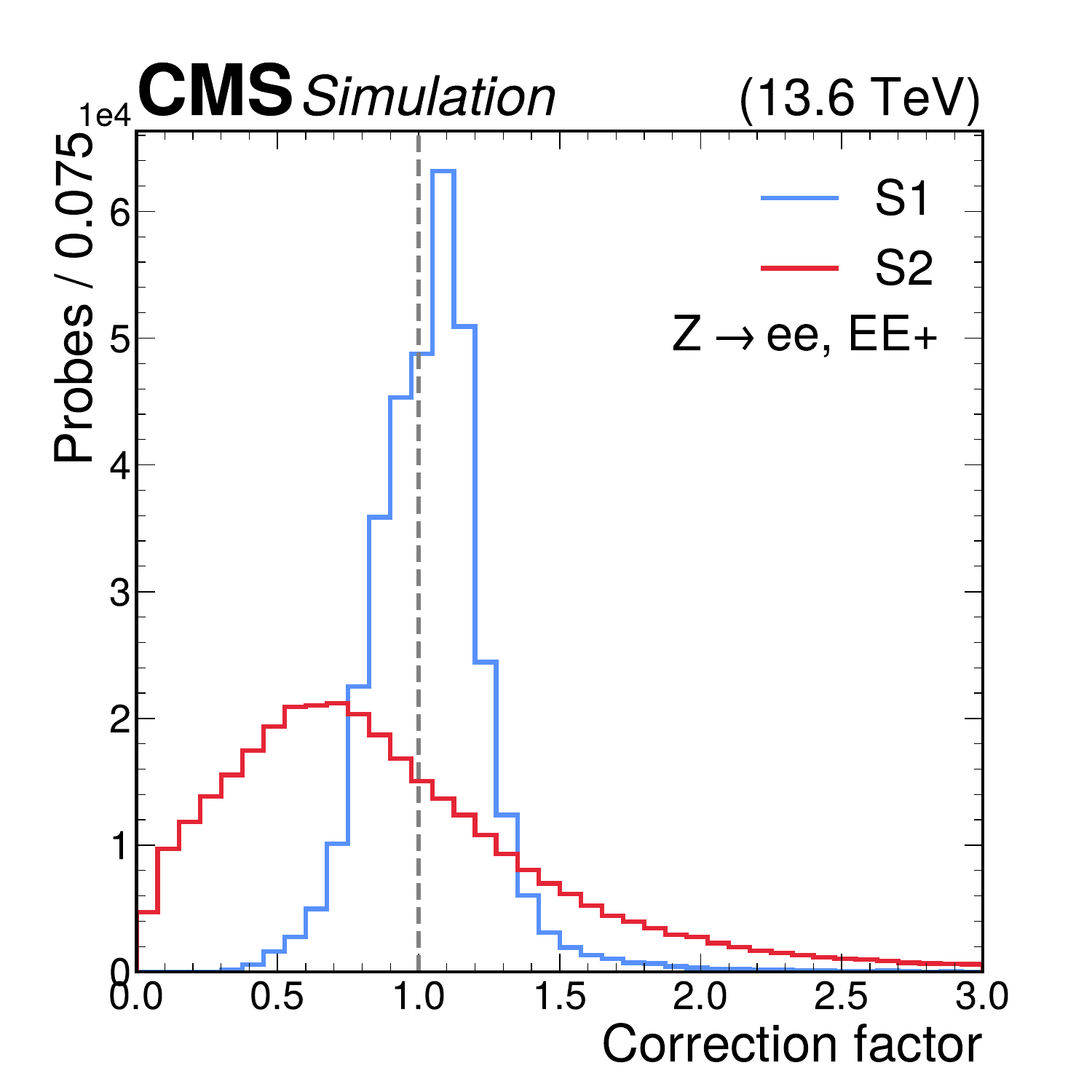}
    \includegraphics[width=.325\linewidth]{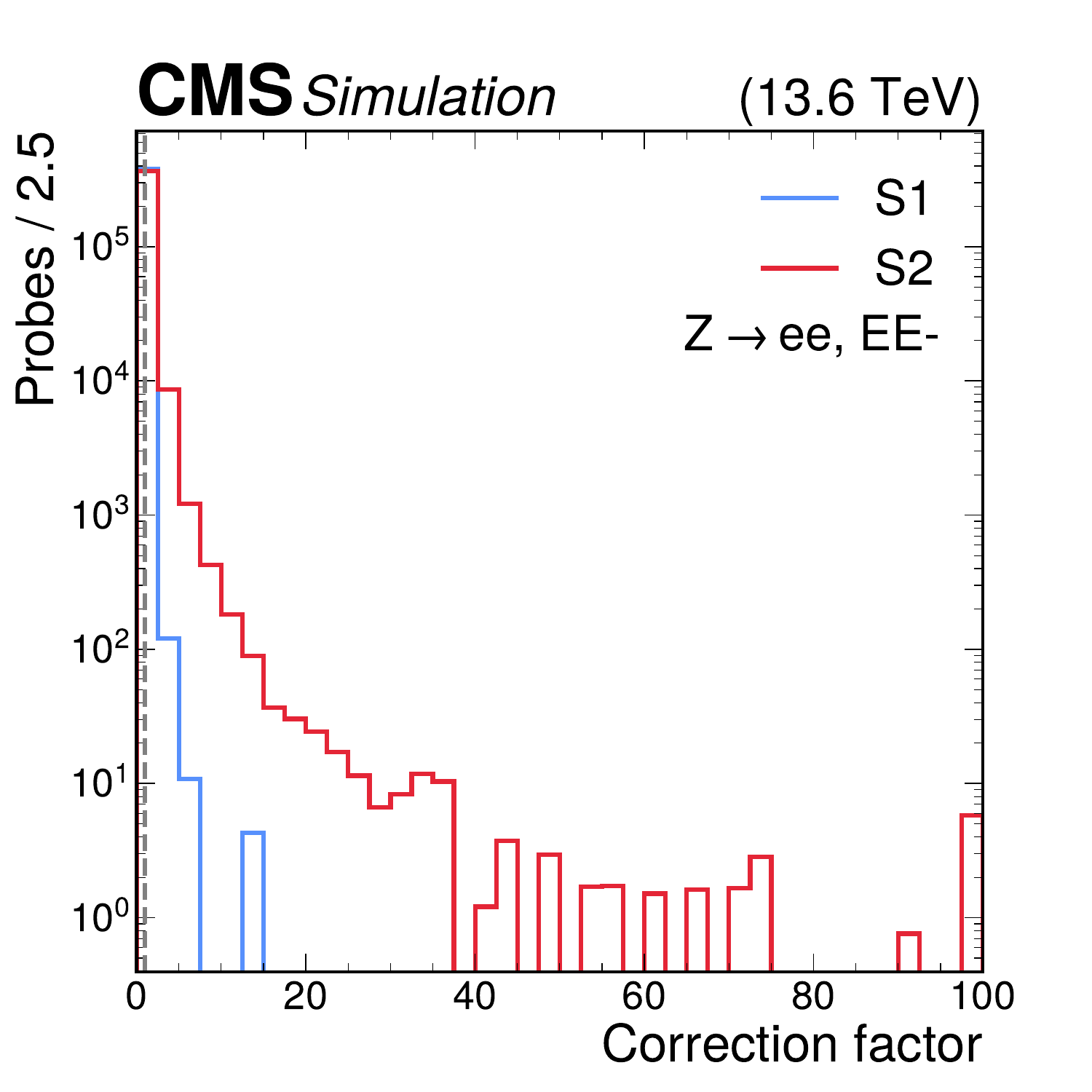}
    \includegraphics[width=.325\linewidth]{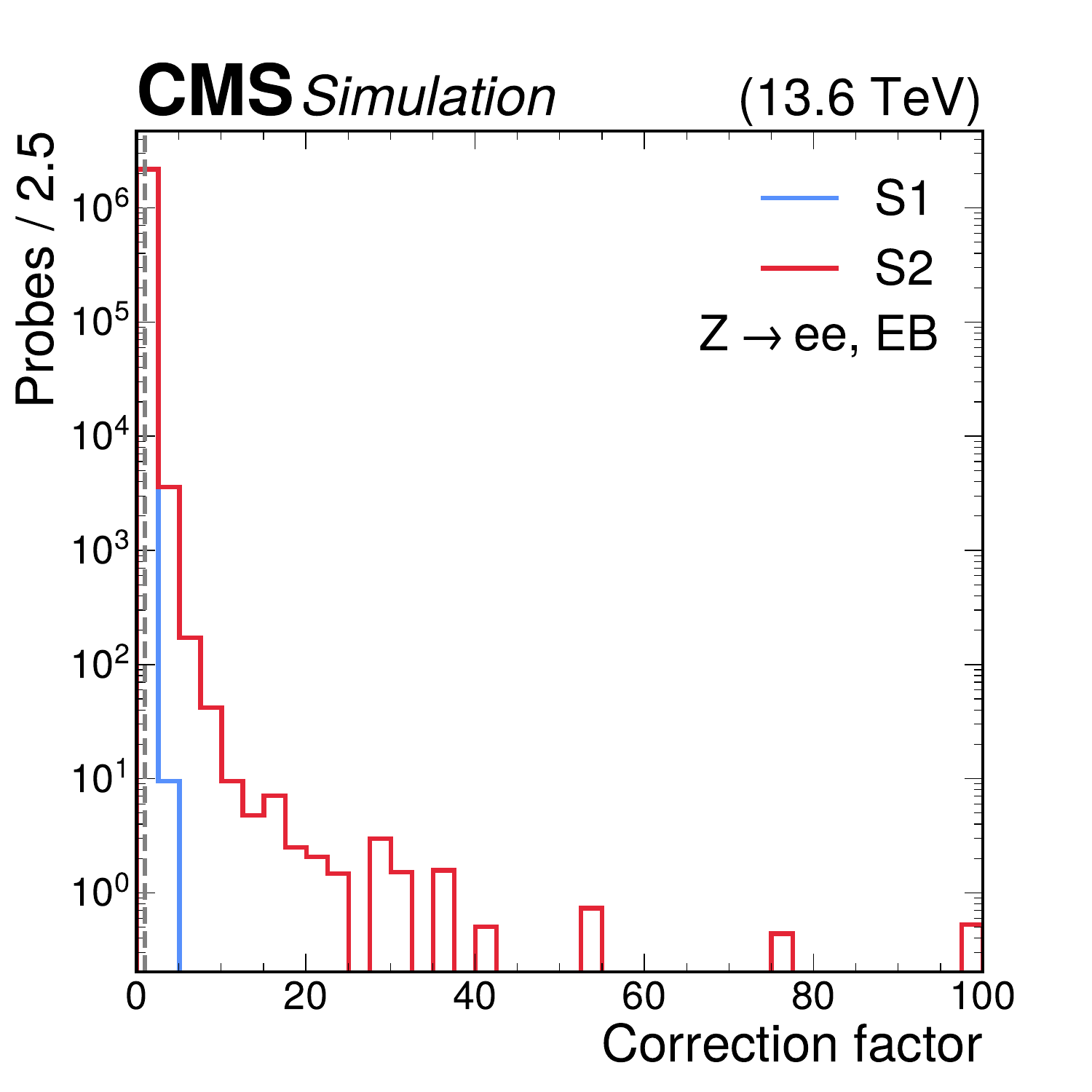}
    \includegraphics[width=.325\linewidth]{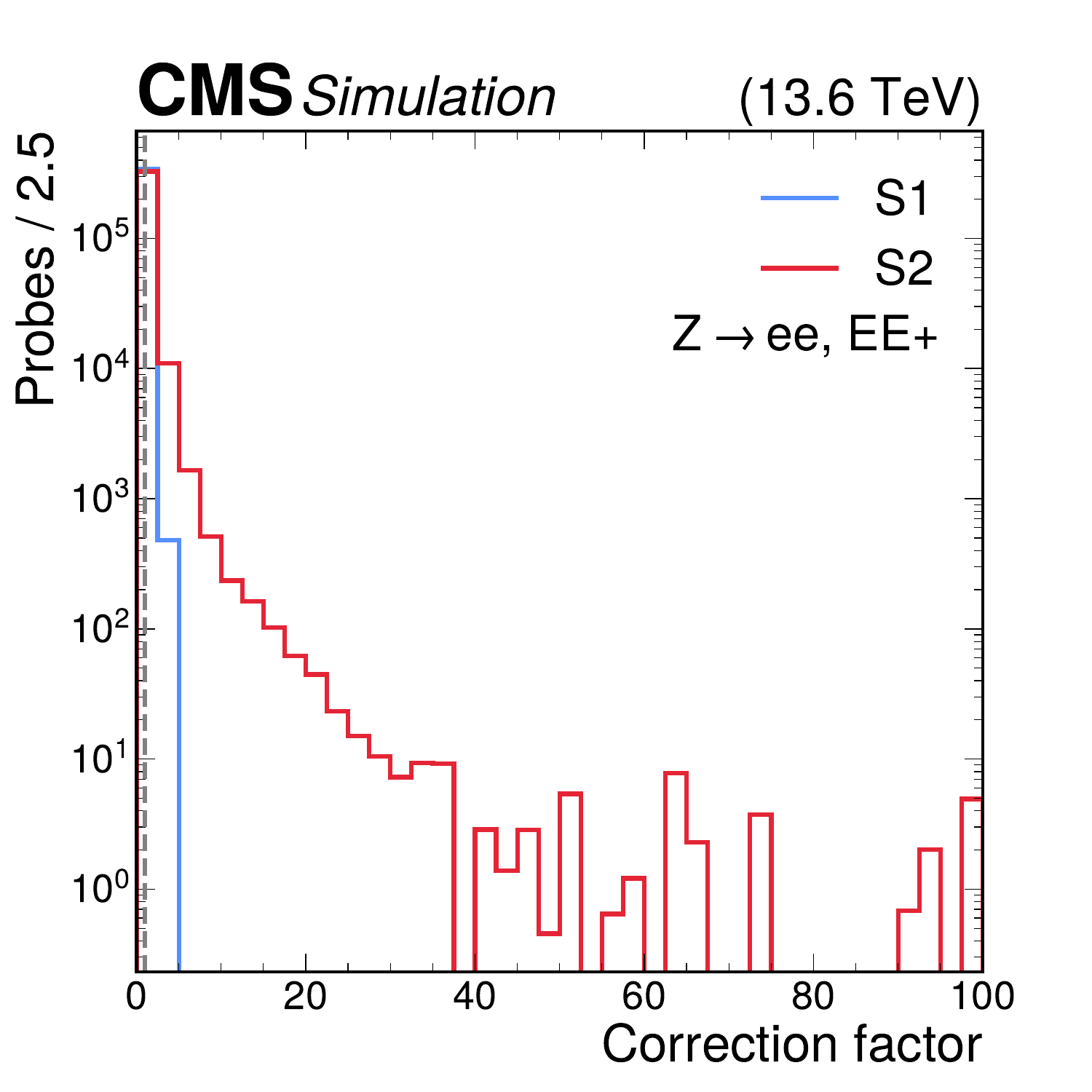}
    \caption{Weight correction distributions for the S1 (blue) and S2 (red) models, shown for the EE$-$ (left), EB (centre), and EE$+$ (right) subdetector regions. Linear and logarithmic y-scale plots are shown in the upper and lower rows, respectively. The lower plots extend up to $w_{\text{ceil}}$, used in the weight-clipping procedure.}
    \label{fig:rwgt_dist}
\end{figure}

Figure~\ref{fig:rwgt_dist} shows the obtained weight correction distributions for the test data set, for each subdetector region, with S1 ($w_{\mathrm{S1}}(x)$) and S2 ($w_{\mathrm{S2}}(x)$) models in blue and red, respectively. Linear-scale plots (upper) emphasise the bulk behaviour, while logarithmic-scale plots (lower) highlight the tails. The EE$+$ and EE$-$ regions exhibit larger variance in the S2 weight corrections, leading to a reduced statistical power of the samples (lower $R_\varepsilon$) after S2 reweighting.

\subsection{Normalising-flow method}\label{sec:norm_flow}
Normalising flows form a class of generative models that learn and sample from multivariate probability distributions by parameterising an invertible transformation between the complex input feature distribution and a simple, tractable base distribution of the same dimensionality.

The transformation $f$ must be invertible and hence ensure a one-to-one correspondence between the input pdf $p(x)$, and the base pdf $q(z)$, \ie~$f: \mathbb{R}^{d} \to \mathbb{R}^{d}$, for a $d$-dimensional feature space.
Given that the composition of multiple invertible functions remains invertible, it is common practice to construct the transformation $f$ as a composition of $N$ invertible transformations to increase the expressiveness of the model~\cite{papamakarios2021normalizing_review, Kobyzev_2021}: ${f = f_{1} \circ \ldots \circ f_{N}}$. If $f$ is differentiable, one can express $p(x)$ as a function of $q(z)$ and the Jacobian matrix of the transformation, $J_f$, with the change of variables formula as
\begin{equation}
    p(x) = q(f(x)) \left| \det J_{f}(x)  \right|.
    \label{eq:pdfs}
\end{equation}
For a composition of transformations, the log-likelihood function becomes
\begin{equation}
    \log p(x) = \log q(f(x)) + \sum_{i=1}^{N} \log \left| \det J_{f_{i}}(y_i) \right|,
    \label{eq:loss2}
\end{equation}
where $y_i = f_{i+1} \circ \dots \circ f_N$ represents the output of intermediate flows.
With a suitable choice of $q(z)$ and the transformation $f$, Eq.~\ref{eq:loss2} has a closed form and can be used to train the normalising flow with a negative log-likelihood loss function.
The transformations $f_{i}$ are typically analytically invertible maps whose parameters are predicted by neural networks, referred to as ``conditioner networks".

The normalising-flow model used here adopts the coupling block architecture of \textsc{RealNVP}~\cite{DBLP:journals/corr/DinhSB16} and \textsc{Glow}~\cite{kingma2018glow}. 
Each input, described as a vector of the shower shape and isolation features $x_s\in\mathbb{R}^n$ is split into two complementary subsets: $(x^A_s,x^B_s)$.  
The subset $x^B_s$ undergoes an invertible neural-spline coupling transformation, parameterised by monotonic rational-quadratic splines~\cite{Spline_flows,muller2019neural_linear_quadratic_splines}, while $x^A_s$ remains unchanged. 
The subsets are permuted during the $N$ transformations so all input features are transformed. 
This coupling block design yields a block triangular Jacobian, which admits a closed form and computationally efficient evaluation of the log-determinant. 
Simultaneously, the adoption of spline-based coupling functions preserves the expressive capacity characteristic of autoregressive flows, and achieves substantially reduced inference time~\cite{Spline_flows}.
Using splines in normalising flows has the advantage that they can approximate more complex distributions compared to affine transformations, while still being simple to invert. 
In addition, the monotonicity of the transformations is important for morphing distributions, because this ensures the quantiles of the distributions are preserved during the transformation.
When a transformation is monotonic, the order of data points is maintained. This means that the same proportion of data points will fall below any given threshold in both the original and transformed distributions. Consequently, the quantiles, which represent these thresholds, remain unchanged. 

To perform the simulation-to-data distribution morphing, the one-flow method of Ref.~\cite{Daumann:2024kfd} is employed. In this approach, a normalising flow is trained simultaneously on both simulation and data samples. The normalising flow is conditioned on the kinematic features $x_k = \{\pt, \eta, \phi, \rho\}$, and an additional Boolean variable $\mathbb{I}_{\text{data}}$~\cite{serra2019blow}, which allows the model to simultaneously learn distinct bijections from the simulation and data domains to the common base distribution. 
Events from data are assigned $\mathbb{I}_{\text{data}}=1$, while simulated events take $\mathbb{I}_{\text{data}}=0$. In this way, the normalising flow learns to approximate the conditional density $p(x_s|x_k,\mathbb{I}_{\text{data}})$.

Once trained, the normalising flow can be used to calibrate the simulation by first mapping a simulated event ($x_{s,i},x_{k,i}$) to the base distribution, 
\begin{equation}
    x_{s,i} \longrightarrow z_i = f(x_{s,i}|x_{k,i},\mathbb{I}_{\text{data}}=0),
\end{equation}
flipping the Boolean flag to $\mathbb{I}_{\text{data}}=1$, and then applying the inverse transformation to project the events into the data domain,
\begin{equation}
    z_i \longrightarrow \widetilde{x}_{s,i} = f^{-1}(z_i|x_{k,i},\mathbb{I}_{\text{data}}=1).
\end{equation}
After this transformation, the conditional pdf over the morphed features in simulation approximates the conditional pdf in data.
Owing to the monotonicity of the bijective transformations, this procedure effectively performs a quantile morphing between the simulation and data distributions via the shared base distribution \cite{Daumann:2024kfd}. Since each transformation is invertible, the mapping is fully deterministic: the same input always yields the same output, with no stochasticity involved. 

Analogous to the reweighting method, the simulation is first reweighted to data in the $x_k$ features using the S1 model. This step prevents the normalising flow from learning the differences in the $x_s$ distributions that arise from discrepancies between data and simulation in the $x_k$ features.
The total log-likelihood loss is obtained by summing the log-likelihood for each event in the training sample, where the simulated events are weighted according to $w^{\text{nom}}_i w_{\mathrm{S1}}(x_{k,i})$ in the sum.
We note that the normalising-flow architecture cannot naturally handle negatively weighted events, and therefore only positively weighted events are used in the training.
This approach is justified by the same arguments discussed in the reweighting method, \ie that the conditional density $p_{\mathrm{MC}}(x_s|x_k)$ is approximately equal for both positively and negatively weighted events.

The normalising flow is implemented using the neural spline architecture of the \textsc{Zuko} package~\cite{zuko} in \textsc{PyTorch}~\cite{paszke2019pytorch}. 
Unlike the reweighting method, 
a single-flow model (conditional in probe $\eta$) was found to be sufficient to achieve good performance across the different subdetector regions.
Training is performed with an early stopping criterion with a patience of 15 epochs. A learning rate scheduler is employed that reduces the initial learning rate of $1\times10^{-3}$ by a factor of 0.1 whenever the validation loss fails to improve for 10 consecutive epochs. The architecture comprises eight spline transformations, each with ten bins, while the conditioner networks consist of two hidden layers of 256 units each. The model was trained with a batch size of 1024. It was found that the performance of the model was robust and did not vary significantly across different hyperparameter settings; therefore no extensive hyperparameter optimisation was required. Moreover, the model showed no signs of overfitting based on comparing the training and validation loss curves, and thus no additional regularisation was applied. 
The normalising-flow model was trained on an NVIDIA A40 GPU, requiring up to 48 GB of VRAM, and took approximately one hour to complete.

\begin{figure}[htb!]
    \centering
    \includegraphics[width=.49\linewidth]{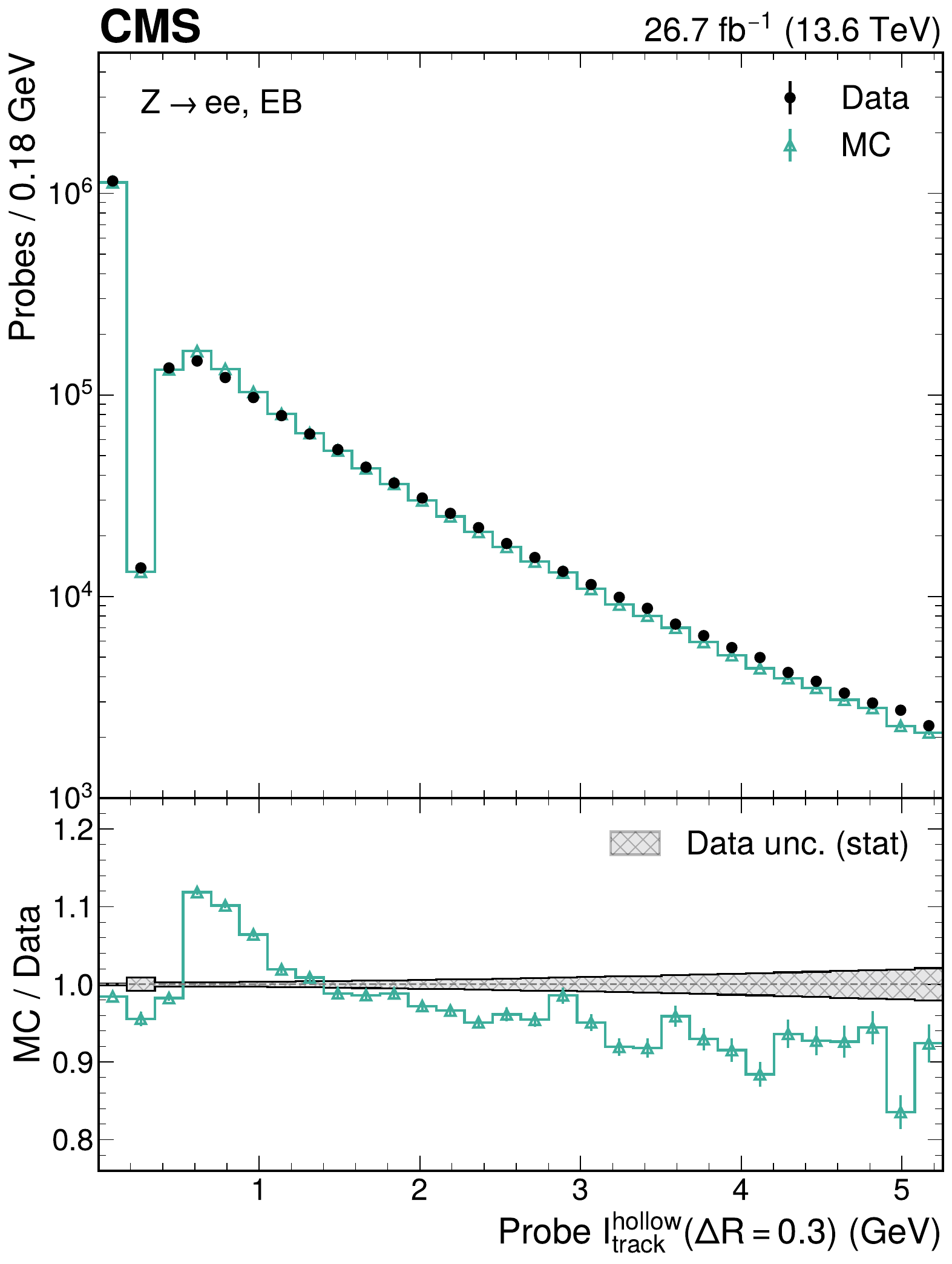}
    \includegraphics[width=.49\linewidth]{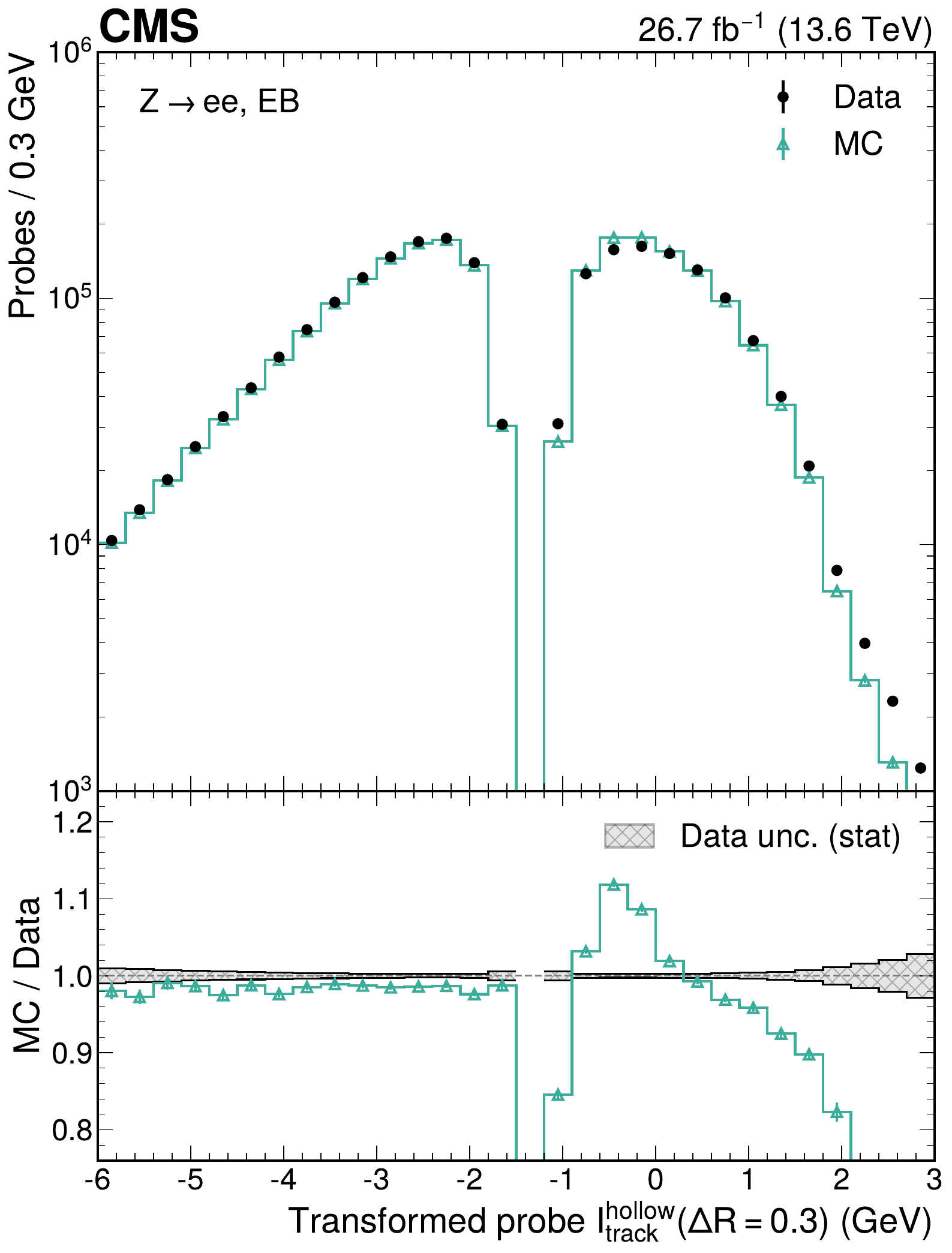}
    \caption{The probe $\mathcal{I}^{\text{hollow}}_{\text{track}}(\Delta R = 0.3)$ distribution in simulation (teal points) and data (black histogram), before (left) and after (right) the transformation of this isolation-related feature, used in the normalising-flow model. The lower panels show the ratio of simulation to data, where the statistical uncertainties in the simulation and data are shown by the teal error bars and the grey hatched bands, respectively.}
    \label{fig:iso_flows}
\end{figure}

A number of challenges also arise in the application of the normalising-flow method to a real-world scenario. 
In particular, complex input feature distributions require special treatment, as normalising flows assume that the distributions of the input features are continuous and differentiable \cite{Kobyzev_2021, papamakarios2021normalizing_review}. 
The isolation-related features contain a discontinuity due to detector thresholds. 
For these variables, the transformation introduced in Ref.~\cite{Daumann:2024kfd} is employed: 
the peak at zero is mitigated by replacing the zero values with values sampled from a triangular distribution.
To ensure a clear separation between the zero values and the continuous part of the distribution, 
the values from the continuous part are first shifted by a parameter $a$. 
The triangular distribution is subsequently defined on the interval $[0,a]$, 
such that its range remains disjoint from the shifted continuous region. 
The zero values are then replaced with values sampled from the pdf,
\begin{equation}
    p_{\mathrm{triangular}}(x) = \begin{cases}
        \frac{2(a-x)}{a^2}, & 0 < x < a \\
        0, & \text{otherwise}
    \end{cases}.
\end{equation}
The parameter $a$ is set to 0.02 for H/E and 0.2 for all other isolation-related variables to reflect the different scales of these features.
Since this sampling procedure involves random sampling, a small seed dependence is in principle introduced when evaluating the normalising flow. 
To quantify the corresponding effect, the evaluation was repeated using different random seeds, and the observed variations were found to be negligible. 
After sampling, a log-transformation is applied to the feature, producing a two-peaked structure.
Figure~\ref{fig:iso_flows} shows the probe $\mathcal{I}^{\text{hollow}}_{\text{track}}(\Delta R = 0.3)$ distributions in simulation and data, before and after this transformation procedure is applied. 
The agreement in the shape of the left-hand peak in the transformed feature reflects the fact that the values for both data and simulation are sampled from the same triangular distribution.
Ultimately, this procedure was found to improve the performance of the normalising flow in learning the isolation-related features. 
For the two EE-only variables, namely $\sigma_{\mathrm{RR}}$ and $E_{\mathrm{ES}}/E_{\text{raw}}$, values in the EB region are set to zero, after which they undergo the same transformations as the other isolation variables. 
This enables the use of a single flow model across all three ECAL regions.

Additionally, all features are normalised for the training, \ie each feature $y_i$ is replaced with $(y_i-\hat{y})/\sigma_y$, where $\hat{y}$ and $\sigma_y$ are the sample mean and standard deviation, respectively.
When using the model for the final calibration, these aforementioned transformations are inverted after mapping the simulated events to the data domain. 

\begin{figure}[htb!]
    \centering
    \includegraphics[width=.49\linewidth]{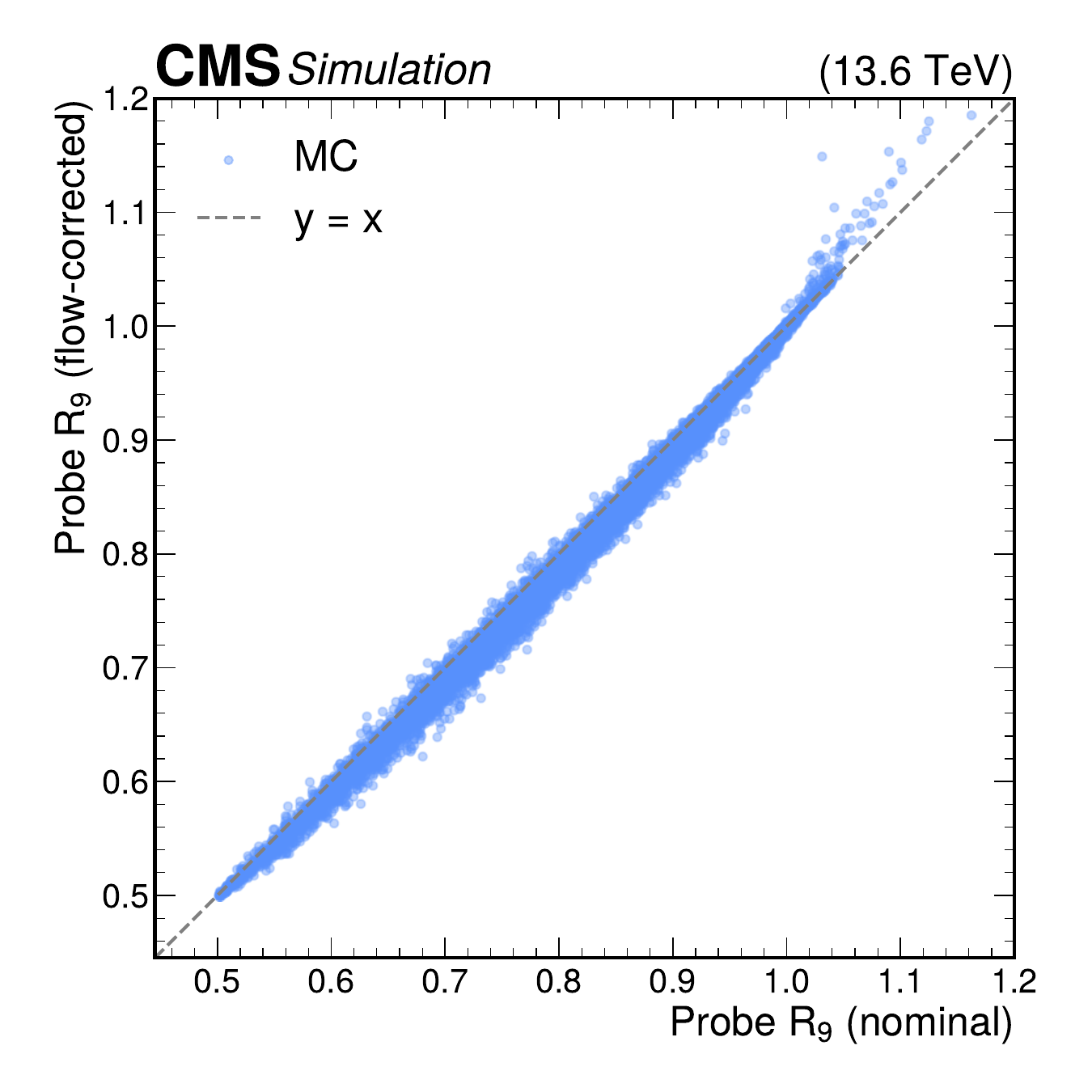}
    \includegraphics[width=.49\linewidth]{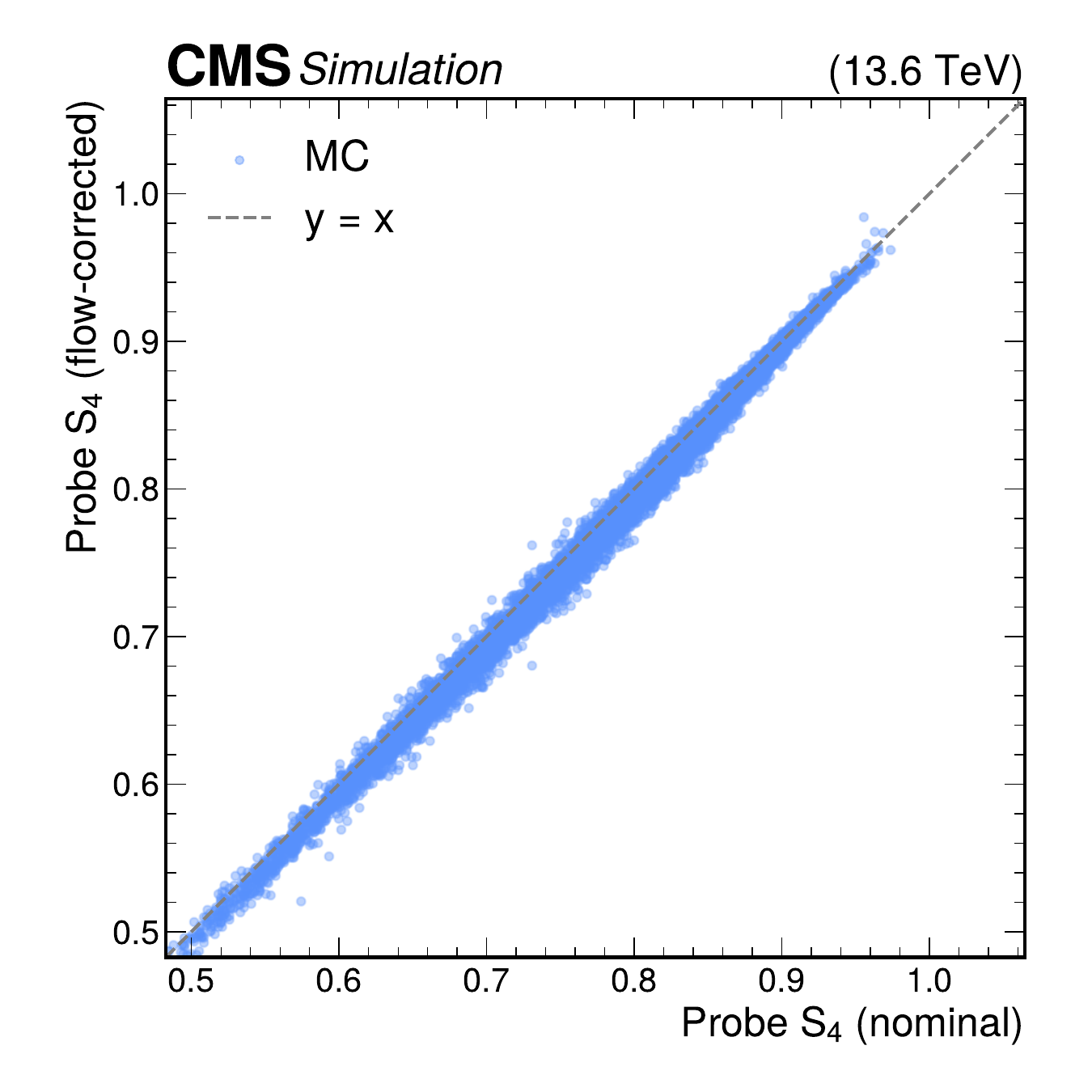}
    \includegraphics[width=.49\linewidth]{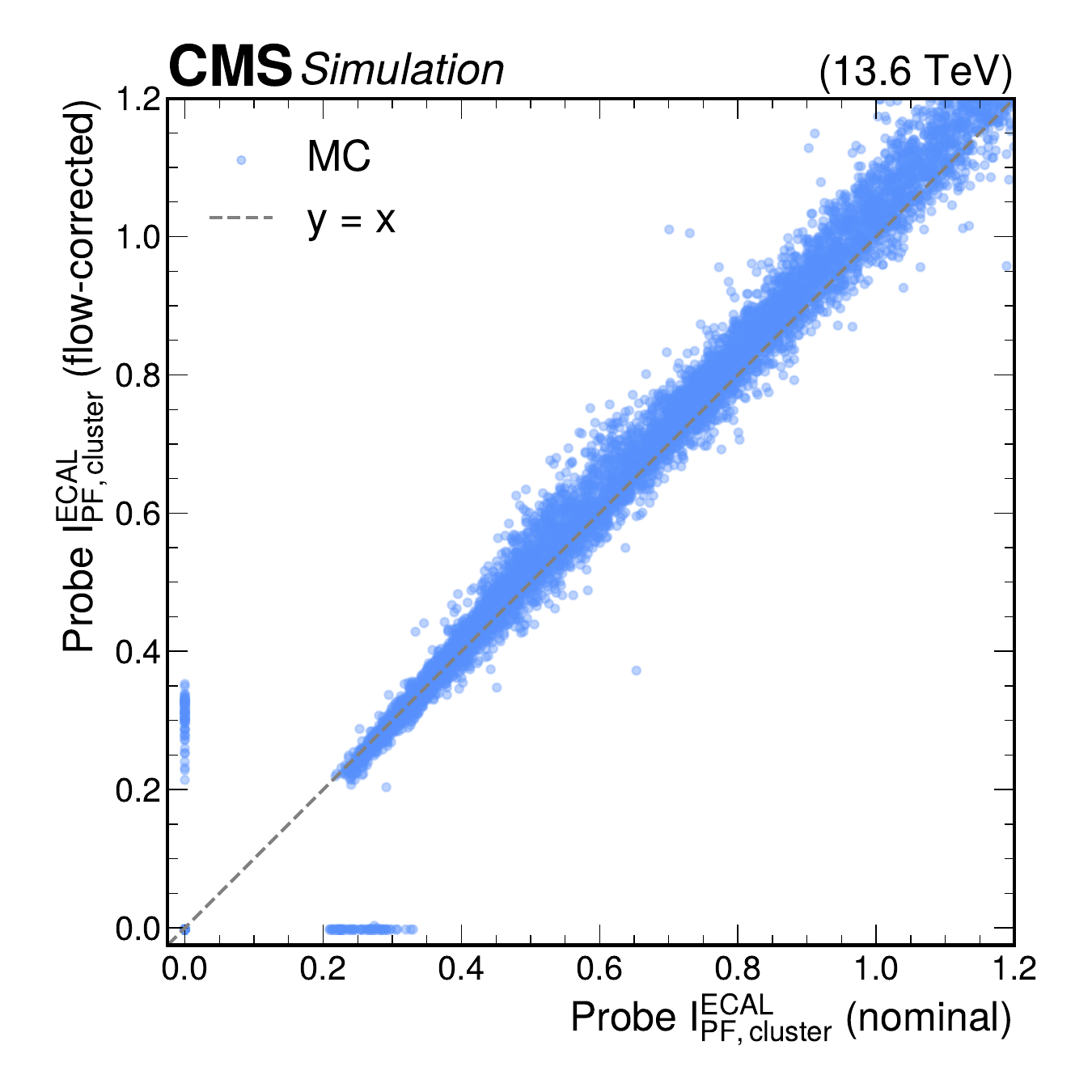}
    \includegraphics[width=.49\linewidth]{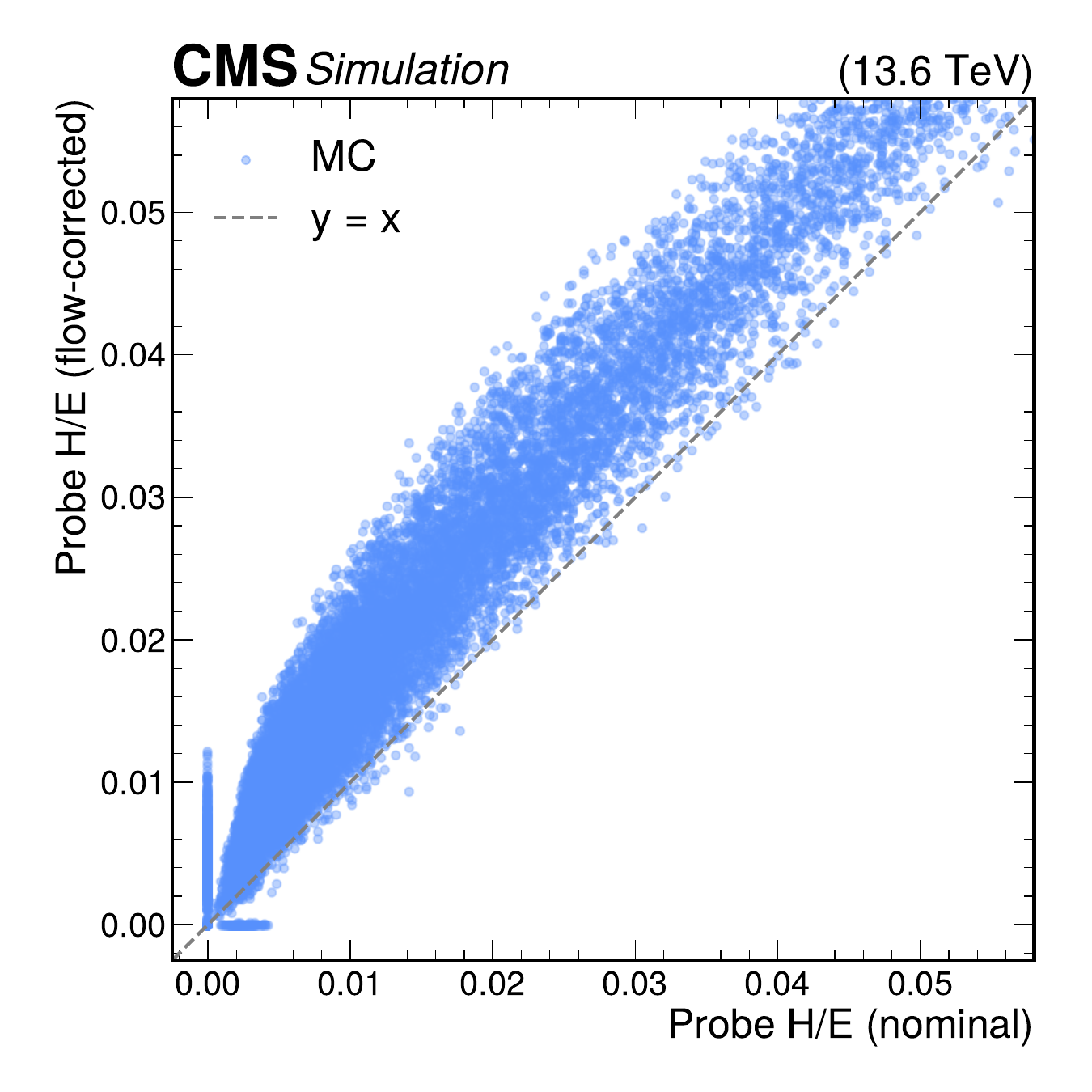}
    \caption{Scatter plots showing the learned high-dimensional mapping projected onto a single dimension: probe $R_9$ (upper left), $S_4$ (upper right), $\mathcal{I}^{\mathrm{ECAL}}_{\text{PF, cluster}}$ (lower left) and H/E (lower right). The $x$ axis shows the nominal values of these features in the simulation sample, while the $y$ axis shows the flow-corrected features. The dashed line in each panel shows the identity line $y=x$, where points on the line are unchanged by the flow corrections in this dimension.}
    \label{fig:flow_scatter}
\end{figure}

The behaviour of the normalising flow can be visualised by projecting its high-dimensional mapping onto a single dimension, using a scatter plot of the nominal simulation values versus the flow-corrected values. As discussed above, the model effectively performs a quantile morphing between the simulation and data distributions, and therefore the transformations are not expected to map events across large distances in the input feature space. For example, an event with a low-$R_{9}$ value should not be corrected to an extremely high-$R_{9}$ value. Figure~\ref{fig:flow_scatter} shows these 1D projections for two features related to the shower-shape of the probes: $R_9$ and $S_4$, and two features related to the isolation: $\mathcal{I}^{\mathrm{ECAL}}_{\text{PF, cluster}}$ and H/E. The diagonal line $y=x$ indicates zero correction in a given dimension, whilst points above (below) the line are shifted to higher (lower) values. The features $R_9$, $S_4$, and $\mathcal{I}^{\mathrm{ECAL}}_{\text{PF, cluster}}$ exhibit minor corrections, with most points clustered close to the line $y=x$. In contrast, H/E shows sizeable corrections, with the majority of points significantly above the line, reflecting the large initial discrepancy between data and simulation in this feature (as shown in Fig.~\ref{fig:results_shower_2}). A further effect is visible in the isolation features (lower panels), where points clustered at zero in just one axis represent events that migrate across the gap separating the peak at zero from the smoothly falling tail.

\section{Results}\label{sec:results}
The performance of the calibration methods is assessed using the independent test sample of probe EM showers, 
by applying the corrections and comparing the level of agreement between data and simulation. 
As this sample contains both positively weighted and negatively weighted events, 
an additional S1 model is first trained to reweight the mixed weight simulation sample to data in the $x_k=\{\pt, \eta, \phi, \rho\}$ features. 
The S2-reweighting and normalising-flow models, trained exclusively on positively weighted events, are then applied. 
This procedure ensures that any residual discrepancies in the calibration performance cannot be accounted for by differences in the $x_k$ distributions between data and the mixed weight simulation samples. 

Simulated distributions are labelled according to the following naming scheme: ``MC" denotes the nominal (mixed weight) sample, while ``MC (S1)" refers to the sample after applying $w_{\mathrm{S1}}(x)$ to align the $x_k$ distributions with data. 
The calibrated sample from the reweighting method is referred to as ``MC (S1$\times$S2)", where the total weight correction is given by the product $w_{\mathrm{S1}}(x)w_{\mathrm{S2}}(x)$, 
and the calibrated sample from the normalising-flow method is referred to as ``MC (S1+NF)", where the normalising flow is applied on top of the S1 reweighting.

\subsection{Invariance of the kinematic and event-level features}\label{sec:results_kinematic}
One of the key objectives of the calibration methods is to maintain kinematic invariance after the corrections are applied. 
In other words, the distributions of $x_k$ should remain unchanged. 
This property is guaranteed by construction for the normalising-flow method, as $x_k$ are used as conditional features in the flow. 
However, the reweighting method provides a global event weight correction, which can therefore lead to deformations of the kinematic distributions. 
The two-stage training procedure was introduced to mitigate this effect.

\begin{figure}[htb!]
    \centering
    \includegraphics[width=.325\linewidth]{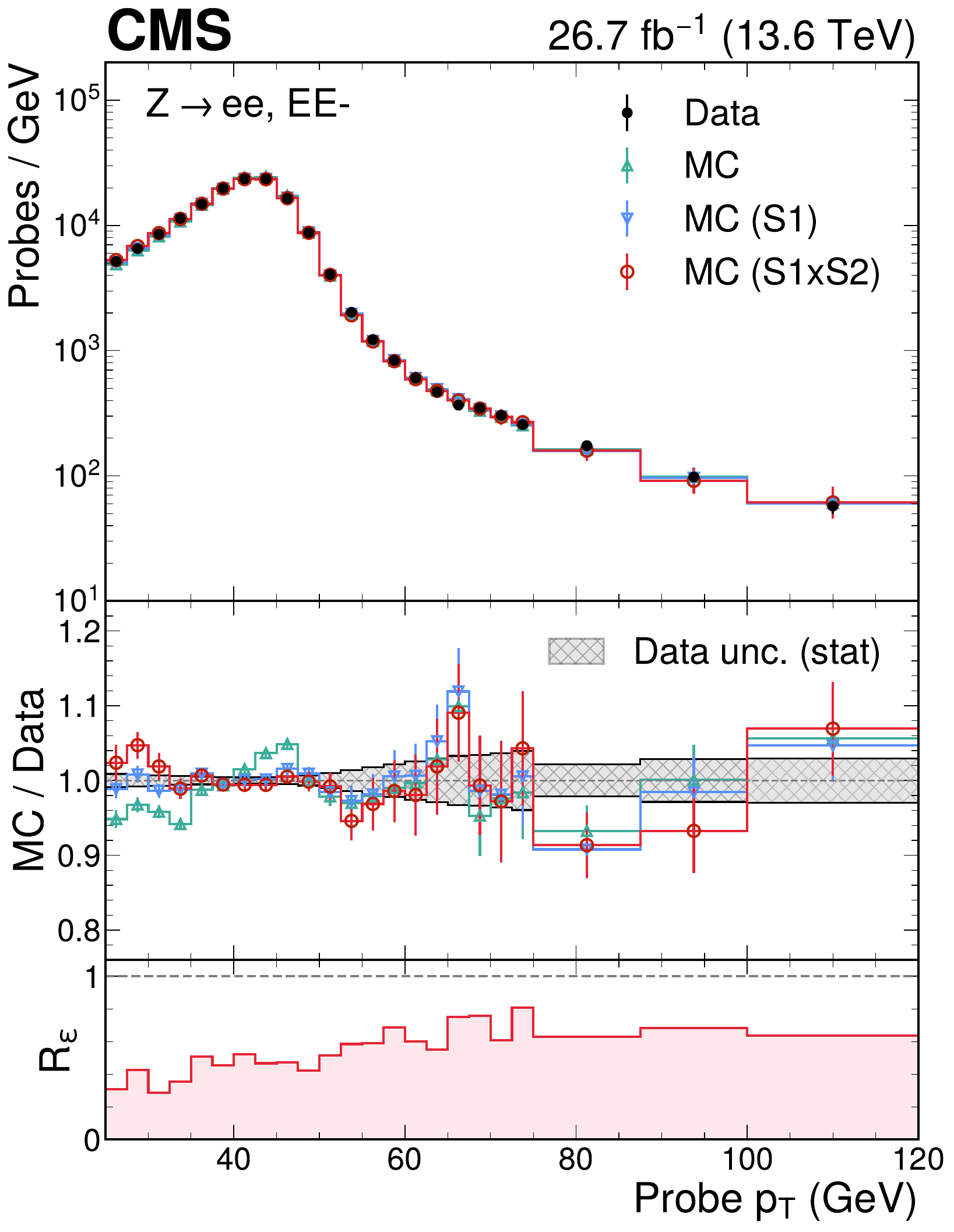}
    \includegraphics[width=.325\linewidth]{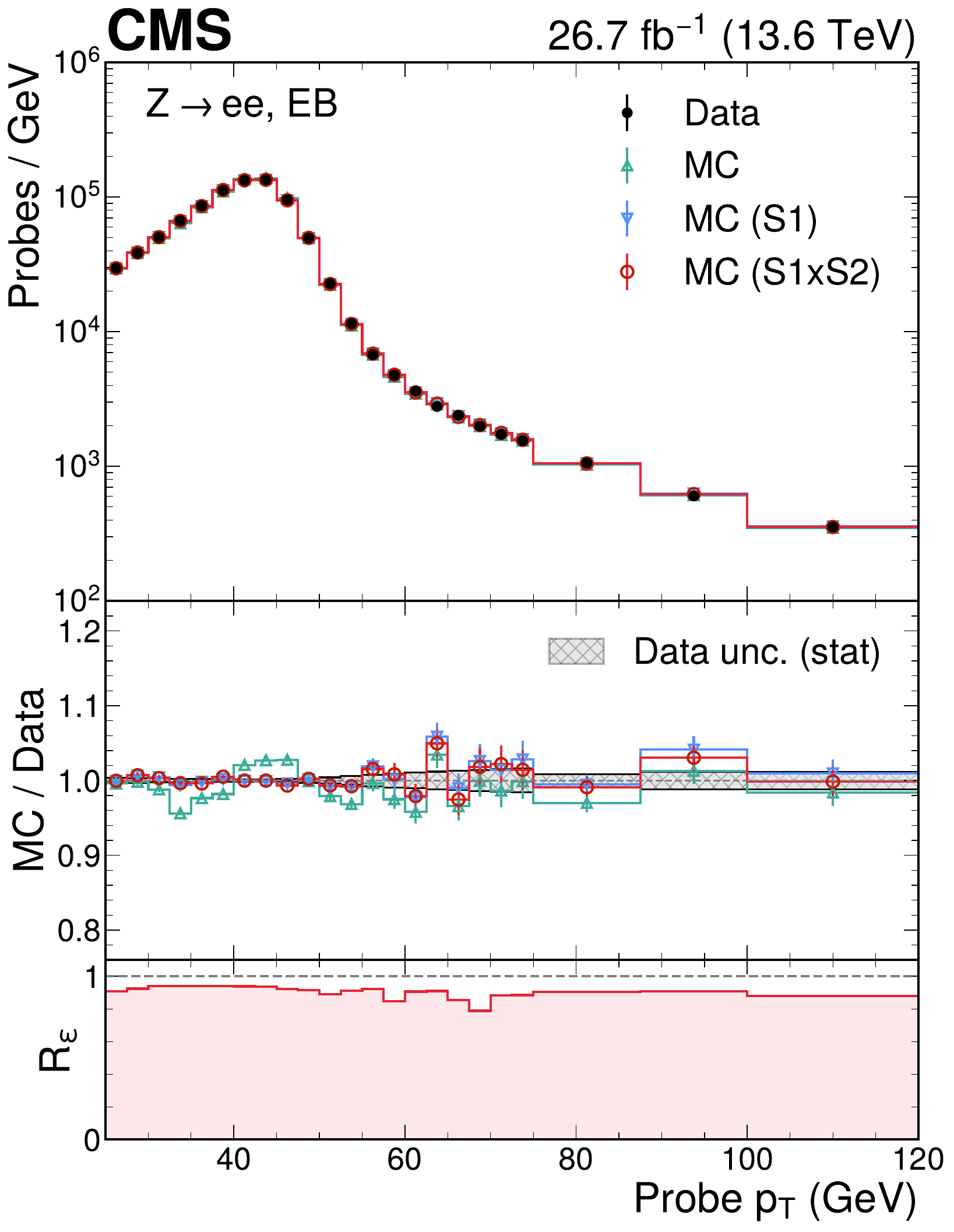}
    \includegraphics[width=.325\linewidth]{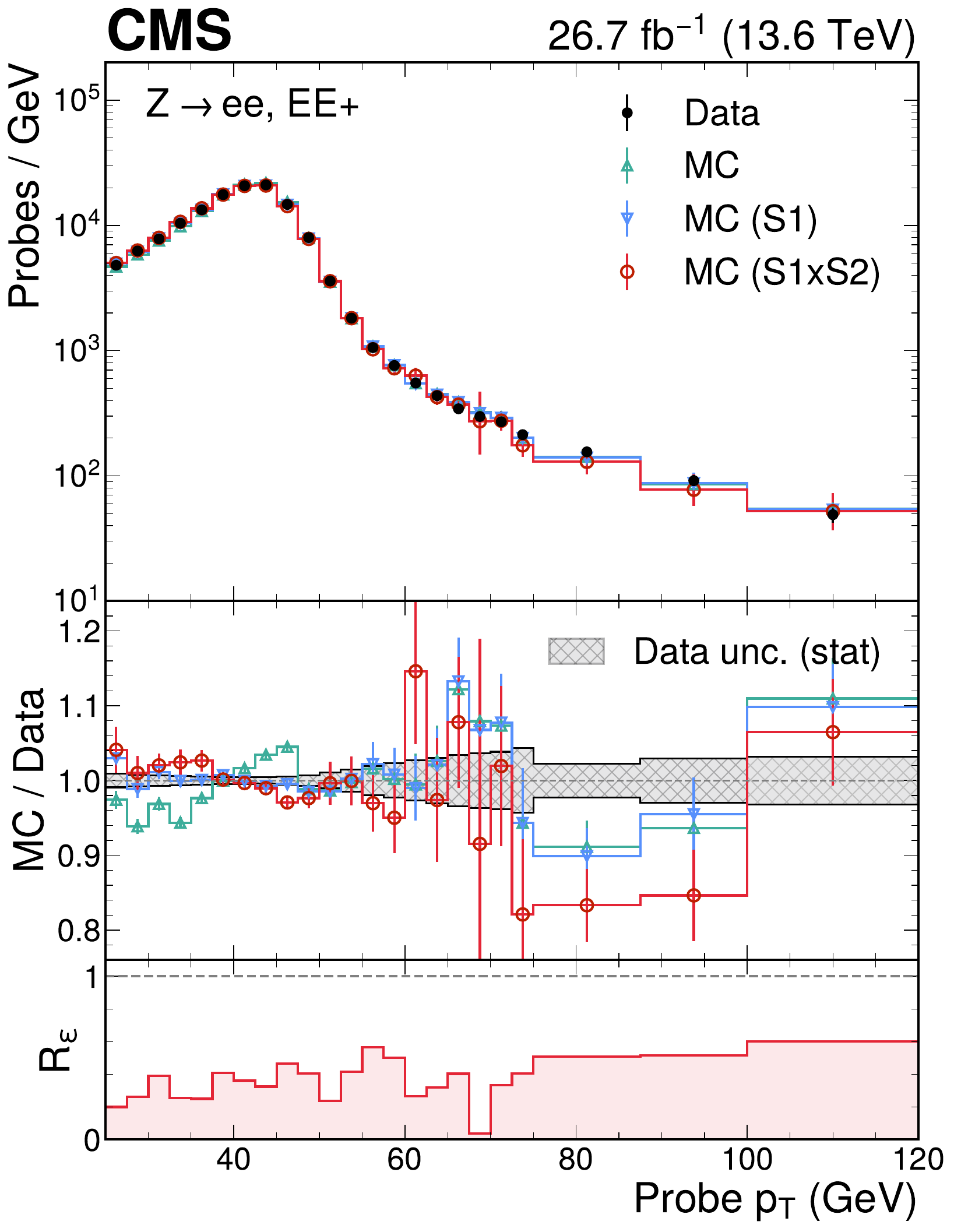}
    \includegraphics[width=.325\linewidth]{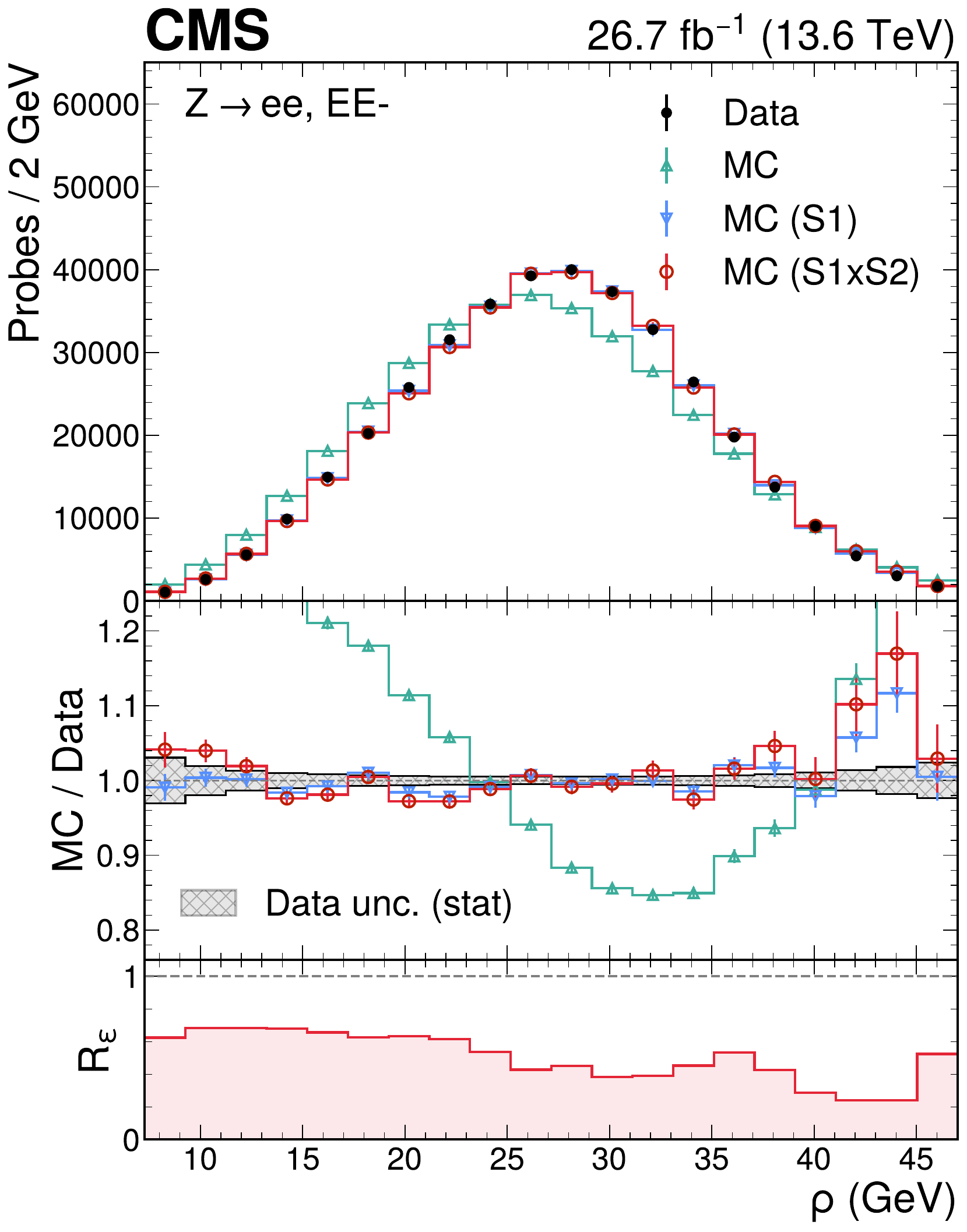}
    \includegraphics[width=.325\linewidth]{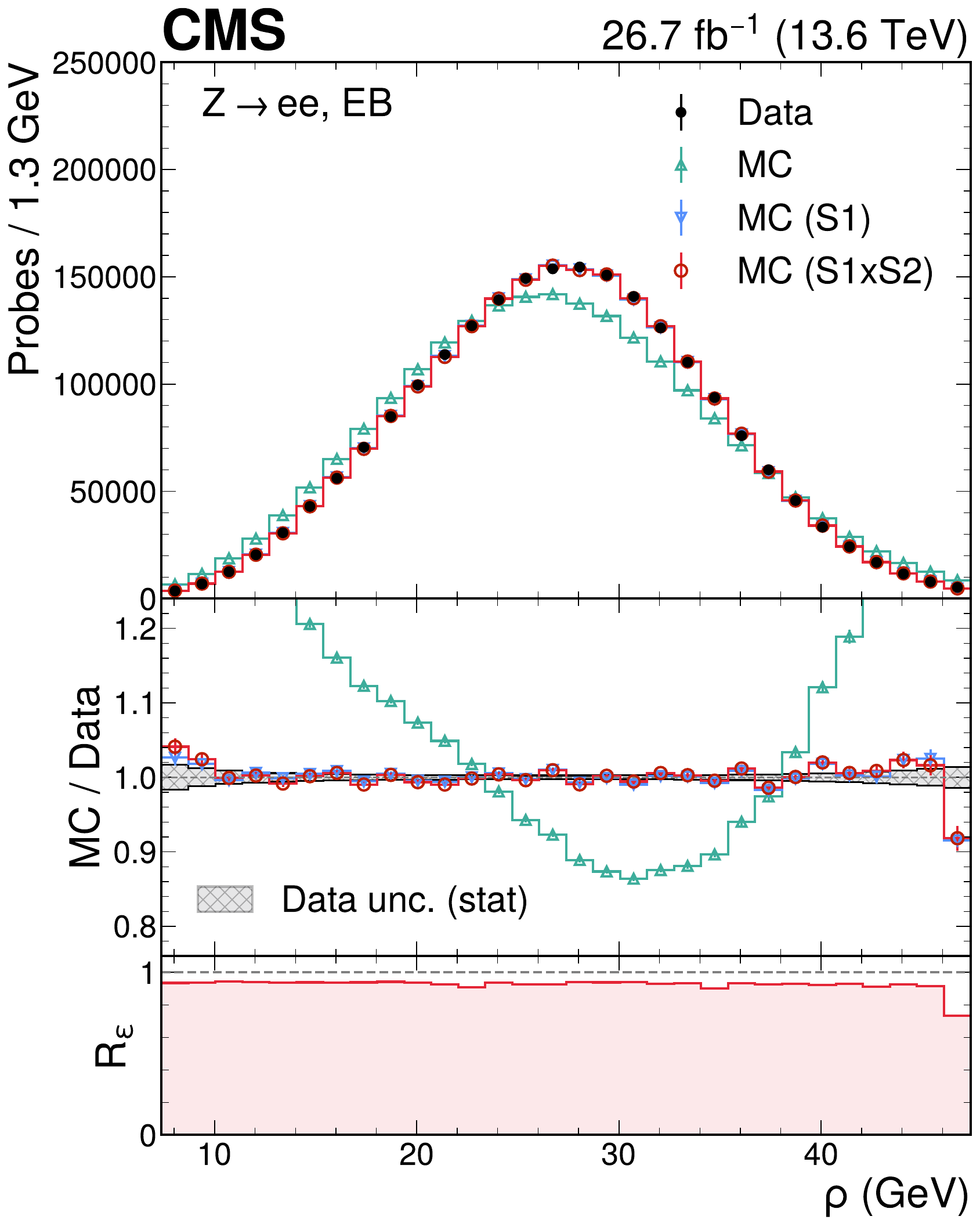}
    \includegraphics[width=.325\linewidth]{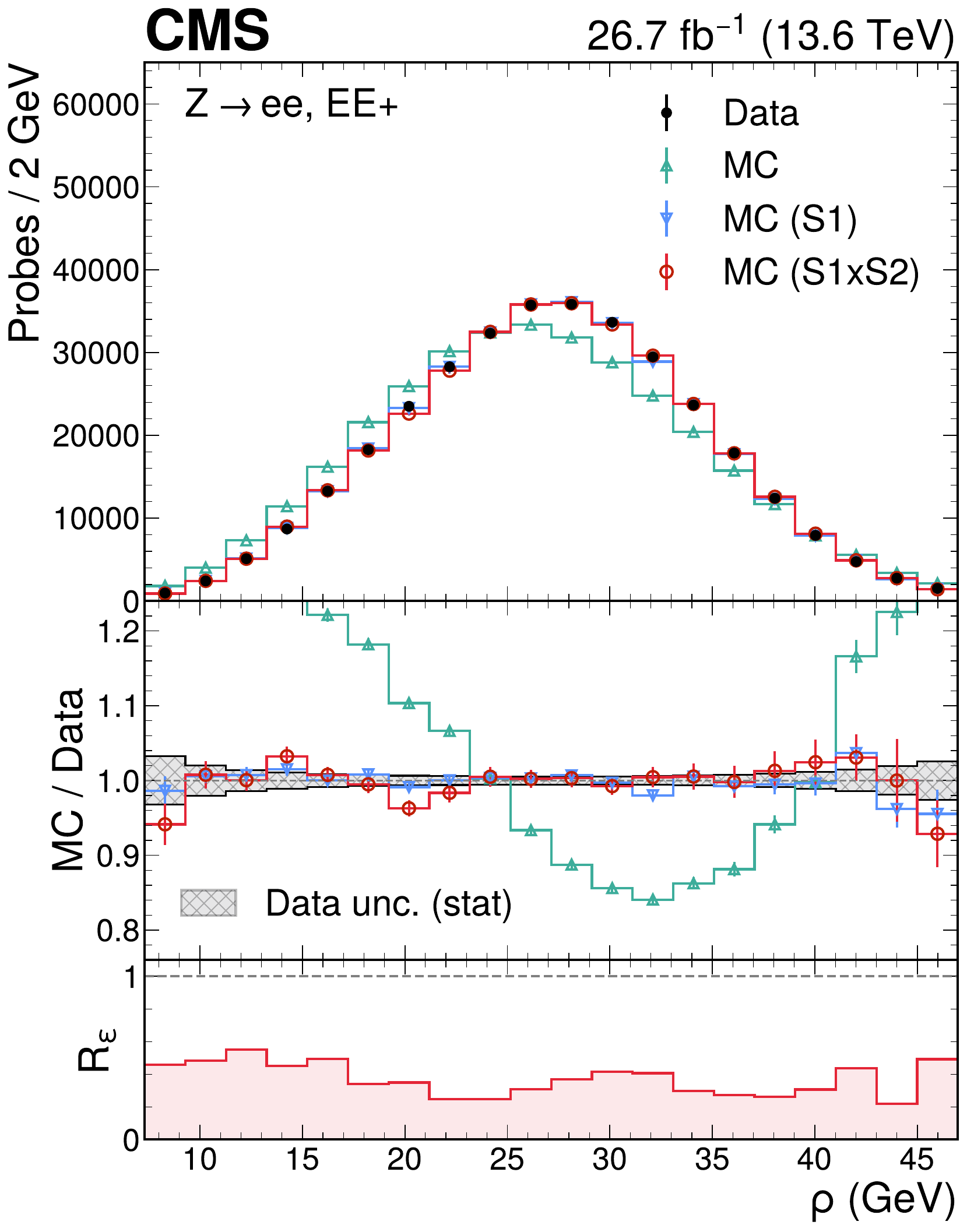}
    \caption{The probe $\pt$ (upper row) and $\rho$ (lower row) distributions in data (black points) and simulation (coloured histograms with points), for the EE$-$ (left), EB (centre), and EE$+$ (right) subdetector regions. 
    The statistical uncertainties in the simulation and data are shown by the error bars on the points.
    The simulation is shown for three scenarios. 
    The teal histograms show the nominal simulation normalised to the data, the blue histograms show the simulation after S1 reweighting, and the red histograms show the simulation after S2 reweighting. 
    The middle panels show the ratio of simulation to data, where the statistical uncertainty in the data is indicated by the grey hatched band. 
    The lower panels show the reduction in statistical power $R_\varepsilon$, in bins of each observable, after applying the S2 reweighting.}
    \label{fig:results_kinematic_pt_rho}
\end{figure}

\begin{figure}[htb!]
    \centering
    \includegraphics[width=.325\linewidth]{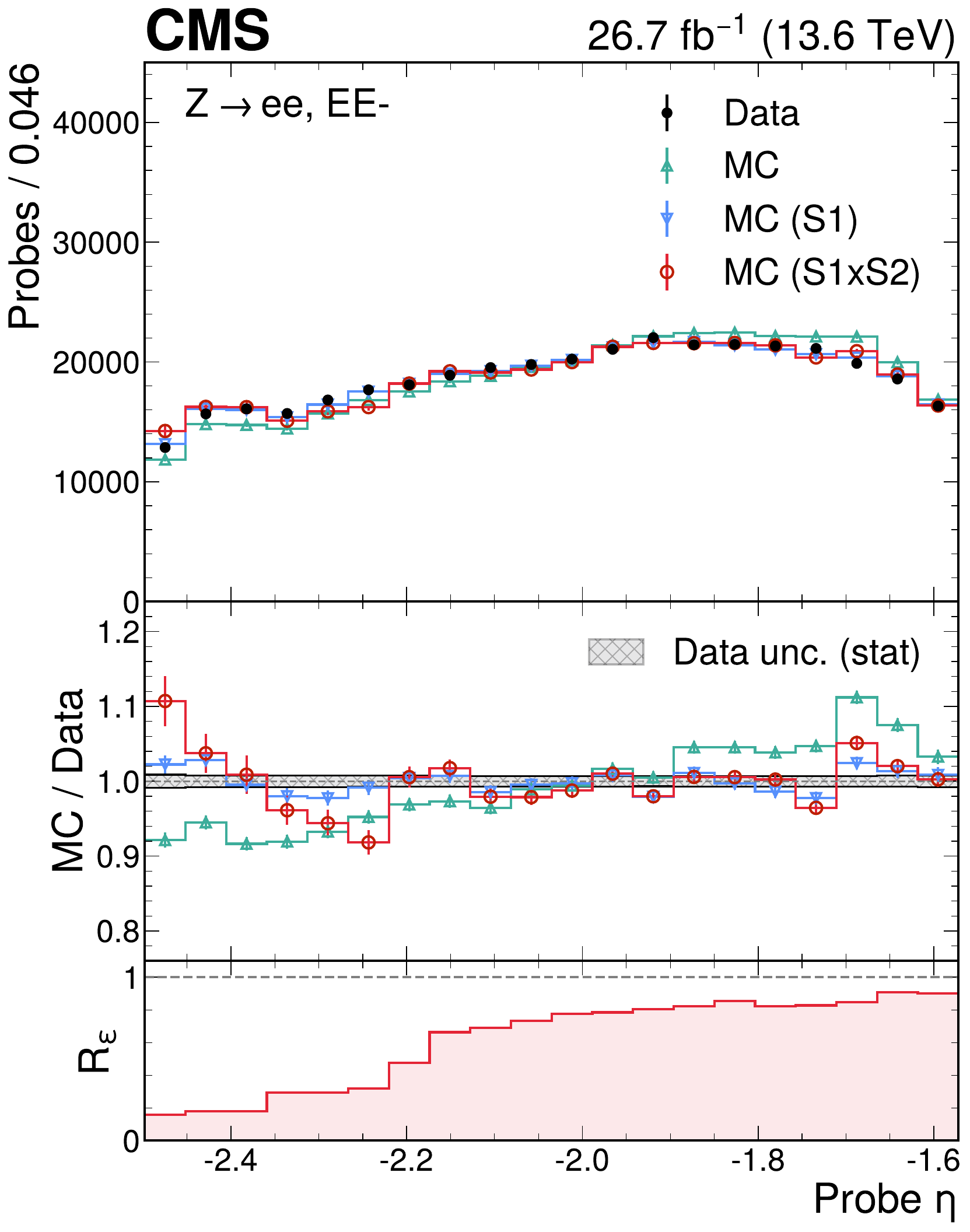}
    \includegraphics[width=.325\linewidth]{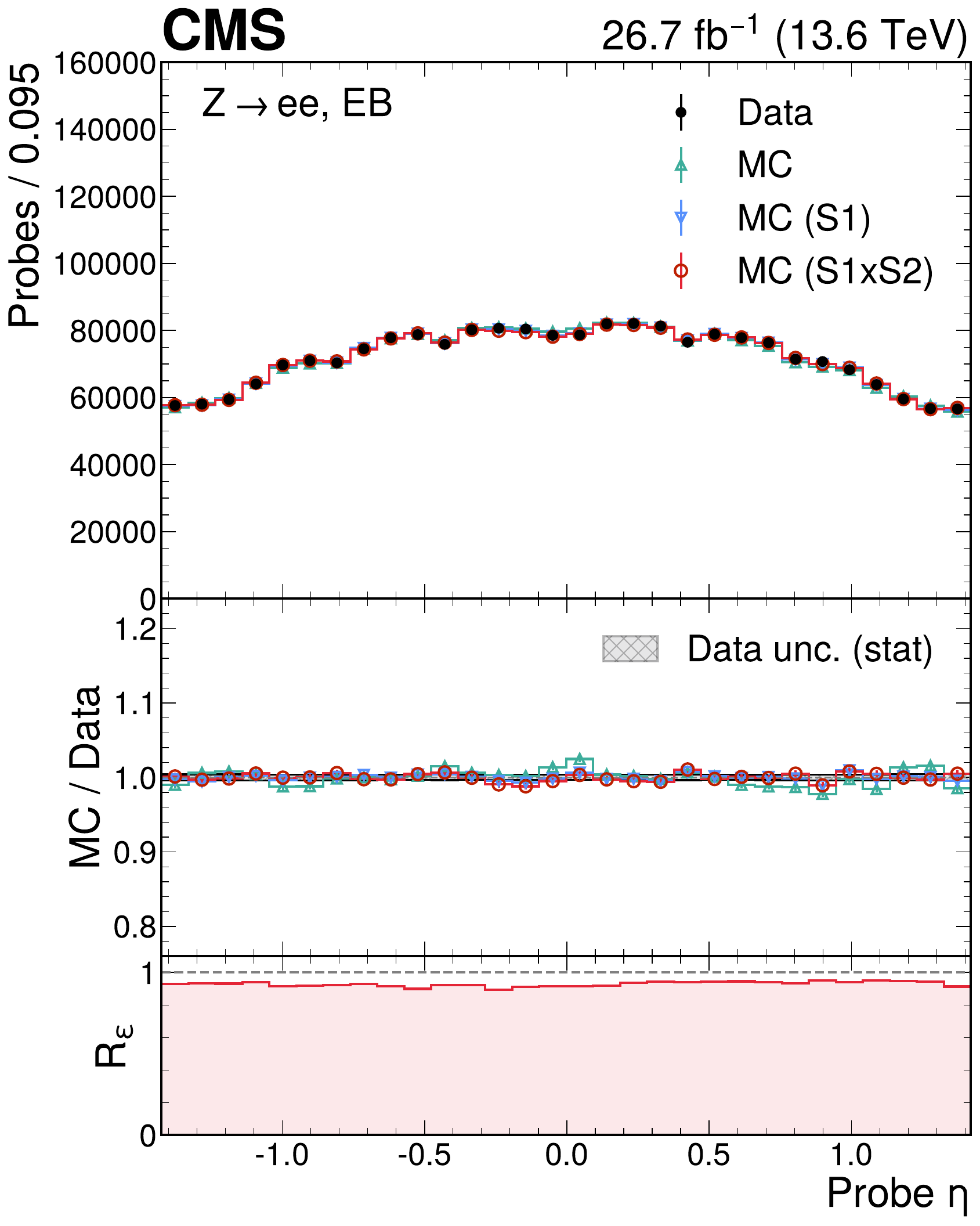}
    \includegraphics[width=.325\linewidth]{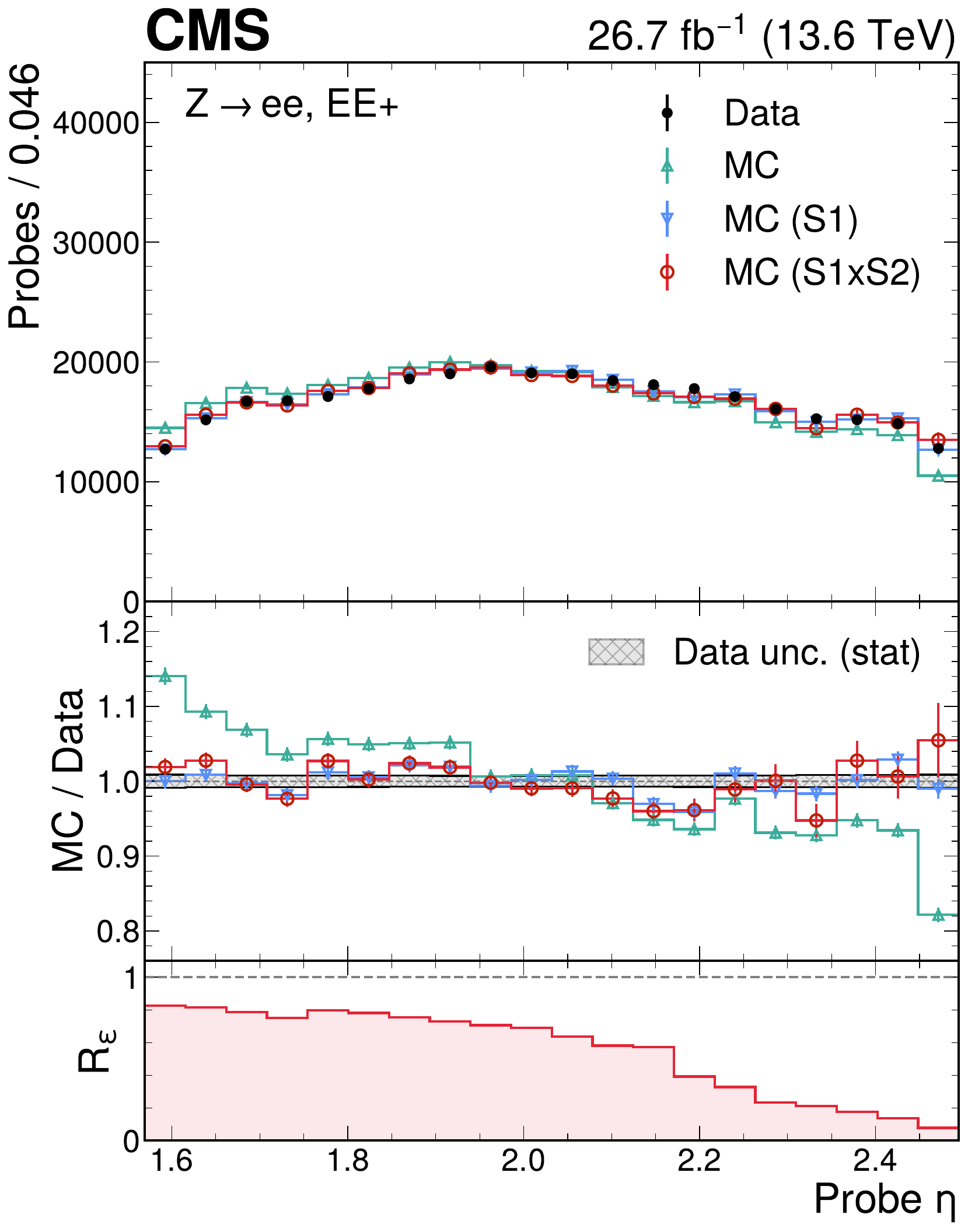}
    \includegraphics[width=.325\linewidth]{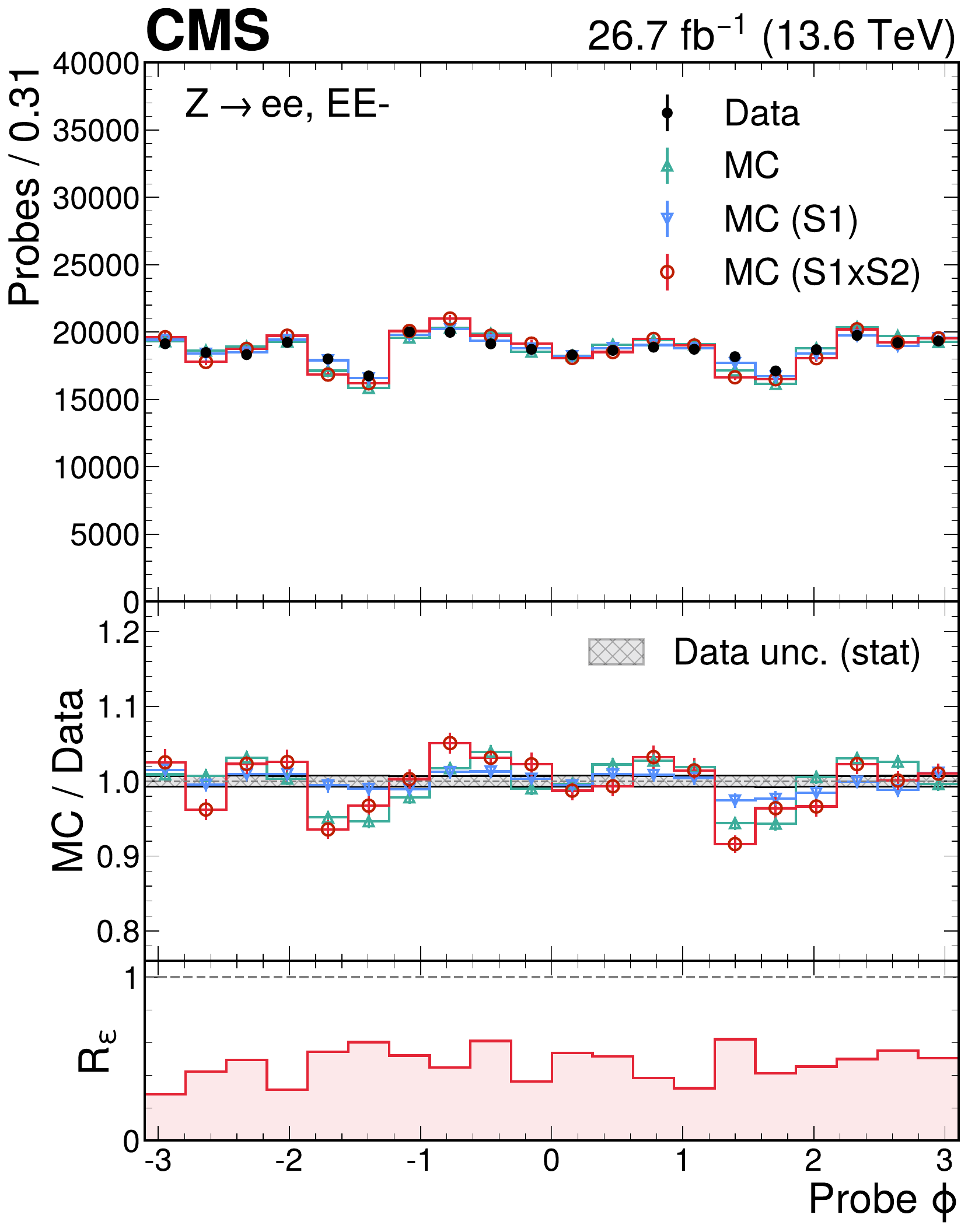}
    \includegraphics[width=.325\linewidth]{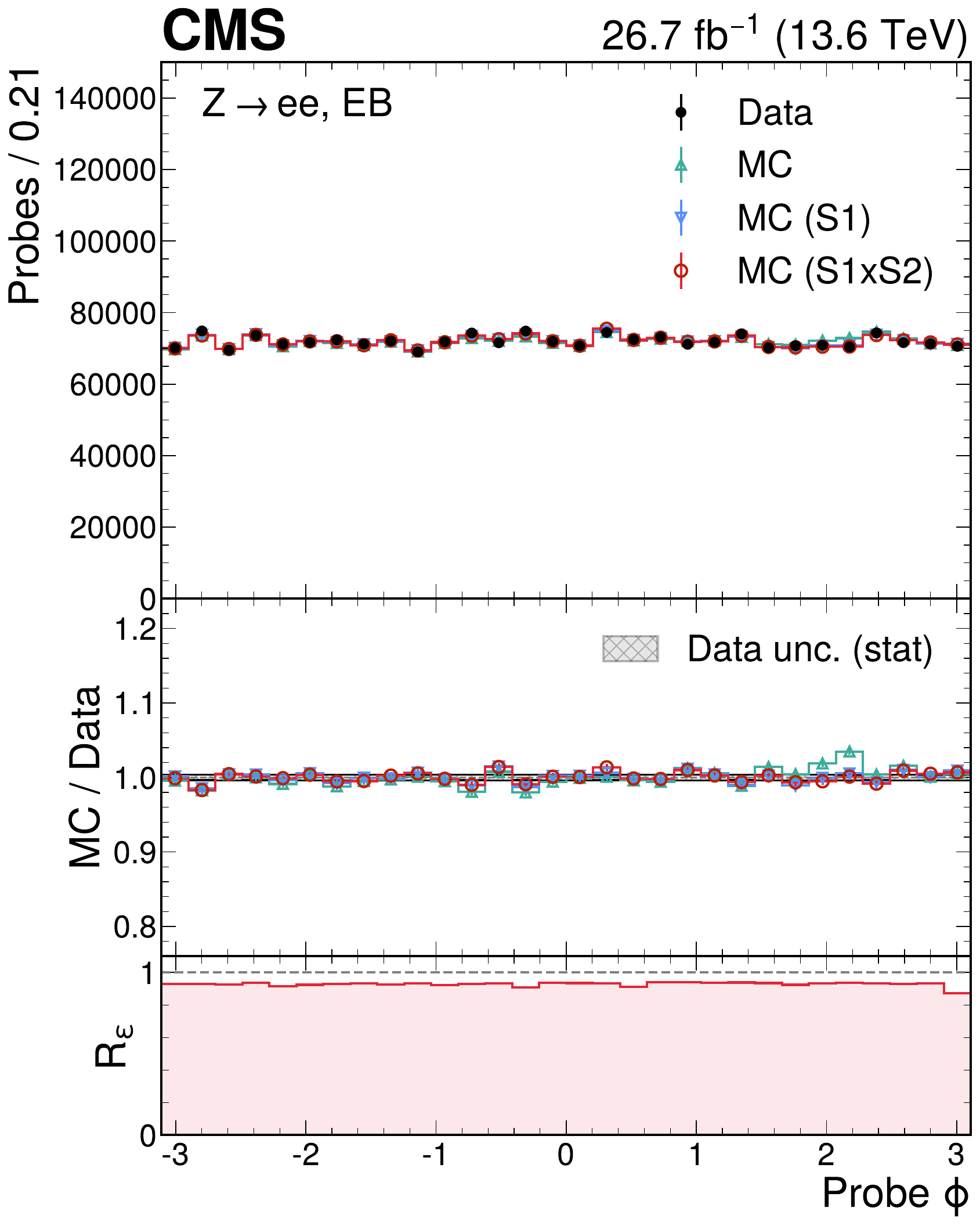}
    \includegraphics[width=.325\linewidth]{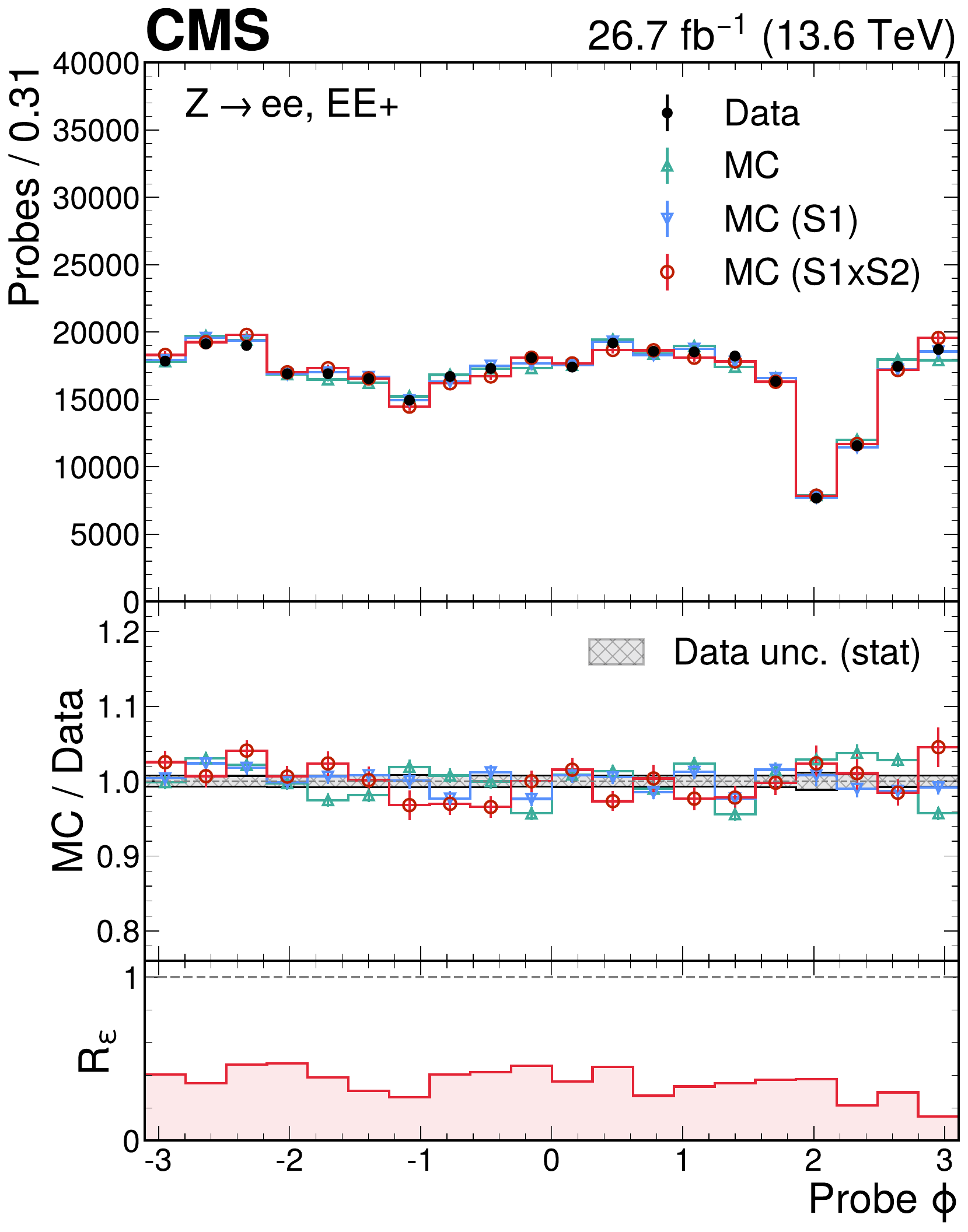}
    \caption{The probe $\eta$ (upper row) and $\phi$ (lower row) distributions in data and simulation, 
    for the EE$-$ (left), EB (centre), and EE$+$ (right) subdetector regions. 
    The plotting conventions are the same as in Fig.~\ref{fig:results_kinematic_pt_rho}.}
    \label{fig:results_kinematic_eta_phi}
\end{figure}

Figures~\ref{fig:results_kinematic_pt_rho} and \ref{fig:results_kinematic_eta_phi} show 1D histograms for the $x_k$ features in data and simulation, 
split into the EE$-$ (left), EB (centre), and EE$+$ (right) subdetector regions. 
The simulation is shown for three reweighting scenarios: MC, MC (S1), and MC (S1$\times$S2) in teal, blue, and red, respectively. 
The middle panels in each plot show the ratio of the simulation to data. 
The blue histograms show excellent agreement with data, demonstrating the good performance of the S1 reweighting.
Comparing the red histograms to the blue histograms shows the deformations in the kinematic distributions that are incurred after applying the S2 reweighting. 
The deformations, whilst small (a few percent) and much smaller than the initial data-simulation differences prior to S1 reweighting, are non-zero.
This indicates that perfect kinematic invariance has not been achieved with the reweighting method.
When applying the reweighting method on an independent sample for an analysis,
the deformations in the kinematic distributions induced by the calibration should be explicitly checked. 
If the analysis is sensitive to the size of the kinematic variations, then a dedicated systematic uncertainty should be introduced to account for this effect.
As mentioned above, the normalising-flow method does not change $x_k$ and therefore maintains the distributions shown by the blue histograms in Figs.~\ref{fig:results_kinematic_pt_rho} and \ref{fig:results_kinematic_eta_phi}.

The lower panels show the reduction in statistical power $R_{\varepsilon}$ (defined in Eq.~\ref{eq:rwgt_statpower}), in bins of each observable after applying the S2 reweighting. 
The EB model retains high statistical power across the full $x_k$ range, with $R_{\varepsilon}$ values close to unity. 
In contrast, the EE$+$ and EE$-$ models exhibit a significant degradation in the statistical power at large $\abs{\eta}$ and high $\rho$ (high pileup). 
This reflects the substantial data-simulation discrepancies in these challenging regions of feature space. 
Importantly this behaviour, while undesired, can be regarded as a feature of the method: 
when carefully propagated through an analysis, the inflated MC statistical uncertainties naturally highlight regions of feature space in which the EM showering is difficult to model.

\subsection{Calibration of the shower shape and isolation features}
\begin{figure}[htb!]
    \centering
    \includegraphics[width=.325\linewidth]{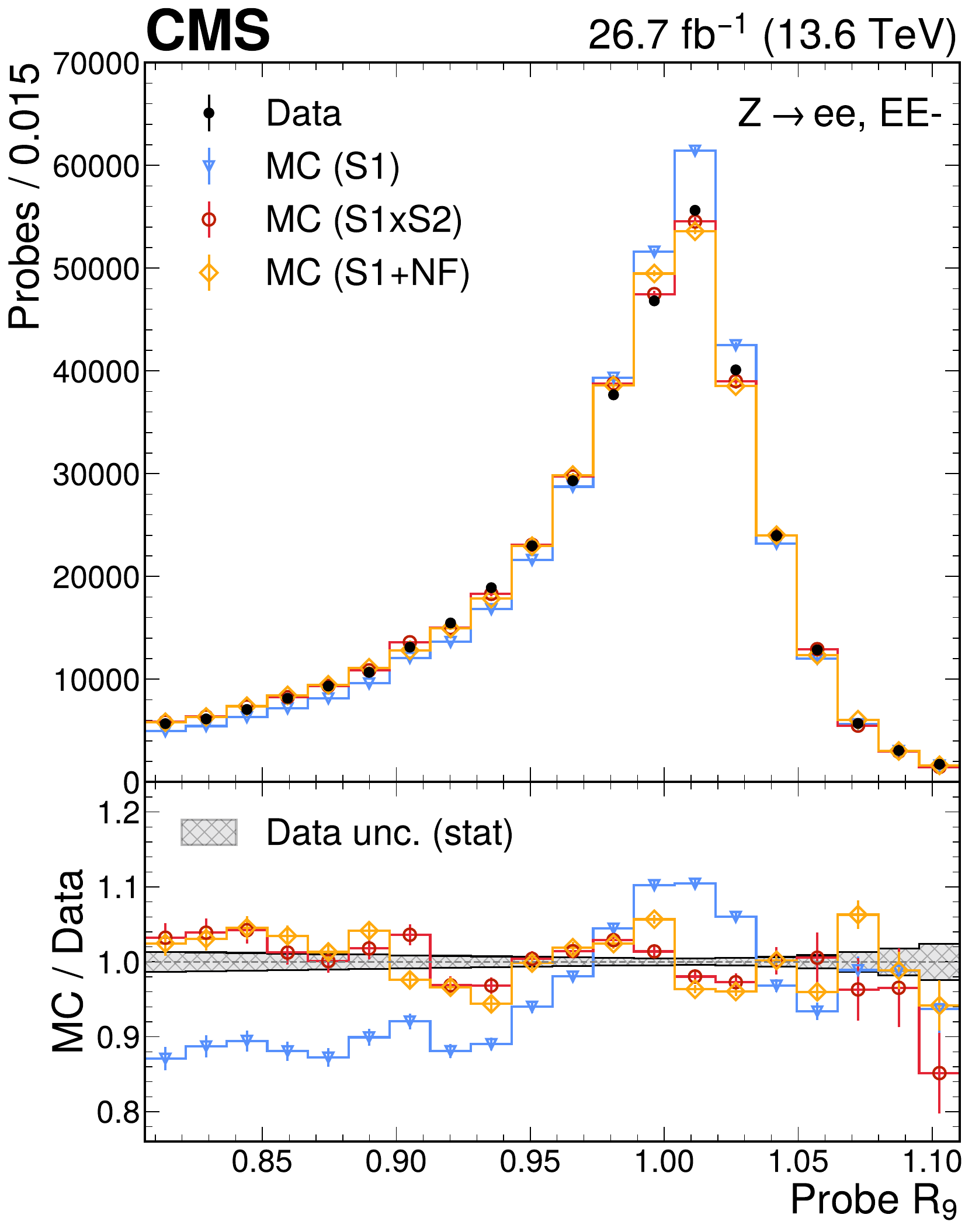}
    \includegraphics[width=.325\linewidth]{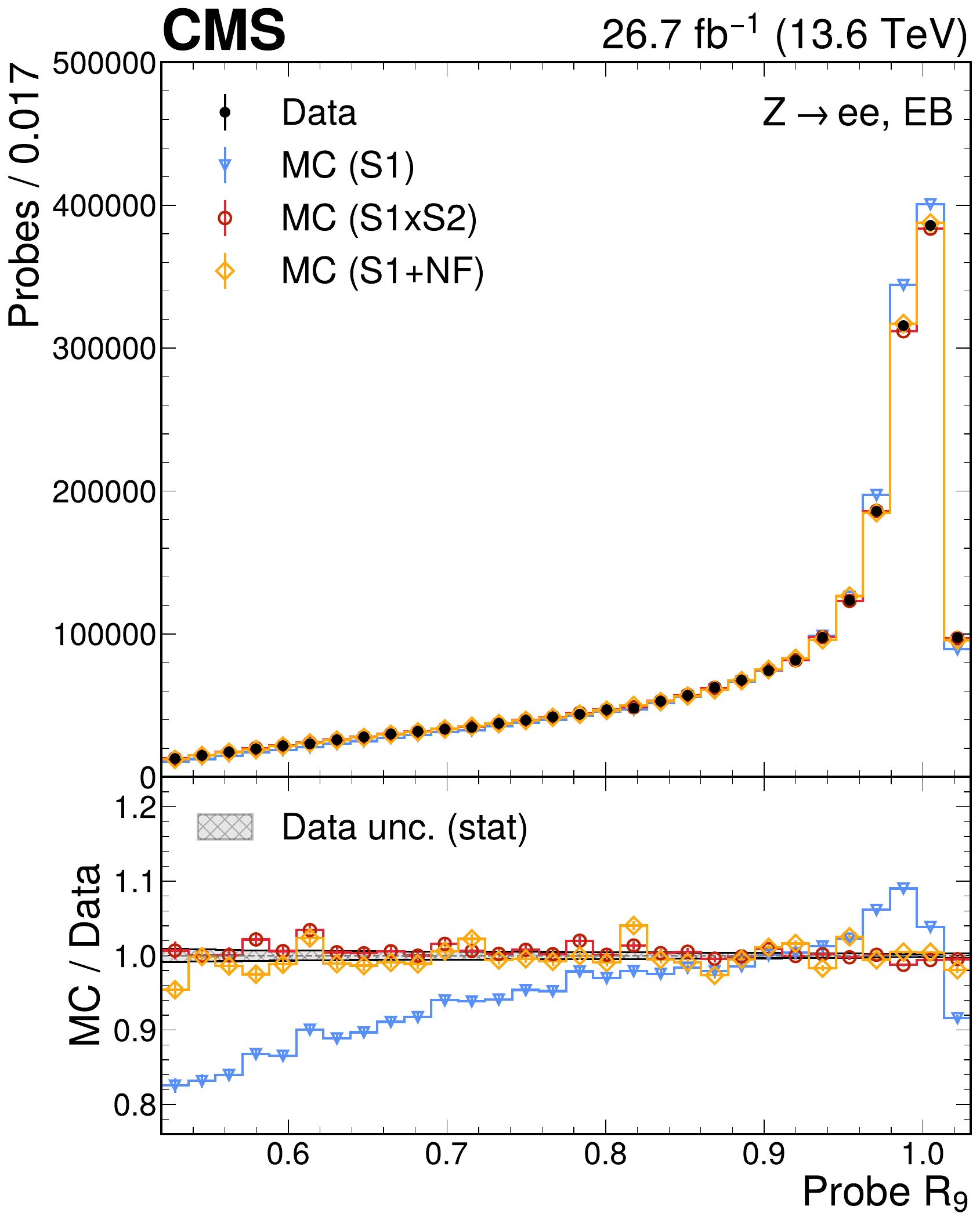}
    \includegraphics[width=.325\linewidth]{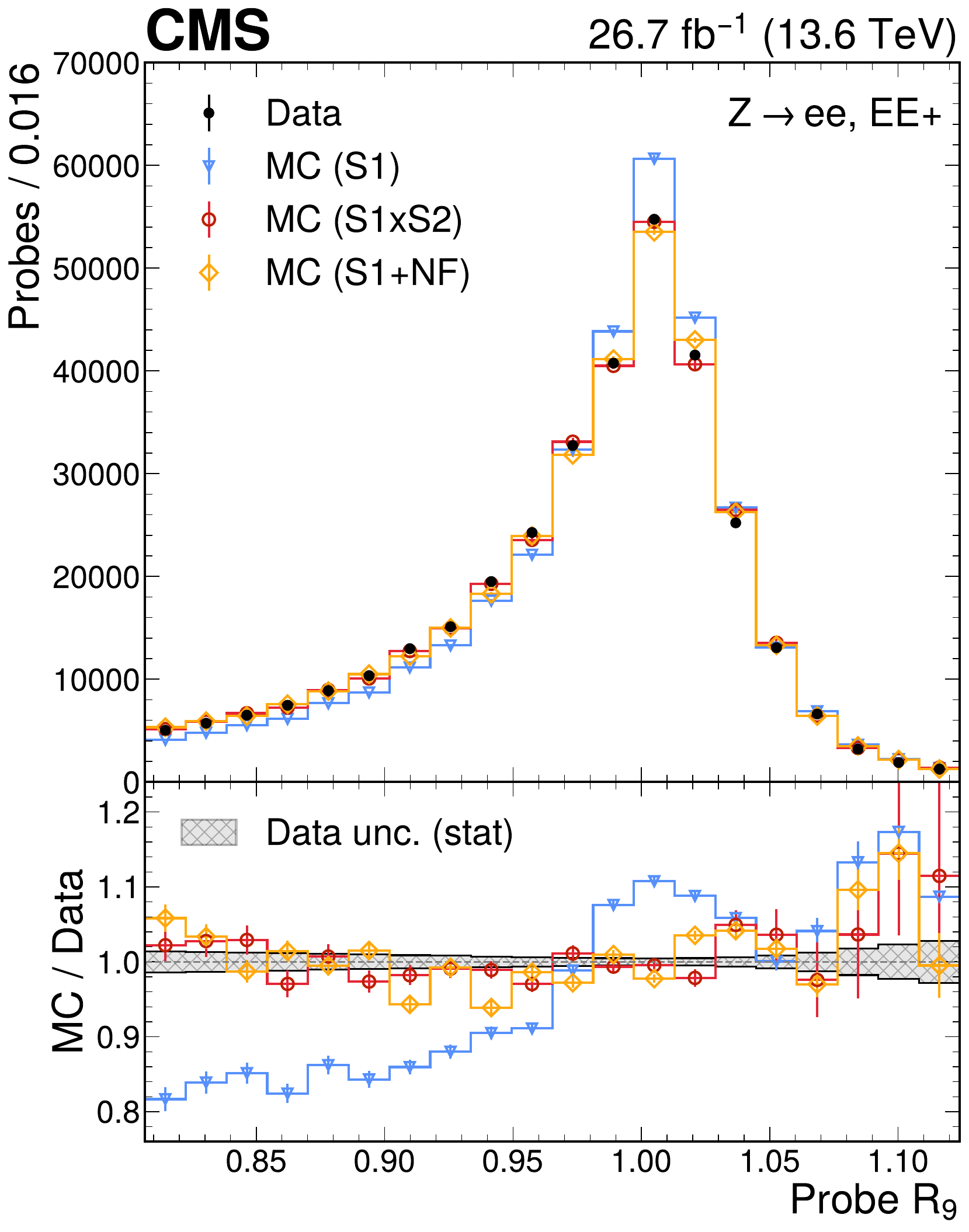}
    \includegraphics[width=.325\linewidth]{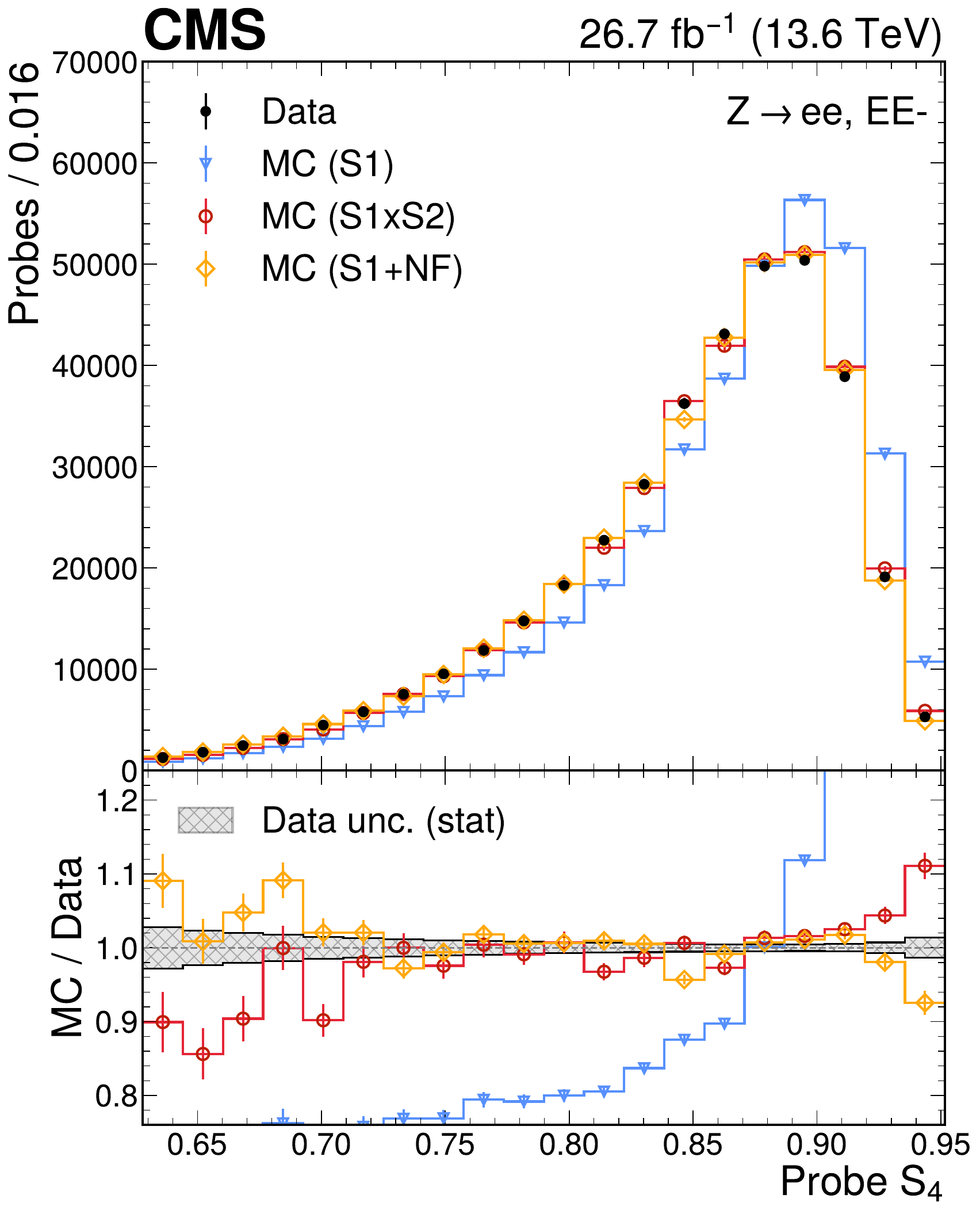}
    \includegraphics[width=.325\linewidth]{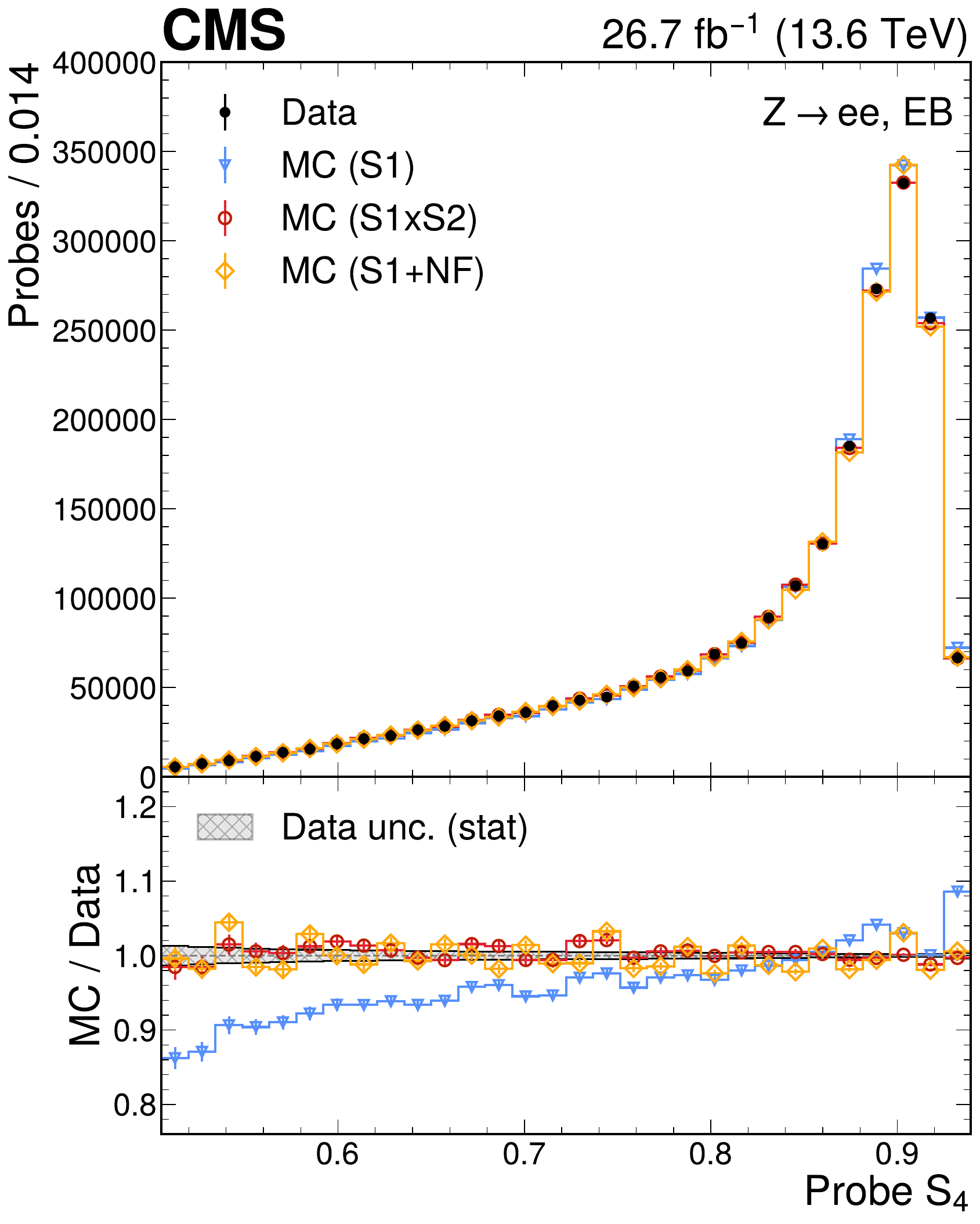}
    \includegraphics[width=.325\linewidth]{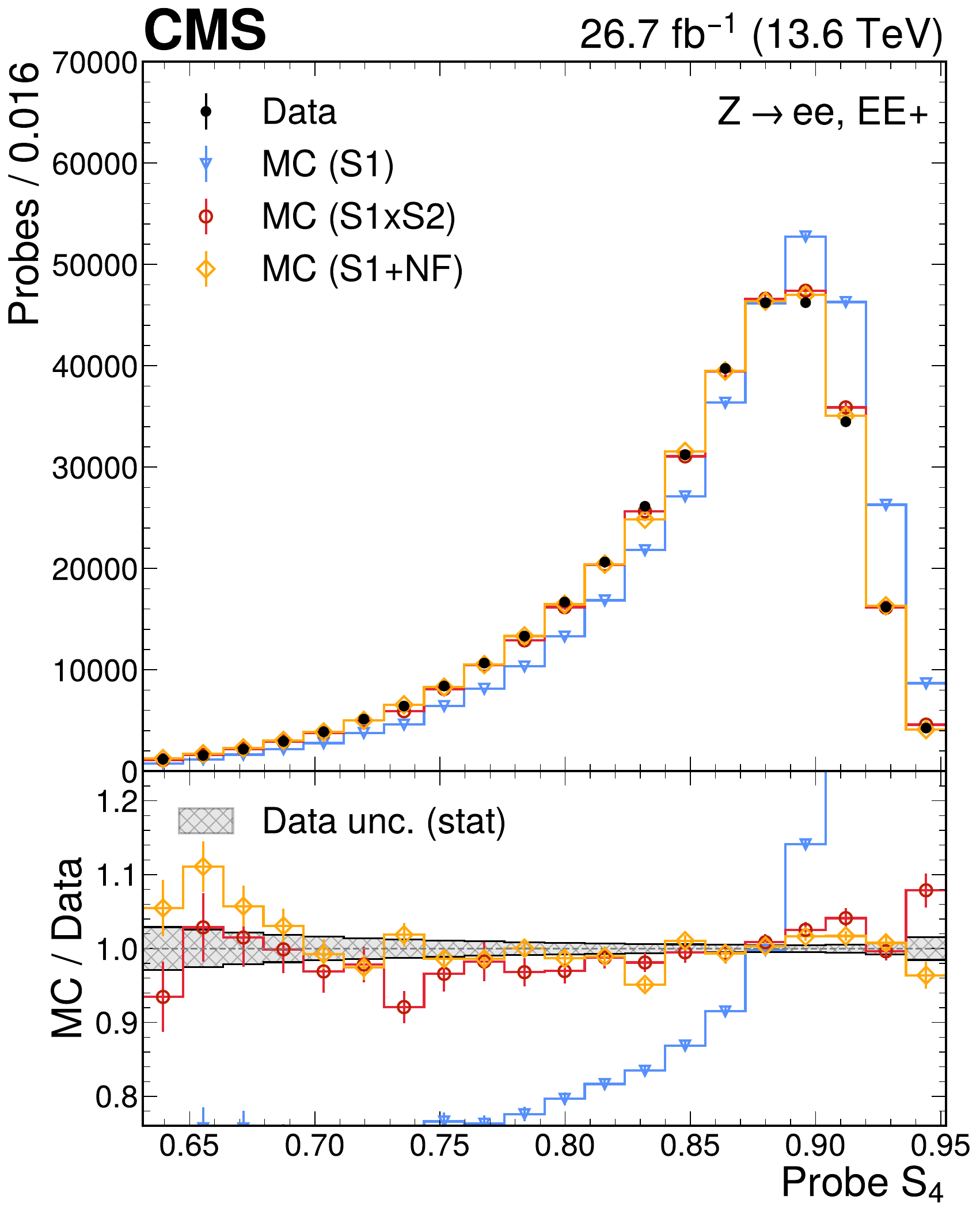}
    \caption{The probe $R_9$ (upper row) and $S_4$ (lower row) distributions in data (black points) and simulation (coloured histograms with points), 
    for the EE$-$ (left), EB (centre), and EE$+$ (right) subdetector regions. 
    The statistical uncertainties in the simulation and data are shown by the error bars on the points.
    The simulation is shown for three scenarios after first normalising the nominal simulation to the data. 
    The blue histograms show the simulation after the S1 reweighting to align the kinematic distributions with data. 
    The red and orange histograms show the calibrated simulation after the S2 reweighting and the flow corrections are applied, respectively. 
    The lower panels show the ratio of simulation to data, where the statistical uncertainty in the data is indicated by the grey hatched band.}
    \label{fig:results_shower_0}
\end{figure}

\begin{figure}[htb!]
    \centering
    \includegraphics[width=.325\linewidth]{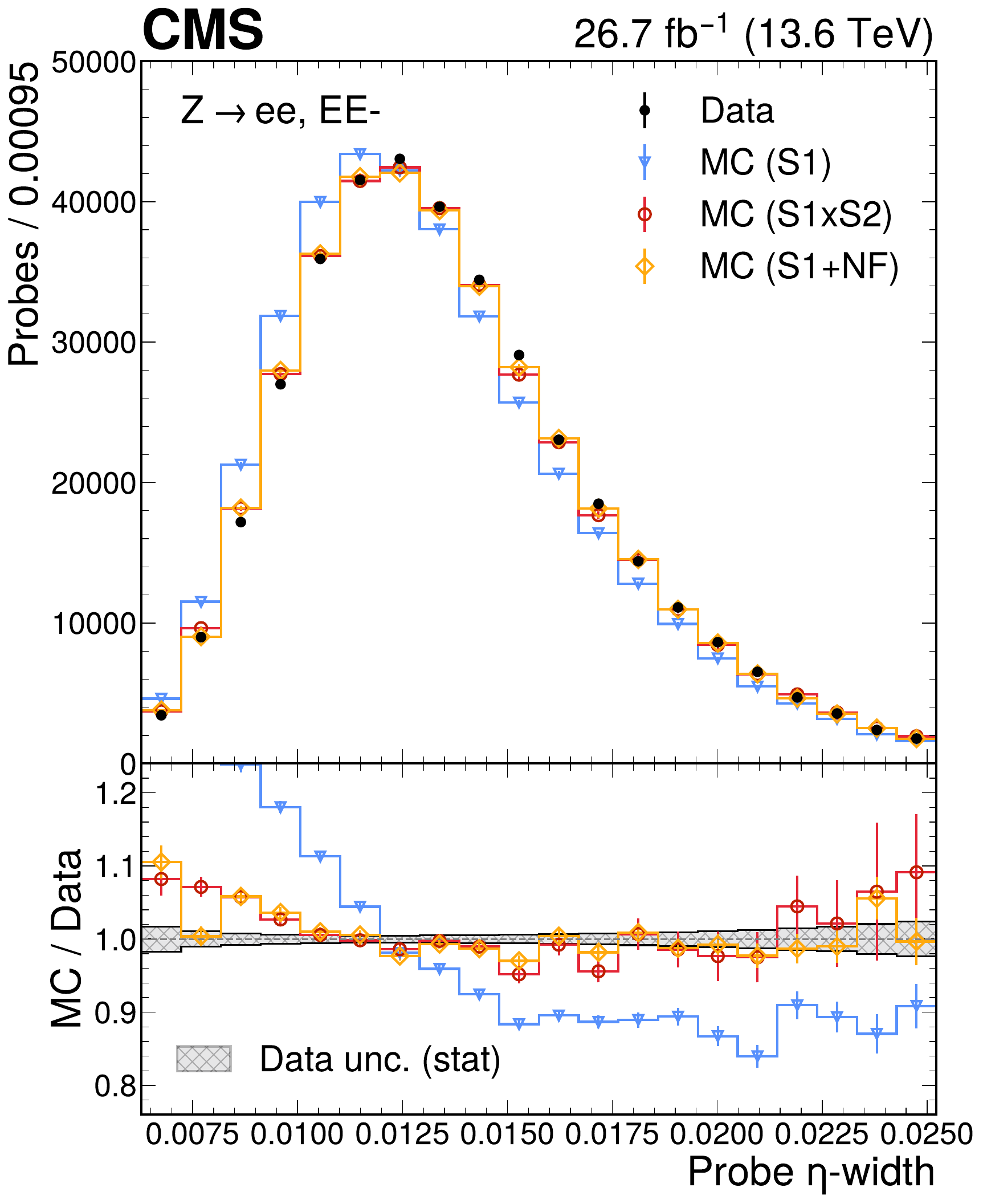}
    \includegraphics[width=.325\linewidth]{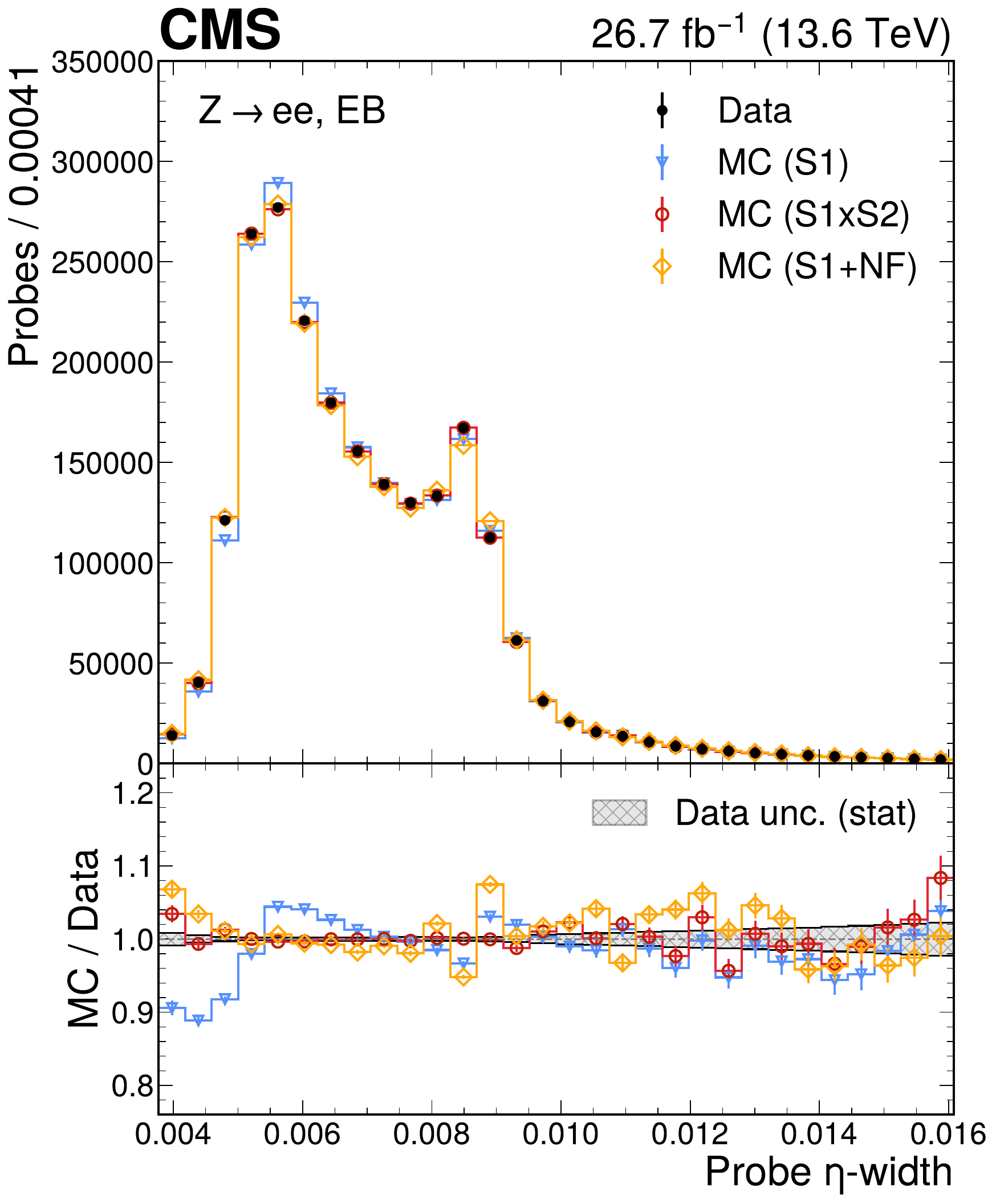}
    \includegraphics[width=.325\linewidth]{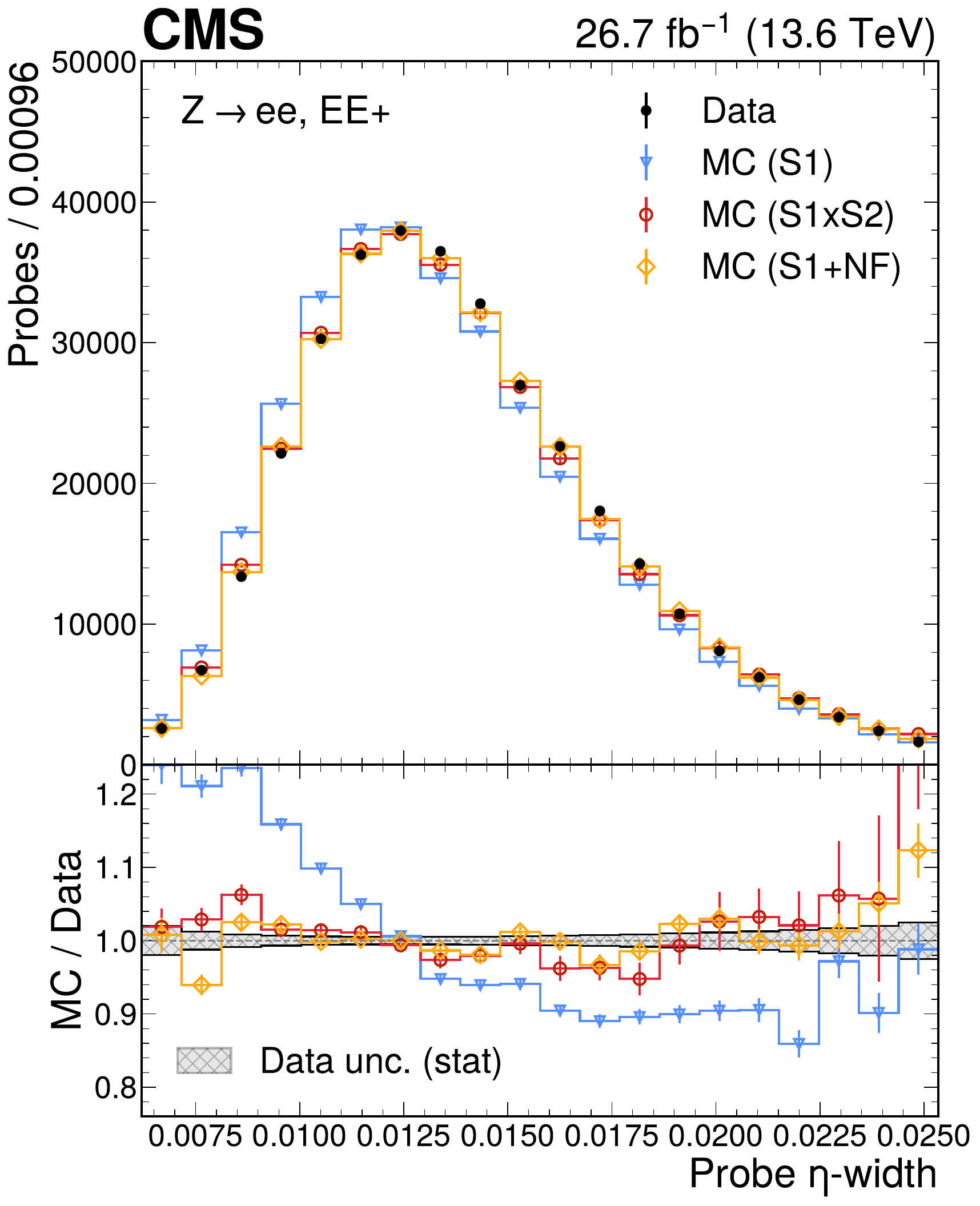}
    \includegraphics[width=.325\linewidth]{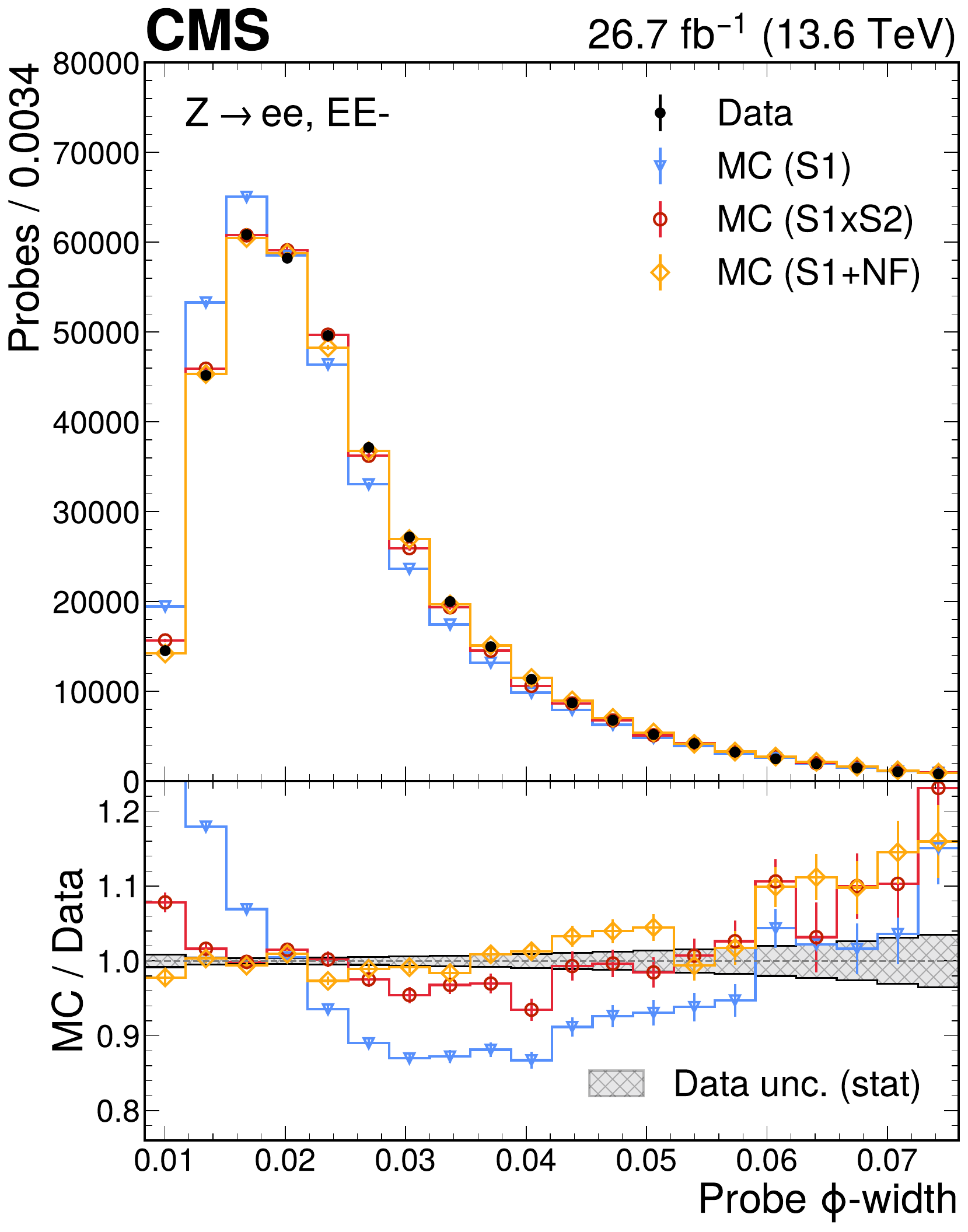}
    \includegraphics[width=.325\linewidth]{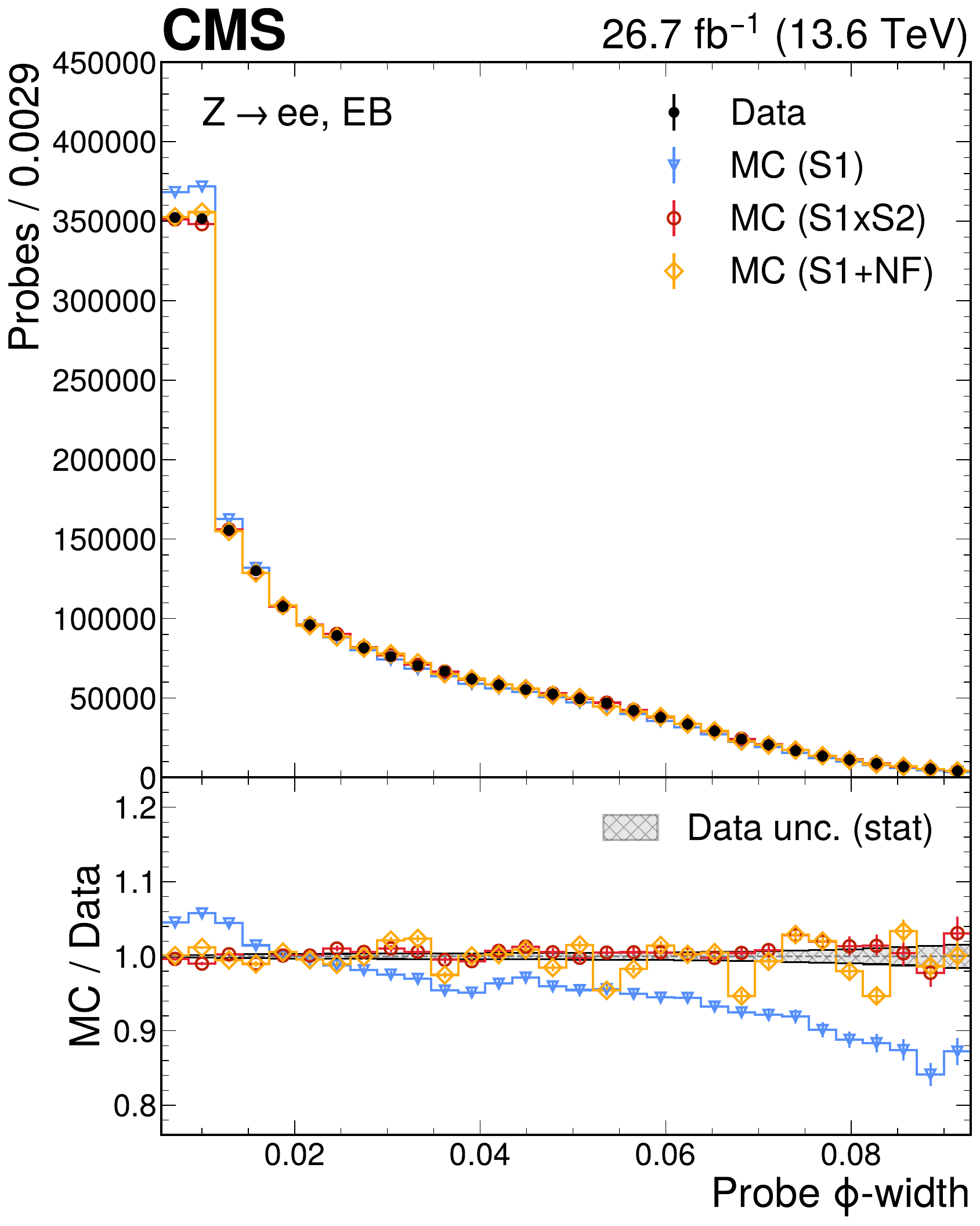}
    \includegraphics[width=.325\linewidth]{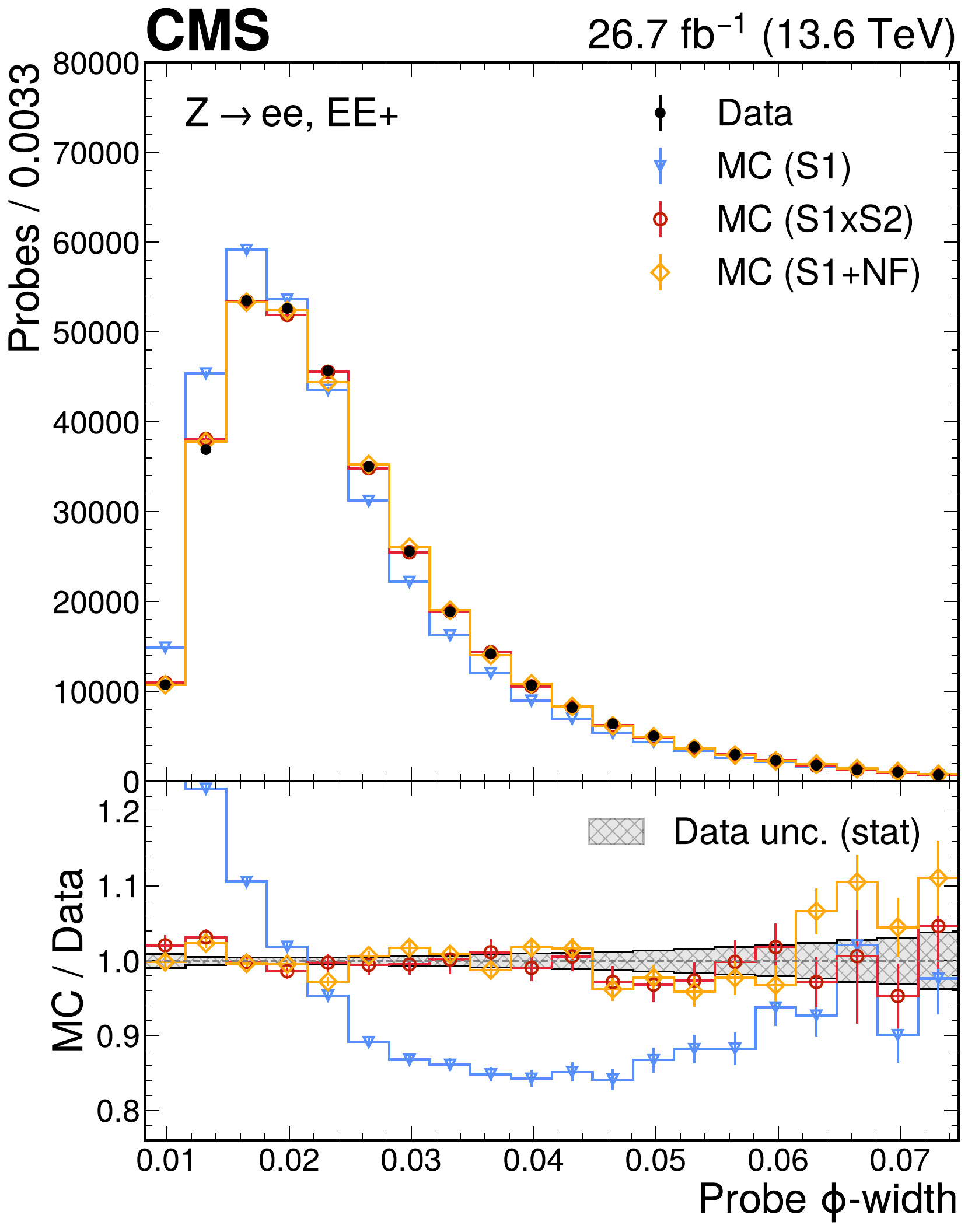}
    \includegraphics[width=.325\linewidth]{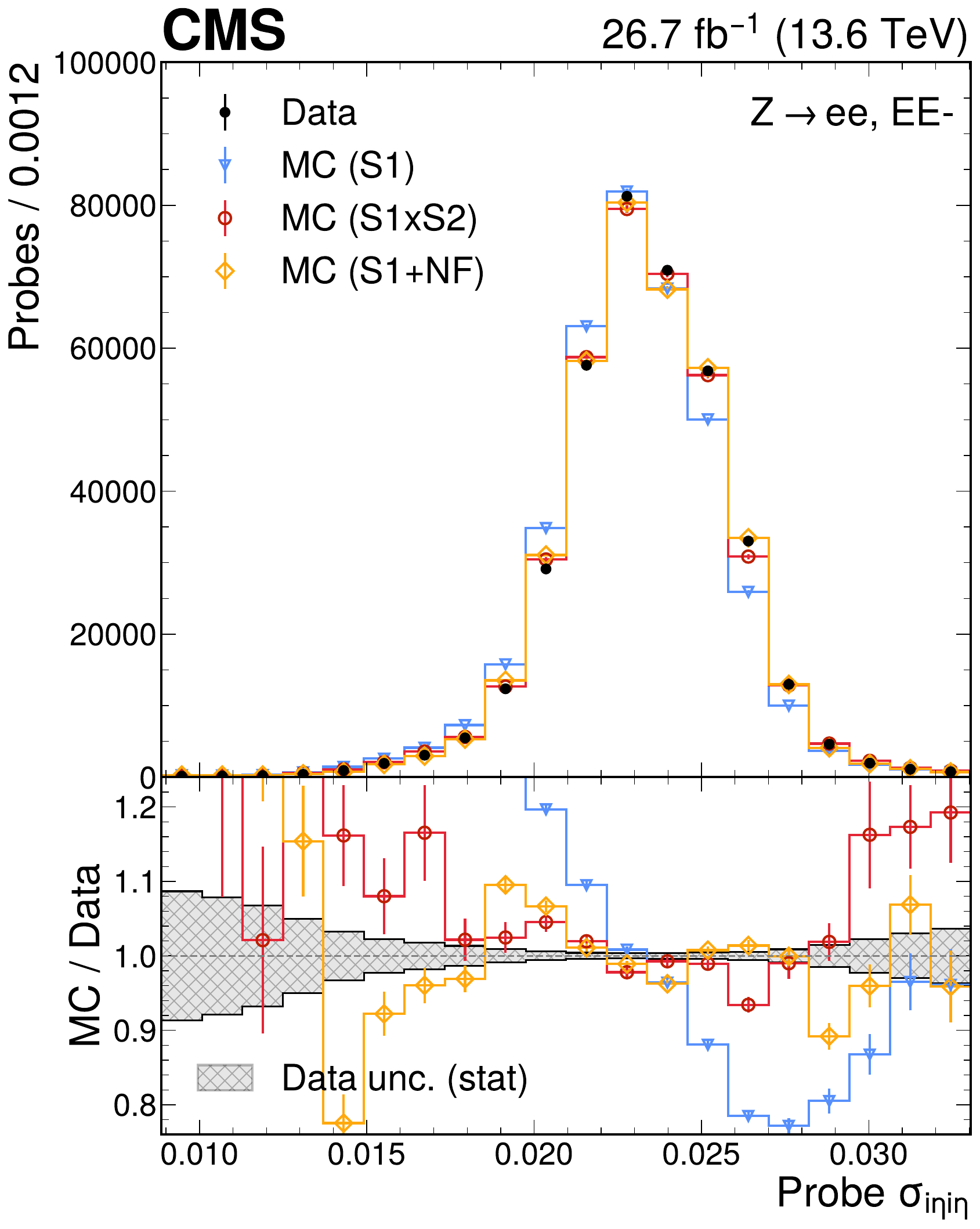}
    \includegraphics[width=.325\linewidth]{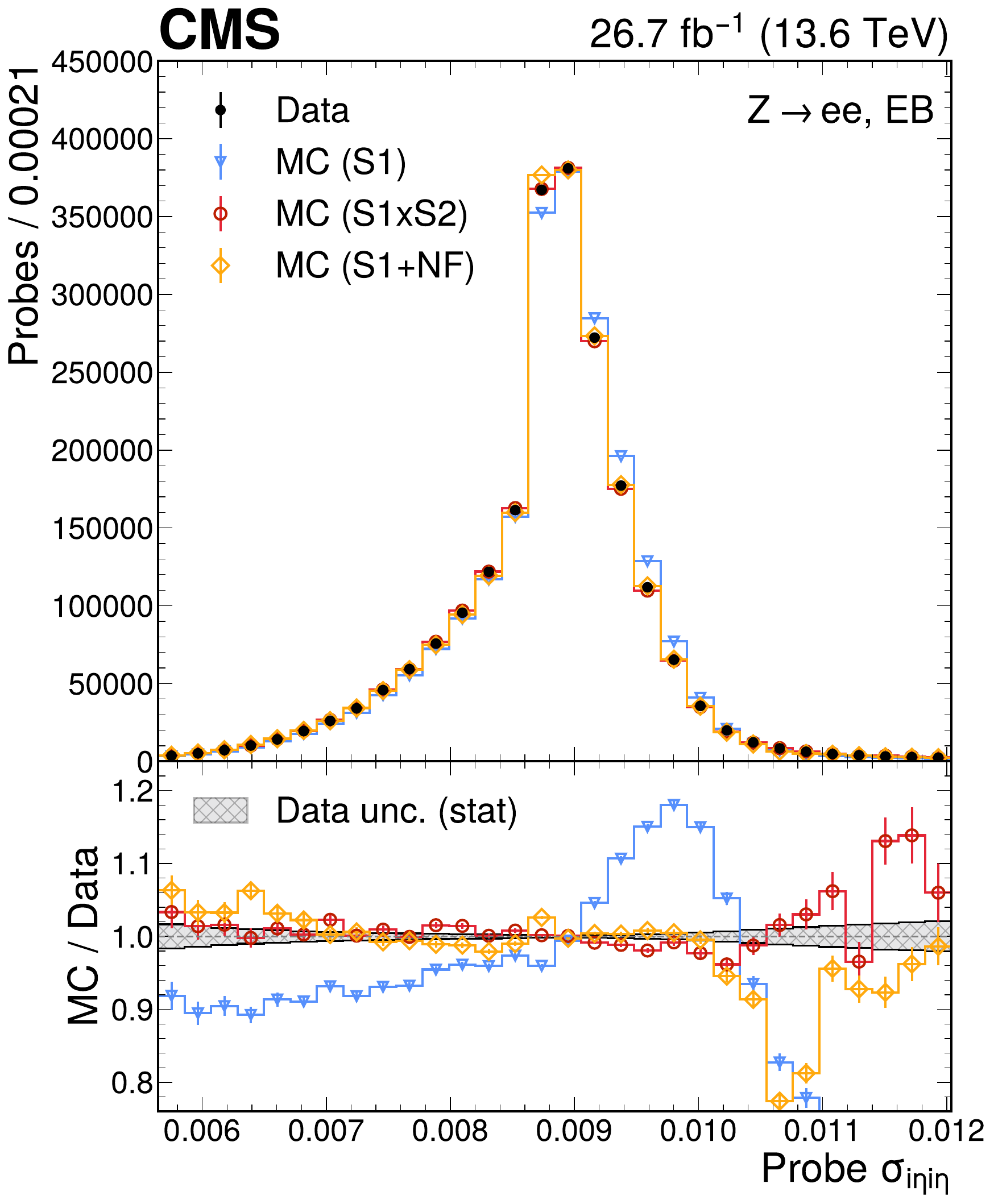}
    \includegraphics[width=.325\linewidth]{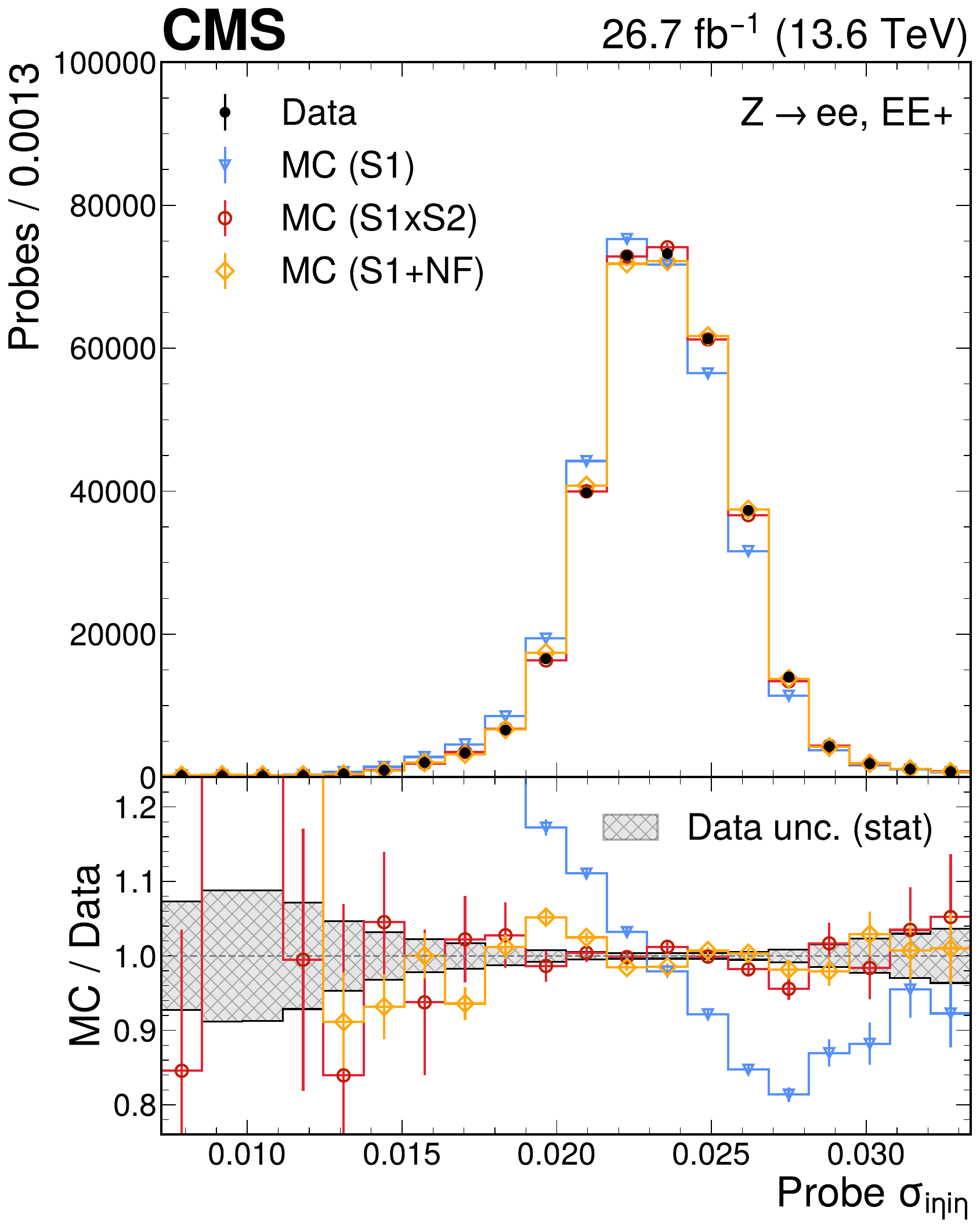}
    \caption{The probe $\eta$-width (upper row), $\phi$-width (middle row), and $\sigma_{i\eta i\eta}$ (lower row) distributions in data and simulation,
    for the EE$-$ (left), EB (centre), and EE$+$ (right) subdetector regions. 
    The plotting conventions are the same as in Fig.~\ref{fig:results_shower_0}.}     
    \label{fig:results_shower_1}
\end{figure}

\begin{figure}[htb!]
    \centering
    \includegraphics[width=.325\linewidth]{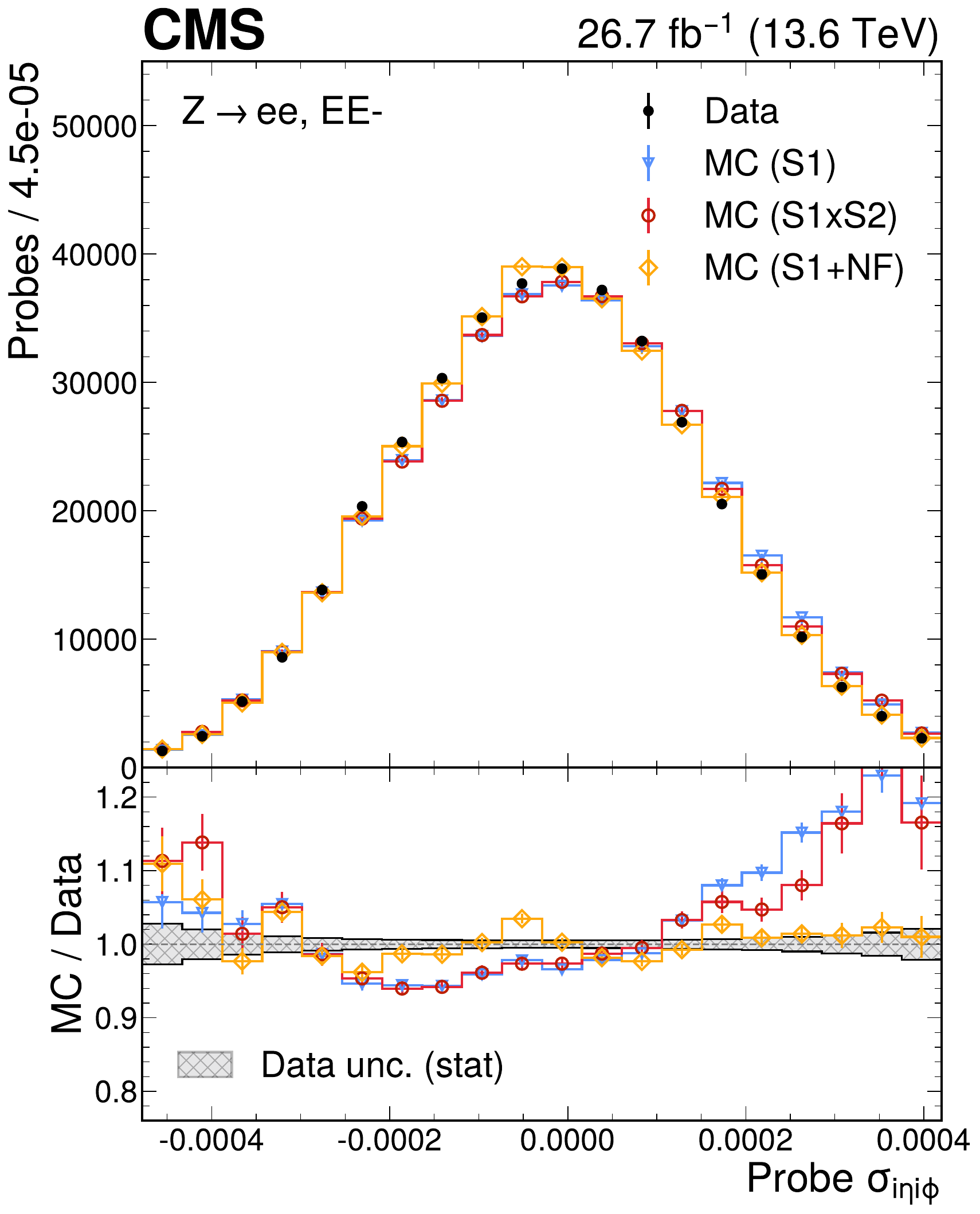}
    \includegraphics[width=.325\linewidth]{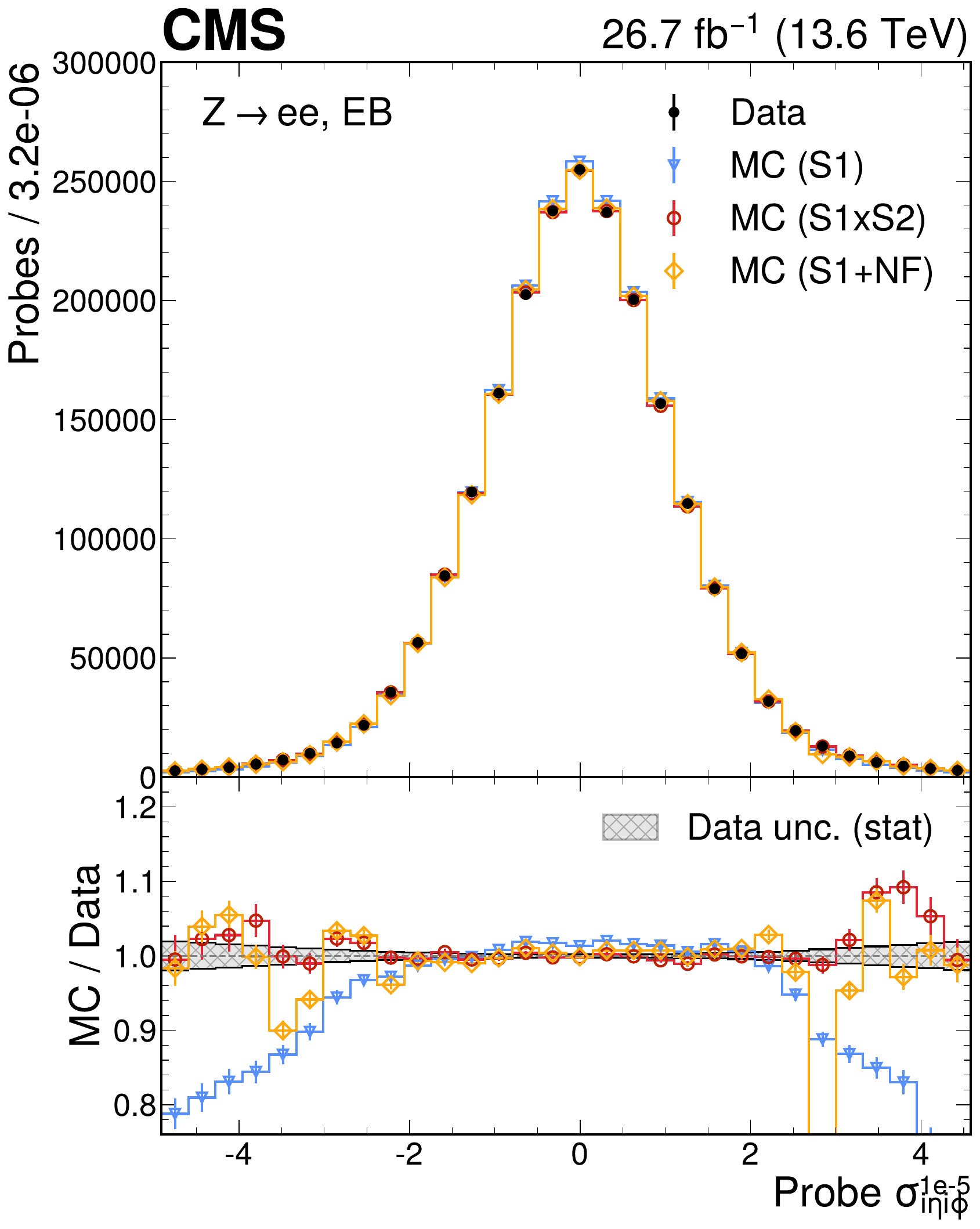}
    \includegraphics[width=.325\linewidth]{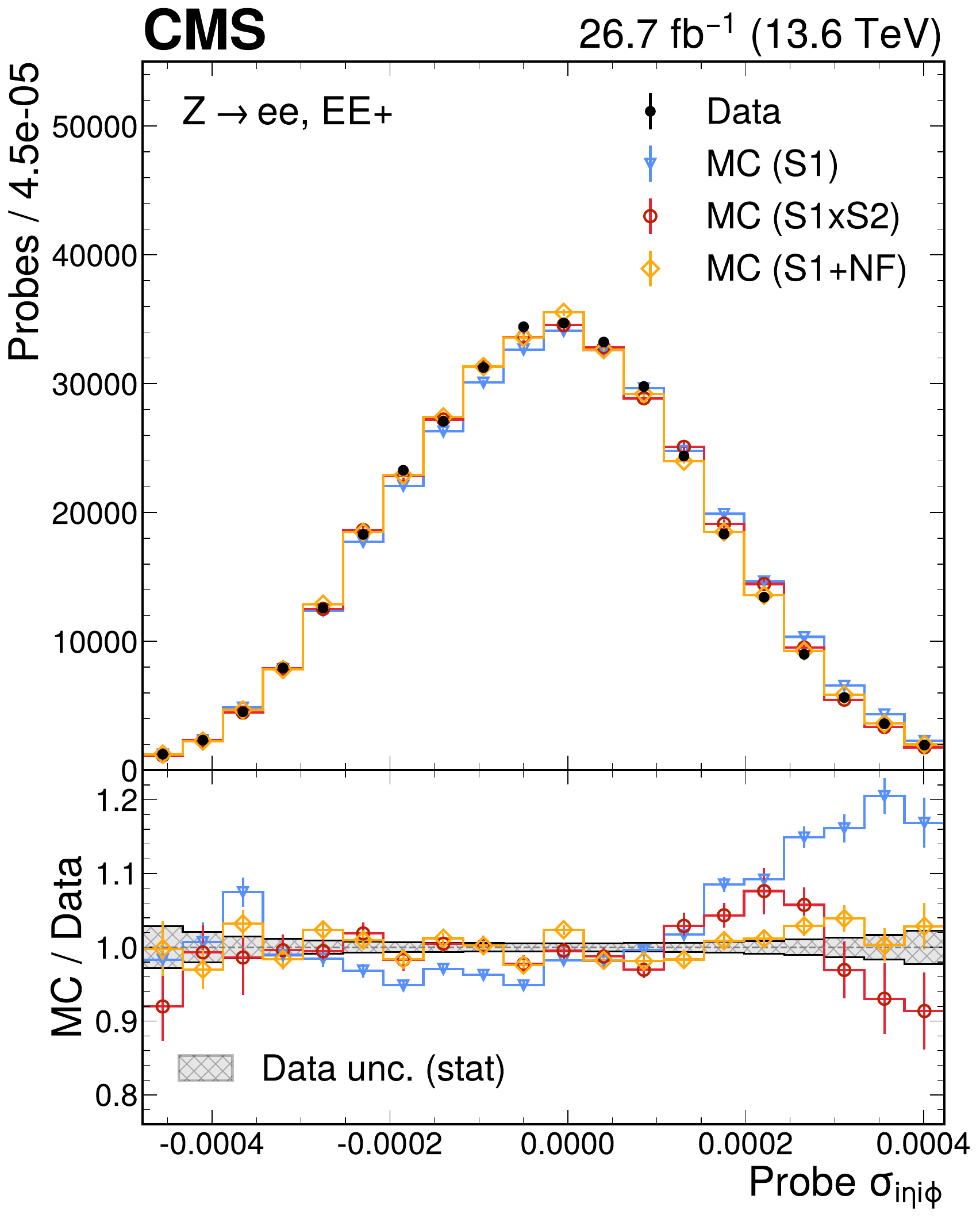}
    \includegraphics[width=.325\linewidth]{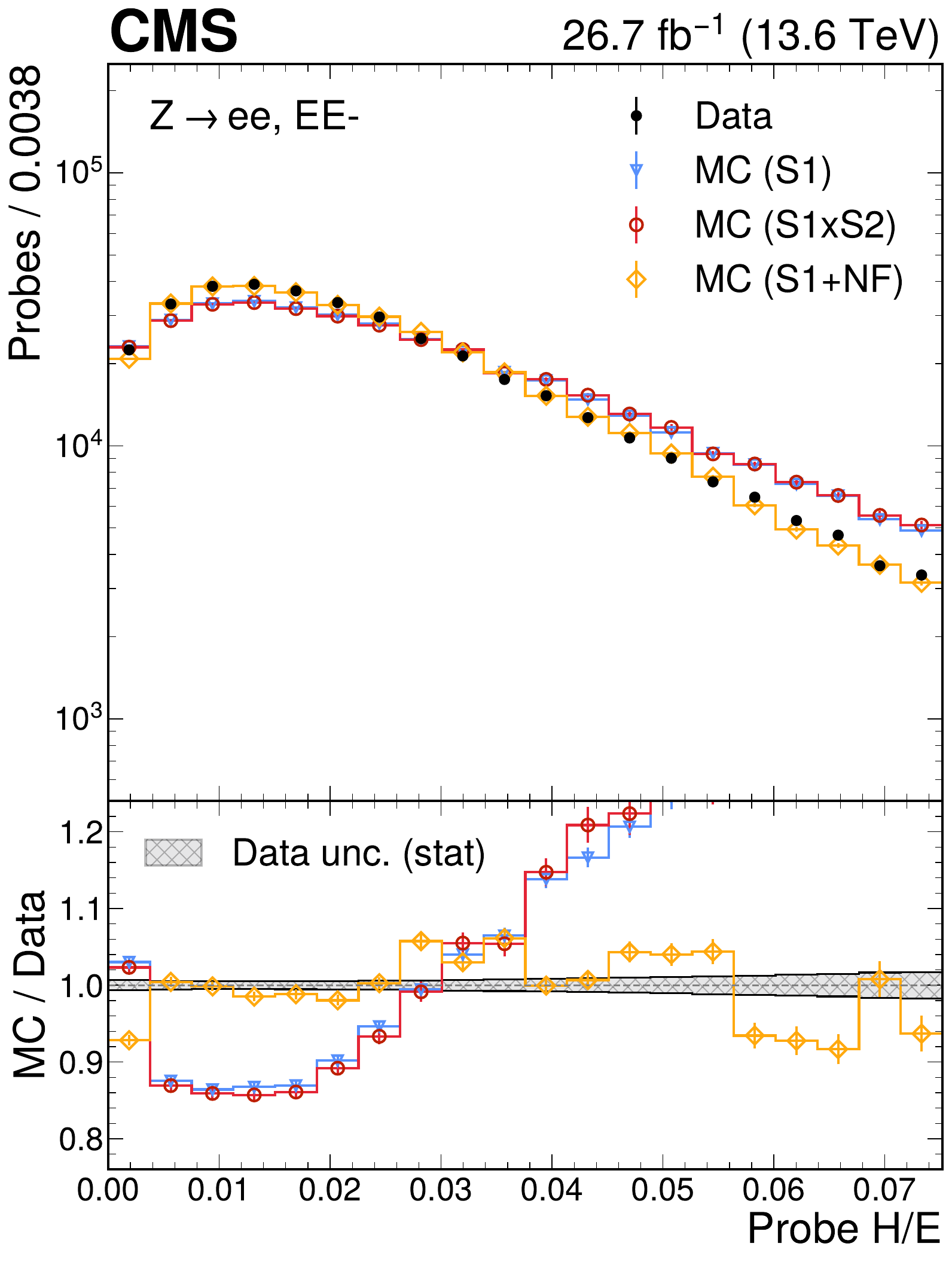}
    \includegraphics[width=.325\linewidth]{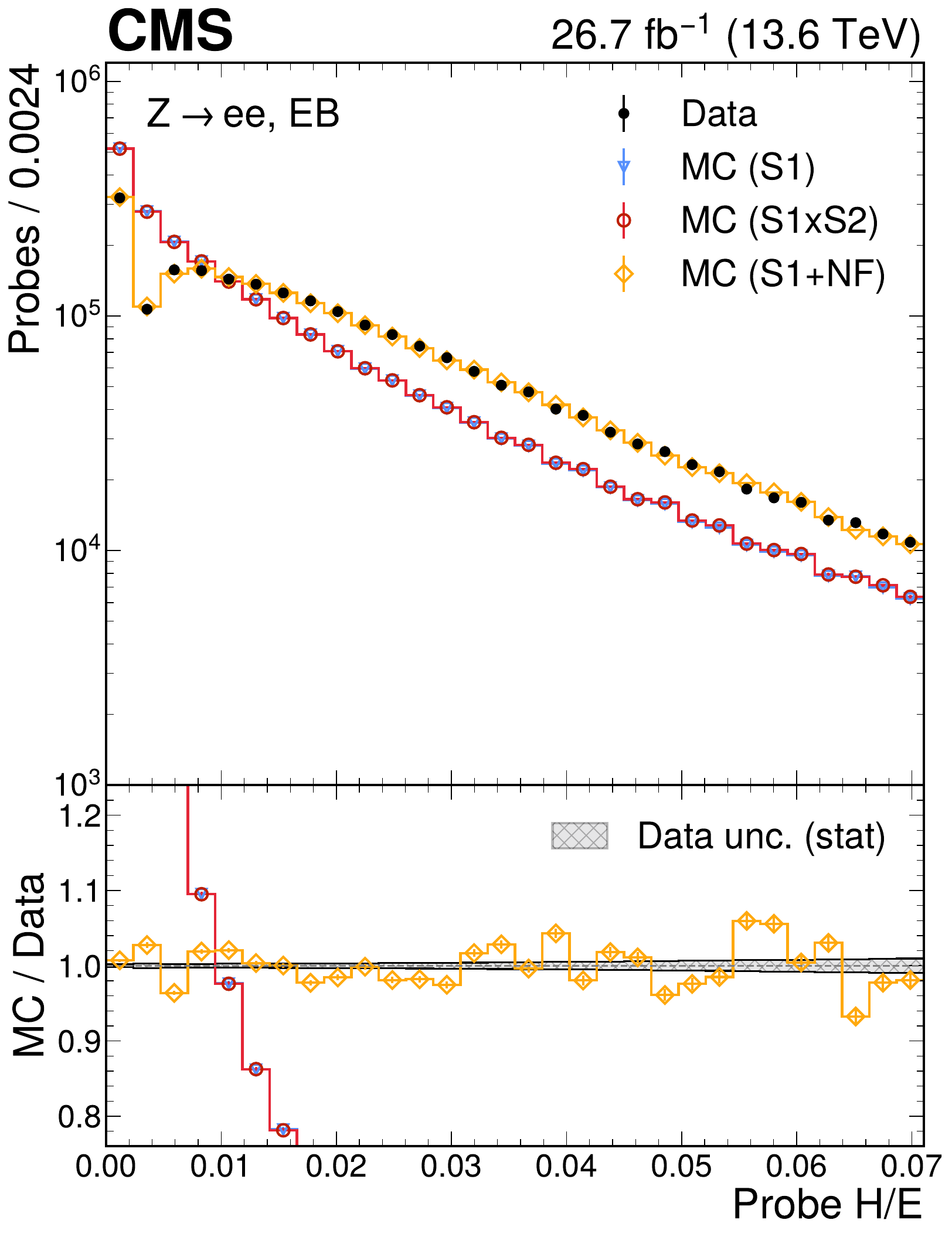}
    \includegraphics[width=.325\linewidth]{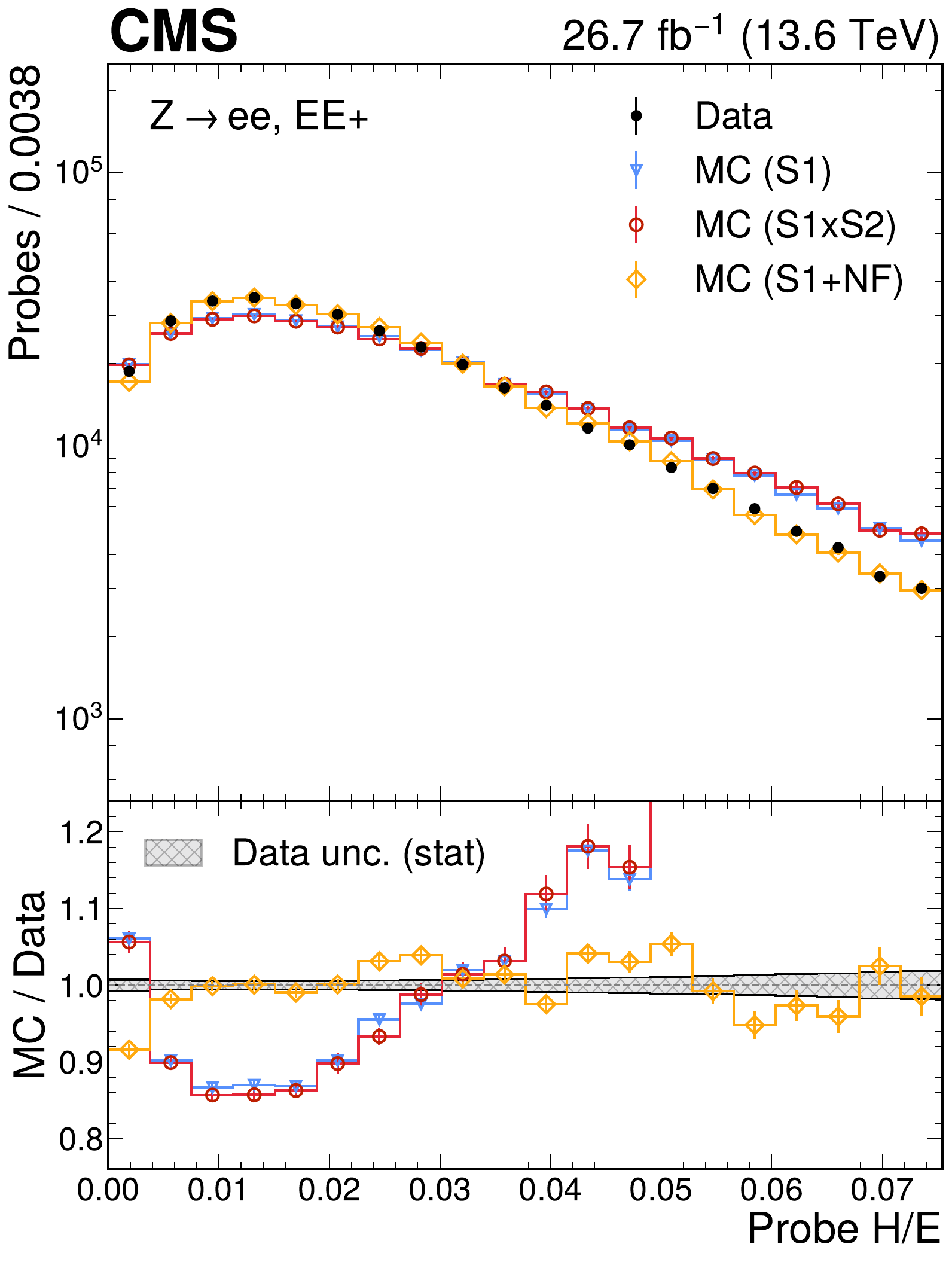}
    \includegraphics[width=.325\linewidth]{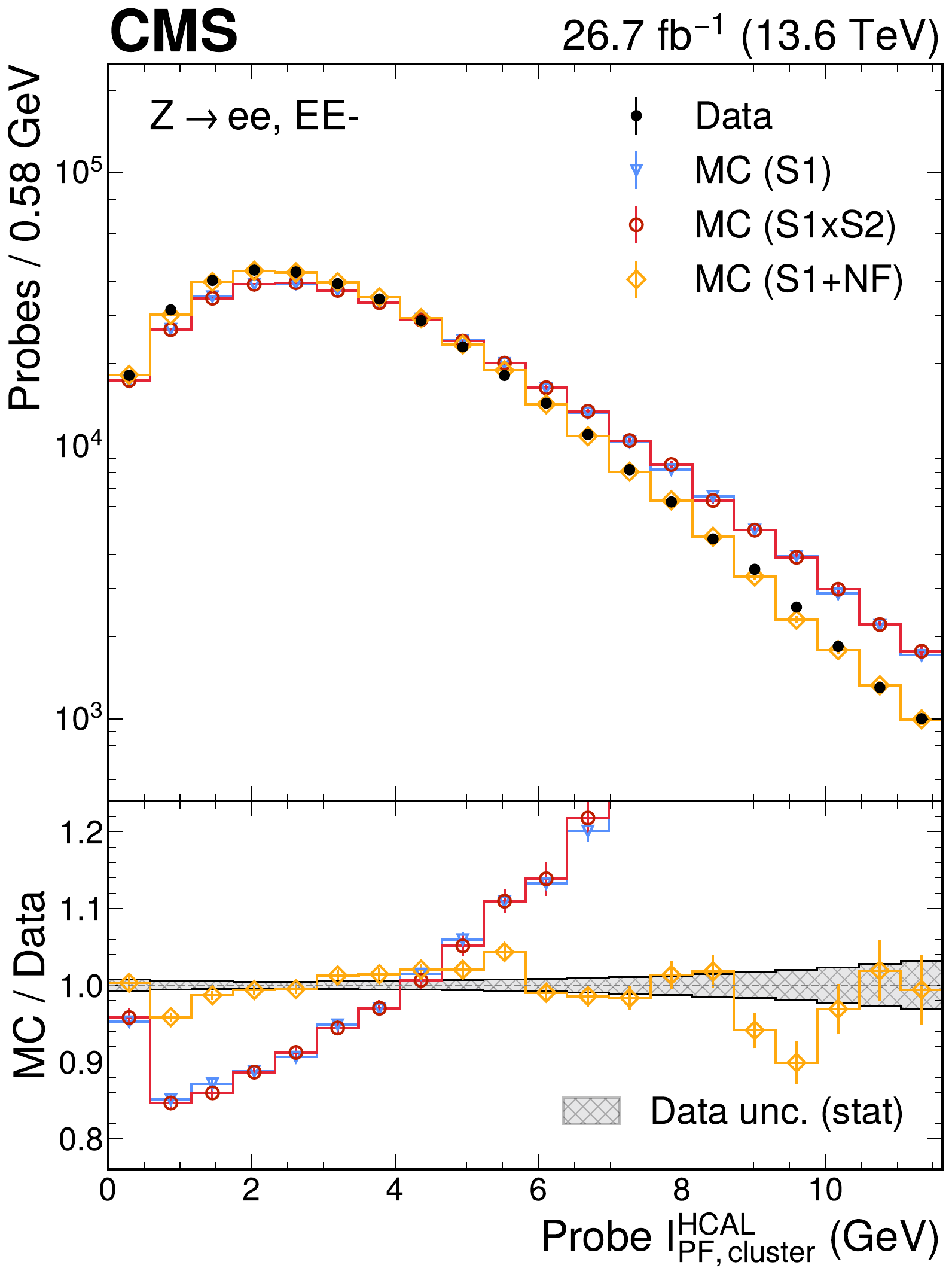}
    \includegraphics[width=.325\linewidth]{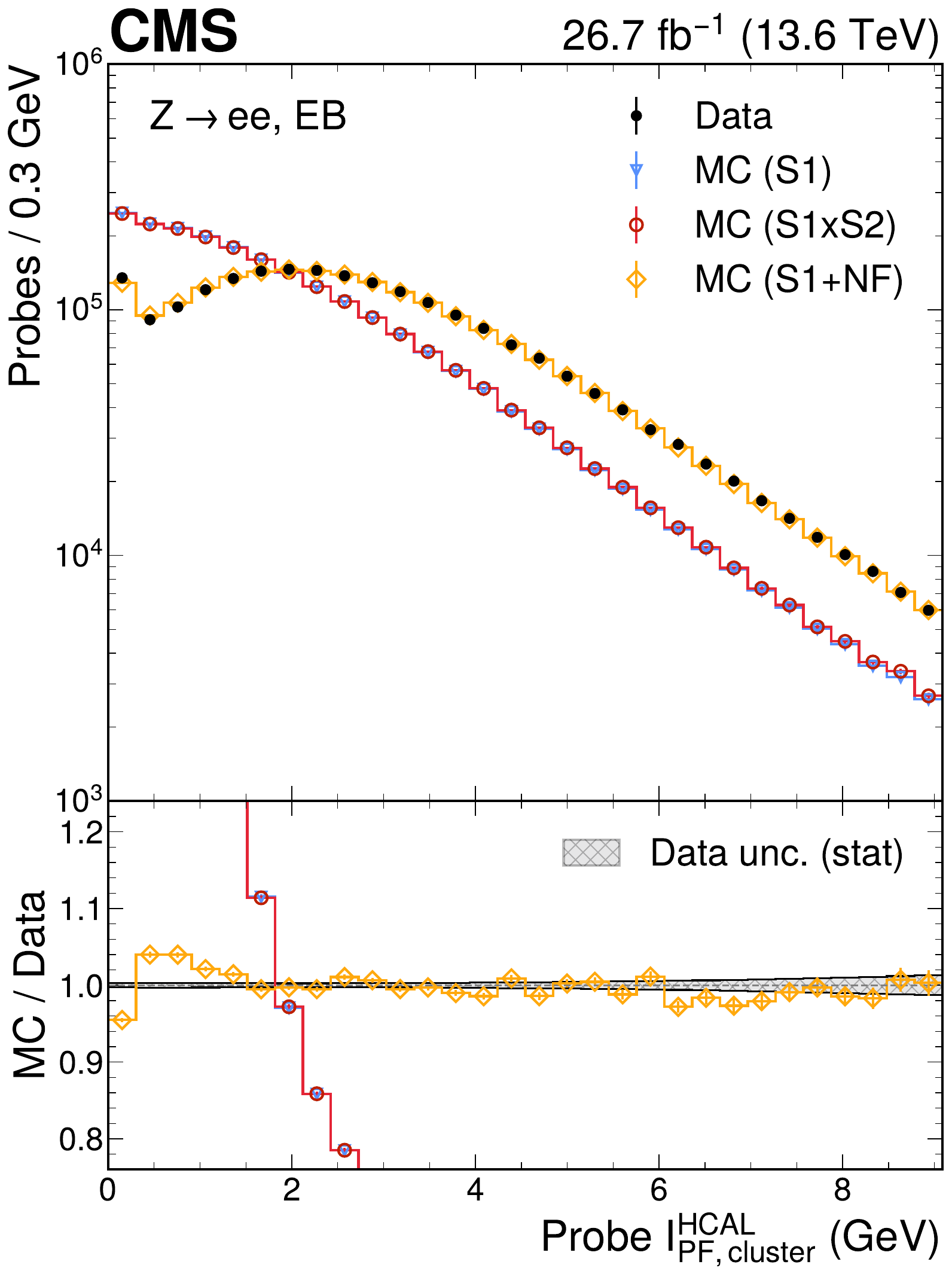}
    \includegraphics[width=.325\linewidth]{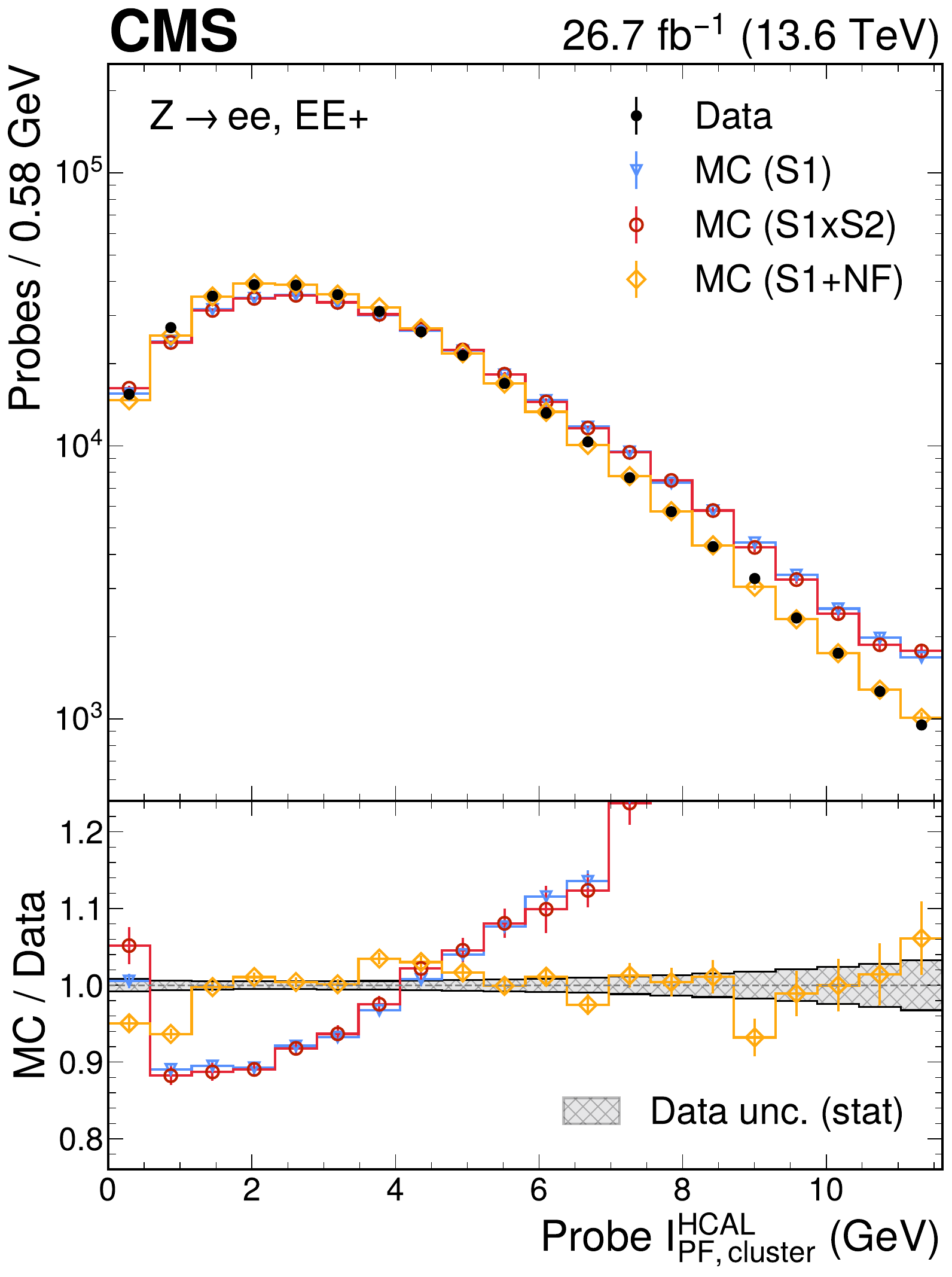}
    \caption{The probe $\sigma_{i\eta i\phi}$ (upper row), H/E (middle row), and $\mathcal{I}^{\mathrm{HCAL}}_{\text{PF, cluster}}$ (lower row) distributions in data and simulation,
    for the EE$-$ (left), EB (centre), and EE$+$ (right) subdetector regions. 
    The plotting conventions are the same as in Fig.~\ref{fig:results_shower_0}.
    The H/E and $\mathcal{I}^{\mathrm{HCAL}}_{\text{PF, cluster}}$ features were not included in the S2 reweighting method training, and therefore the S2 reweighting has a negligible effect on the distributions.}
    \label{fig:results_shower_2}
\end{figure}

\begin{figure}[htb!]
    \centering
    \includegraphics[width=.325\linewidth]{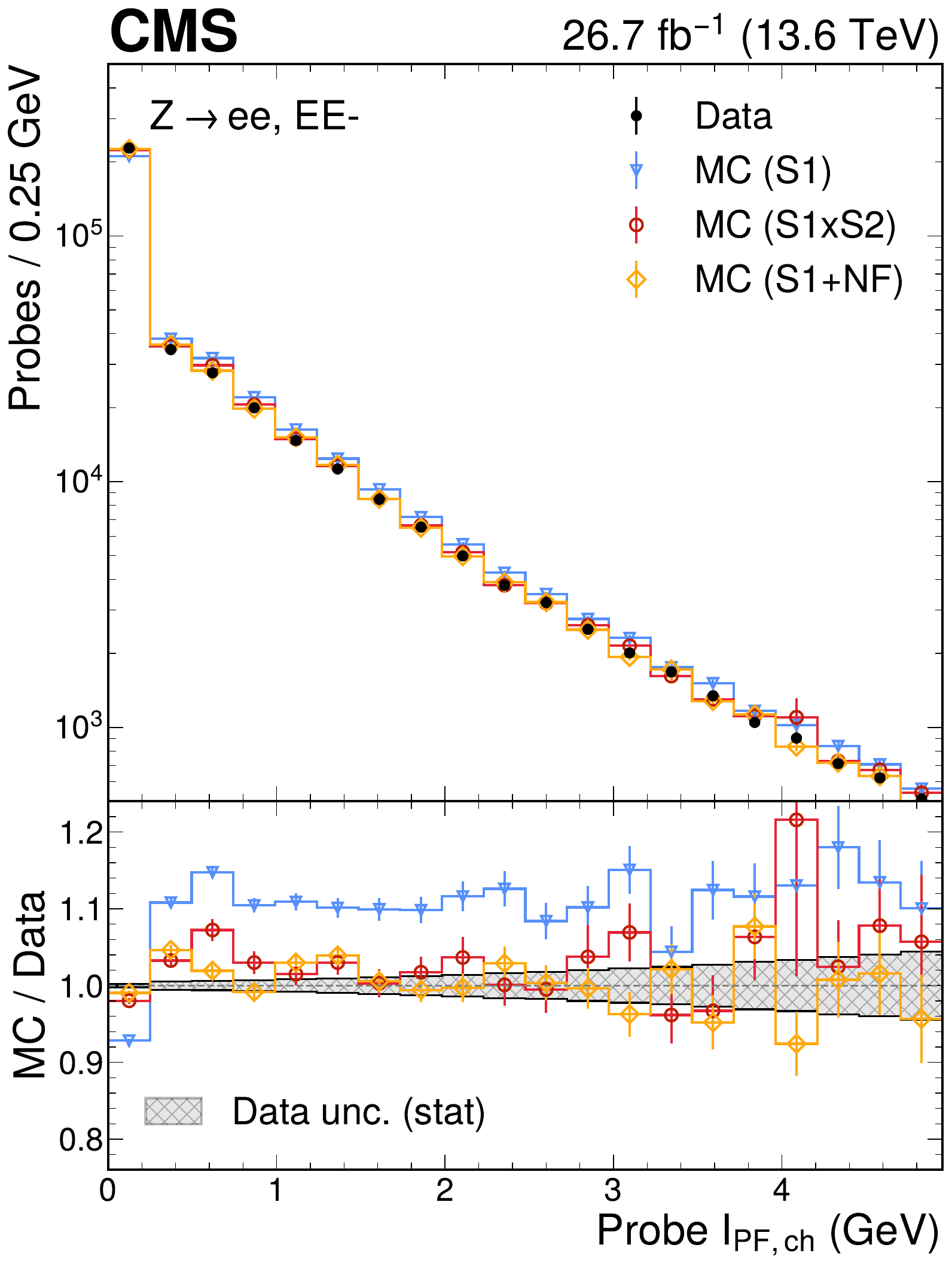}
    \includegraphics[width=.325\linewidth]{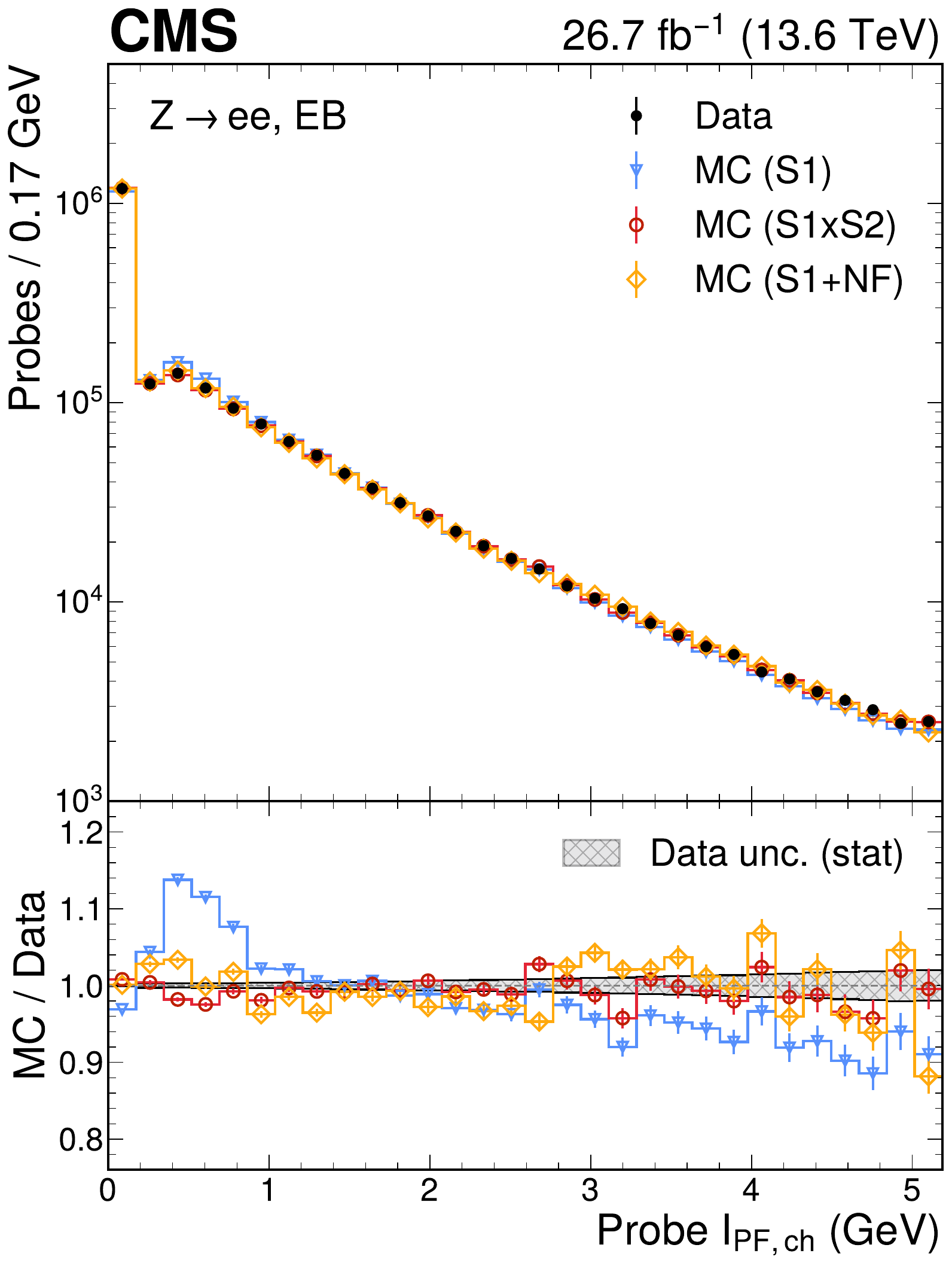}
    \includegraphics[width=.325\linewidth]{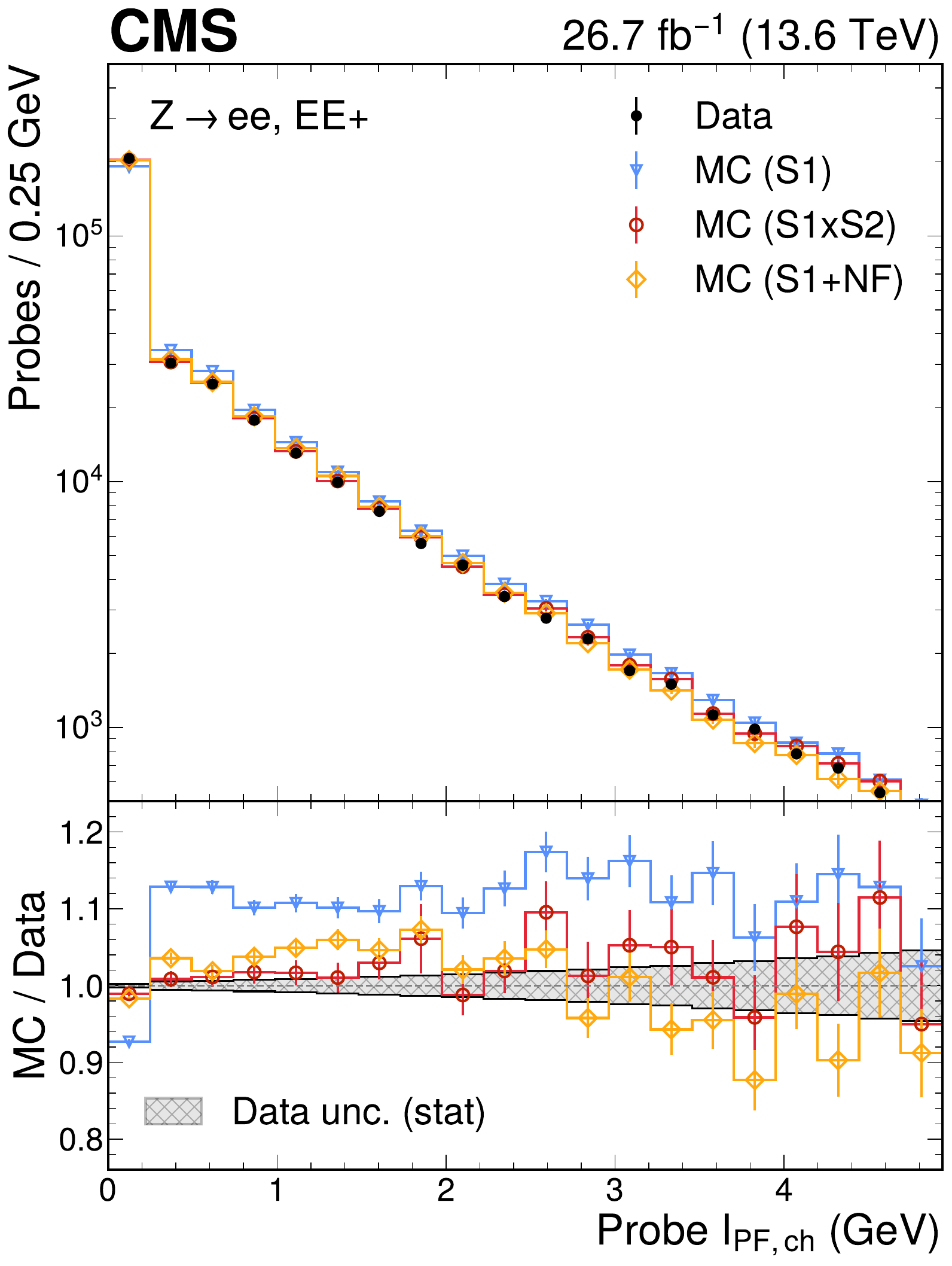}
    \includegraphics[width=.325\linewidth]{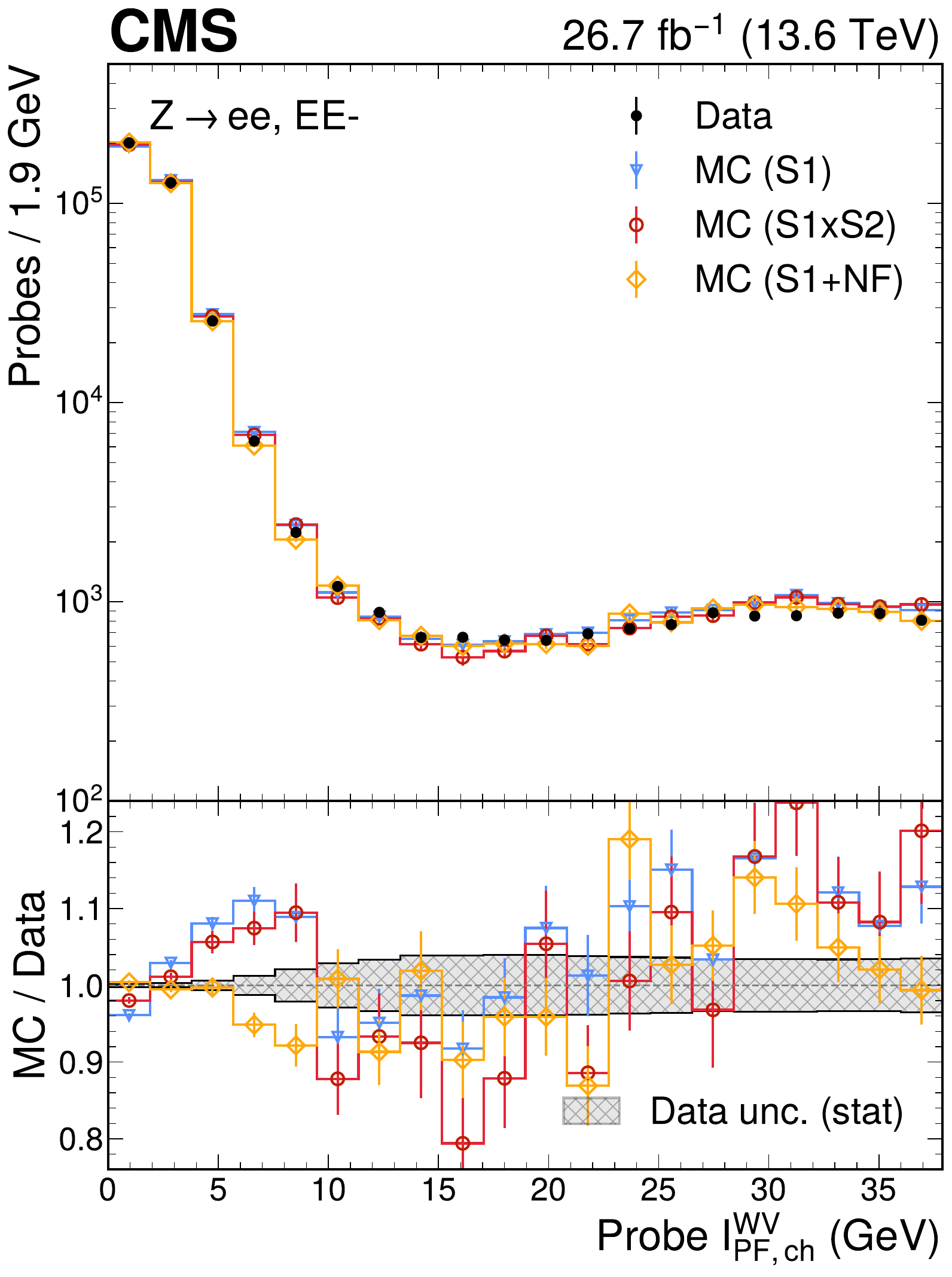}
    \includegraphics[width=.325\linewidth]{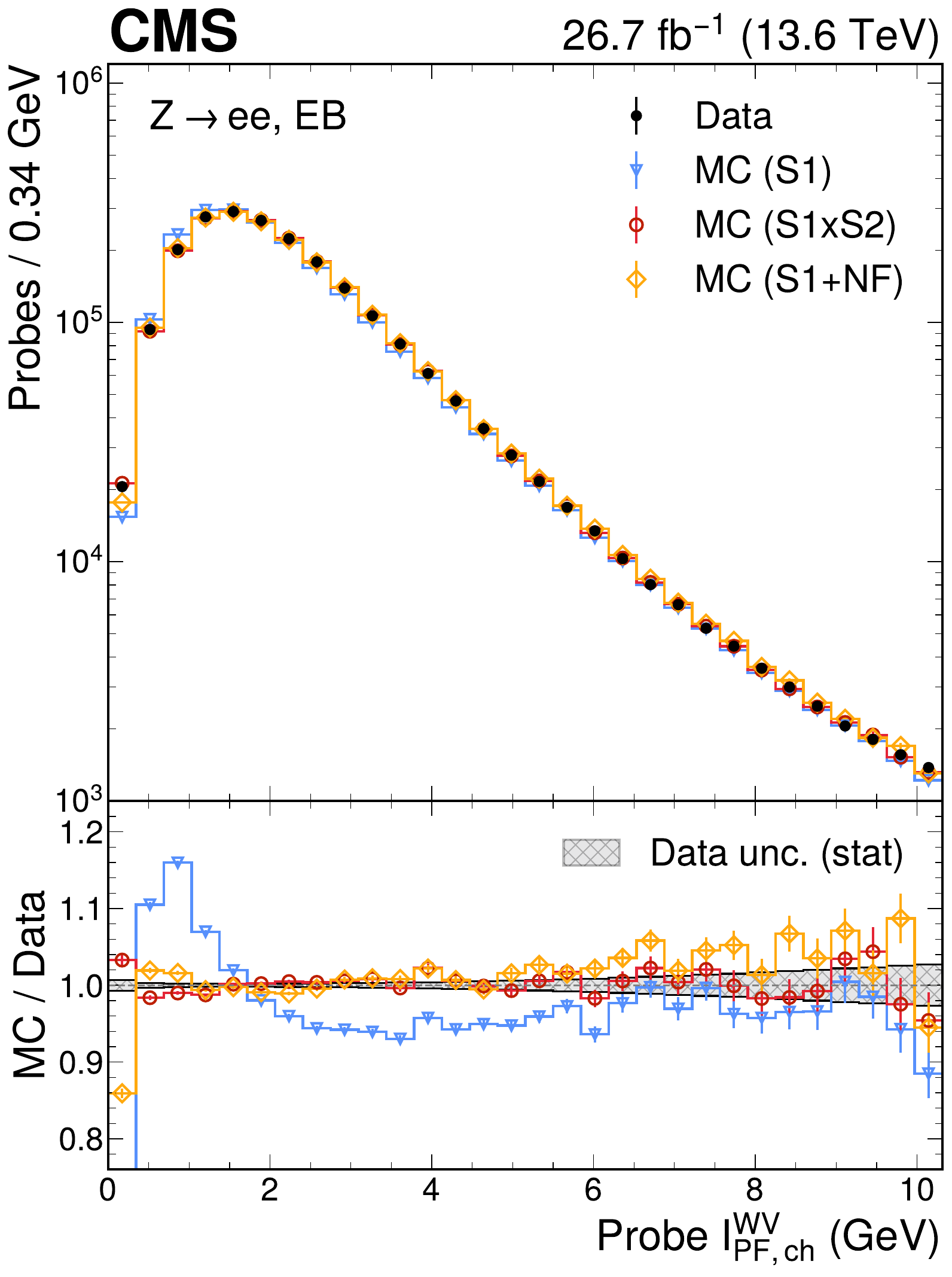}
    \includegraphics[width=.325\linewidth]{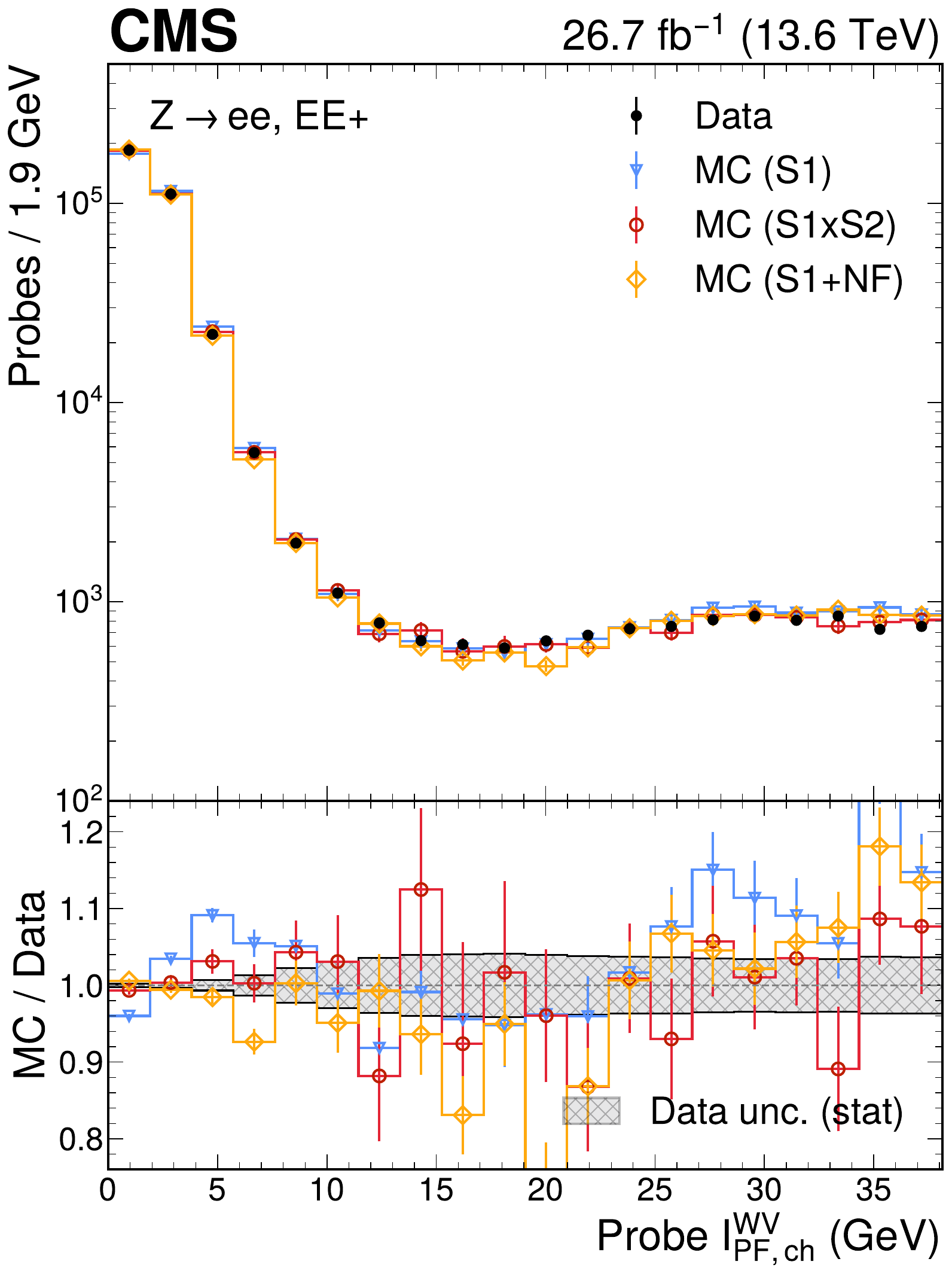}
    \includegraphics[width=.325\linewidth]{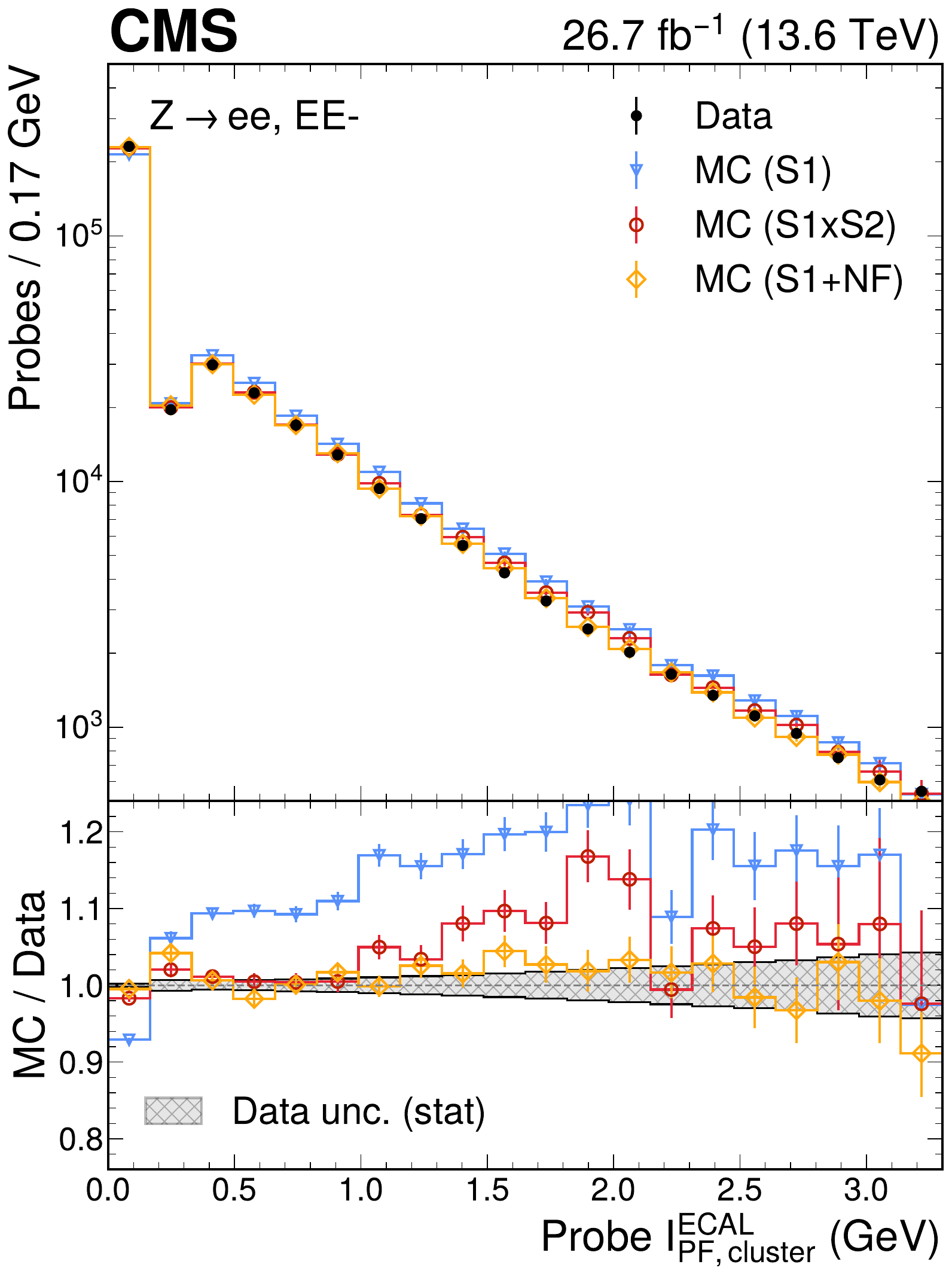}
    \includegraphics[width=.325\linewidth]{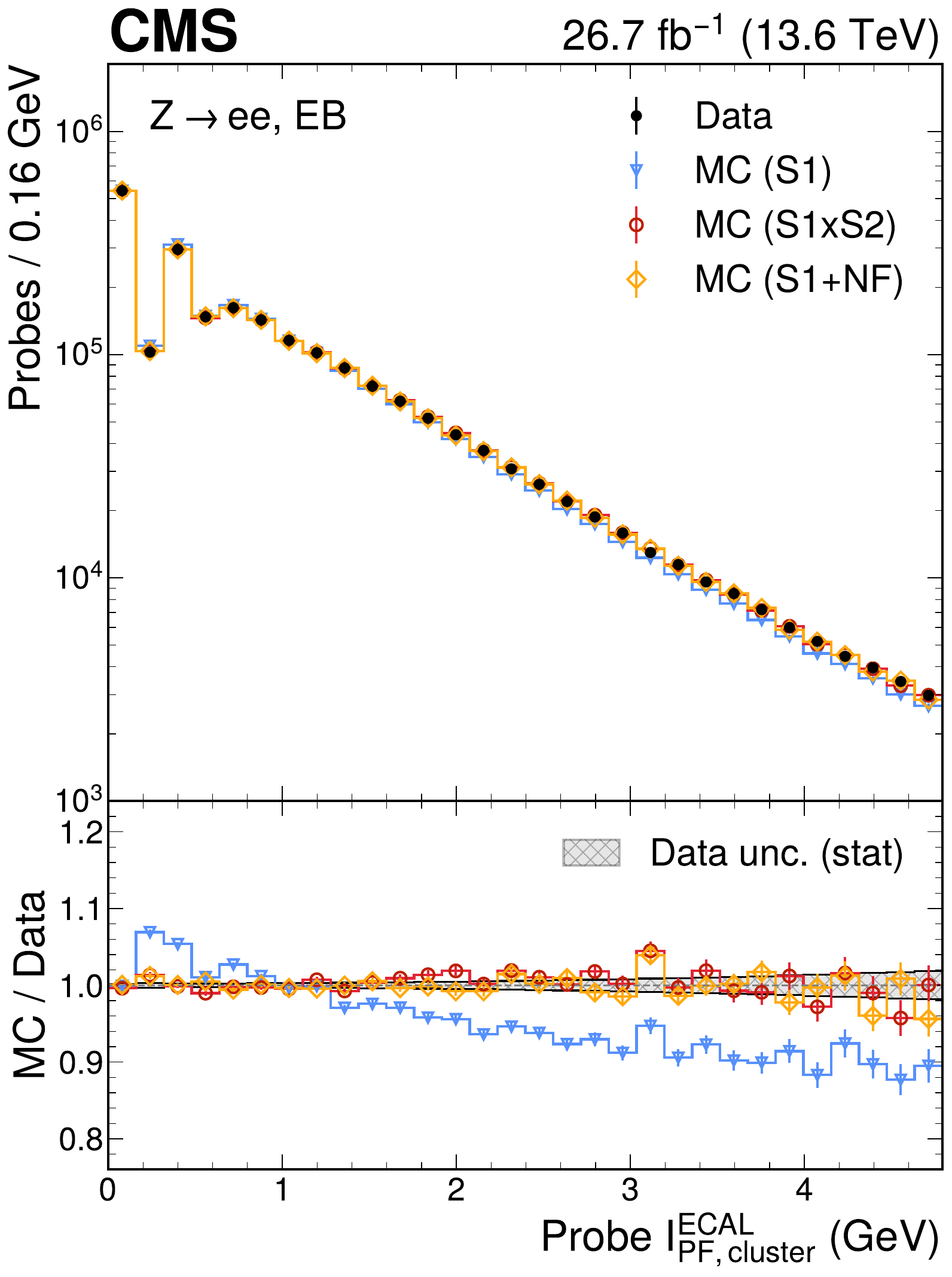}
    \includegraphics[width=.325\linewidth]{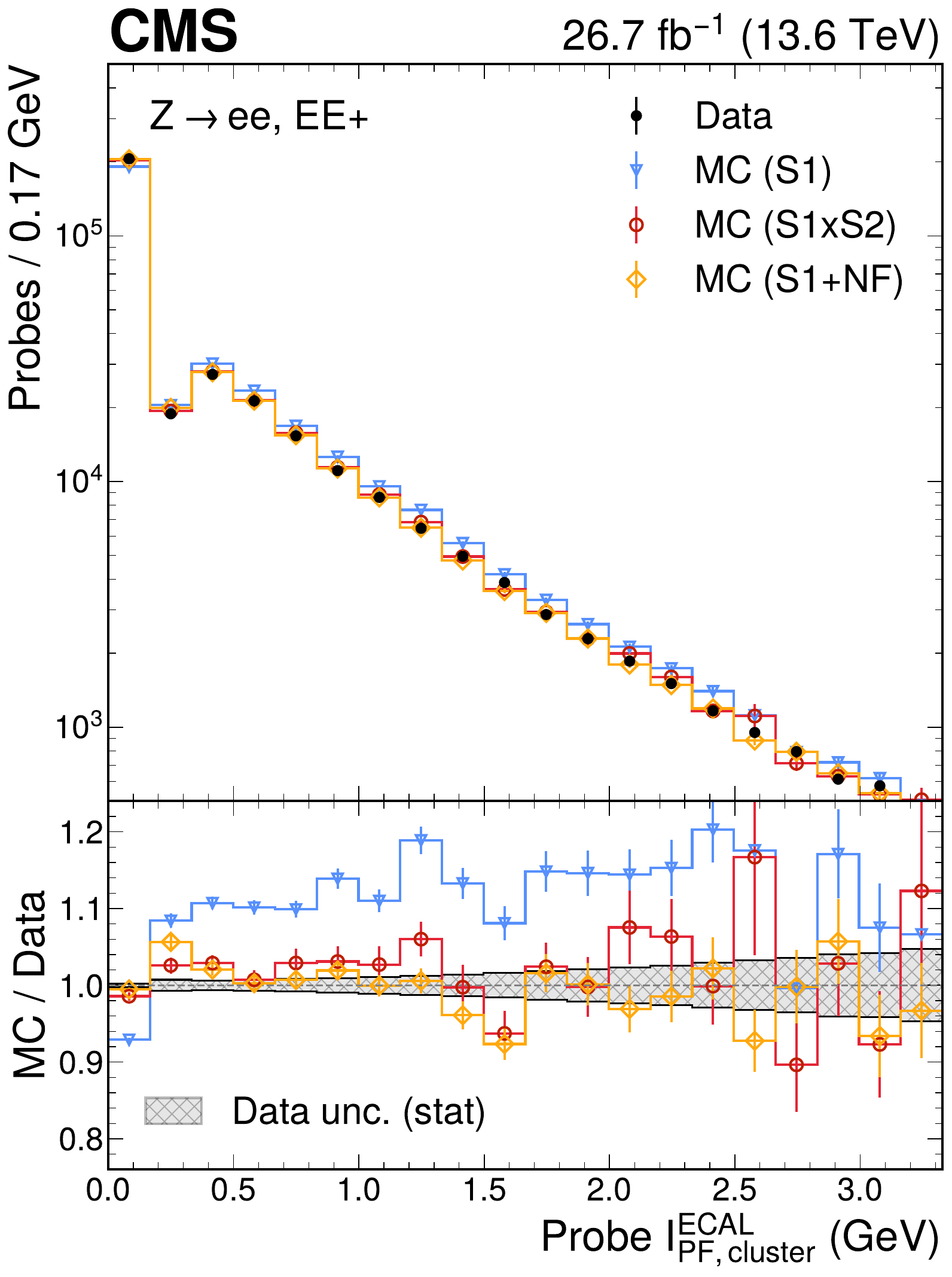}
    \caption{The probe $\mathcal{I}_{\mathrm{PF,ch}}$ (upper row), $\mathcal{I}^{\mathrm{WV}}_{\mathrm{PF,ch}}$ (middle row), and $\mathcal{I}^{\mathrm{ECAL}}_{\text{PF, cluster}}$ (lower row) distributions in data and simulation,
    for the EE$-$ (left), EB (centre), and EE$+$ (right) subdetector regions. 
    The plotting conventions are the same as in Fig.~\ref{fig:results_shower_0}.}
    \label{fig:results_shower_3}
\end{figure}

\begin{figure}[htb!]
    \centering
    \includegraphics[width=.325\linewidth]{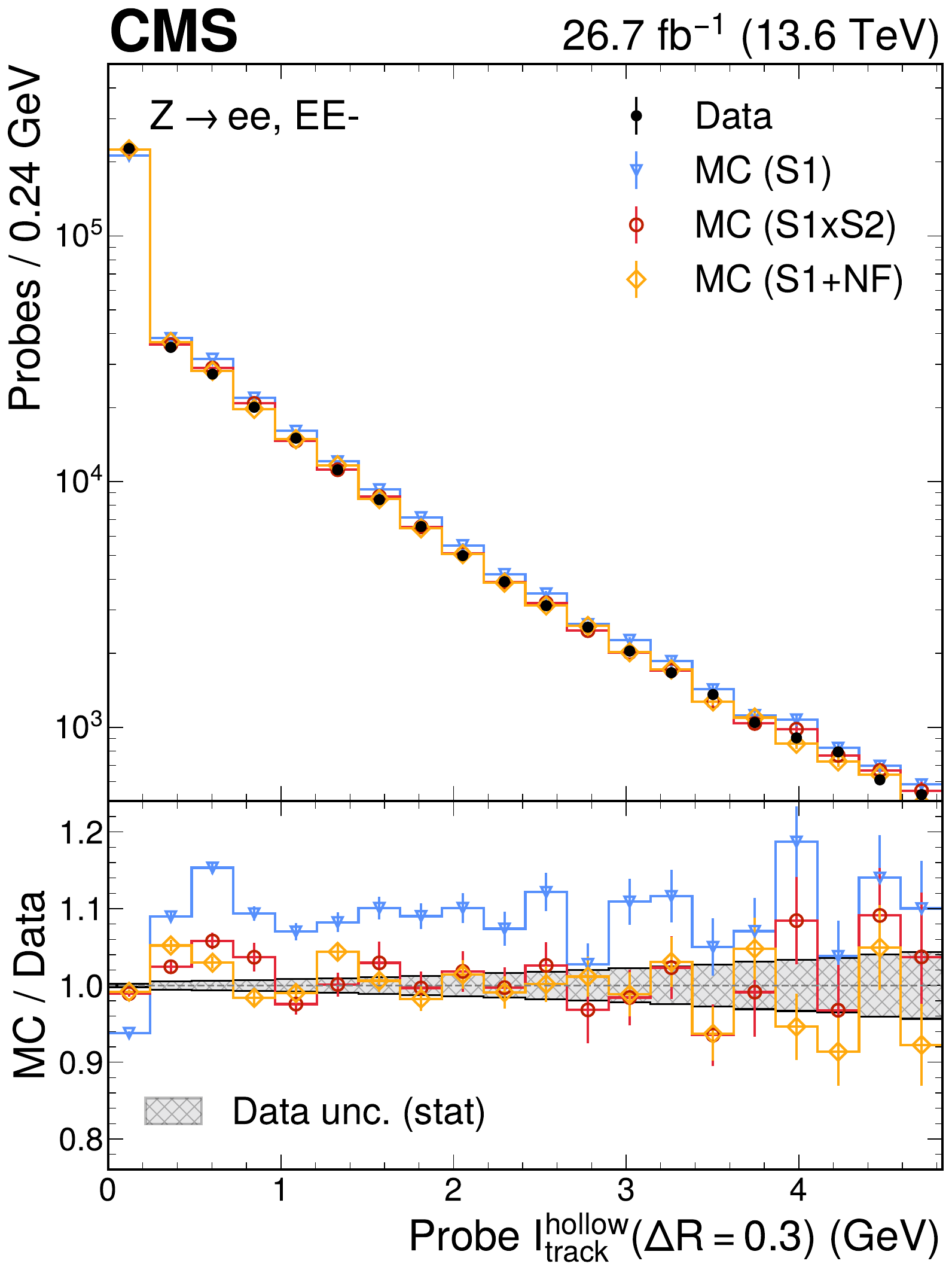}
    \includegraphics[width=.325\linewidth]{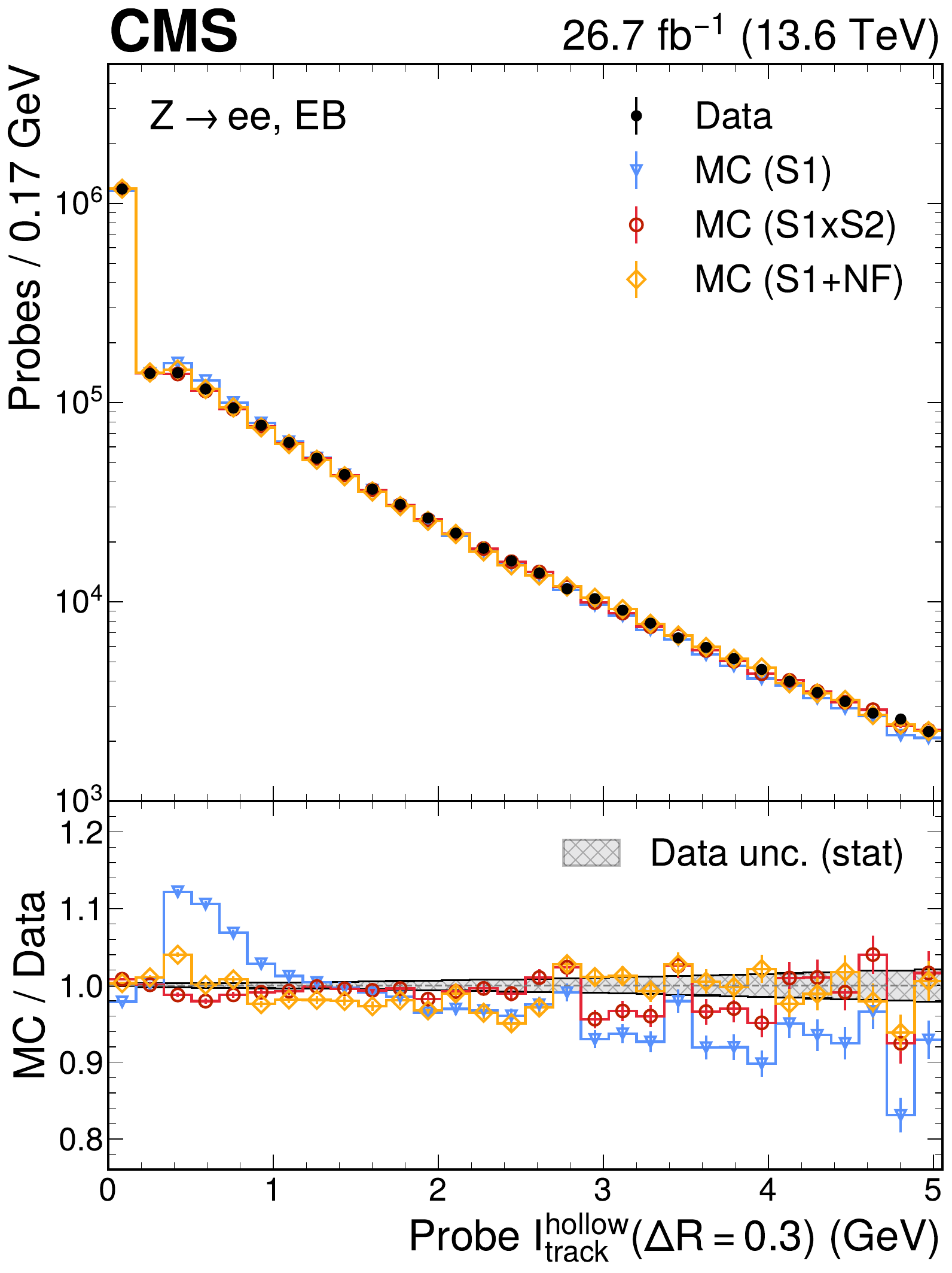}
    \includegraphics[width=.325\linewidth]{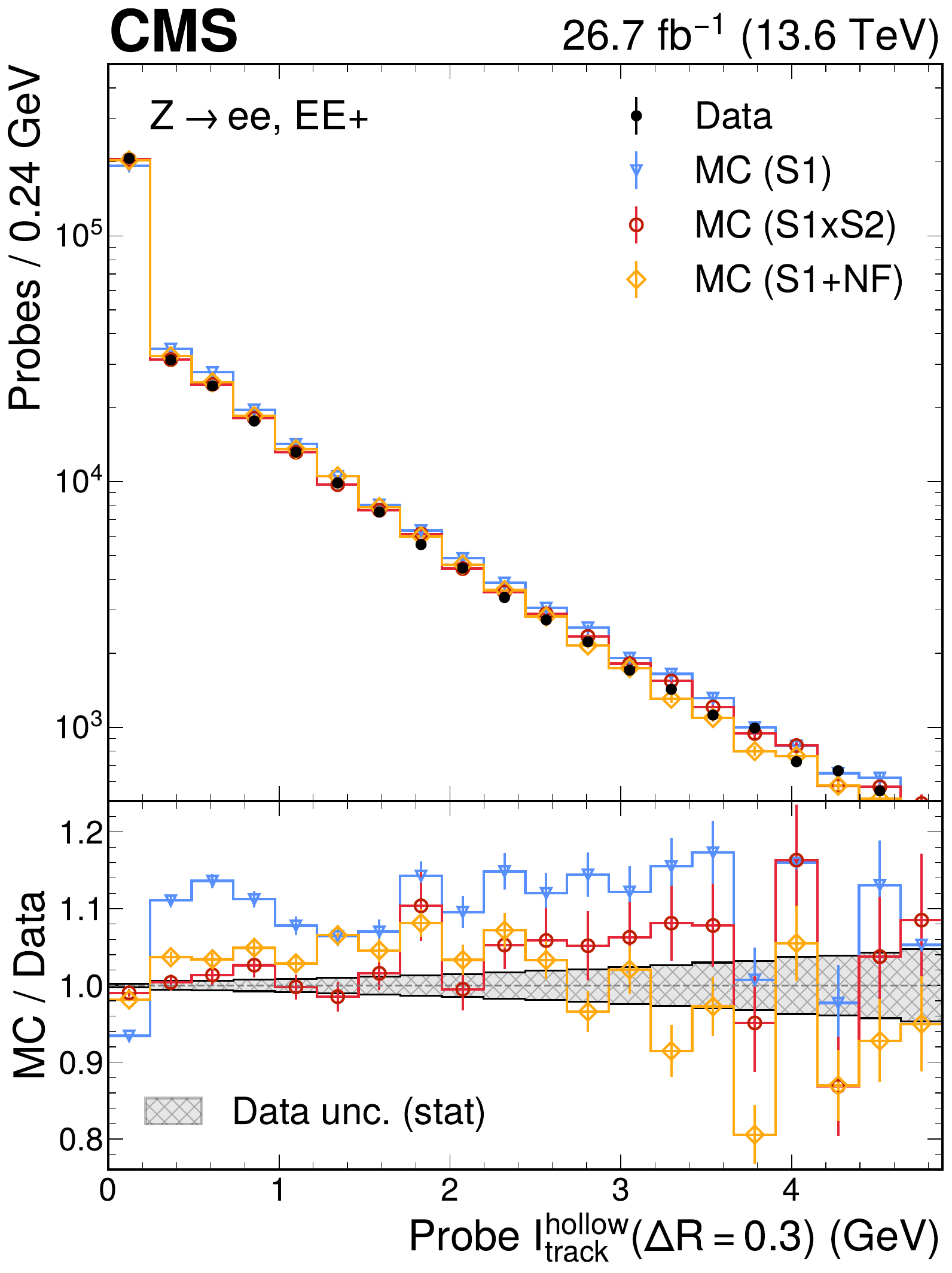}
    \includegraphics[width=.325\linewidth]{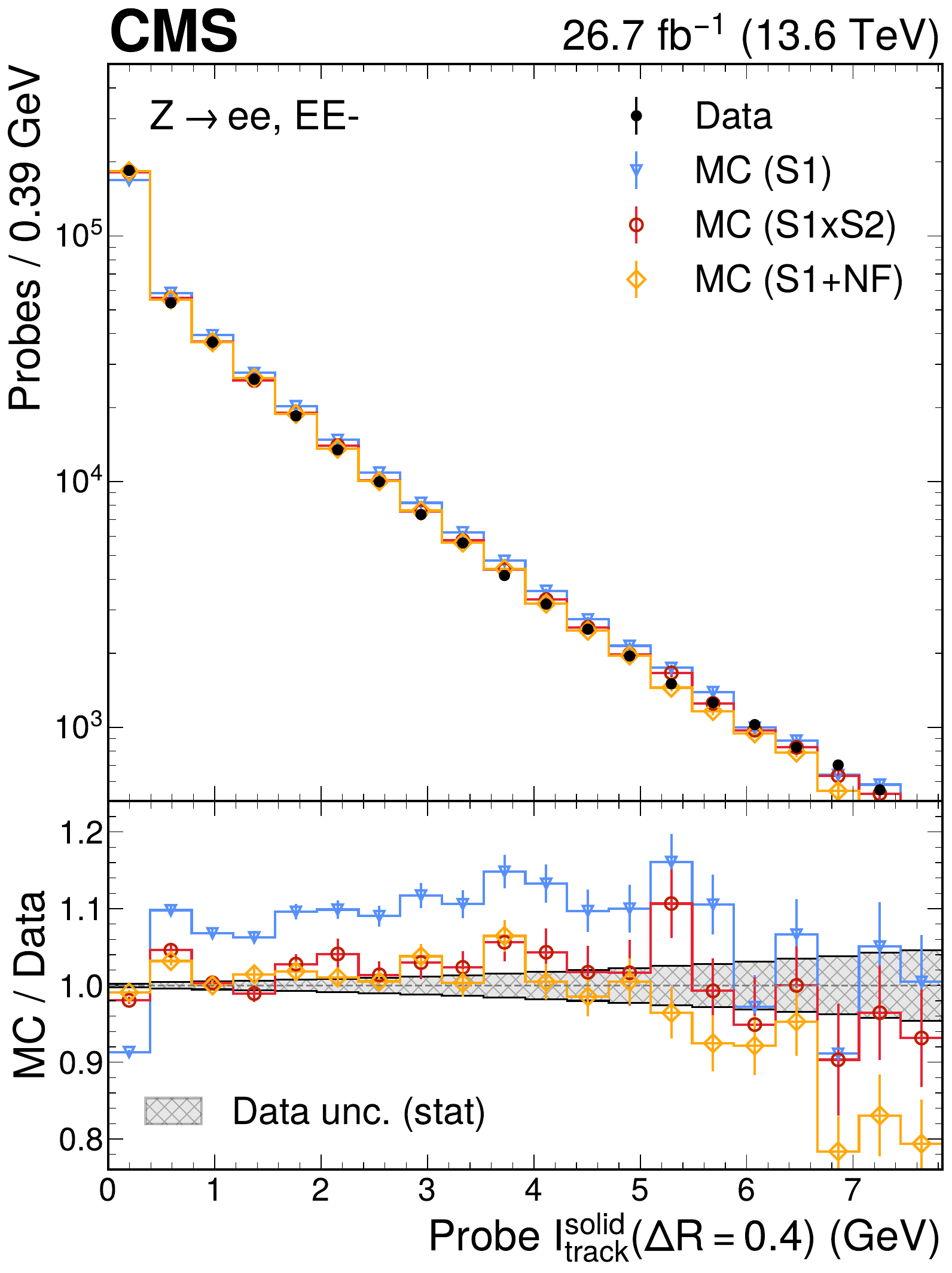}
    \includegraphics[width=.325\linewidth]{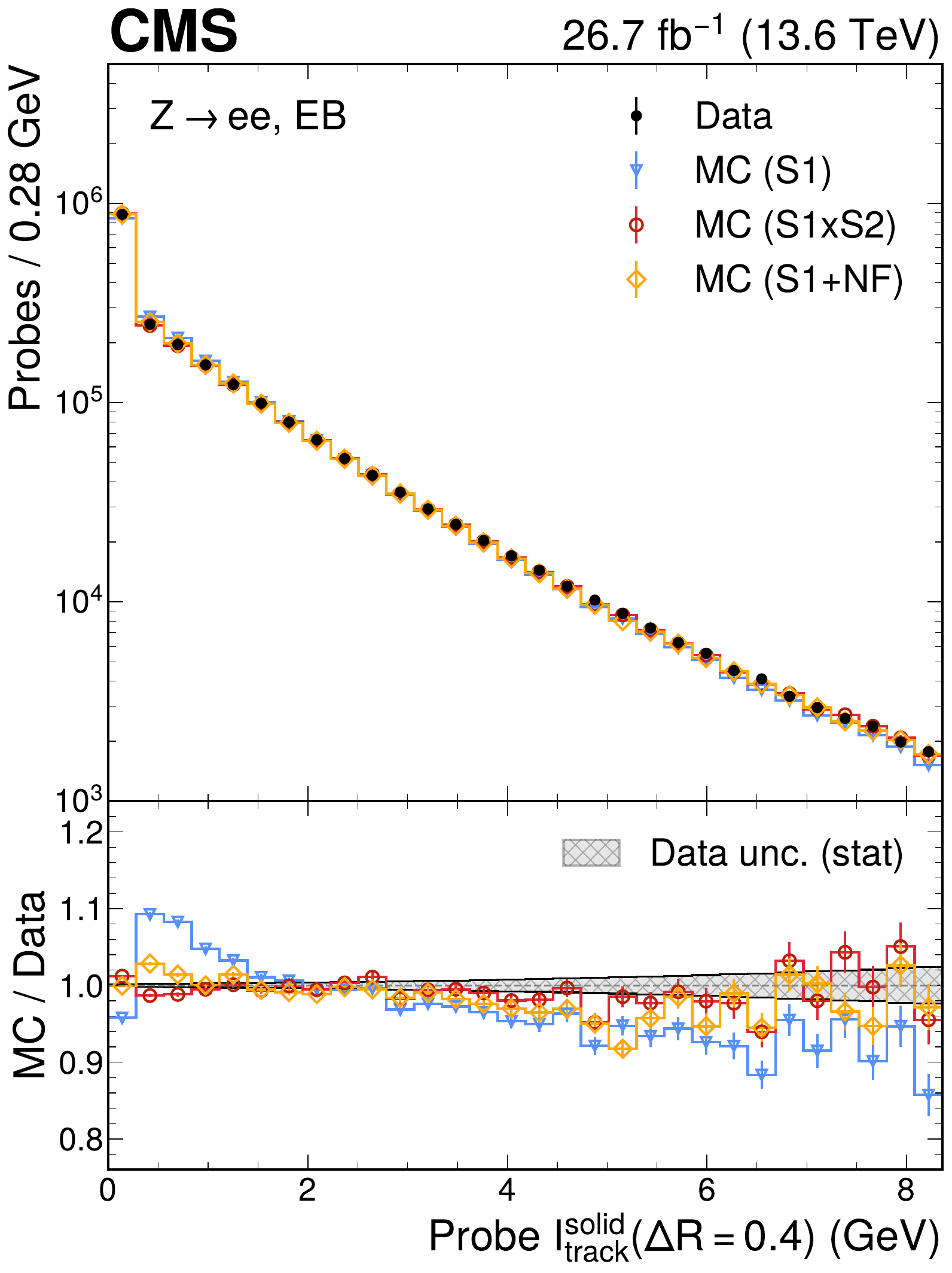}
    \includegraphics[width=.325\linewidth]{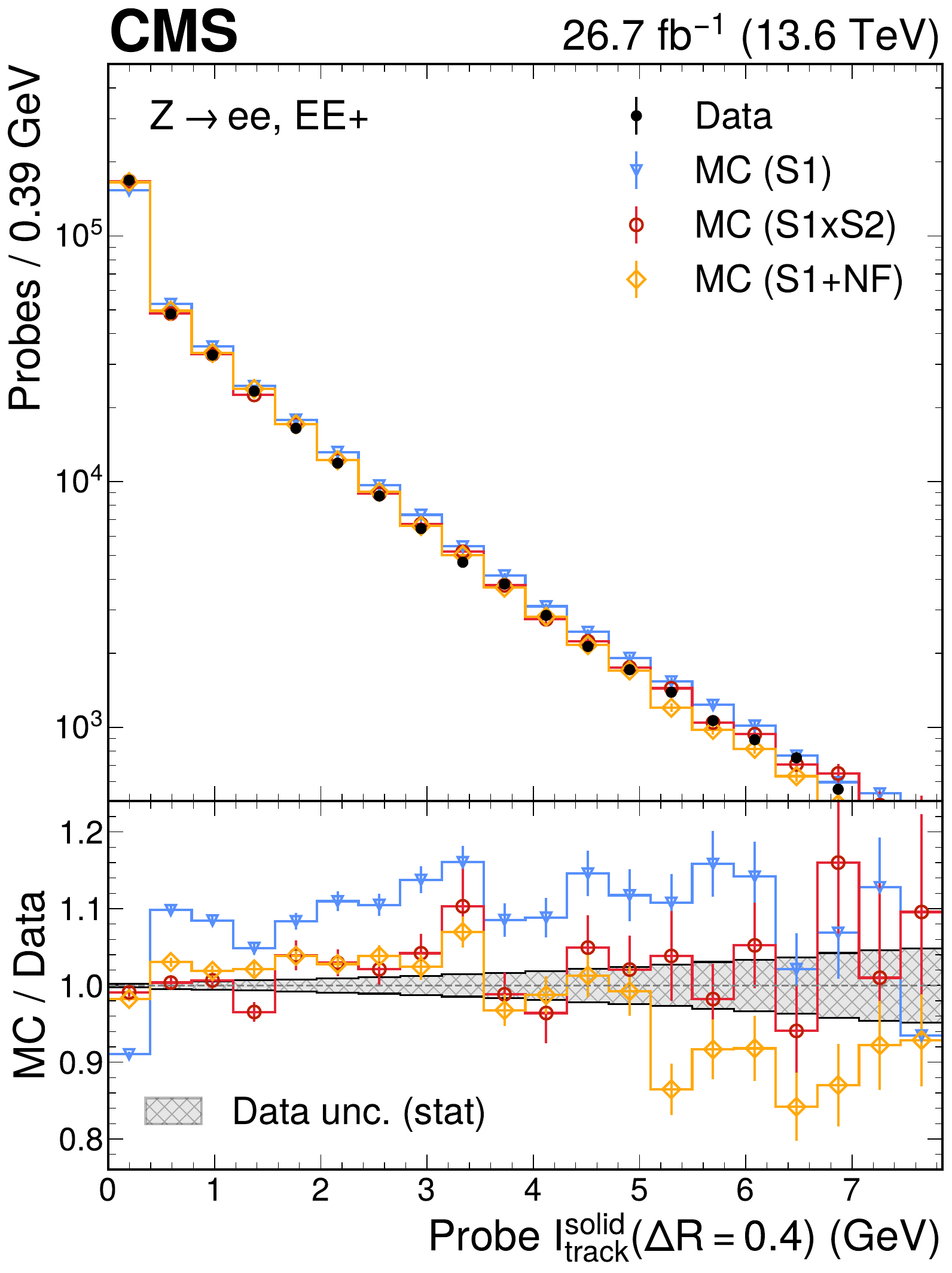}
    \caption{The probe $\mathcal{I}^{\text{hollow}}_{\text{track}}(\Delta R = 0.3)$ (upper row) and $\mathcal{I}^{\text{solid}}_{\text{track}}(\Delta R = 0.4)$ (lower row) distributions in data and simulation,
    for the EE$-$ (left), EB (centre), and EE$+$ (right) subdetector regions. 
    The plotting conventions are the same as in Fig.~\ref{fig:results_shower_0}.}
    \label{fig:results_shower_4}
\end{figure}

\begin{figure}[htb!]
    \centering
    \includegraphics[width=.49\linewidth]{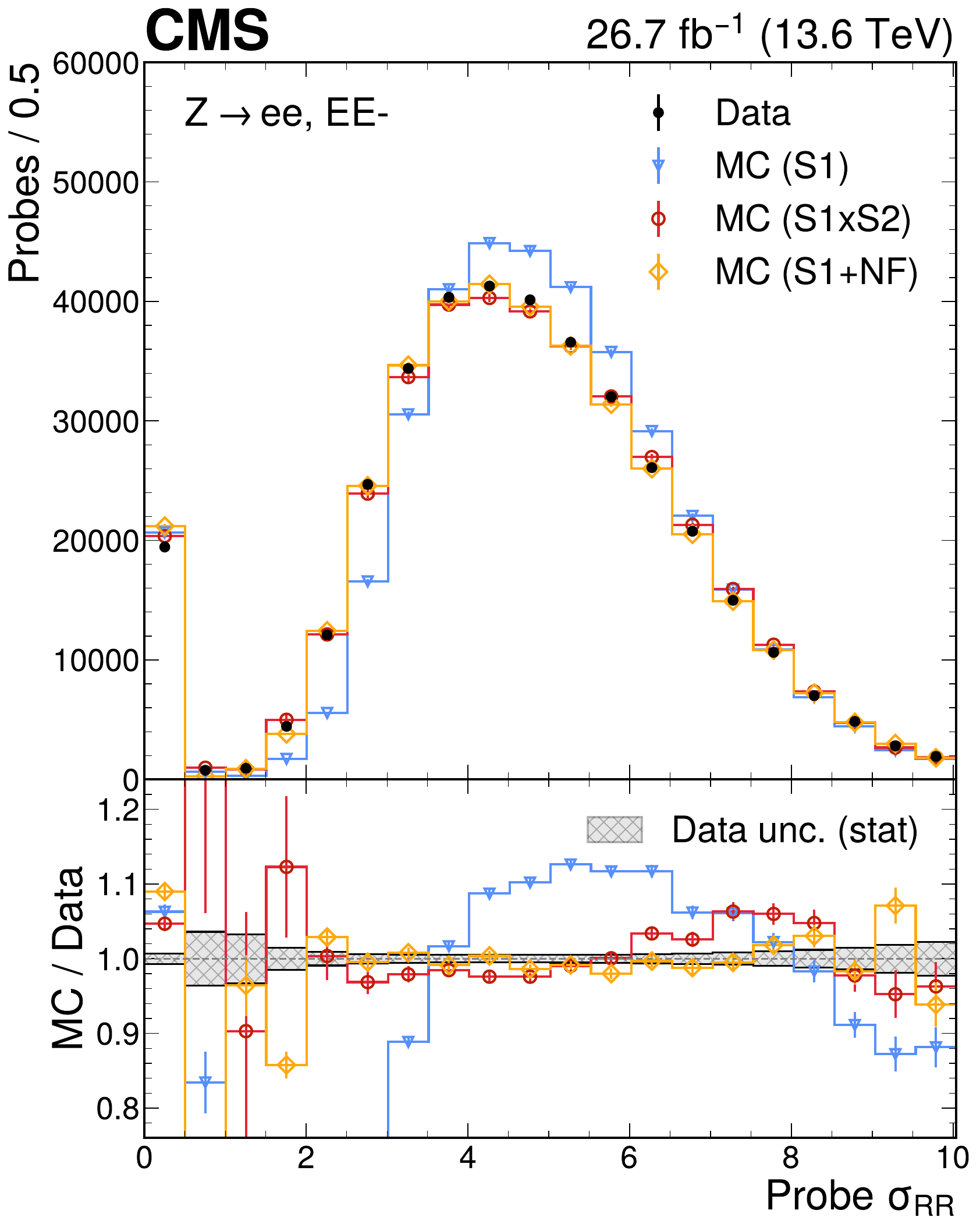}
    \includegraphics[width=.49\linewidth]{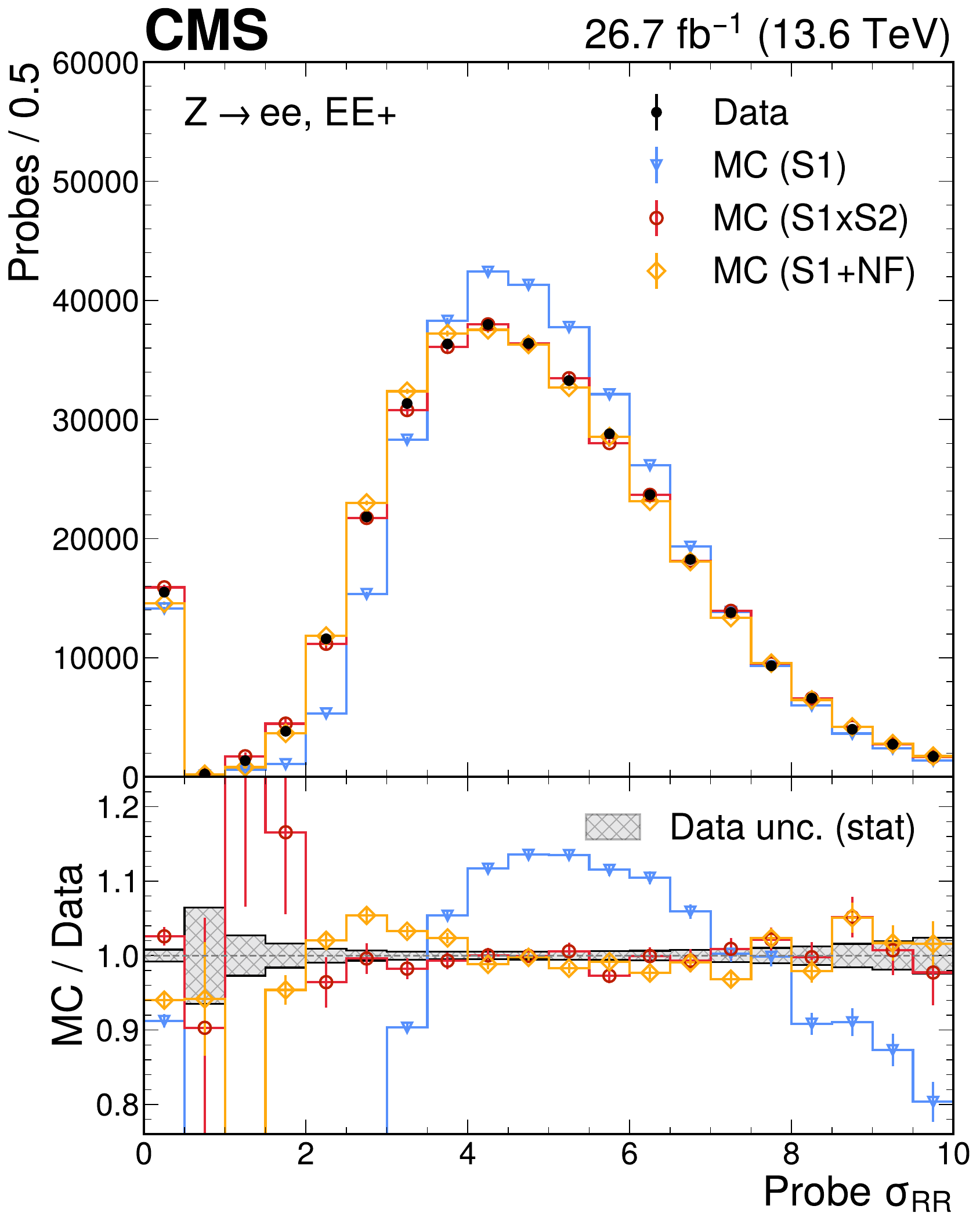}
    \\
    \includegraphics[width=.49\linewidth]{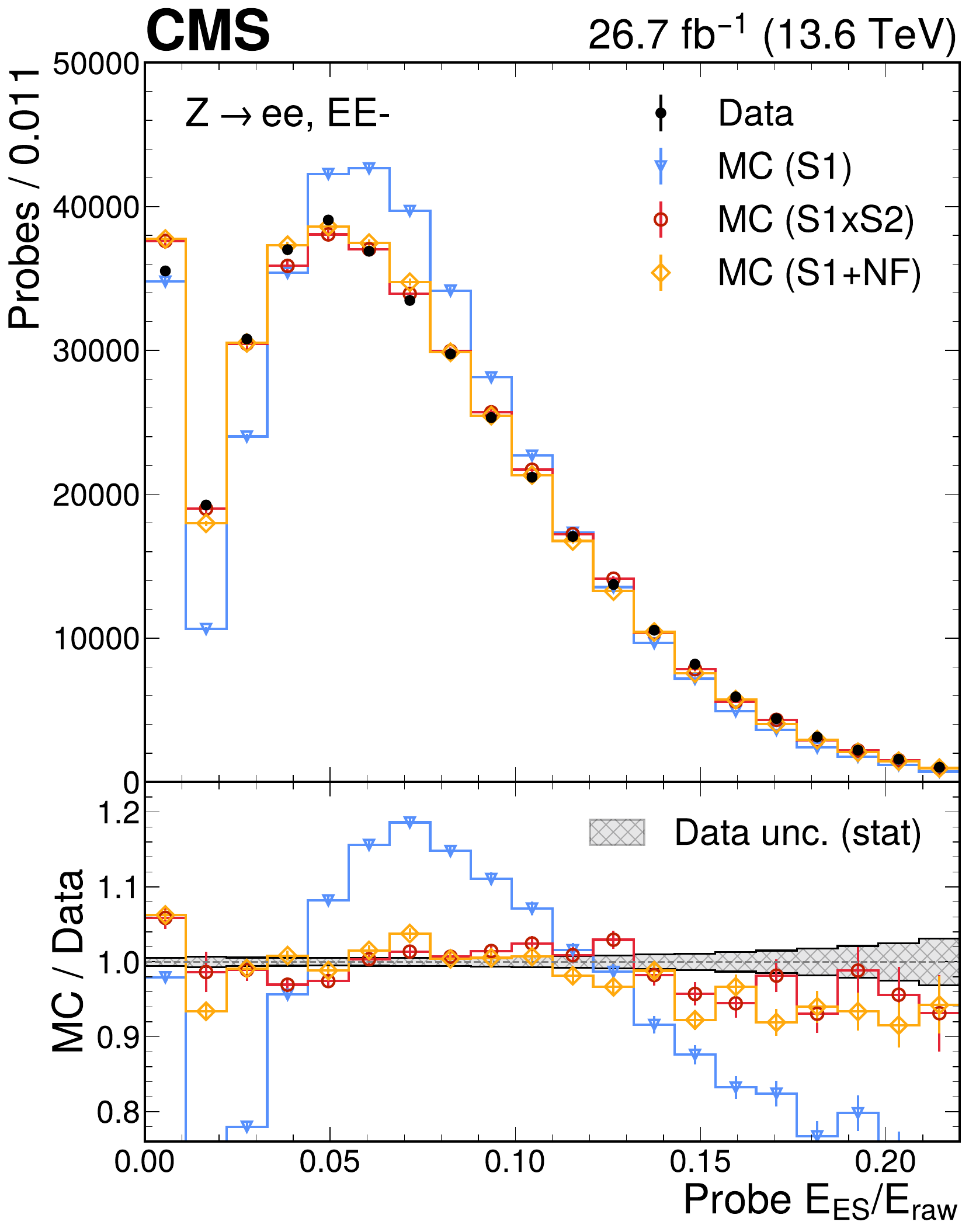}
    \includegraphics[width=.49\linewidth]{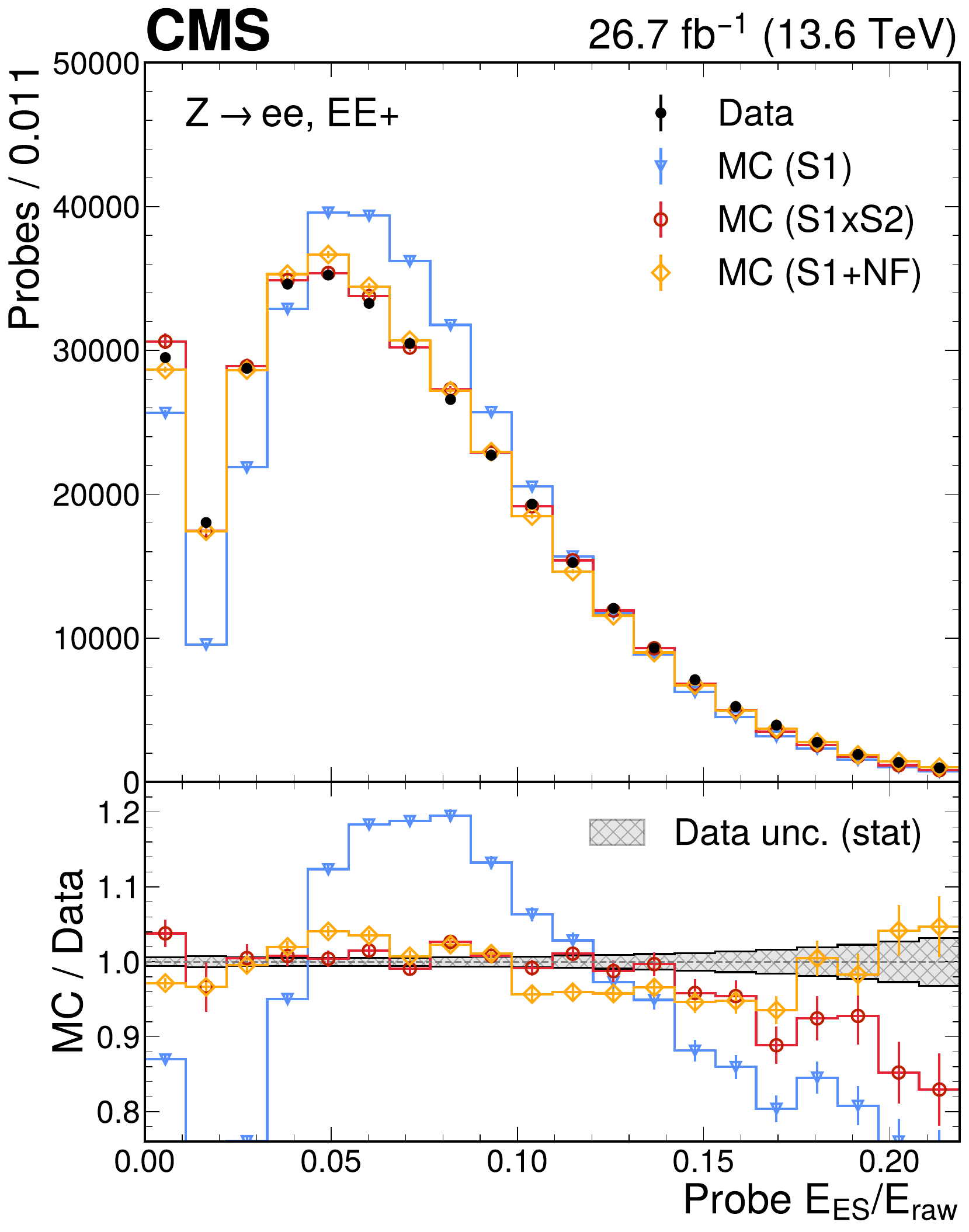}
    \caption{The probe features related to the preshower detector, $\sigma_{\mathrm{RR}}$ (upper row) and $E_{\mathrm{ES}}/E_{\text{raw}}$ (lower row), are shown. These variables are only defined for probes in the endcaps, where the preshower detectors are located.
    The plots show the distributions in data and simulation,
    for the EE$-$ (left) and EE$+$ (right) subdetector regions. 
    The plotting conventions are the same as in Fig.~\ref{fig:results_shower_0}.}
    \label{fig:results_shower_5}
\end{figure}

The 1D histograms for the $x_s$ features are shown for data and simulation in Figs.~\ref{fig:results_shower_0}--\ref{fig:results_shower_5}. 
These features are the targets of the calibration methods. 
As with the previous figures, the plots are separated into the different subdetector regions: EE$-$ (left), EB (centre), and EE$+$ (right). 
The simulation is shown for three scenarios. 
The blue histograms represent the baseline, showing the simulation after the initial S1 reweighting. 
The red and orange histograms show the corrected simulation after applying the S2 reweighting and the flow corrections, respectively. 
The improved agreement with data after applying the calibration methods is evident from the lower panels in each plot, which show the ratio of simulation to data. 

To quantify this improvement, two complementary metrics are evaluated for each feature.
The first is the reduced $\chi^2/n_{\mathrm{dof}}$, between the data and simulation histograms, as defined in Eq.~\ref{eq:chi2_reduced}.
This metric has a clear statistical interpretation, with values close to unity expected for statistically compatible distributions when all relevant uncertainties are correctly accounted for.
Since the uncertainties used in Eq.~\ref{eq:chi2_reduced} are purely statistical, and do not account for additional systematic effects, 
values greater than unity are generally observed.
The metric is sensitive to localised discrepancies between data and simulation, 
such as those in the tails of the distributions.
However, because it is based on bin-by-bin comparisons, 
the $\chi^2/n_{\mathrm{dof}}$ does not effectively capture correlated discrepancies that extend across multiple bins.
To address this limitation,
the 1D EMD~\cite{inbook_villani} between the data and simulation samples is used as a second metric.
This unbinned metric is more sensitive to global differences between distributions,
such as shifts in the mean or an overall broadening.
For the EMD calculation, the features are first normalised to place them on a comparable scale.
It is important to note that the $\chi^2/n_{\mathrm{dof}}$ values obtained for the reweighting method include the inflated statistical uncertainties in the simulation, often leading to smaller values compared to the normalising-flow method. 
This makes it difficult to directly compare the performance of the two methods using the $\chi^2/n_{\mathrm{dof}}$ metric alone.
In contrast, the EMD metric does not directly depend on the statistical uncertainties, and therefore provides a useful relative measure for comparing the improvement achieved by the two calibration methods.
The results for both metrics across all input features are summarised in Tables~\ref{tab:results_chi2} and \ref{tab:results_emd}.

As discussed in Section~\ref{sec:reweighting}, the H/E and $\mathcal{I}^{\mathrm{HCAL}}_{\text{PF, cluster}}$ features were not included in the reweighting method training due to large initial discrepancies between data and simulation. 
This leads to weight corrections that vary substantially from unity, and therefore decrease the statistical power of the simulation sample. 
The choice to omit these features from the training is evident from Fig.~\ref{fig:results_shower_2}, and Tables~\ref{tab:results_chi2} and \ref{tab:results_emd}, as the S2 reweighting has a negligible effect on the distributions. 
The normalising-flow method however manages to correct these large initial discrepancies. 

\begin{figure}[htb!]
    \centering
    \includegraphics[width=.49\linewidth]{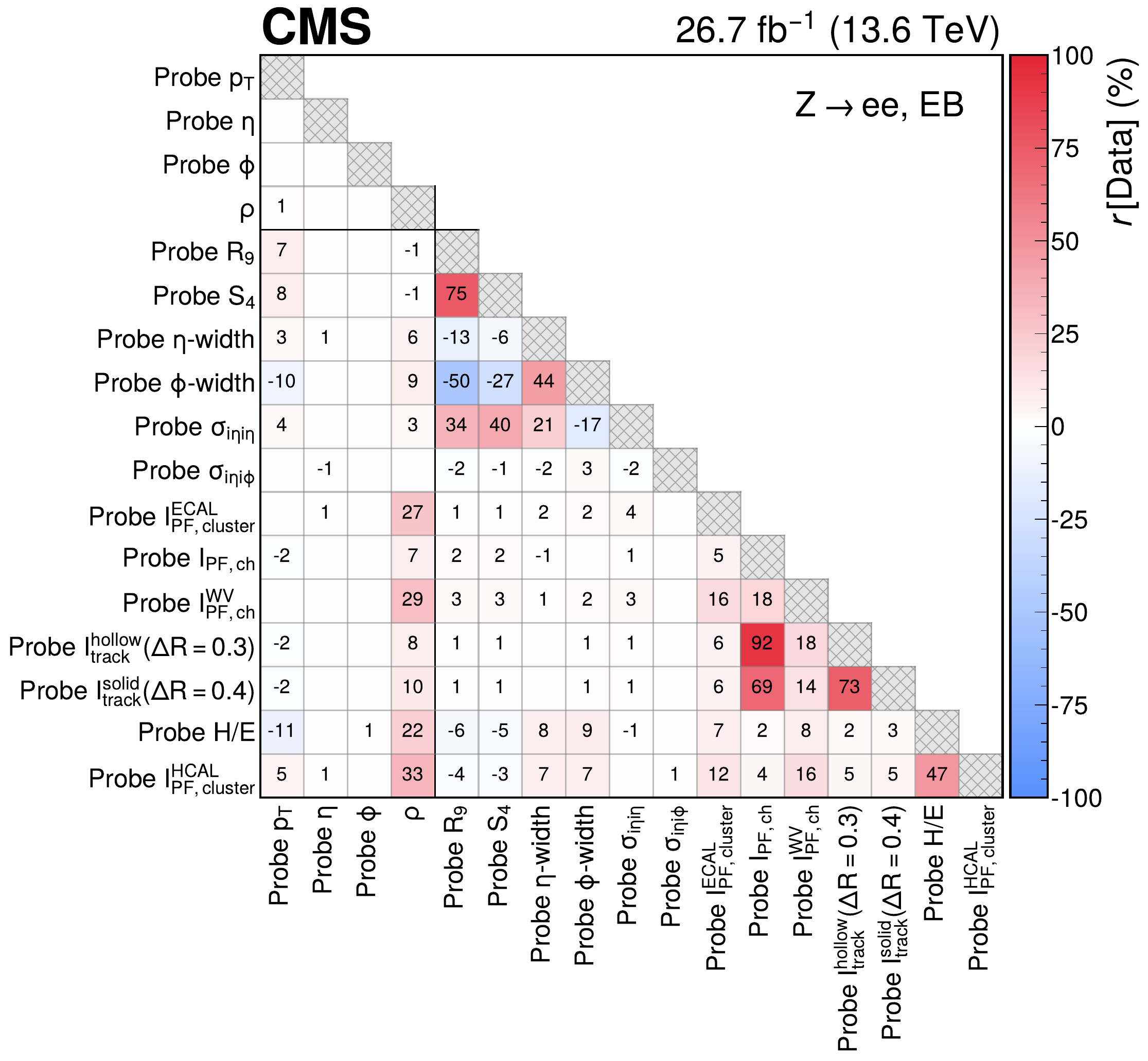}
    \includegraphics[width=.49\linewidth]{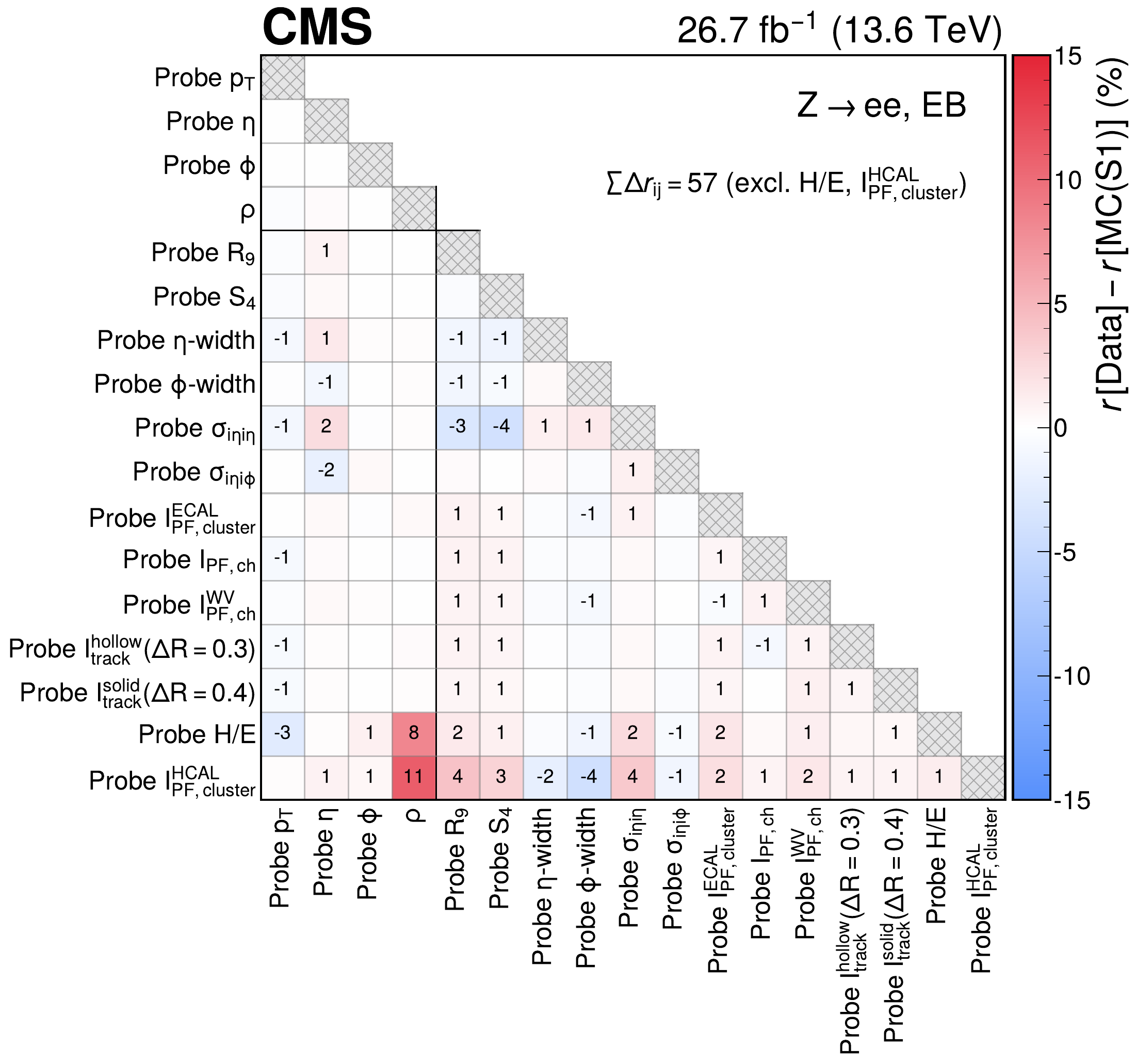}
    \includegraphics[width=.49\linewidth]{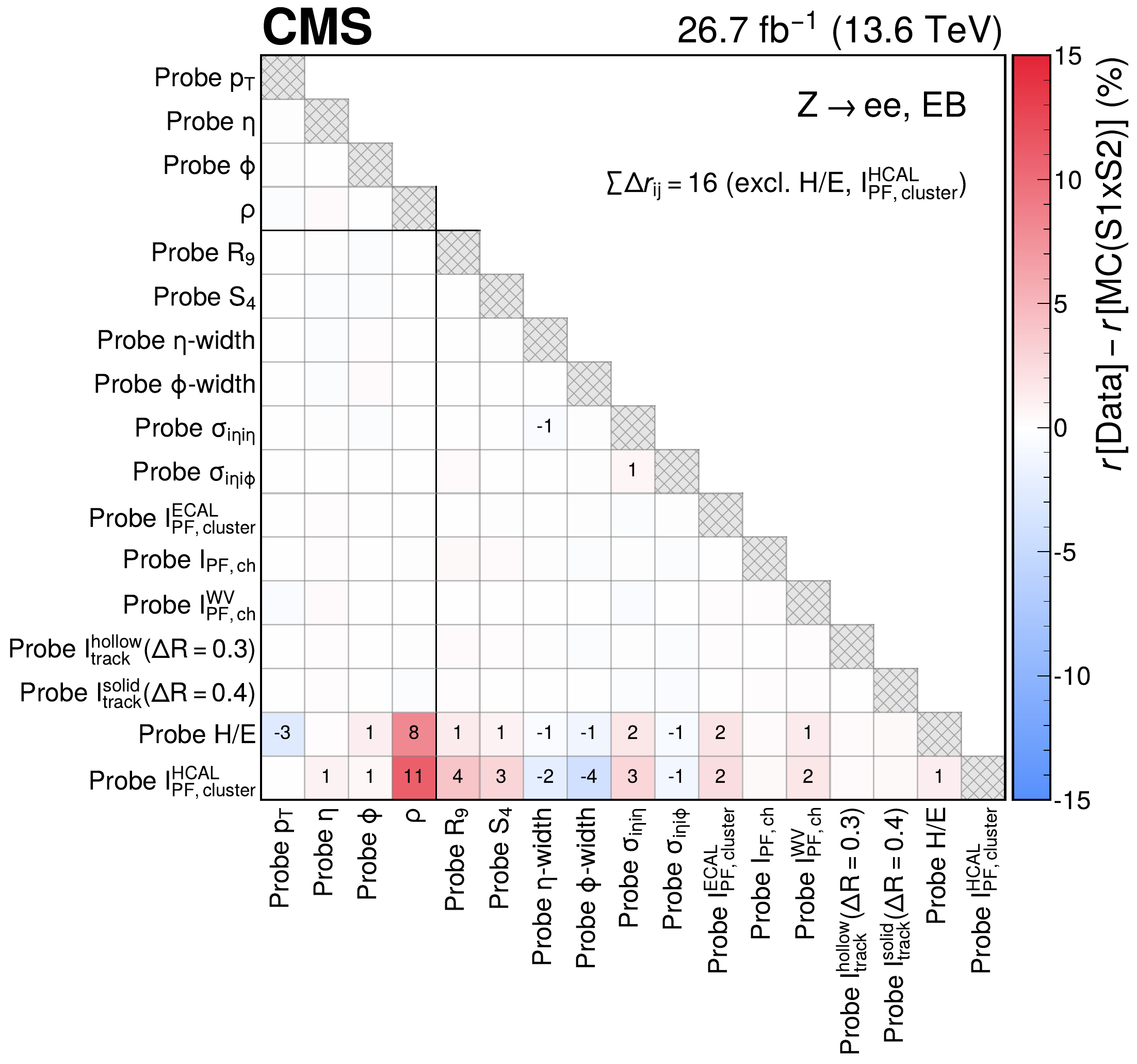}
    \includegraphics[width=.49\linewidth]{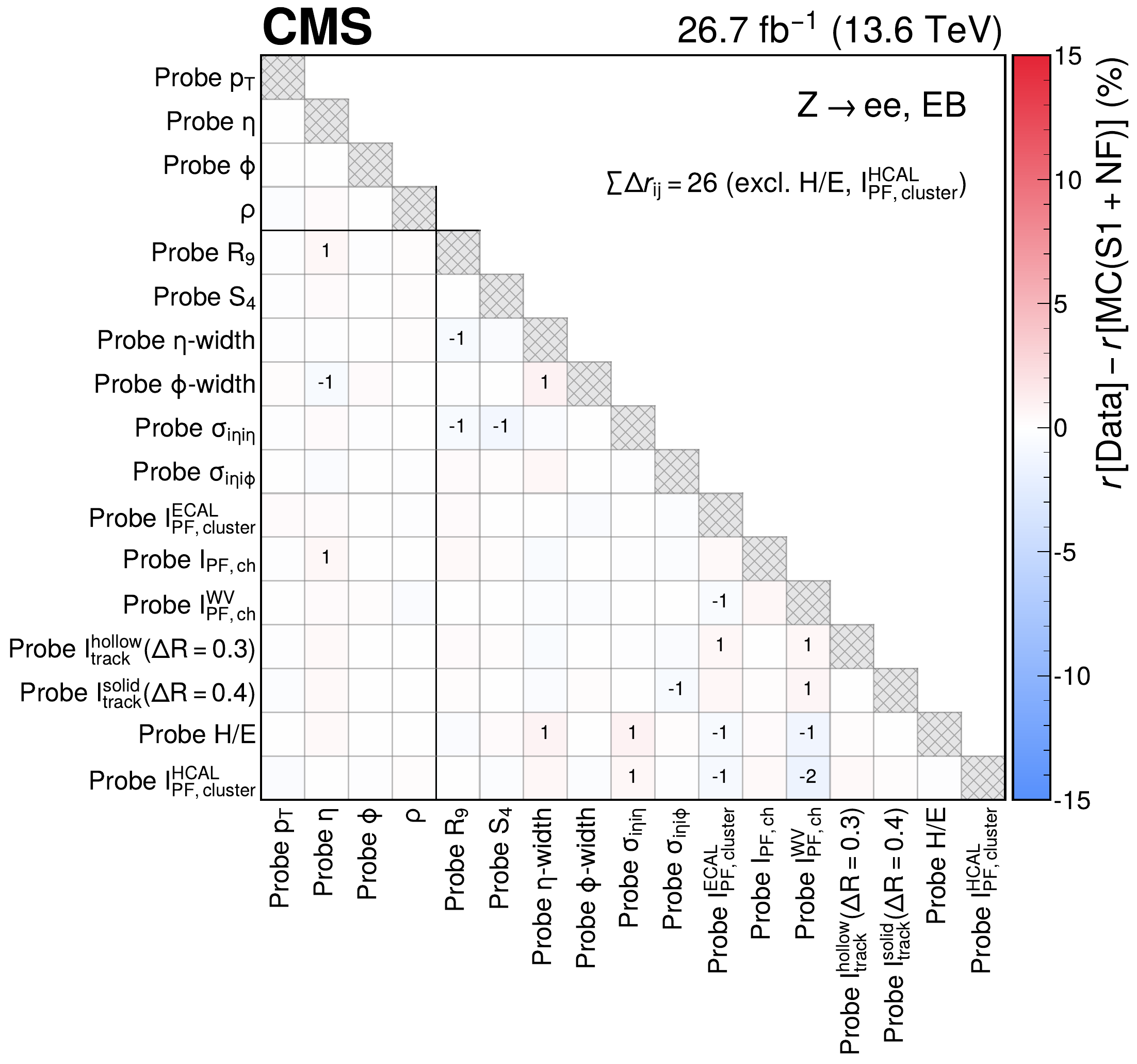}
    \caption{The upper left plot shows the Pearson correlation coefficients $r_{ij}$, 
    between pairs of features in data for the EB subdetector. 
    The other three plots show the difference in correlation coefficients $\Delta r_{ij}$ between data and simulation for three scenarios: the upper right plot corresponds to simulation after the initial S1 reweighting to align the kinematic distributions with data, while the lower left and lower right plots show the calibrated simulation after the S2 reweighting and the flow corrections are applied, respectively. All $r_{ij}$ and $\Delta r_{ij}$ values are rounded to the nearest percent, and values of absolute size less than 0.5\% are not shown. The sum of absolute $\Delta r_{ij}$ terms are shown in each of the relevant plots, where the sum excludes the probe H/E and $\mathcal{I}^{\mathrm{HCAL}}_{\text{PF, cluster}}$ features, since these are not included in the reweighting method. A solid line is used to separate the $x_k$ and $x_s$ features.}
    \label{fig:results_pearson_EB}
\end{figure}

\begin{figure}[htb!]
    \centering
    \includegraphics[width=.49\linewidth]{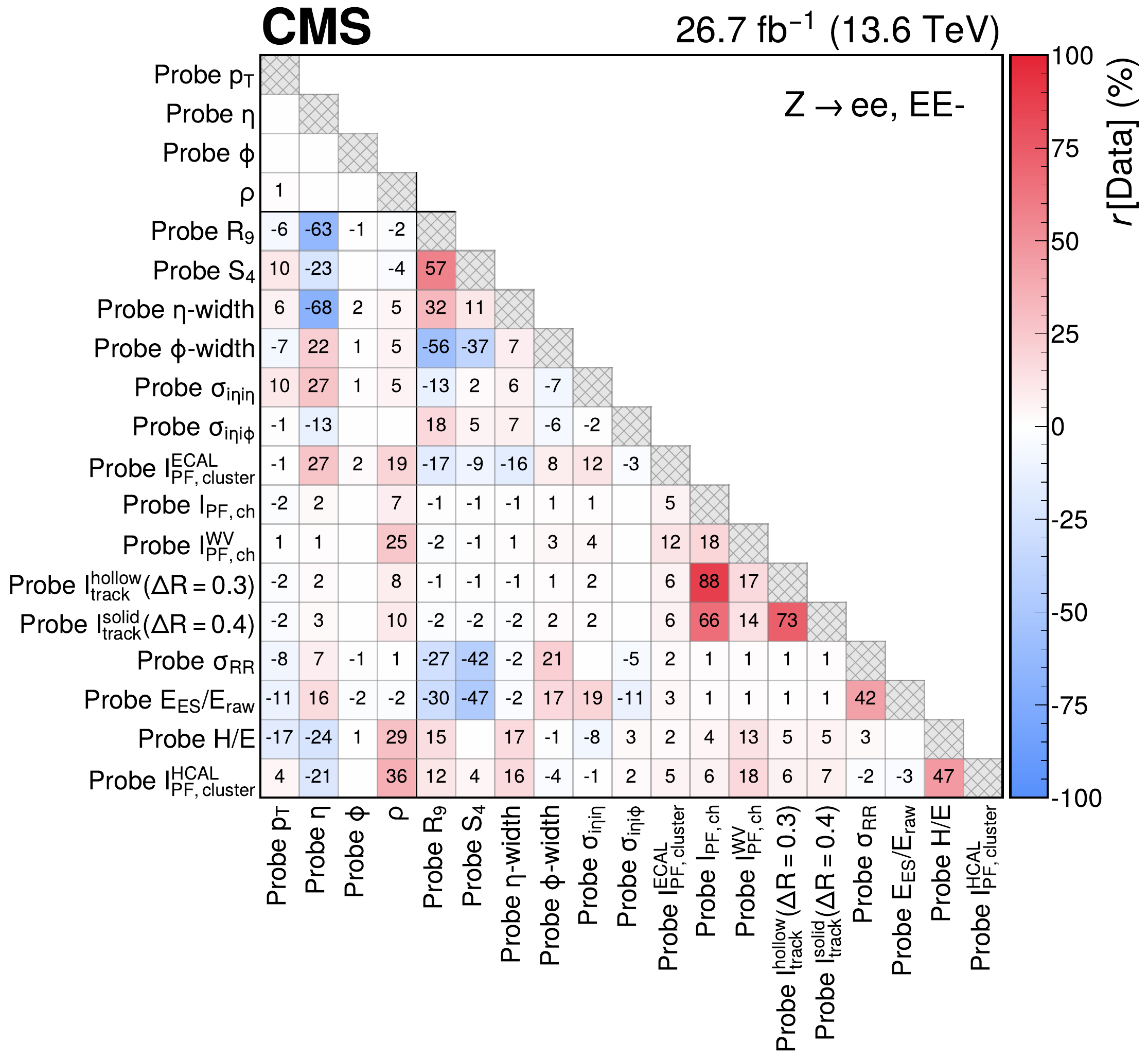}
    \includegraphics[width=.49\linewidth]{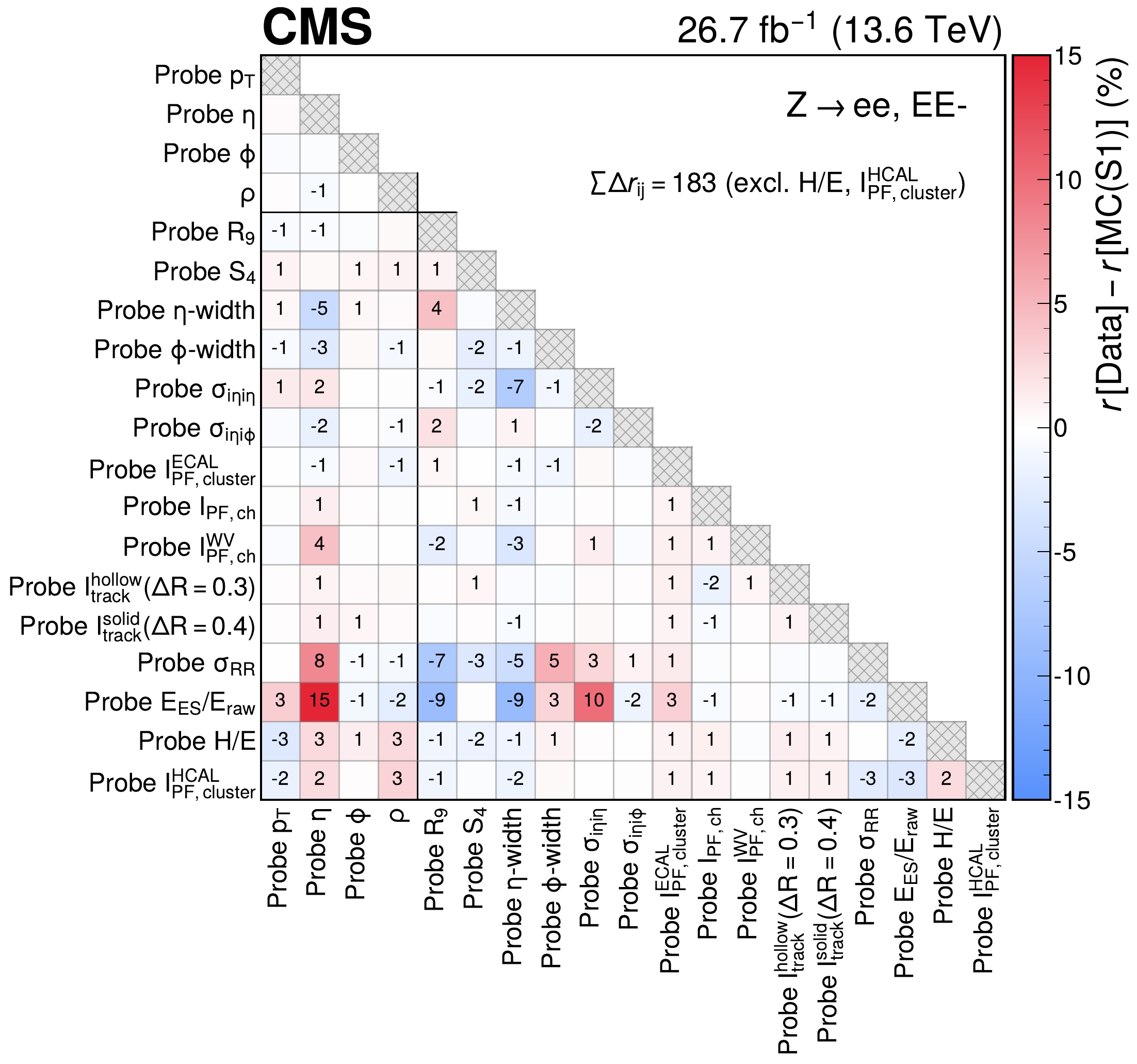}
    \includegraphics[width=.49\linewidth]{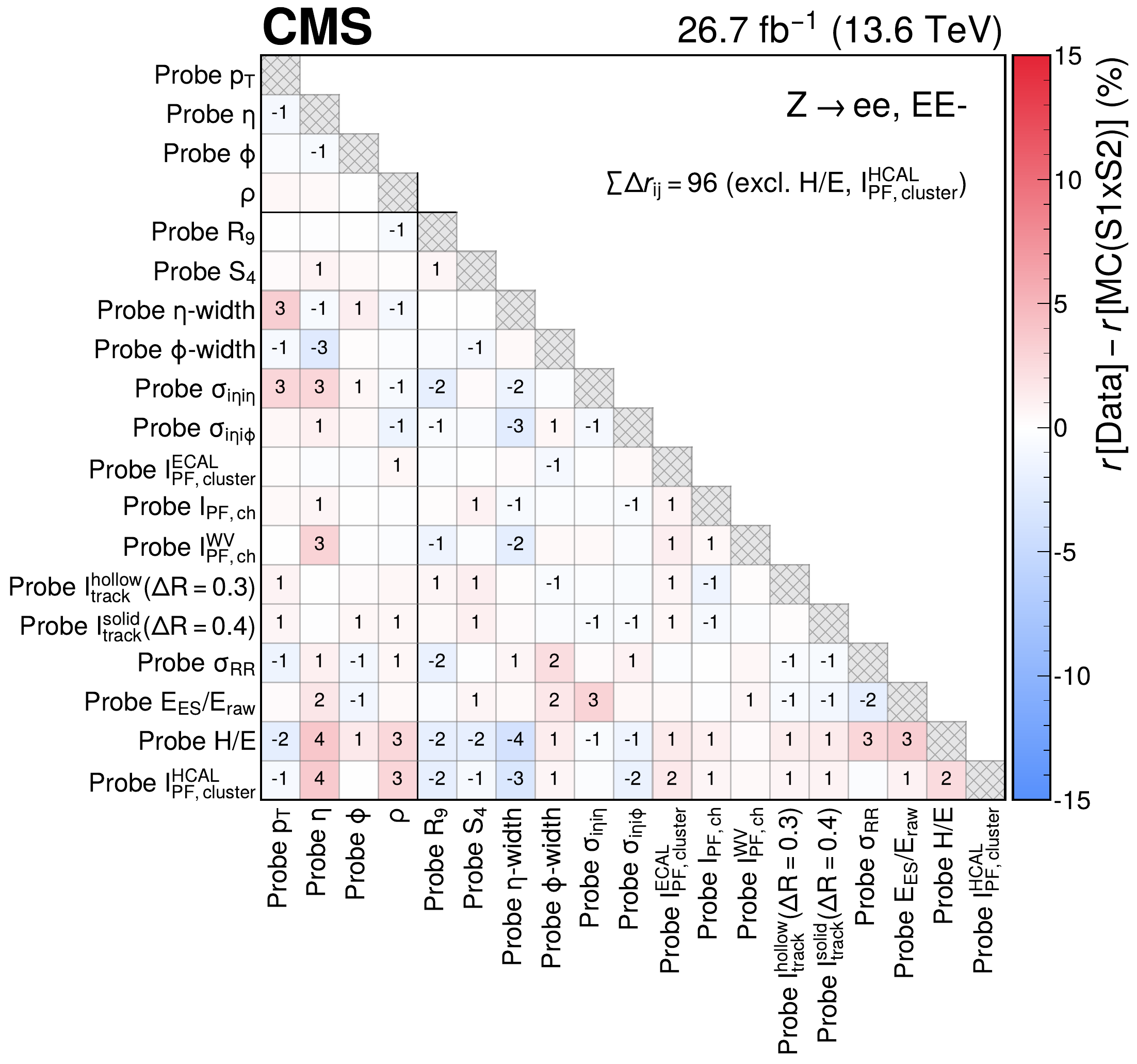}
    \includegraphics[width=.49\linewidth]{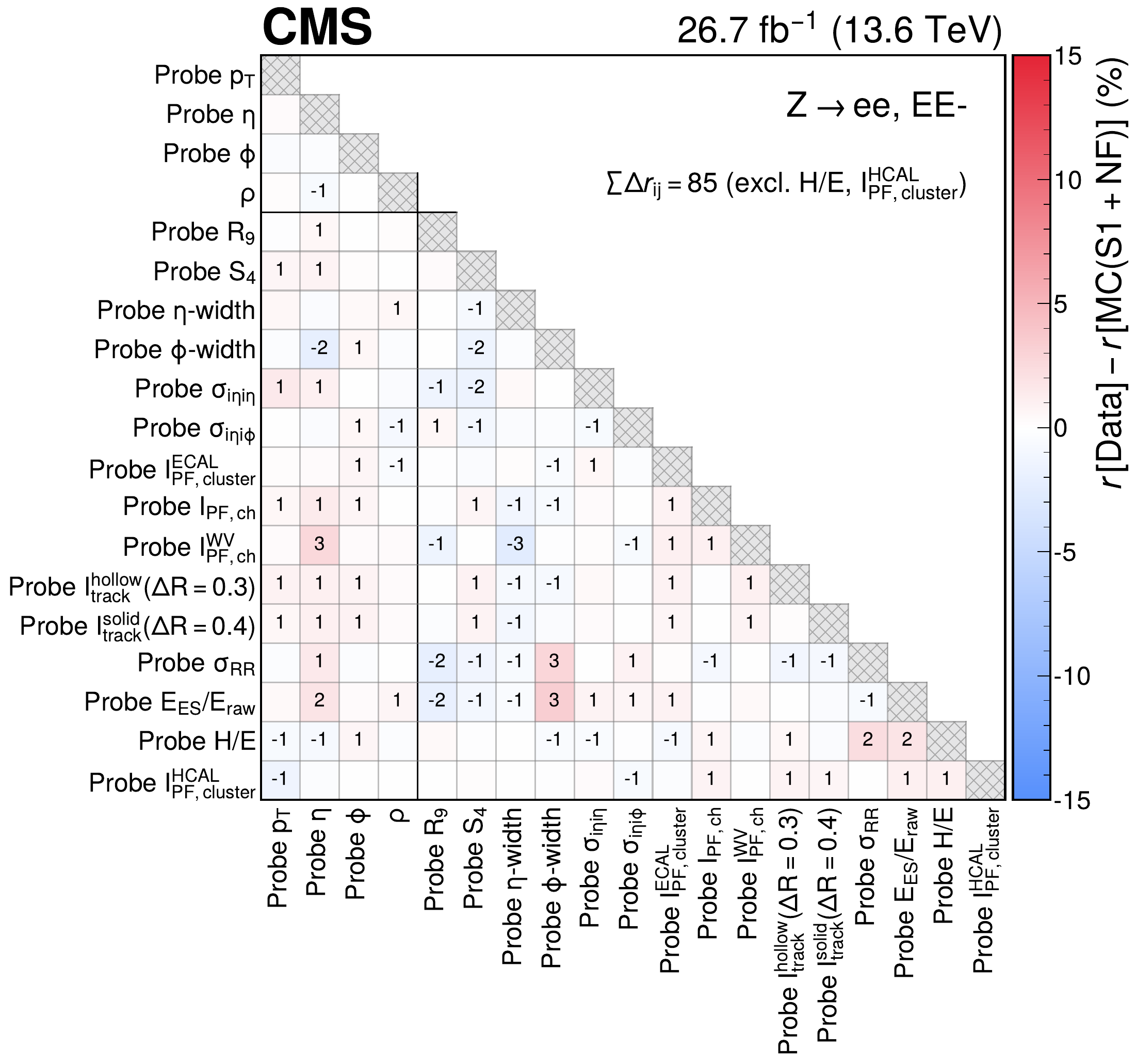}
    \caption{The upper left plot shows the Pearson correlation coefficients $r_{ij}$, between pairs of features in data for the EE$-$ subdetector. The other three plots show the difference in correlation coefficients $\Delta r_{ij}$ between data and simulation for three scenarios: the upper right plot corresponds to simulation after the initial S1 reweighting to align the kinematic distributions with data, while the lower left and lower right plots show the calibrated simulation after the S2 reweighting and the flow corrections are applied, respectively. All $r_{ij}$ and $\Delta r_{ij}$ values are rounded to the nearest percent, and values of absolute size less than 0.5\% are not shown. The sum of absolute $\Delta r_{ij}$ terms are shown in each of the relevant plots, where the sum excludes the probe H/E and $\mathcal{I}^{\mathrm{HCAL}}_{\text{PF, cluster}}$ features, since these are not included in the reweighting method. A solid line is used to separate the $x_k$ and $x_s$ features.}
    \label{fig:results_pearson_EEm}
\end{figure}

\begin{figure}[htb!]
    \centering
    \includegraphics[width=.49\linewidth]{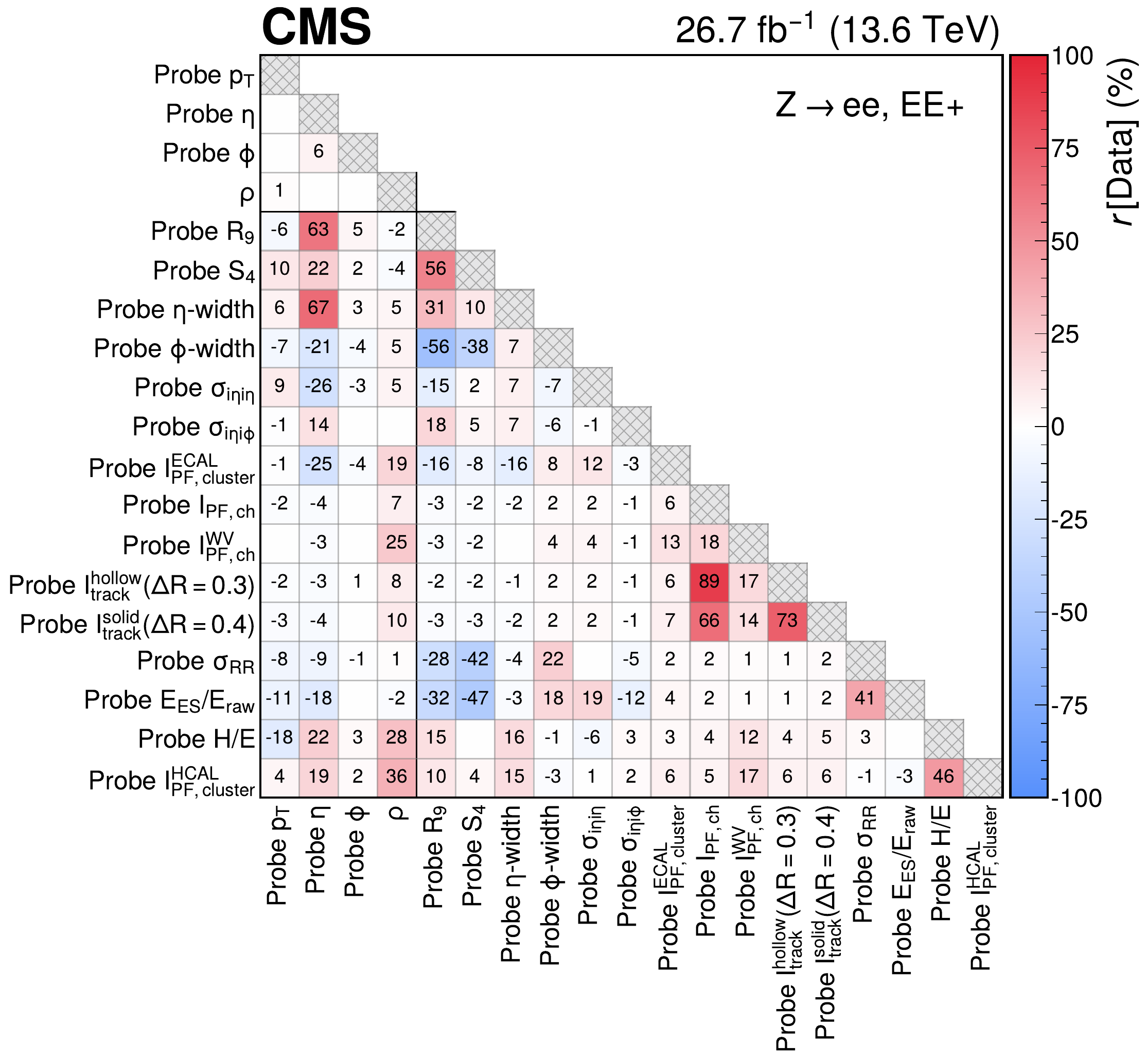}
    \includegraphics[width=.49\linewidth]{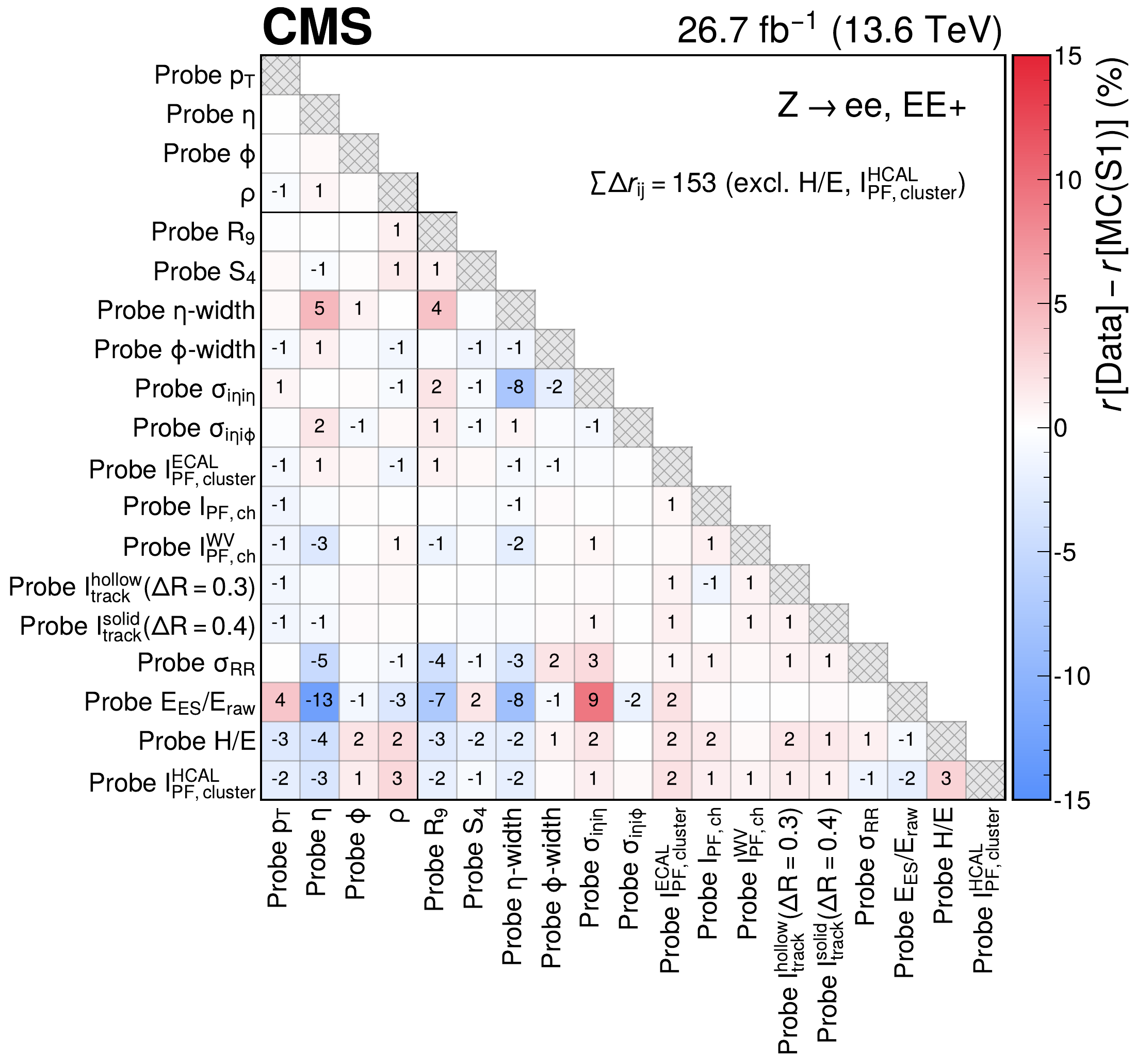}
    \includegraphics[width=.49\linewidth]{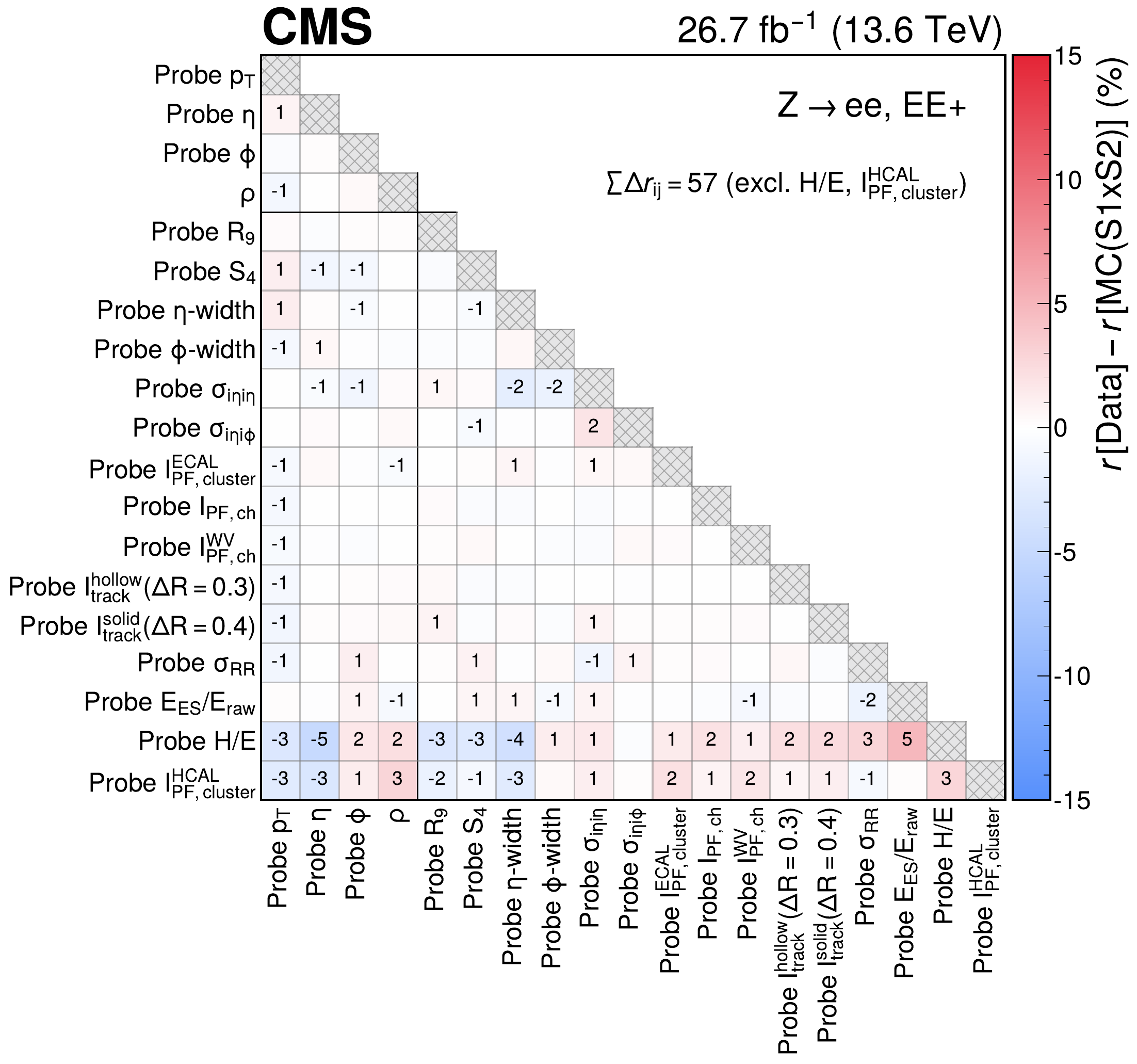}
    \includegraphics[width=.49\linewidth]{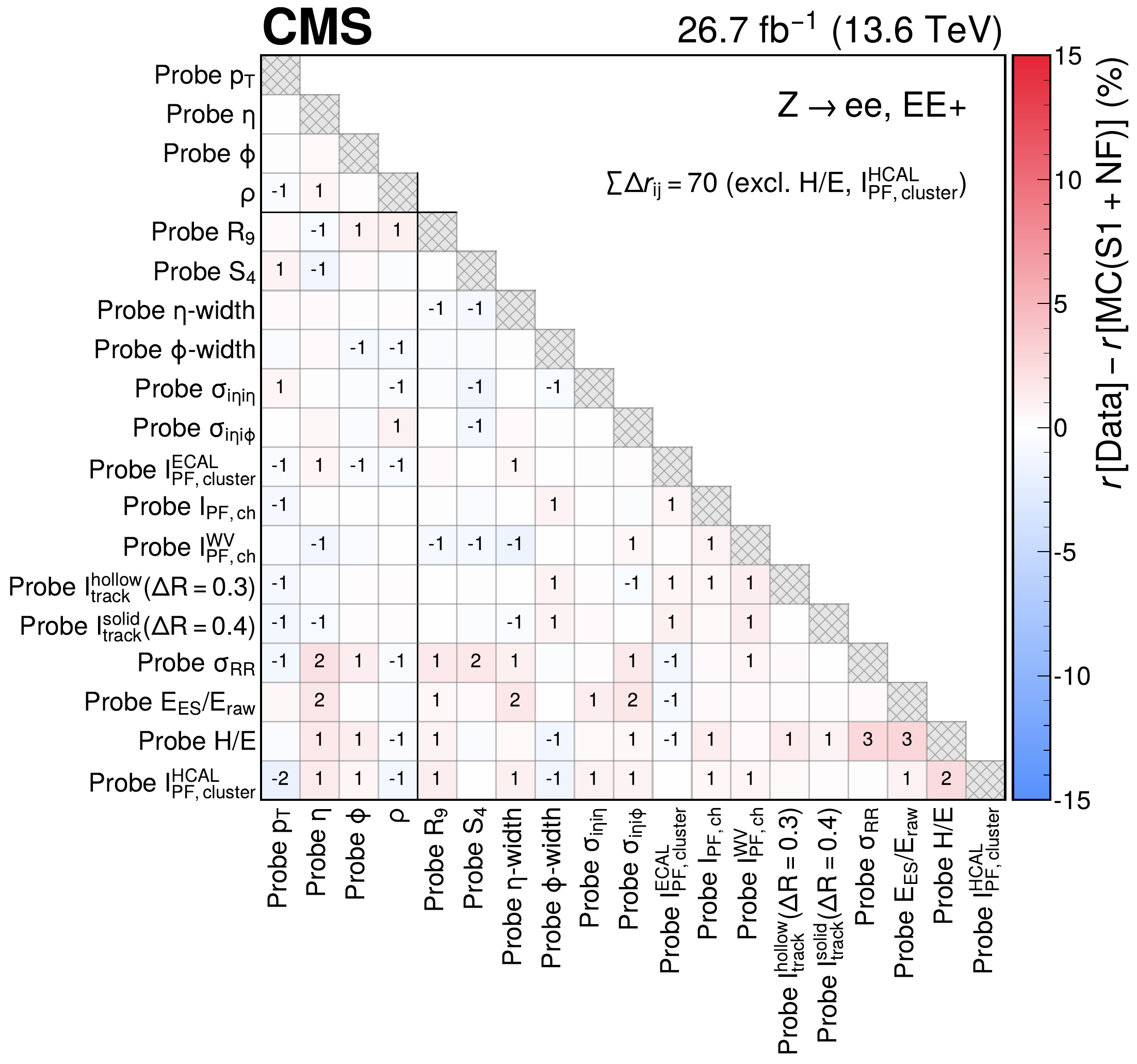}
    \caption{The upper left plot shows the Pearson correlation coefficients $r_{ij}$, between pairs of features in data for the EE$+$ subdetector. The other three plots show the difference in correlation coefficients $\Delta r_{ij}$ between data and simulation for three scenarios: the upper right plot corresponds to simulation after the initial S1 reweighting to align the kinematic distributions with data, while the lower left and lower right plots show the calibrated simulation after the S2 reweighting and the flow corrections are applied, respectively. All $r_{ij}$ and $\Delta r_{ij}$ values are rounded to the nearest percent, and values of absolute size less than 0.5\% are not shown. The sum of absolute $\Delta r_{ij}$ terms are shown in each of the relevant plots, where the sum excludes the probe H/E and $\mathcal{I}^{\mathrm{HCAL}}_{\text{PF, cluster}}$ features, since these are not included in the reweighting method. A solid line is used to separate the $x_k$ and $x_s$ features.}
    \label{fig:results_pearson_EEp}
\end{figure}

While the metrics discussed above provide a good indication of the agreement in 1D, 
they fail to capture the agreement in higher dimensions. 
Pearson correlation coefficients $r_{ij}$ are computed between all input features to assess whether the calibration methods can also capture the correlations in data correctly. 
Figures~\ref{fig:results_pearson_EB}--\ref{fig:results_pearson_EEp} show $r_{ij}$ in data, as well as the differences $\Delta r_{ij}$, between data and simulation, for each subdetector region. 
Results are presented for the three simulation scenarios discussed above: MC (S1), MC (S1$\times$S2), and MC (S1+NF).
A notable improvement is observed after the corrections are applied. 
To quantify this, the sum of absolute $\Delta r_{ij}$ values across all feature pairs is computed, excluding the probe H/E and $\mathcal{I}^{\mathrm{HCAL}}_{\text{PF, cluster}}$ features, since these are not included in the reweighting method, and would otherwise bias the metric. The $\abs{\Delta r_{ij}}$ sums decrease for the reweighting method (normalising-flow method) by factors of 3.6 (2.2), 1.9 (2.2), and 2.7 (2.2), in the EB, EE$-$, and EE$+$ subdetectors, respectively. This demonstrates that both calibration methods are capable of learning the complex correlation structure among input features in this application.

Four additional metrics are provided to quantify the higher-dimensional agreement between data and the corrected simulation samples. 
First, the sum of 1D $\chi^2/n_{\mathrm{dof}}$ over $x_k$ and over $x_s$ are provided in Table~\ref{tab:results_chi2}.
Although the sums do not account for correlations between features and therefore have limited statistical meaning, they nevertheless provide useful information as lower values  correspond to better overall agreement.
The sum of $\chi^2/n_{\mathrm{dof}}$ over $x_k$ serves as a global metric for the level of kinematic invariance achieved after applying the calibration methods, 
while the sum over $x_s$ quantifies the overall improvement in the shower shape and isolation features.
The H/E and $\mathcal{I}^{\mathrm{HCAL}}_{\text{PF, cluster}}$ features are omitted from the $x_s$ sum as these features were not included in the S2 reweighting, and would therefore skew the results. 
In a similar vein,
the SWD is also evaluated between data and simulation over $x_k$ ($x_s$), such that the EMD is calculated for 500 randomly sampled slices of the 4-dimensional (13-dimensional) feature space and averaged over. 
Again, the H/E and $\mathcal{I}^{\mathrm{HCAL}}_{\text{PF, cluster}}$ features are omitted from the SWD calculation over $x_s$. 
The SWD values, provided in Table~\ref{tab:results_emd}, offer a measure of the agreement that accounts for the correlations between features, and
does not depend on the inflated statistical uncertainties in the simulation after the S2 reweighting is applied.
Other metrics for assessing higher-dimensional agreement exist in the literature~\cite{Kansal:2022spb}; 
however, since the primary focus of this work is the calibration performance of the PID output, they are not considered here.

\begin{table*}[ht!]
    \centering
    \topcaption{A comparison of the 1D $\chi^2/n_{\mathrm{dof}}$ metric for the three scenarios: MC (S1), MC (S1$\times$S2) and MC (S1+NF). 
    The first four rows correspond to the kinematic and event-level features $x_k$, to highlight the level of kinematic invariance achieved with the S2 reweighting model. The normalising-flow method does not change the $x_k$ distributions, and therefore the values for S1 and S1+NF are identical. 
    The sum of $\chi^2/n_{\mathrm{dof}}$ values calculated over the $x_k$ features are also provided as a global metric for the level of kinematic invariance achieved.
    The following rows correspond to the shower shape and isolation features $x_s$, which show the improved agreement after the calibration methods are applied. 
    This is further quantified by the sum of $\chi^2/n_{\mathrm{dof}}$ values calculated over the $x_s$ features.
    The asterisk indicates that H/E and $\mathcal{I}^{\mathrm{HCAL}}_{\text{PF, cluster}}$ are omitted from the sums,
    as these features are not used as inputs into the S2 reweighting model.
    The final row shows the reduction in the 1D $\chi^2/n_{\mathrm{dof}}$ values for the PID output.
    Boldface indicates the best performing calibration method for each of the corrected $x_s$ features, the summed $x_s$ metric, and the PID output.
    We note that the $\chi^2/n_{\mathrm{dof}}$ values for the reweighting method (S1$\times$S2) include the inflated statistical uncertainties in the simulation,
    often leading to smaller values than the normalising-flow method (S1+NF).
    }
    \centering
    \renewcommand{\arraystretch}{1.4}
   \cmsTableAlways{
        \begin{tabular}{rccc{c}@{\hspace*{5pt}}ccc{c}@{\hspace*{5pt}}ccc}
             & \multicolumn{3}{c}{EB} && \multicolumn{3}{c}{EE+} && \multicolumn{3}{c}{EE-} \\
            & S1 & S1$\times$S2 & S1+NF && S1 & S1$\times$S2 & S1+NF && S1 & S1$\times$S2 & S1+NF \\
            \cline{2-4}\cline{6-8}\cline{10-12}
            Probe \pt & $1.2$ & $1.1$ & $1.2$ && $1.1$ & $1.9$ & $1.1$ && $1.0$ & $0.99$ & $1.0$ \\
            Probe $\phi$ & $1.6$ & $1.4$ & $1.6$ && $1.4$ & $1.7$ & $1.4$ && $1.1$ & $5.7$ & $1.1$ \\
            Probe $\eta$ & $0.55$ & $0.87$ & $0.55$ && $2.0$ & $2.2$ & $2.0$ && $1.7$ & $4.6$ & $1.7$ \\
            $\rho$ & $2.2$ & $2.3$ & $2.2$ && $0.70$ & $1.1$ & $0.70$ && $2.2$ & $2.7$ & $2.2$ \\[\cmsTabSkip]
            $\sum\chi^2/n_{\mathrm{dof}}$ over $x_k$ & $5.6$ & $5.8$ & $5.6$ && $5.2$ & $6.9$ & $5.2$ && $6.0$ & $14$ & $6.0$ \\[\cmsTabSkip]
            Probe $R_{9}$ & $124$ & \textbf{$2.1$} & $5.9$ && $70$ & \textbf{$1.6$} & $7.4$ && $54$ & \textbf{$3.0$} & $10$ \\
            Probe $S_{4}$ & $45$ & \textbf{$1.7$} & $11$ && $385$ & \textbf{$2.4$} & $2.9$ && $424$ & $5.1$ & \textbf{$3.8$} \\
            Probe $\eta$-width & $44$ & \textbf{$1.5$} & $22$ && $66$ & \textbf{$2.5$} & $2.6$ && $103$ & $4.3$ & \textbf{$4.3$} \\
            Probe $\phi$-width & $65$ & \textbf{$1.7$} & $6.7$ && $126$ & \textbf{$0.81$} & $2.6$ && $103$ & $5.3$ & \textbf{$3.9$} \\
            Probe $\sigma_{i{\eta}i\eta}$ & $176$ & \textbf{$4.1$} & $19$ && $90$ & \textbf{$1.2$} & $7.7$ && $148$ & \textbf{$5.8$} & $13$ \\
            Probe $\sigma_{i{\eta}i\phi}$ & $36$ & \textbf{$1.9$} & $21$ && $20$ & \textbf{$2.0$} & $2.3$ && $27$ & $9.7$ & \textbf{$3.7$} \\
            Probe H/E & $6302$ & $5937$ & \textbf{$14$} && $134$ & $64$ & \textbf{$6.5$} && $156$ & $97$ & \textbf{$9.2$} \\
            Probe $\mathcal{I}^{\mathrm{HCAL}}_{\text{PF, cluster}}$ & $8141$ & $7602$ & \textbf{$13$} && $95$ & $43$ & \textbf{$5.4$} && $142$ & $84$ & \textbf{$3.6$} \\
            Probe $\mathcal{I}_{\mathrm{PF,ch}}$ & $89$ & \textbf{$4.0$} & $9.9$ && $57$ & \textbf{$1.1$} & $5.7$ && $60$ & $3.5$ & \textbf{$3.1$} \\
            Probe $\mathcal{I}_{\mathrm{PF,ch}}^{\mathrm{WV}}$ & $174$ & \textbf{$2.7$} & $13$ && $15$ & \textbf{$0.86$} & $3.4$ && $16$ & $4.6$ & \textbf{$2.3$} \\
            Probe $\mathcal{I}^{\mathrm{ECAL}}_{\text{PF, cluster}}$ & $44$ & $1.9$ & \textbf{$1.2$} && $54$ & \textbf{$2.1$} & $2.5$ && $62$ & $4.4$ & \textbf{$1.5$} \\
            Probe $\mathcal{I}^{\text{hollow}}_{\text{track}}({\Delta}R=0.3)$ & $67$ & \textbf{$3.9$} & $7.5$ && $49$ & \textbf{$1.6$} & $7.5$ && $48$ & \textbf{$2.5$} & $3.7$ \\
            Probe $\mathcal{I}^{\text{solid}}_{\text{track}}({\Delta}R=0.4)$ & $84$ & \textbf{$4.1$} & $7.0$ && $57$ & \textbf{$1.4$} & $5.3$ && $59$ & \textbf{$3.2$} & $3.9$ \\
            Probe $\sigma_{RR}$ & \NA & \NA & \NA && $296$ & \textbf{$1.1$} & $11$ && $295$ & \textbf{$4.8$} & $17$ \\
            Probe $E_{\mathrm{ES}} / E_{\text{raw}}$ & \NA & \NA & \NA && $251$ & \textbf{$3.3$} & $6.6$ && $221$ & \textbf{$3.3$} & $8.7$ \\[\cmsTabSkip]
            $\sum\chi^2/n_{\mathrm{dof}}$ over $x_s$ (*) & $948$ & \textbf{$30$} & $124$ && $1537$ & \textbf{$22$} & $67$ && $1620$ & \textbf{$60$} & $79$ \\[\cmsTabSkip]
            Probe PID output & $284$ & \textbf{$9.1$} & $13$ && $94$ & \textbf{$0.48$} & $7.0$ && $81$ & \textbf{$1.1$} & $4.5$ \\
        \end{tabular}
    }
    \label{tab:results_chi2}
\end{table*}

\begin{table*}[ht!]
    \centering
    \topcaption{A comparison of the EMD metric for the three scenarios: MC (S1), MC (S1$\times$S2) and MC (S1+NF). 
    The EMD values are calculated for the normalised feature distributions, 
    and scaled by a factor of 100 for readability.
    The first four rows correspond to the kinematic and event-level features $x_k$, to highlight the level of kinematic invariance achieved with the S2 reweighting model. The normalising-flow method does not change the $x_k$ distributions, and therefore the values for S1 and S1+NF are identical. 
    The SWD values calculated for 500 randomly sampled slices of the $x_k$ feature space and averaged over are also provided as a global metric for the level of kinematic invariance achieved. 
    The following rows correspond to the shower shape and isolation features $x_s$, which show the improved agreement after the calibration methods are applied. 
    This is further quantified by the SWD values calculated over the $x_s$ feature space using the same number of slices as for $x_k$.
    The asterisk indicates that H/E and $\mathcal{I}^{\mathrm{HCAL}}_{\text{PF, cluster}}$ are omitted from the SWD calculations,
    as these features are not used as inputs into the S2 reweighting model.
    The final row shows the reduction in the EMD values for the PID output.
    Boldface indicates the best performing calibration method for each of the corrected $x_s$ features, the SWD over $x_s$ metric, and the PID output.
    }
    \centering
    \renewcommand{\arraystretch}{1.4}
   \cmsTableAlways{
        \begin{tabular}{rccc{c}@{\hspace*{5pt}}ccc{c}@{\hspace*{5pt}}ccc}
             & \multicolumn{3}{c}{EB} && \multicolumn{3}{c}{EE+} && \multicolumn{3}{c}{EE-} \\
            & S1 & S1$\times$S2 & S1+NF && S1 & S1$\times$S2 & S1+NF && S1 & S1$\times$S2 & S1+NF \\
            \cline{2-4}\cline{6-8}\cline{10-12}
            Probe \pt & $0.26$ & $0.20$ & $0.26$ && $0.65$ & $1.4$ & $0.65$ && $0.45$ & $0.75$ & $0.45$ \\
            Probe $\phi$ & $0.14$ & $0.17$ & $0.14$ && $0.65$ & $0.71$ & $0.65$ && $0.43$ & $0.68$ & $0.43$ \\
            Probe $\eta$ & $0.07$ & $0.11$ & $0.07$ && $0.47$ & $0.99$ & $0.47$ && $0.39$ & $1.0$ & $0.39$ \\
            $\rho$ & $0.19$ & $0.22$ & $0.19$ && $0.32$ & $0.79$ & $0.32$ && $1.0$ & $1.6$ & $1.0$ \\[\cmsTabSkip]
            SWD over $x_k$ & $0.20$ & $0.20$ & $0.20$ && $0.57$ & $0.90$ & $0.57$ && $0.56$ & $0.91$ & $0.56$ \\[\cmsTabSkip]
            Probe $R_{9}$ & $5.3$ & $0.59$ & \textbf{$0.29$} && $8.5$ & \textbf{$1.0$} & $1.5$ && $6.0$ & $1.2$ & \textbf{$1.2$} \\
            Probe $S_{4}$ & $3.7$ & $0.44$ & \textbf{$0.29$} && $20$ & $2.0$ & \textbf{$0.94$} && $20$ & $2.6$ & \textbf{$0.71$} \\
            Probe $\eta$-width & $1.2$ & \textbf{$0.13$} & $0.51$ && $8.6$ & $2.0$ & \textbf{$0.78$} && $11$ & $1.7$ & \textbf{$1.4$} \\
            Probe $\phi$-width & $4.5$ & \textbf{$0.45$} & $0.46$ && $10$ & $0.75$ & \textbf{$0.53$} && $7.3$ & $1.8$ & \textbf{$1.7$} \\
            Probe $\sigma_{i{\eta}i\eta}$ & $5.4$ & \textbf{$0.68$} & $0.95$ && $11$ & $1.0$ & \textbf{$0.98$} && $13$ & $2.6$ & \textbf{$1.5$} \\
            Probe $\sigma_{i{\eta}i\phi}$ & $3.2$ & \textbf{$0.37$} & $0.61$ && $4.0$ & $0.87$ & \textbf{$0.62$} && $4.5$ & $4.4$ & \textbf{$0.78$} \\
            Probe H/E & $35$ & $35$ & \textbf{$0.34$} && $14$ & $15$ & \textbf{$1.2$} && $15$ & $16$ & \textbf{$1.3$} \\
            Probe $\mathcal{I}^{\mathrm{HCAL}}_{\text{PF, cluster}}$ & $42$ & $41$ & \textbf{$0.75$} && $14$ & $13$ & \textbf{$1.2$} && $17$ & $18$ & \textbf{$1.0$} \\
            Probe $\mathcal{I}_{\mathrm{PF,ch}}$ & $3.1$ & \textbf{$0.71$} & $1.5$ && $4.6$ & \textbf{$0.89$} & $2.2$ && $4.4$ & $1.6$ & \textbf{$1.4$} \\
            Probe $\mathcal{I}_{\mathrm{PF,ch}}^{\mathrm{WV}}$ & $3.3$ & \textbf{$0.59$} & $0.67$ && $3.1$ & \textbf{$0.42$} & $1.1$ && $3.0$ & $1.7$ & \textbf{$1.00$} \\
            Probe $\mathcal{I}^{\mathrm{ECAL}}_{\text{PF, cluster}}$ & $2.6$ & $0.39$ & \textbf{$0.35$} && $4.8$ & $2.3$ & \textbf{$1.3$} && $5.6$ & $1.9$ & \textbf{$0.95$} \\
            Probe $\mathcal{I}^{\text{hollow}}_{\text{track}}({\Delta}R=0.3)$ & $1.4$ & \textbf{$0.57$} & $0.60$ && $0.85$ & \textbf{$0.39$} & $0.50$ && $1.9$ & \textbf{$0.86$} & $0.92$ \\
            Probe $\mathcal{I}^{\text{solid}}_{\text{track}}({\Delta}R=0.4)$ & $0.26$ & \textbf{$0.14$} & $0.19$ && $1.5$ & $1.2$ & \textbf{$1.1$} && $2.6$ & $1.8$ & \textbf{$1.6$} \\
            Probe $\sigma_{\mathrm{RR}}$ & \NA & \NA & \NA && $9.0$ & \textbf{$0.69$} & $1.4$ && $8.0$ & $1.5$ & \textbf{$1.2$} \\
            Probe $E_{\mathrm{ES}} / E_{\text{raw}}$ & \NA & \NA & \NA && $8.7$ & $1.9$ & \textbf{$1.5$} && $7.4$ & \textbf{$1.1$} & $1.8$ \\[\cmsTabSkip]
            SWD over $x_s$ (*) & $4.2$ & \textbf{$0.49$} & $0.69$ && $7.4$ & $1.4$ & \textbf{$1.1$} && $7.3$ & $1.8$ & \textbf{$1.3$} \\[\cmsTabSkip]
            Probe PID output & $8.9$ & $0.56$ & \textbf{$0.48$} && $10$ & \textbf{$0.22$} & $1.4$ && $9.1$ & \textbf{$0.55$} & $0.77$ \\
        \end{tabular}
    }
    \label{tab:results_emd}
\end{table*}

\clearpage
\newpage
\subsection{Calibration of the PID output}
\begin{figure}[htb!]
    \centering
    \includegraphics[width=.325\linewidth]{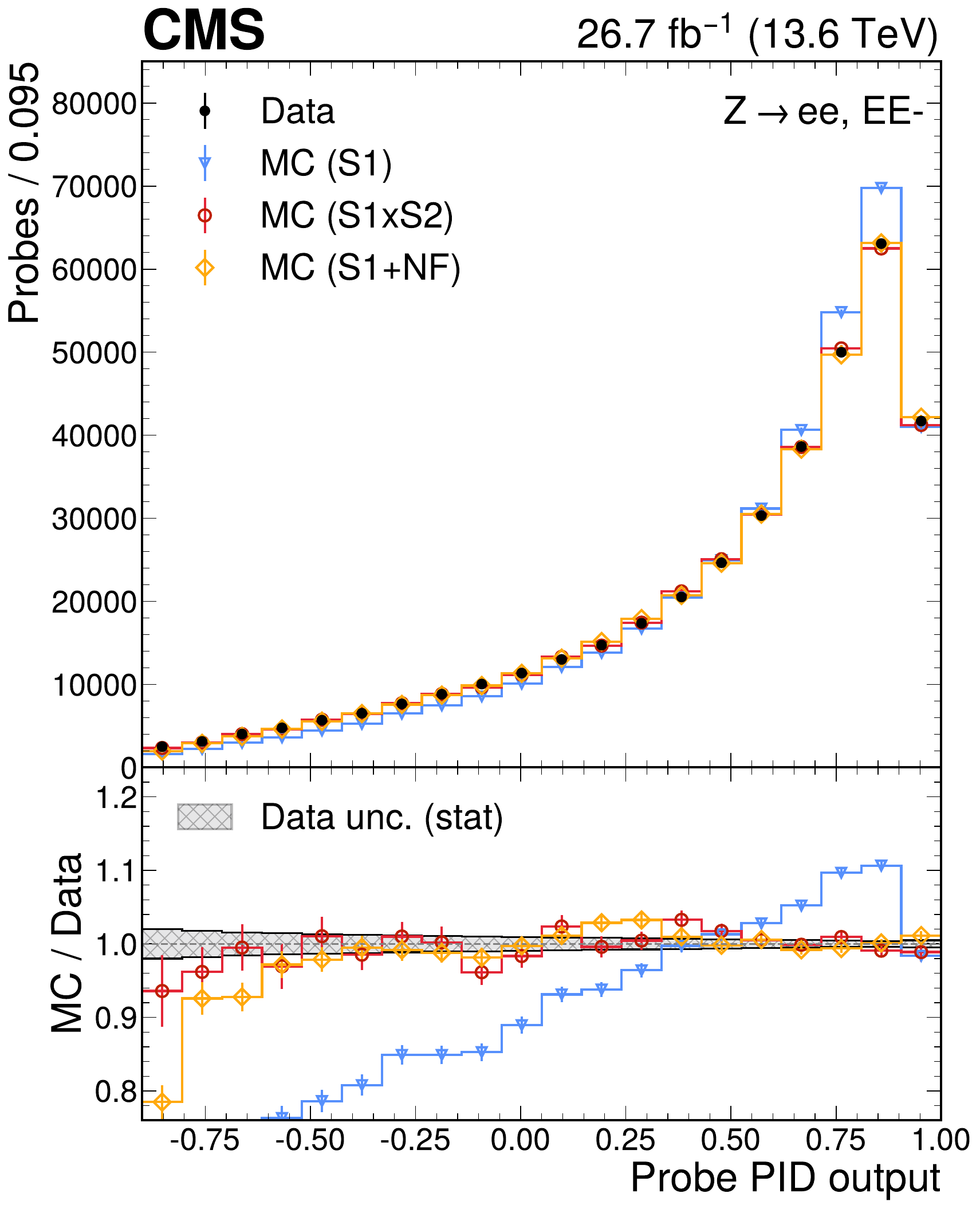}
    \includegraphics[width=.325\linewidth]{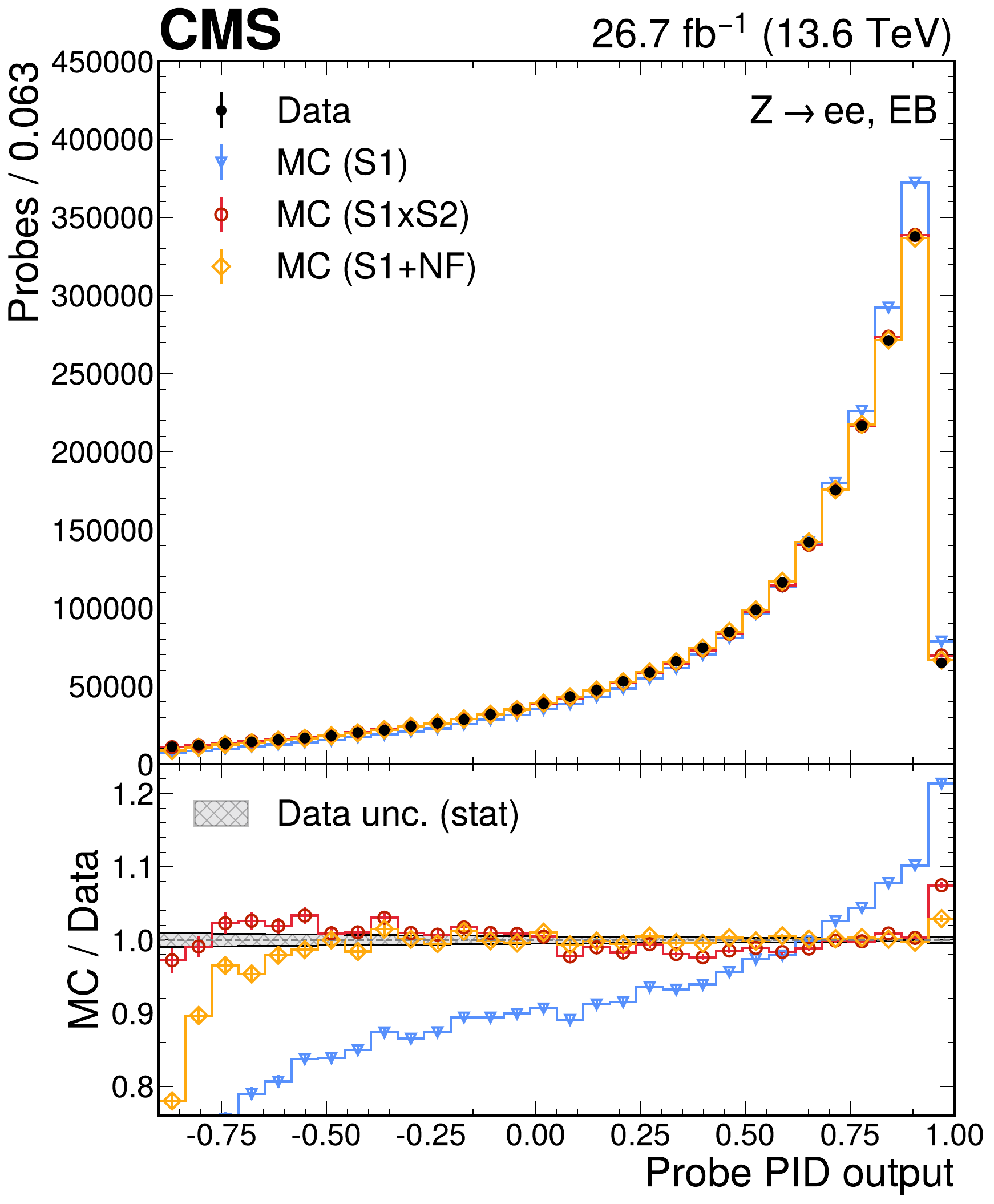}
    \includegraphics[width=.325\linewidth]{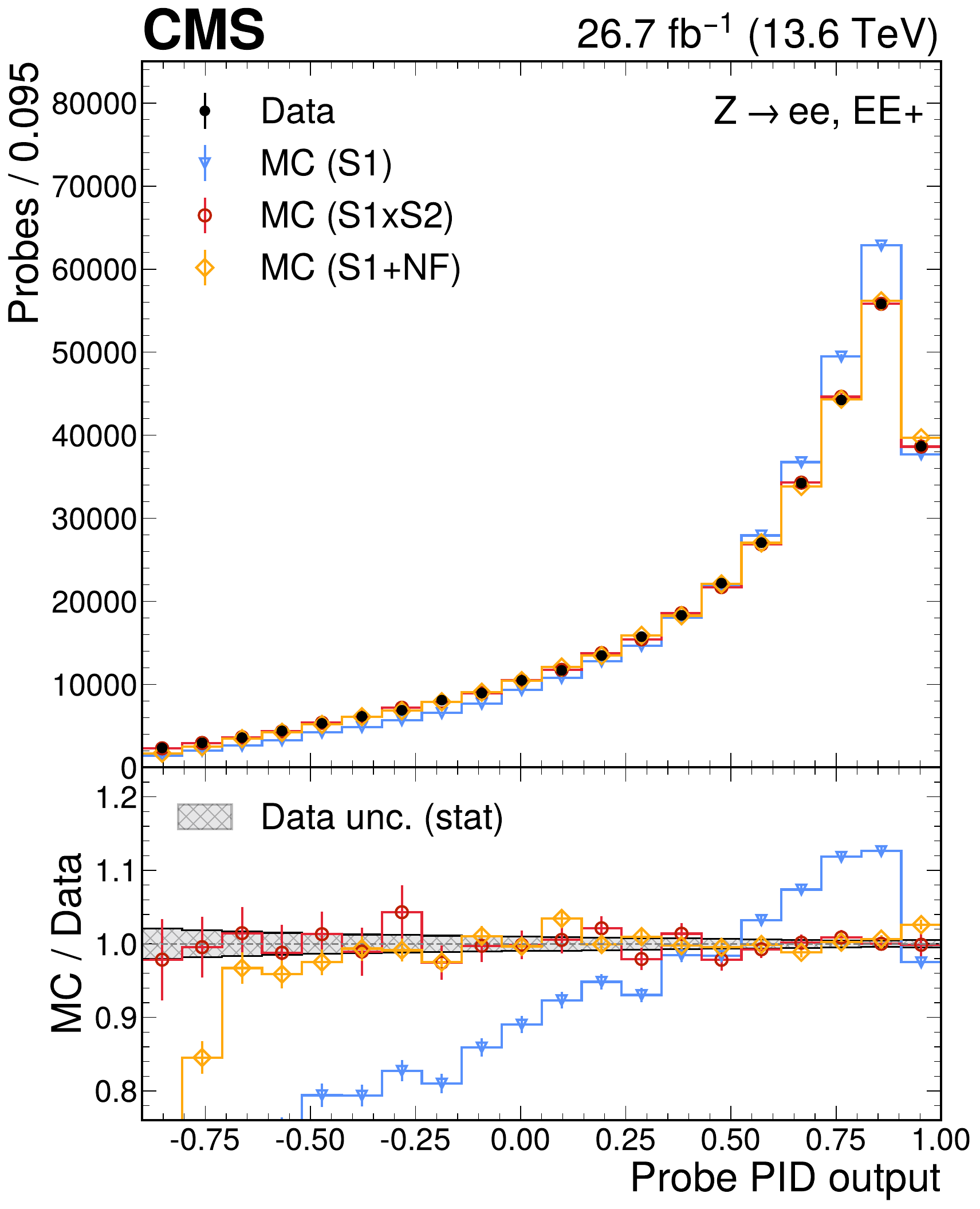}
    \caption{The probe PID output distributions in data (black points) and simulation (coloured histograms with points),
    for the EE$-$ (left), EB (centre), and EE$+$ (right) subdetector regions. 
    The statistical uncertainties in the simulation and data are shown by the error bars on the points.
    The simulation is shown for three scenarios. The blue histograms show the simulation after the initial S1 reweighting to align the kinematic distributions with data. 
    The red and orange histograms show the calibrated simulation after the S2 reweighting and the flow corrections are applied, respectively. 
    The lower panels show the ratio of simulation to data, where the statistical uncertainty in the data is indicated by the grey hatched band.}   
    \label{fig:results_id}
\end{figure}

The ultimate aim of the calibration methods is to improve the agreement in the PID output distribution. 
Since the PID output is a complex, non-linear function of the input features,
it serves as a useful summary statistic for assessing the high-dimensional agreement between data and the corrected simulation samples.
To obtain the corrected PID output, the normalising-flow corrections must be propagated through by re-evaluating the PID on the corrected input features. 
This is not required in the reweighting method as the weight corrections apply globally to the event.
Figure~\ref{fig:results_id} shows the PID output distributions for data and simulation, where simulation is shown for the same three scenarios described above. 
After applying the calibration methods, the agreement between data and simulation is significantly improved.
This can be seen by the drastic reduction in the $\chi^2/n_{\mathrm{dof}}$ and EMD values for the PID output in Tables~\ref{tab:results_chi2} and \ref{tab:results_emd}, respectively.

\begin{figure}[htb!]
    \centering
    \includegraphics[width=.49\linewidth]{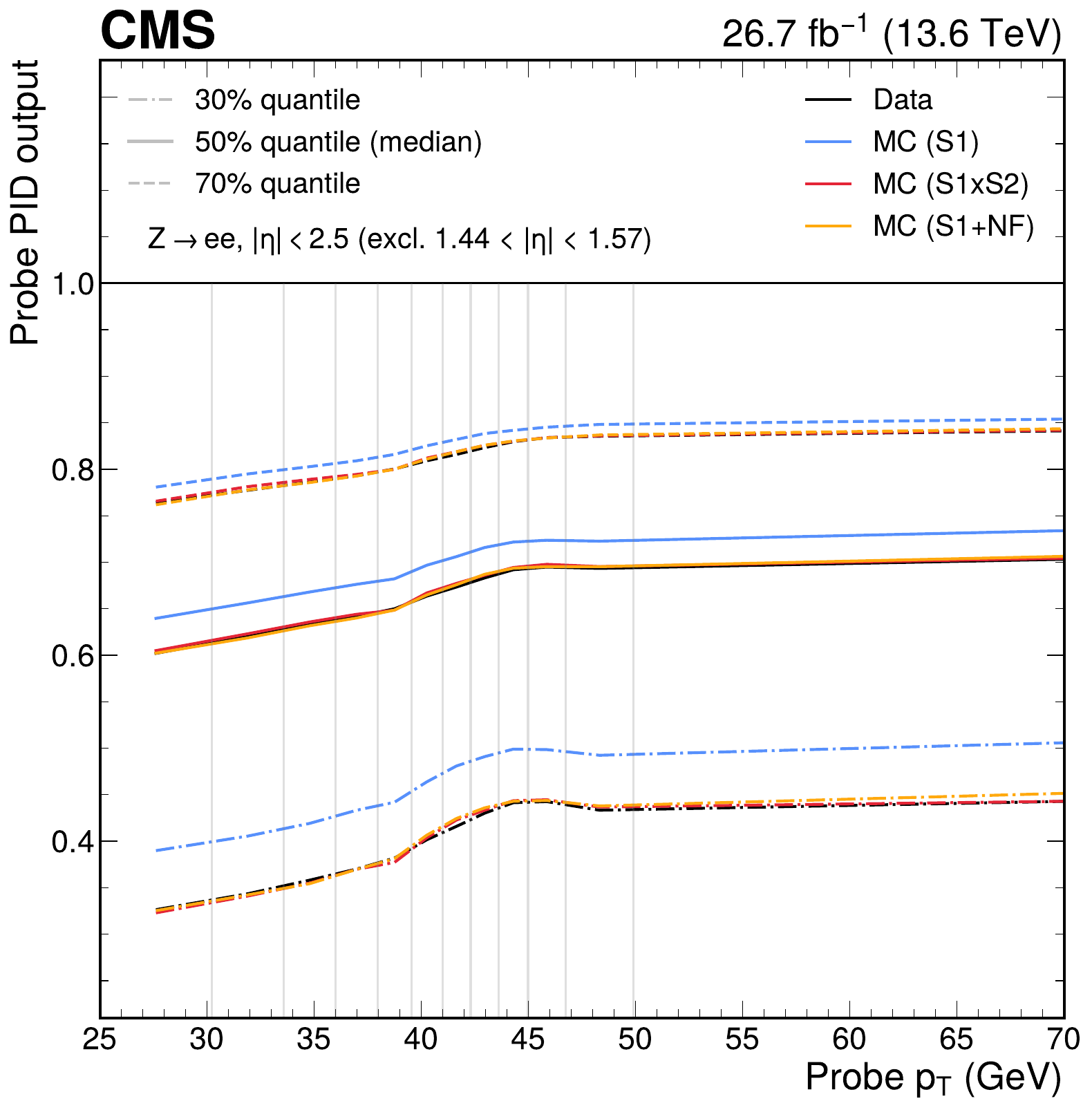}
    \includegraphics[width=.49\linewidth]{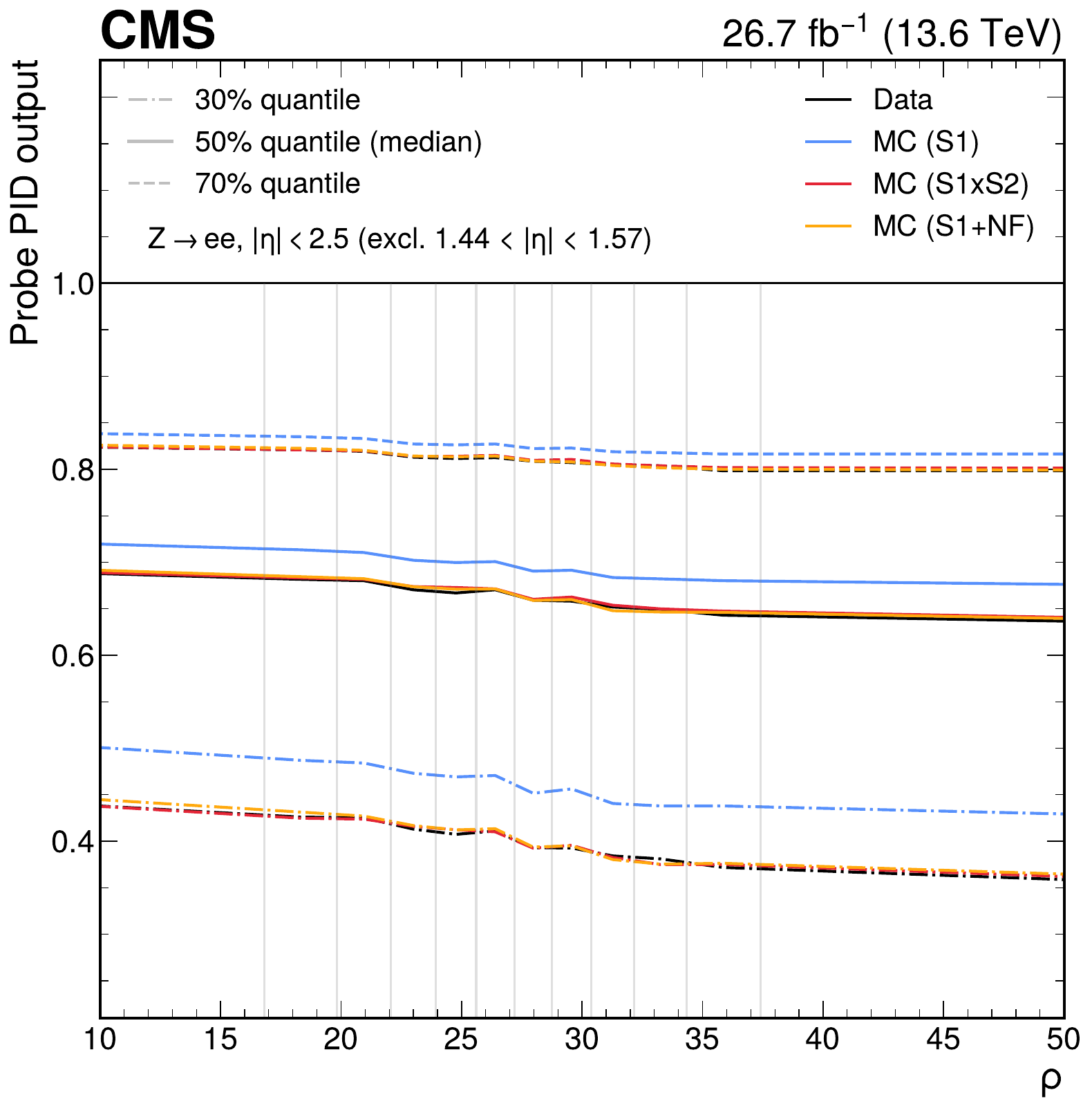}
    \includegraphics[width=.49\linewidth]{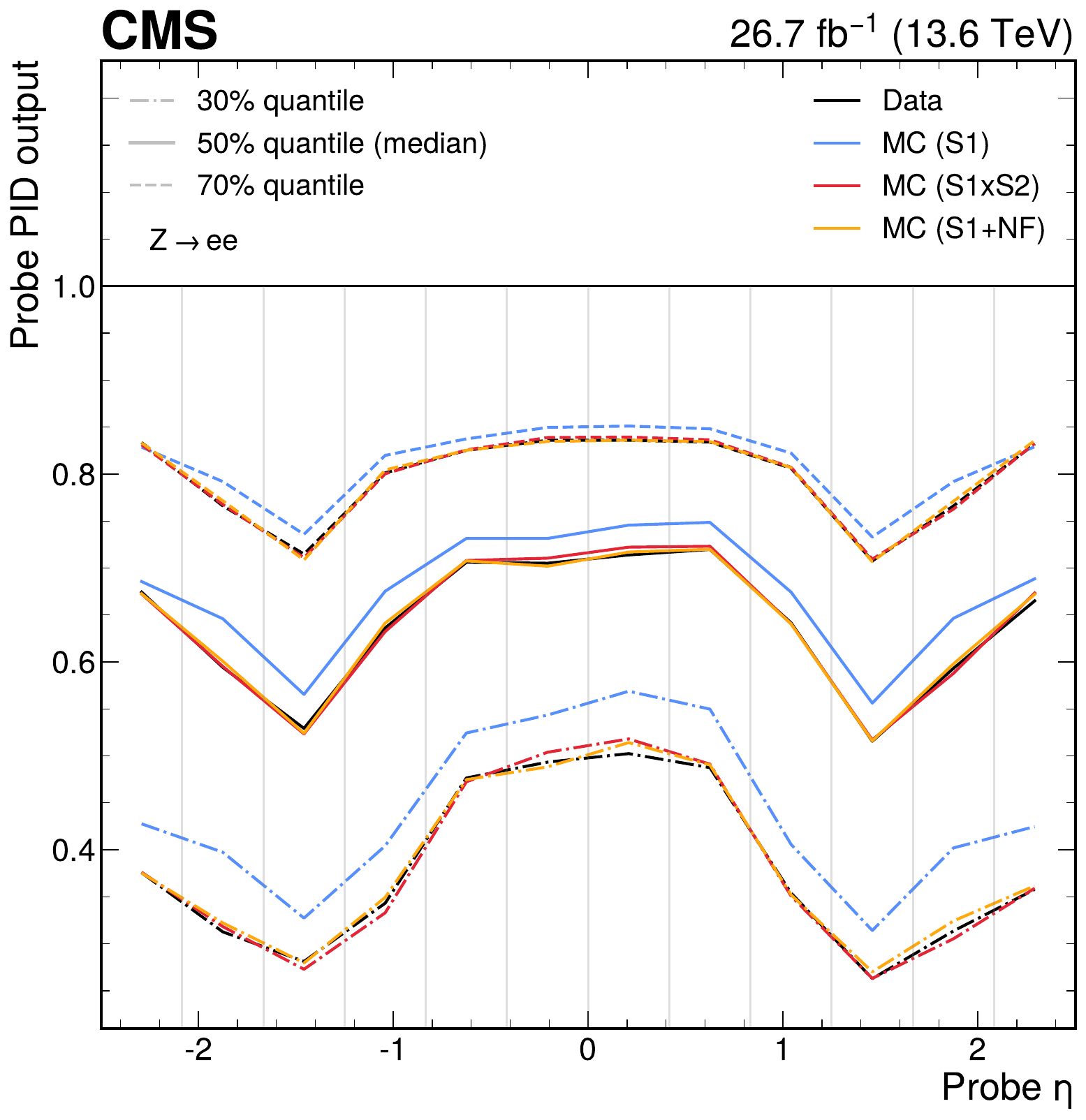}
    \includegraphics[width=.49\linewidth]{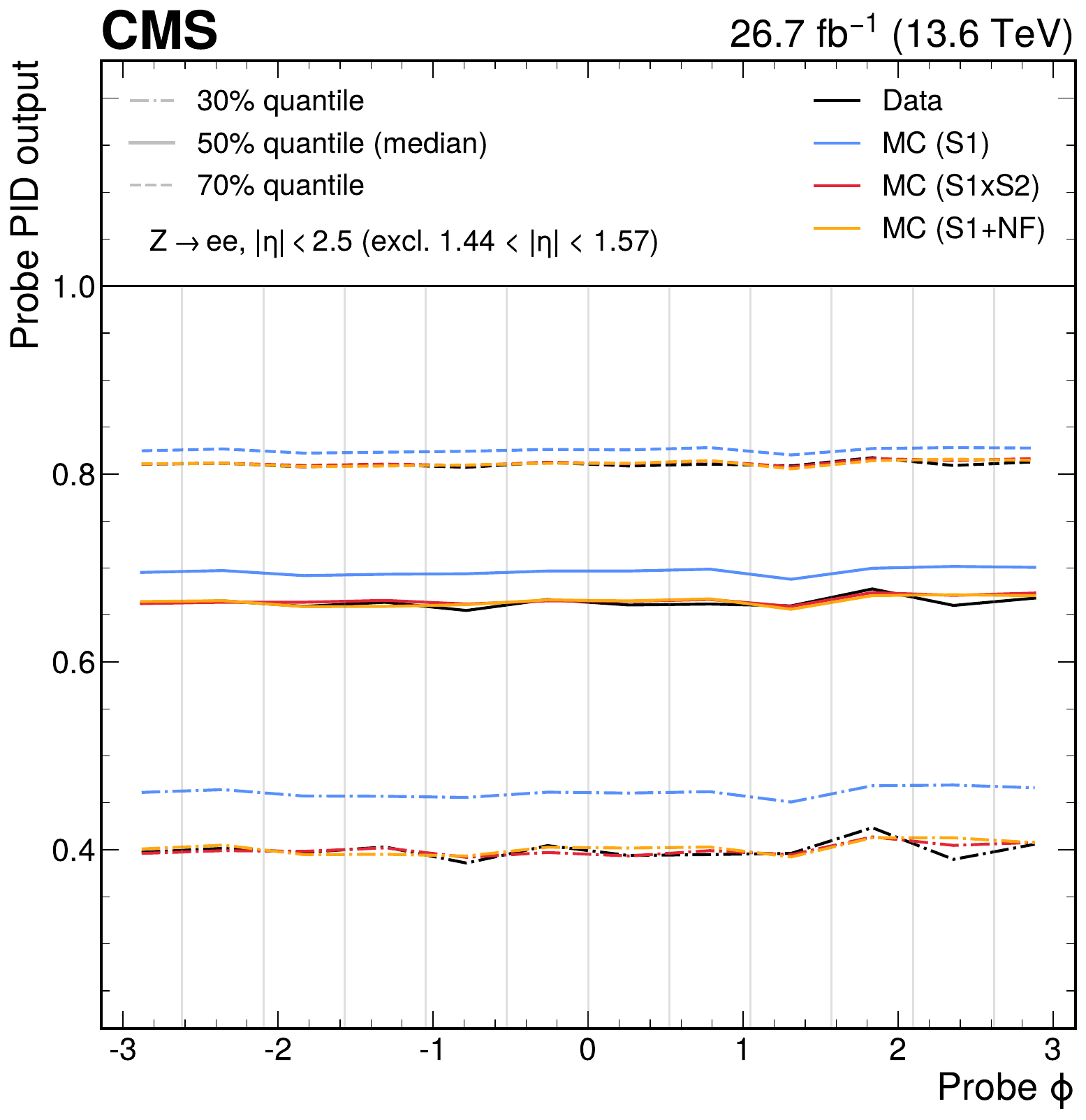}
    \caption{Quantiles of the PID output distribution as a function of the $x_k$ features: $\pt$ (upper left), $\rho$ (upper right), $\eta$ (lower left), and $\phi$ (lower right). The probe EM showers are divided into 12 bins for each feature. For $\pt$ and $\rho$, bin edges are chosen such that each bin contains an equal number of events in data, while for $\eta$ and $\phi$ the bins are linearly spaced. Within each bin, the 30\%, 50\%, and 70\% quantiles of the PID output are shown as dot-dashed, solid, and dashed lines, respectively. The quantiles are shown for data (black), simulation after the initial S1 reweighting (blue), and the calibrated simulation after applying the S2 reweighting (red) or the normalising flow (orange). The quantiles for $\pt$, $\rho$, and $\phi$ are computed inclusively in $\eta$, across all subdetector regions.}
    \label{fig:results_qprofile}
\end{figure}

Finally, the calibration performance is studied as a function of the kinematic and event-level features $x_k$, 
to ensure that the corrections are successful across the full range. 
This is important when the corrections are applied to an independent sample with a different distribution over $x_k$. 
The probes from the test data set are partitioned into 12 bins in each of the four $x_k$ features: $\pt$, $\eta$, $\phi$, and $\rho$. 
For $\pt$ and $\rho$, bin edges are chosen such that each bin contains an equal number of events in data, 
while for $\eta$ and $\phi$ the bins are linearly spaced. 
Within each bin, the 30\%, 50\% and 70\% quantiles of the PID output distribution are computed. 
Figure~\ref{fig:results_qprofile} compares the quantiles for data, and the nominal and corrected simulation samples, as functions of the four $x_k$ features. 
After the calibration methods are applied, the quantile curves are in good agreement, demonstrating that the methods are successful over the full range of $x_k$.

\clearpage
\newpage

\newpage
\clearpage
\section{Discussion of the methods}\label{sec:discussion}
This paper has presented two novel ML calibration methods, which operate by morphing the multidimensional distribution from simulation to that observed in data. 
Both methods are shown to perform well,
significantly improving the agreement in the PID output distribution (as shown in Fig.~\ref{fig:results_id}).
Nevertheless, the two methods are based on fundamentally different principles: the reweighting method exemplifies vertical morphing, modifying the per-event probability via weight corrections, whereas the normalising flow method is a horizontal morphing technique that shifts the values of the quantities of interest. 
As a result, each method requires different considerations and may be optimal under different conditions in physics analyses.
In this section, we compare the strengths and limitations of these methods against more traditional calibration approaches, as well as against each other. 

\begin{itemize}
    \item \textit{Corrections in a high-dimensional feature space}: traditional calibration methods, such as binned scale factors, are typically limited to a low-dimensional feature space by the finite number of samples available for their derivation. After applying binned scale factors, the improved agreement between data and simulation is limited to the few dimensions. Both methods presented in this paper employ ML to correct a high-dimensional feature space. This simultaneous correction ensures that any derived quantity based on the input features (\eg PID output) is also corrected. Furthermore, the two methods use a single ML model in the calibration procedure, which is trained on the full feature space concurrently. This drastically simplifies the model training compared to previous ML calibration approaches, such as chained quantile regression~\cite{CMS:2022wpo}, which required training a large number of BDTs.
    
    \item \textit{Continuous (unbinned) corrections}: the use of ML removes the need for binning the samples, instead providing a correction that is continuous across the input feature space. Consequently, the full shape of the multivariate input feature distribution (along with any derived feature distribution) is corrected. By accurately modelling the shape of the PID output distribution, an analysis is no longer constrained to fixed working points with dedicated efficiency scale factors. Moreover, the PID output can be reliably propagated through the analysis chain and used, for example, as an input feature to an event-level classifier designed to distinguish signal from background. 

    \item \textit{Sample normalisation}: the reweighting method modifies the per-event probability and can therefore affect the overall normalisation of an event sample. When the S2 reweighting function is applied to the same phase space as used for training (\eg to the test data set), the overall normalisation is found to change by less than 1\% across all subdetector regions. However, when applied to an independent data set with a different selection phase space, a change in the normalisation is expected. In principle, this should bring the normalisation of the simulation sample in the selected phase space closer to data. We note that the selection phase space should always be a subset of the full training phase space, ensuring that the algorithm is not applied in regions unseen during training. 
    In contrast, the normalising flow method preserves the sample normalisation. Nevertheless, it must not be applied after selection criteria have been imposed on any of the corrected features, since events may migrate across the selection boundaries and thereby alter the post-selection normalisation. 
    
    For both methods, we recommend to still derive data-simulation efficiency scale factors for any selection criteria on the corrected features. This ensures that the sample normalisation is accurately modelled, with the resulting efficiency scale factors expected to be significantly closer to unity than those obtained without calibration.

    \item \textit{Kinematic invariance}: similarly, modifications to the per-event probabilities via the reweighting method can induce distortions in the $x_k$ distributions. The two-stage training procedure helps to mitigate this effect by learning the conditional density ratio, albeit some per-cent level kinematic variations remain (as shown in Figs.~\ref{fig:results_kinematic_pt_rho} and~\ref{fig:results_kinematic_eta_phi}). If an analysis is sensitive to these kinematic variations, a dedicated systematic uncertainty should be introduced to account for the shape distortions of $p_{\mathrm{MC}}(x_k)$. Alternatively, one may reweight the corrected simulation in $x_k$ to recover the nominal $p_{\mathrm{MC}}(x_k)$, although this may degrade the performance of the calibration. The normalising flow method does not have this issue, as $x_k$ are provided as conditional features and therefore remain unchanged.
    
    \item \textit{Preserving physical laws among features}: since the normalising flow alters the values of the input features, it may break underlying physical constraints. This would pose a more significant concern if the method were applied to correct particle four-momentum vectors, as energy-momentum conservation laws may be broken. For the application in this paper, however, the calibrated features are related to the shower formation and detector modelling, so this property of the method is less critical. In contrast, the reweighting method inherently preserves physical laws, as the input feature values remain unchanged for each event. Moreover, quantities derived from, or even correlated with, the input features are implicitly corrected (to some extent) in the reweighting method, since the weight correction is applied globally at the event level.
    
    \item \textit{Poor overlap between distributions}: a common challenge for all reweighting procedures is a limited overlap between the two pdfs being compared. The ML reweighting method detailed in this paper is no exception; in such scenarios the obtained weight corrections can deviate significantly from unity. This, in turn, reduces the statistical power of the simulation sample, which can be problematic if MC statistical uncertainties are dominant in an analysis. It is this feature of the method that led to the omission of H/E and $\mathcal{I}^{\mathrm{HCAL}}_{\text{PF, cluster}}$ in the model training. Whilst this choice helps preserve the statistical power of the sample, these variables remain uncorrected, which may result in residual discrepancies in the PID output distribution. This issue is also particularly pronounced in the tails of distributions. Nevertheless, as discussed in Section~\ref{sec:results_kinematic}, this limitation of the reweighting method can (if handled carefully) be turned into a feature: the inflated MC statistical uncertainties that are propagated through the analysis coincide with regions of feature space with poor overlap. 

    Another challenge for the reweighting method arises from inactive detector regions, which can introduce ``holes" in the training feature space (as discussed in Section~\ref{sec:reweighting}). 
    A misalignment between these regions in data and simulation can lead to fluctuations in the learned weight corrections near the hole boundaries.
    For the application considered in this paper, the reweighting method was found to be sufficiently robust, with a negligible impact on the overall calibration performance. However, in other applications, inactive detector regions may require additional treatment to ensure the robustness of the method.
    Further studies are needed to determine an optimal strategy for handling such regions, which is left to future work.

    \item \textit{Truncation of corrections}: as discussed in Section~\ref{sec:reweighting}, a weight-clipping procedure was introduced to the reweighting method to prevent extremely large weight corrections. The hyperparameter $w_{\text{ceil}}$ should be chosen according to the requirements of the analysis: smaller values help preserve the statistical power of the simulation sample, but worsen the data-simulation agreement. A similar clipping technique could, in principle, be applied in the normalising flow method to prevent mapping events across large distances in the multidimensional feature space, although this was not required in this application. 
    
    \item \textit{Unphysical regions}: certain regions of the feature space are unphysical and are not populated by either simulation or data, \eg the gap between the tail and peak of an isolation-related feature. Since the normalising flow method modifies the feature values, it can occasionally map events into such unphysical regions. Depending on the use case, it may be necessary to implement a post-calibration cleaning procedure to prevent events from residing in the unphysical regions. In this application, using the probe $\mathcal{I}^{\mathrm{ECAL}}_{\text{PF, cluster}}$ distribution as an illustrative example, only $0.015\%$ of the events were mapped into such regions. Given the negligible impact of this fraction on the overall analysis, no post-calibration cleaning procedure was implemented.

    \item \textit{Propagating corrections through derived features}: this occurs naturally with the reweighting method as the weight corrections are applied globally at the event level. In contrast, when using the normalising flow method, derived features such as the PID must be reevaluated using the corrected input features. As a result, the reweighting method can be applied after using the input features in an analysis, \eg for selection criteria or classifier training, whereas the normalising flow method must be applied beforehand.
    
    \item \textit{Treatment of backgrounds}: traditional calibration techniques, such as binned scale factors, account for background contributions through subtraction methods. These backgrounds are typically estimated using either parametric fits (as in the derivation of electron reconstruction scale factors in Ref.~\cite{CMS:2020uim}) or from simulated templates or templates derived from control samples in data (as used in the derivation of hadronically decaying $\tau$ lepton identification scale factors in Ref.~\cite{CMS:2018jrd}).
    Background subtraction in the context of ML-based calibration methods is more challenging, as these algorithms operate directly on event-level data rather than on aggregated distributions. As discussed in Section~\ref{sec:selection}, the data sample used in this application is extremely pure in \zee decays, allowing background contributions to be safely neglected. However, other reconstructed objects in CMS that require calibration often lack access to such pure control regions in data. In these cases, sophisticated background subtraction techniques within the model training become necessary. 
    A detailed investigation of these techniques is left to future work.
\end{itemize}

The properties outlined above illustrate key considerations when selecting a calibration method for a given analysis. Due to the complementary strengths and limitations of each approach, the optimal choice depends on the specific requirements of the application. If needed, the methods can be used in combination to ensure that each feature is calibrated using the most appropriate technique.

One of the key benefits of both calibration methods is their ability to identify aspects of the simulation that are poorly modelled.
This allows analysts to direct their attention, diagnose the underlying issues, and ultimately improve the simulation for future iterations.
In the reweighting method, large weight corrections highlight regions of the feature space that are poorly modelled.
Additionally, the feature importance ranking of the classifier can be used to identify the features driving the discrepancies between data and simulation.
For the normalising flow approach, the magnitude of the learned transformation provides a direct measure of mismodelling.
The studies presented in this paper helped reveal several issues in CMS simulation,
most notably in the mismodelling of the HCAL noise thresholds that led to significant discrepancies in the H/E and $\mathcal{I}^{\mathrm{HCAL}}_{\text{PF, cluster}}$ distributions.
These issues were subsequently fixed for the next iteration of the simulation.

Before concluding, we outline how these calibration methods should be integrated into a physics analysis. For the reweighting method, only the S2 function $w_{\mathrm{S2}}(x)$ is applied, serving as a multiplicative weight correction for each object to be calibrated within an event. Given the statistical independence of the shower shape and isolation features between different EM showers, it is valid to compute the total event weight correction as the product of individual object weight corrections. Similarly, in the case of the normalising flow method, the calibrated features of all relevant objects should be shifted according to the learned transformation. 

Although the methods yield significantly improved agreement between data and simulation (as shown in Fig.~\ref{fig:results_id}), residual discrepancies remain and should be accounted for by introducing systematic uncertainties if an analysis is sensitive to them. 
Importantly, these uncertainties are significantly smaller than the initial disagreement between data and the uncalibrated simulation (as shown in Fig.~\ref{fig:selection}, right), underscoring the potential of ML-based calibrations to enhance the precision of physics measurements.

\section{Summary}\label{sec:summary}
Particle physics measurements depend on comparisons between Monte Carlo simulations and data. Discrepancies arising from imperfect simulation can introduce biases, which are accounted for by assigning systematic uncertainties to cover the mismodelling effects. 
Improving the calibration of simulation can help mitigate these biases, thereby reducing the associated systematic uncertainties and enhancing the precision of measurements and the sensitivity of searches for new physics.

In this paper, two machine-learning (ML) based calibration methods are presented. The methods are employed to correct the features of simulated electromagnetic (EM) showers, as measured with the CMS detector. A subset of the proton-proton collision data collected by the CMS experiment in 2022 at $\sqrt{s}=13.6$\TeV is used, corresponding to an integrated luminosity of 26.7\fbinv. The tag-and-probe technique is applied to \zee decays to obtain a pure sample of EM showers in both simulation and data. The probe EM showers are then used to train the ML calibration methods to correct a high-dimensional feature space characterising the shower shape and isolation properties. The corrections are conditioned on the kinematic properties of the EM shower, as well as a variable sensitive to the pileup conditions. The final performance is evaluated by comparing the agreement between simulation and data in the output of a particle identification algorithm based on the corrected features.

The two methods follow fundamentally different approaches. The reweighting method is an example of vertical morphing. An ML classifier is trained to distinguish between samples drawn from simulation and data, effectively learning the ratio of conditional densities between the two classes. The classifier output is then translated to a weight, which acts as a multiplicative correction to the total event weight. In contrast, the normalising flow method demonstrates horizontal morphing. A high-dimensional transformation is learned that directly modifies the input feature values of the simulation to match the data distribution.
Each method offers different advantages and limitations, which are important to consider when applying them in particle physics analyses.

Both methods offer a significant advantage over traditional calibration techniques by providing continuous corrections in high-dimensional feature spaces. 
The ML methods demonstrate excellent performance in correcting the input features, as well as the particle identification output distribution.
This work documents the first application of these advanced ML calibration techniques to collision data in CMS, and serves as both motivation and guidance for future applications of these methods in particle physics and related fields.

\begin{acknowledgments}
We congratulate our colleagues in the CERN accelerator departments for the excellent performance of the LHC and thank the technical and administrative staffs at CERN and at other CMS institutes for their contributions to the success of the CMS effort. In addition, we gratefully acknowledge the computing centres and personnel of the Worldwide LHC Computing Grid and other centres for delivering so effectively the computing infrastructure essential to our analyses. Finally, we acknowledge the enduring support for the construction and operation of the LHC, the CMS detector, and the supporting computing infrastructure provided by the following funding agencies: SC (Armenia), BMFWF and FWF (Austria); FNRS and FWO (Belgium); CNPq, CAPES, FAPERJ, FAPERGS, and FAPESP (Brazil); MES and BNSF (Bulgaria); CERN; CAS, MoST, and NSFC (China); MINCIENCIAS (Colombia); MSES and CSF (Croatia); RIF (Cyprus); SENESCYT (Ecuador); ERC PRG and PSG, TARISTU24-TK10 and MoER TK202 (Estonia); Academy of Finland, MEC, and HIP (Finland); CEA and CNRS/IN2P3 (France); SRNSF (Georgia); BMFTR, DFG, and HGF (Germany); GSRI (Greece); MATE and NKFIH (Hungary); DAE and DST (India); IPM (Iran); SFI (Ireland); INFN (Italy); MSIT and NRF (Republic of Korea); MES (Latvia); LMTLT (Lithuania); MOE and UM (Malaysia); BUAP, CINVESTAV, CONACYT, LNS, SEP, and UASLP-FAI (Mexico); MOS (Montenegro); MBIE (New Zealand); PAEC (Pakistan); MSHE, NSC, and NAWA (Poland); FCT (Portugal);  MESTD (Serbia); MICIU/AEI and PCTI (Spain); MOSTR (Sri Lanka); Swiss Funding Agencies (Switzerland); MST (Taipei); MHESI (Thailand); TUBITAK and TENMAK (T\"{u}rkiye); NASU (Ukraine); STFC (United Kingdom); DOE and NSF (USA).

\begin{sloppypar}
\setlength\emergencystretch{\hsize}
\hyphenation{Rachada-pisek} Individuals have received support from the Marie-Curie programme and the European Research Council and Horizon 2020 Grant, contract Nos.\ 675440, 724704, 752730, 758316, 765710, 824093, 101115353, 101002207, 101001205, and COST Action CA16108 (European Union); the Leventis Foundation; the Alfred P.\ Sloan Foundation; the Alexander von Humboldt Foundation; the Science Committee, project no. 22rl-037 (Armenia); the Fonds pour la Formation \`a la Recherche dans l'Industrie et dans l'Agriculture (FRIA) and Fonds voor Wetenschappelijk Onderzoek contract No. 1228724N (Belgium); the Beijing Municipal Science \& Technology Commission, No. Z191100007219010, the Fundamental Research Funds for the Central Universities, the Ministry of Science and Technology of China under Grant No. 2023YFA1605804, the Natural Science Foundation of China under Grant No. 12535004, and USTC Research Funds of the Double First-Class Initiative No.\ YD2030002017 (China); the Ministry of Education, Youth and Sports (MEYS) of the Czech Republic; the Shota Rustaveli National Science Foundation (Georgia); the Deutsche Forschungsgemeinschaft (DFG), among others, under Germany's Excellence Strategy -- EXC 2121 ``Quantum Universe" -- 390833306, and under project number 400140256 - GRK2497; the Hellenic Foundation for Research and Innovation (HFRI), Project Number 2288 (Greece); the Hungarian Academy of Sciences, the New National Excellence Program - \'UNKP, the NKFIH research grants K 131991, K 138136, K 143460, K 143477, K 147557, K 146913, K 146914, K 147048, TKP2021-NKTA-64, and 2025-1.1.5-NEMZ\_KI-2025-00004, and MATE KKP and KKPCs Research Excellence and Flagship Research Groups grants (Hungary); the Council of Science and Industrial Research, India; ICSC -- National Research Centre for High Performance Computing, Big Data and Quantum Computing, FAIR -- Future Artificial Intelligence Research, and CUP I53D23001070006 (Mission 4 Component 1), funded by the NextGenerationEU program, the Italian Ministry of University and Research (MUR) under Bando PRIN 2022 -- CUP I53C24002390006, PRIN PRIMULA 2022RBYK7T (Italy); the Latvian Council of Science; the Ministry of Science and Higher Education, project no. 2022/WK/14, and the National Science Centre, contracts Opus 2021/41/B/ST2/01369, 2021/43/B/ST2/01552, 2023/49/B/ST2/03273, and the NAWA contract BPN/PPO/2021/1/00011 (Poland); the Funda\c{c}\~ao para a Ci\^encia e a Tecnologia (Portugal); the National Priorities Research Program by Qatar National Research Fund; MICIU/AEI/10.13039/501100011033, ERDF/EU, ``European Union NextGenerationEU/PRTR", projects PID2022-142604OB-C21, PID2022-139519OB-C21, PID2023-147706NB-I00, PID2023-148896NB-I00, PID2023-146983NB-I00, PID2023-147115NB-I00, PID2023-148418NB-C41, PID2023-148418NB-C42, PID2023-148418NB-C43, PID2023-148418NB-C44, PID2024-158190NB-C22, RYC2021-033305-I, RYC2024-048719-I, CNS2023-144781, CNS2024-154769 and Plan de Ciencia, Tecnolog{\'i}a e Innovaci{\'o}n de Asturias, Spain; the Chulalongkorn Academic into Its 2nd Century Project Advancement Project, the National Science, Research and Innovation Fund program IND\_FF\_68\_369\_2300\_097, and the Program Management Unit for Human Resources \& Institutional Development, Research and Innovation, grant B39G680009 (Thailand); the Eric \& Wendy Schmidt Fund for Strategic Innovation through the CERN Next Generation Triggers project under grant agreement number SIF-2023-004; the Kavli Foundation; the Nvidia Corporation; the SuperMicro Corporation; the Welch Foundation, contract C-1845; and the Weston Havens Foundation (USA).
\end{sloppypar}
\end{acknowledgments}\section*{Data availability} Release and preservation of data used by the CMS Collaboration as the basis for publications is guided by the  \href{https://doi.org/10.7483/OPENDATA.CMS.1BNU.8V1W}{CMS data preservation, re-use and open access policy}.

\bibliography{auto_generated} 
\cleardoublepage \appendix\section{The CMS Collaboration \label{app:collab}}\begin{sloppypar}\hyphenpenalty=5000\widowpenalty=500\clubpenalty=5000\cmsinstitute{Yerevan Physics Institute, Yerevan, Armenia}
{\tolerance=6000
A.~Gevorgyan\cmsorcid{0000-0003-2751-9489}, A.~Hayrapetyan, V.~Makarenko\cmsorcid{0000-0002-8406-8605}, A.~Tumasyan\cmsAuthorMark{1}\cmsorcid{0009-0000-0684-6742}
\par}
\cmsinstitute{Institut f\"{u}r Hochenergiephysik, Vienna, Austria}
{\tolerance=6000
P.S.~Hussain\cmsorcid{0000-0002-4825-5278}, M.~Sonawane\cmsorcid{0000-0003-0510-7010}
\par}
\cmsinstitute{Marietta Blau Institute for Particle Physics, Vienna, Austria}
{\tolerance=6000
W.~Adam\cmsorcid{0000-0001-9099-4341}, L.~Benato\cmsorcid{0000-0001-5135-7489}, T.~Bergauer\cmsorcid{0000-0002-5786-0293}, M.~Dragicevic\cmsorcid{0000-0003-1967-6783}, S.~Gundacker\cmsorcid{0000-0003-2087-3266}, A.K.~Guven\cmsorcid{0009-0004-5670-5138}, M.~Jeitler\cmsAuthorMark{2}\cmsorcid{0000-0002-5141-9560}, N.~Krammer\cmsorcid{0000-0002-0548-0985}, A.~Li\cmsorcid{0000-0002-4547-116X}, D.~Liko\cmsorcid{0000-0002-3380-473X}, M.~Matthewman, A.~Pfeiffer\cmsorcid{0000-0001-5328-448X}, J.~Schieck\cmsAuthorMark{2}\cmsorcid{0000-0002-1058-8093}, R.~Sch\"{o}fbeck\cmsAuthorMark{2}\cmsorcid{0000-0002-2332-8784}, M.~Shooshtari\cmsorcid{0009-0004-8882-4887}, N.~Van~Den~Bossche\cmsorcid{0000-0003-2973-4991}, W.~Waltenberger\cmsorcid{0000-0002-6215-7228}, C.E.~Wulz\cmsAuthorMark{2}\cmsorcid{0000-0001-9226-5812}
% ...
\par}
$^{1}$Also at Yerevan State University, Yerevan, Armenia\\
$^{2}$Also at Technische Universit\"{a}t Wien, Vienna, Austria\\
% ...
\end{sloppypar}
\end{document}